\documentclass[12pt]{article}

\pdfoutput = 1

\usepackage{times}
\usepackage{authblk}
\usepackage{amsmath}
\usepackage{amssymb}
\usepackage{amsthm}
\usepackage{bm}
\usepackage{braket}
\usepackage{slashed}
\usepackage[svgnames,psnames]{xcolor}
\usepackage[nottoc]{tocbibind}
\usepackage{cite}
\usepackage{here}
\usepackage[height = 8.8 in, width = 6.45 in]{geometry}
\usepackage{graphicx}
\usepackage{adjustbox}
\usepackage{tikz-cd}
\usepackage[titletoc]{appendix}
\usepackage{comment}
\usepackage[colorlinks, linktocpage, hypertexnames = false]{hyperref}

\hypersetup{linkcolor = Crimson, citecolor = MediumBlue, urlcolor = MediumBlue}

\numberwithin{equation}{section}

\theoremstyle{definition}

\begin{document}

~\vspace{4cm}~
\begin{center}{\fontsize{16}{0}
\textbf{On lattice models of gapped phases with fusion category symmetries}}
\end{center}

\begin{center}
Kansei Inamura
\end{center}

\begin{center}
{\small Institue for Solid State Physics, University of Tokyo, Kashiwa, Chiba 277-8581, Japan}
\end{center}
~

\begin{abstract}
We construct topological quantum field theories (TQFTs) and commuting projector Hamiltonians for any 1+1d gapped phases with non-anomalous fusion category symmetries, i.e. finite symmetries that admit SPT phases.
The construction is based on two-dimensional state sum TQFT whose input datum is an $H$-simple left $H$-comodule algebra, where $H$ is a finite dimensional semisimple Hopf algebra.
We show that the actions of fusion category symmetries $\mathcal{C}$ on the boundary conditions of these state sum TQFTs are represented by module categories over $\mathcal{C}$.
This agrees with the classification of gapped phases with symmetry $\mathcal{C}$.
We also find that the commuting projector Hamiltonians for these state sum TQFTs have fusion category symmetries at the level of the lattice models and hence provide lattice realizations of gapped phases with fusion category symmetries.
As an application, we discuss the edge modes of SPT phases based on these commuting projector Hamiltonians.
Finally, we mention that we can extend the construction of topological field theories to the case of anomalous fusion category symmetries by replacing a semisimple Hopf algebra with a semisimple pseudo-unitary connected weak Hopf algebra.
\end{abstract}

\setcounter{page}{0}

\thispagestyle{empty}

\newpage

\tableofcontents

\flushbottom

\section{Introduction and summary}
\label{sec: Introduction and summary}
Symmetries of physical systems are characterized by the algebraic relations of topological defects.
For instance, ordinary group symmetries are associated with invertible topological defects with codimension one.
When the codimensions of invertible topological defects are greater than one, the corresponding symmetries are called higher form symmetries \cite{GKSW2015}.
We can generalize these symmetries by relaxing the invertibility of topological defects.
Symmetries associated with such non-invertible topological defects are called non-invertible symmetries, which are studied recently in various contexts \cite{RS2020, HMcNMRRV2021, Johnson-Freyd2020b, JY2021, KTZ2020, KNY2021, FFRS2010, BCP2014a, BCP2014b, BCP2015, CR2016, BT2018, CLSWY2019, TW2019, JSW2020, LS2021, LTLSB2020, KORS2020, CL2020, HL2021, Inamura2021, TW2021, HLOTT2021, HLS2021, VLDWOHV2021, AMF2016, AFM2020, FTLTKWF2007, GATLTW2009, BG2017, NTU2021a, NTU2021b, Sharpe2021, CMS2020, CRS2018, CRS2019, CRS2020, JW2020, Johnson-Freyd2020a, KZ2020, KLWZZ2020}.
The algebraic structures of non-invertible symmetries are in general captured by higher categories \cite{Kapustin2010, DR2018, Johnson-Freyd2020a, KZ2020, KLWZZ2020}.
In particular, non-invertible symmetries associated with finitely many topological defect lines in 1+1 dimensions are described by unitary fusion categories \cite{BT2018}.
These symmetries are called fusion category symmetries \cite{TW2019} and are investigated extensively \cite{FFRS2010, BCP2014a, BCP2014b, BCP2015, CR2016, BT2018, CLSWY2019, TW2019, JSW2020, LS2021, LTLSB2020, KORS2020, CL2020, TW2021, HL2021, Inamura2021, HLOTT2021, HLS2021, VLDWOHV2021, AMF2016, AFM2020, FTLTKWF2007, GATLTW2009, BG2017}.

Fusion category symmetries are ubiquitous in two-dimensional conformal field theories (CFTs).
A basic example is the symmetry of the Ising CFT \cite{KW1941, FFRS2004, AMF2016}: the Ising CFT has a fusion category symmetry generated by the non-invertible Kramers-Wannier duality defect and the invertible $\mathbb{Z}_2$ spin-flip defect.\footnote{The fusion category that describes this symmetry is known as a $\mathbb{Z}_2$ Tambara-Yamagami category \cite{TY1998, BT2018, CLSWY2019, TW2019, JSW2020, LS2021}.} 
More generally, any diagonal RCFTs have fusion category symmetries generated by the Verlinde lines \cite{Ver1988}.
Fusion category symmetries are also studied in other CFTs such as CFTs with central charge $c = 1$ \cite{FGRS2007, CL2020, TW2021} and RCFTs that are not necessarily diagonal \cite{PZ2001, FFRS2007, FRS2002, FRS2004a, FRS2004b, FRS2005, FFRS2006}.\footnote{Precisely, $c=1$ CFTs can have infinitely many topological defect lines labeled by continuous parameters \cite{FGRS2007, CL2020, TW2021}, whose algebraic structure should be described by a mathematical framework beyond fusion categories.}

We can also consider fusion category symmetries in topological quantum field theories (TQFTs).
In particular, it is shown in \cite{TW2019, KORS2020} that unitary TQFTs with fusion category symmetry $\mathcal{C}$ are classified by semisimple module categories over $\mathcal{C}$.
This result will be heavily used in the rest of this paper.
This classification reveals that fusion category symmetries do not always admit SPT phases, i.e. symmetric gapped phases with unique ground states.
If fusion category symmetries do not have SPT phases, they are said to be anomalous \cite{TW2019}, and otherwise non-anomalous.

Fusion category symmetries exist on the lattice as well.
Remarkably, 2d statistical mechanical models with general fusion category symmetries are constructed recently in \cite{AMF2016, AFM2020}.
There are also examples of 1+1d lattice models known as anyonic chains \cite{FTLTKWF2007, GATLTW2009, BG2017}.
These models might cover all the gapped phases with fusion category symmetries.
However, to the best of my knowledge, systematic construction of 1+1d TQFTs and corresponding gapped Hamiltonians with fusion category symmetries is still lacking.

In this paper, we explicitly construct TQFTs and commuting projector Hamiltonians for any 1+1d gapped phases with arbitrary non-anomalous fusion category symmetries.
For this purpose, we first show that a TQFT with fusion category symmetry, which is formulated axiomatically in \cite{BT2018}, is obtained from another TQFT with different symmetry by a procedure that we call pullback.
This is a natural generalization of the pullback of an SPT phase with finite group symmetry by a group homomorphism \cite{WWW2018}.
Specifically, we can pull back topological defects of a TQFT with symmetry $\mathcal{C}^{\prime}$ by a tensor functor $F: \mathcal{C} \rightarrow \mathcal{C}^{\prime}$ to obtain a TQFT with symmetry $\mathcal{C}$.
This corresponds to the fact that given a $\mathcal{C}^{\prime}$-module category $\mathcal{M}$ and a tensor functor $F: \mathcal{C} \rightarrow \mathcal{C}^{\prime}$, we can endow $\mathcal{M}$ with a $\mathcal{C}$-module category structure.
By using this technique, we can construct any TQFTs with non-anomalous fusion category symmetries.\footnote{We can also construct any TQFTs with anomalous fusion category symmetries in the same way, see section \ref{sec: A comment on a generalization to anomalous symmetries}.}
To see this, we first recall that non-anomalous symmetries are described by fusion categories that admit fiber functors \cite{TW2019, KORS2020, Inamura2021}.
Such fusion categories are equivalent to the representation categories $\mathrm{Rep}(H)$ of finite dimensional semisimple Hopf algebras $H$. 
Therefore, TQFTs with non-anomalous fusion category symmetries are classified by semisimple module categories over $\mathrm{Rep}(H)$.
Among these module categories, we are only interested in indecomposable ones because any semisimple module category can be decomposed into a direct sum of indecomposable module categories.
Every indecomposable semisimple module category over $\mathrm{Rep}(H)$ is equivalent to the category ${}_K \mathcal{M}$ of left $K$-modules where $K$ is an $H$-simple left $H$-comodule algebra \cite{AM2007}.
The $\mathrm{Rep}(H)$-module category structure on ${}_K \mathcal{M}$ is represented by a tensor functor from $\mathrm{Rep}(H)$ to the category $\mathrm{End}({}_K \mathcal{M})$ of endofunctors of ${}_K \mathcal{M}$.
Since $\mathrm{End}({}_K \mathcal{M})$ is equivalent to the category ${}_K \mathcal{M}_K$ of $K$-$K$ bimodules \cite{Watts1960}, we have a tensor functor $F_K: \mathrm{Rep}(H) \rightarrow {}_K \mathcal{M}_K$.
We can use this tensor functor to pull back a ${}_K \mathcal{M}_K$ symmetric TQFT to a $\mathrm{Rep}(H)$ symmetric TQFT.
We show in section \ref{sec: Pullback of fusion category TQFTs by tensor functors} that a $\mathrm{Rep}(H)$ symmetric TQFT corresponding to a $\mathrm{Rep}(H)$-module category ${}_K \mathcal{M}$ is obtained as the pullback of a specific ${}_K \mathcal{M}_K$ symmetric TQFT, which corresponds to the same category ${}_K \mathcal{M}$ regarded as a ${}_K \mathcal{M}_K$-module category, by a tensor functor $F_K$.

We also describe the pullback in the context of state sum TQFTs in section \ref{sec: State sum TQFTs and commuting projector Hamiltonians}.
Here, a state sum TQFT is a TQFT obtained by state sum construction \cite{FHK94}, which is a recipe to construct a 2d TQFT from a semisimple algebra.
The pullback of the state sum TQFTs enables us to explicitly construct TQFTs with $\mathrm{Rep}(H)$ symmetry.
Specifically, it turns out that the state sum TQFT obtained from an $H$-simple left $H$-comodule algebra $K$ is a $\mathrm{Rep}(H)$ symmetric TQFT ${}_K \mathcal{M}$.
Any indecomposable TQFTs with $\mathrm{Rep}(H)$ symmetry can be constructed in this way because an $H$-simple left $H$-comodule algebra is always semisimple \cite{Skr2007}.

The existence of the state sum construction suggests that we can realize the $\mathrm{Rep}(H)$ symmetric TQFTs by lattice models.
Indeed, the vacua of a state sum TQFT are in one-to-one correspondence with the ground states of an appropriate commuting projector Hamiltonian \cite{WW2017, Bullivant2018}.
Specifically, when the input algebra of a state sum TQFT is $K$, the commuting projector Hamiltonian $H$ is given by
\begin{equation}
H = \sum_i (1 - h_{i, i+1}), \quad h_{i, i+1} = \Delta_K \circ m_K = \adjincludegraphics[valign = c, width = 2.2cm]{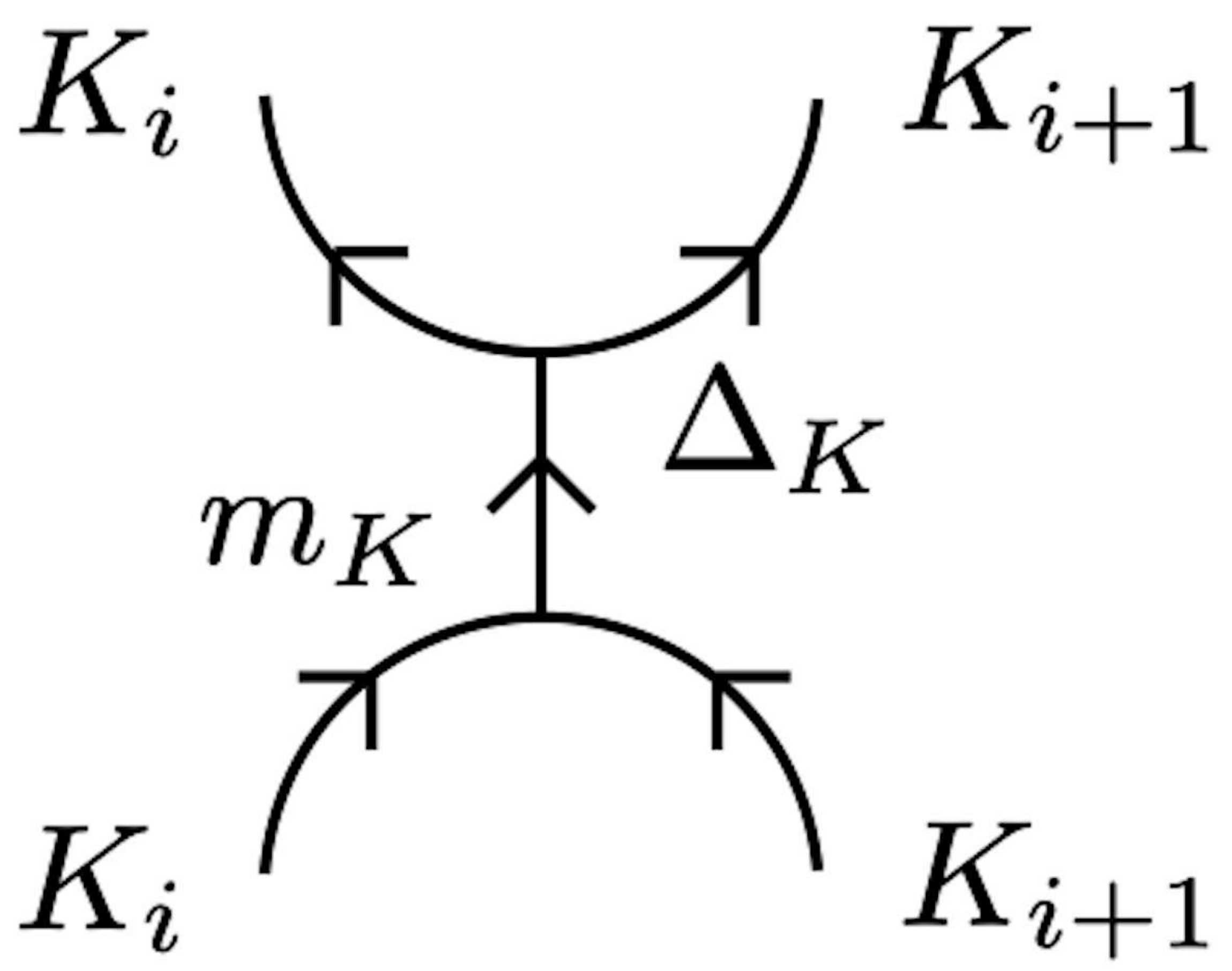}: K_i \otimes K_{i+1} \rightarrow K_i \otimes K_{i+1},
\label{eq: Hamiltonian}
\end{equation}
where $K_i := K$ is the local Hilbert space on the lattice, $m_K: K \otimes K \rightarrow K$ is multiplication on $K$, and $\Delta_K: K \rightarrow K \otimes K$ is comultiplication for the Frobenius algebra structure on $K$.
The diagram in the above equation is the string diagram representation of the linear map $h_{i, i+1}$.
We find that when $K$ is a left $H$-comodule algebra, we can define the action of $\mathrm{Rep}(H)$ on the lattice Hilbert space $\mathcal{H} = \bigotimes_i K_i$ via the left $H$-comodule action on $K$.
Here, we need to choose $K$ appropriately so that the $\mathrm{Rep}(H)$ action becomes faithful on the lattice.
In section \ref{sec: State sum TQFTs and commuting projector Hamiltonians}, we show that the above Hamiltonian has a $\mathrm{Rep}(H)$ symmetry by explicitly computing the commutation relation of the Hamiltonian \eqref{eq: Hamiltonian} and the action of the $\mathrm{Rep}(H)$ symmetry.
Moreover, we will see that the $\mathrm{Rep}(H)$ symmetry action of the lattice model agrees with that of the state sum TQFT when the Hilbert space $\mathcal{H}$ is restricted to the subspace spanned by the ground states.
This implies that the commuting projector Hamiltonian \eqref{eq: Hamiltonian} realizes a $\mathrm{Rep}(H)$ symmetric TQFT ${}_K \mathcal{M}$.

We also examine the edge modes of SPT phases with $\mathrm{Rep}(H)$ symmetry by putting the systems on an interval.
The ground states of the commuting projector Hamiltonian \eqref{eq: Hamiltonian} on an interval are described by the input algebra $K$ itself \cite{LP2007, DKR2011}.
In particular, for SPT phases, $K$ is isomorphic to the endomorphism algebra $\mathrm{End}(M) \cong M^* \otimes M$ of a simple left $K$-module $M$.
We can interpret $M^*$ and $M$ as the edge modes by using the matrix product state (MPS) representation of the ground states.
Thus, the edge modes of the Hamiltonian \eqref{eq: Hamiltonian} for a $\mathrm{Rep}(H)$ SPT phase ${}_K \mathcal{M}$ become either a left $K$-module $M$ or a right $K$-module $M^*$ depending on which boundary they are localized to.
As a special case, we reproduce the well-known result that the edge modes of an SPT phase with finite group symmetry $G$ have anomalies, which take values in the second group cohomology $H^2(G, \mathrm{U}(1))$.
We note that the edge modes of the Hamiltonian \eqref{eq: Hamiltonian} are not necessarily minimal: it would be possible to partially lift the degeneracy on the boundaries by adding symmetric perturbations.

Although we will only consider the fixed point Hamiltonians \eqref{eq: Hamiltonian} in this paper, we can add terms to our models while preserving the $\mathrm{Rep}(H)$ symmetry.
In general, the lattice models still have the $\mathrm{Rep}(H)$ symmetry if the additional terms are $H$-comodule maps.
Since the Hamiltonians with additional terms are generically no longer exactly solvable, one would use numerical calculations to determine the phase diagrams.
For this purpose, we need to write the Hamiltonians in the form of matrices by choosing a basis of the lattice Hilbert space $\mathcal{H}$.
As a concrete example, we will explicitly compute the action of the Hamiltonian \eqref{eq: Hamiltonian} with $\mathrm{Rep}(G)$ symmetry by choosing a specific basis of $\mathcal{H}$.
Here, $\mathrm{Rep}(G)$ is the category of representations of a finite group $G$, which describes the symmetry of $G$ gauge theory.

Before proceeding to the next section, we comment on a relation between the state sum models discussed in this paper and the anyon chain models.\footnote{The author thanks Kantaro Ohmori for a discussion on the anyon chain models.}
As we summarized above, we construct a $\mathrm{Rep}(H)$ symmetric commuting projector Hamiltonian of the state sum model by using a left $H$-comodule algebra $K$ in this paper.
On the other hand, we can also construct a $\mathrm{Rep}((H^{*})^{\mathrm{cop}})$ symmetric commuting projector Hamiltonian of the anyon chain model by using the same algebra $K$,\footnote{Equivalently, a $\mathrm{Rep}(H)$ symmetric commuting projector Hamiltonian of the anyon chain model is obtained from a left $H$-module algebra instead of a left $H$-comodule algebra.} where $(H^{*})^{\mathrm{cop}}$ is the coopposite coalgebra of the dual Hopf algebra $H^*$.
The anyon chain with $\mathrm{Rep}((H^*)^{\mathrm{cop}})$ symmetry is a lattice model whose Hilbert space is spanned by fusion trees in $\mathrm{Rep}((H^*)^{\mathrm{cop}})$.
The commuting projector Hamiltonian of the anyon chain can be written diagrammatically as
\begin{equation}
H = \sum_i (1-h_{i,i+1}), \quad
h_{i, i+1}: \adjincludegraphics[valign = c, width = 2.5cm]{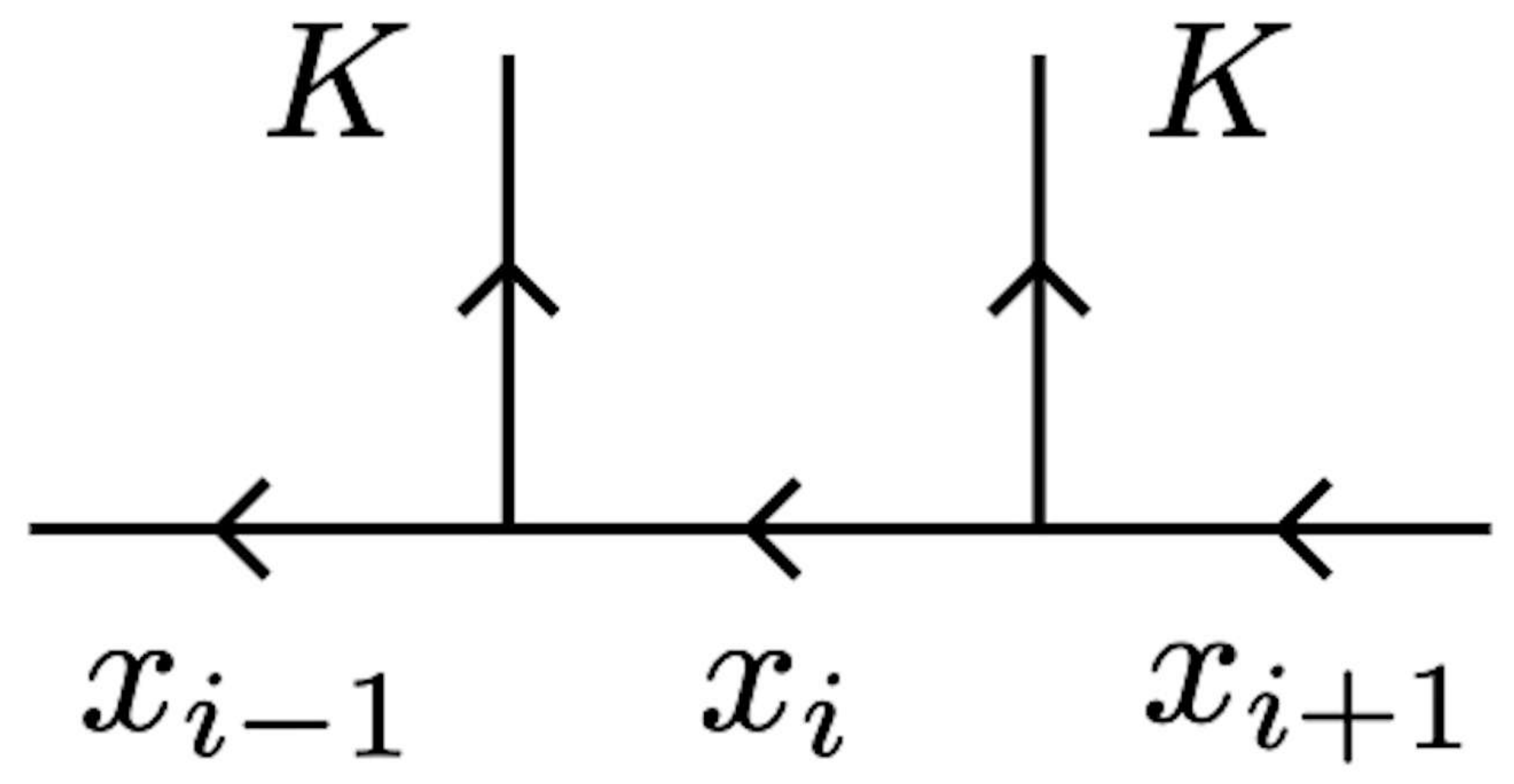} \rightarrow \adjincludegraphics[valign = c, width = 2.5cm]{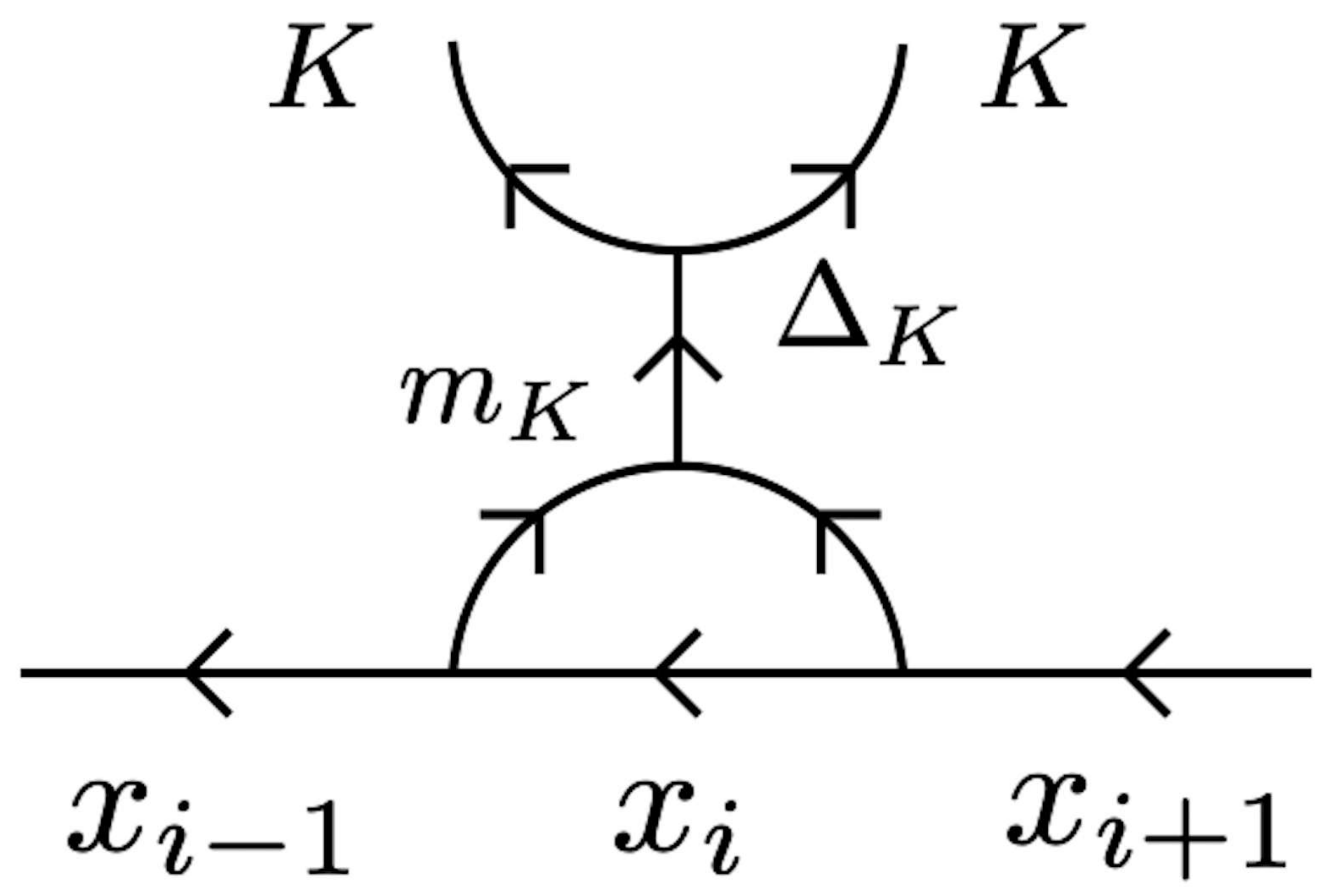},
\label{eq: anyon Hamiltonian}
\end{equation}
where the horizontal edges of the fusion diagrams are labeled by objects in $\mathrm{Rep}((H^*)^{\mathrm{cop}})$.
We note that a left $H$-comodule algebra $K$ is an algebra object in $\mathrm{Rep}((H^*)^{\mathrm{cop}})$.
The right diagram in eq. \eqref{eq: anyon Hamiltonian} can be deformed to a sum of fusion trees via $F$-moves and hence the Hamiltonian can be explicitly written in terms of $F$-symbols.
The above Hamiltonian has ground states represented by the fusion trees all of whose horizontal edges are labeled by a right $K$-module $M \in \mathrm{Rep}((H^*)^{\mathrm{cop}})$.
This suggests, though not prove, that the gapped phase of this anyon chain corresponds to the category of right $K$-modules in $\mathrm{Rep}((H^*)^{\mathrm{cop}})$, which is a $\mathrm{Rep}((H^*)^{\mathrm{cop}})$-module category.
As we will argue in section \ref{sec: State sum TQFTs with defects}, this also suggests that the gapped phase of the anyon chain model constructed from the opposite algebra $K^{\mathrm{op}}$ is obtained by the generalized gauging of the state sum model constructed from $K$, see footnote \ref{foot: mapping}.

The reason why the state sum model \eqref{eq: Hamiltonian} and the anyon chain model \eqref{eq: anyon Hamiltonian} have different symmetries despite the similarity between their Hamiltonians is that the symmetry actions are defined differently due to the different structures of their Hilbert spaces.
Specifically, the $\mathrm{Rep}((H^*)^{\mathrm{cop}})$ symmetry of the anyon chain model is defined via the fusion of topological defect lines and the horizontal edges, whereas the $\mathrm{Rep}(H)$ symmetry of the state sum model is defined via the $H$-comodule structure on the algebra $K$ as we will discuss in section \ref{sec: State sum TQFTs and commuting projector Hamiltonians}, see eq. \eqref{eq: Rep(H) action}.
Since the state sum models do not have counterparts of horizontal edges of fusion trees, the $\mathrm{Rep}((H^*)^{\mathrm{cop}})$ symmetry does not act on the state sum models. 
Conversely, since the $\mathrm{Rep}(H)$ action \eqref{eq: Rep(H) action} is not a morphism in $\mathrm{Rep}((H^*)^{\mathrm{cop}})$ and therefore is not given by a fusion diagram in $\mathrm{Rep}((H^*)^{\mathrm{cop}})$, the $\mathrm{Rep}(H)$ symmetry does not act on the anyon chains.

The rest of the paper is organized as follows.
In section \ref{sec: Preliminaries}, we briefly review some mathematical backgrounds.
In section \ref{sec: Pullback of fusion category TQFTs by tensor functors}, we introduce the notion of pullback of a TQFT and show that every TQFT with non-anomalous fusion category symmetry $\mathrm{Rep}(H)$ is obtained by pulling back a ${}_K \mathcal{M}_K$ symmetric TQFT ${}_K \mathcal{M}$ by a tensor functor $F_K: \mathrm{Rep}(H) \rightarrow {}_K \mathcal{M}_K$.
In section \ref{sec: State sum TQFTs and commuting projector Hamiltonians}, we define state sum TQFTs with $\mathrm{Rep}(H)$ symmetry and show that they are realized by the commuting projector Hamiltonians \eqref{eq: Hamiltonian}.
We emphasize that these Hamiltonians have fusion category symmetries at the level of the lattice models.
These lattice realizations enable us to examine the edge modes of $\mathrm{Rep}(H)$ SPT phases.
We also comment on a generalization to TQFTs and commuting projector Hamiltonians with anomalous fusion category symmetries in the last subsection.
In appendix \ref{sec: State sum TQFTs on surfaces with interfaces}, we describe state sum TQFTs with fusion category symmetries in the presence of interfaces.

\section{Preliminaries}
\label{sec: Preliminaries}

\subsection{Fusion categories, tensor functors, and module categories}
\label{sec: Fusion categories and tensor functors}
We begin with a brief review of unitary fusion categories, tensor functors, and module categories \cite{EGNO2015}.
A unitary fusion category $\mathcal{C}$ is equipped with a bifunctor $\otimes: \mathcal{C} \times \mathcal{C} \rightarrow \mathcal{C}$, which is called a tensor product.
The tensor product of objects $x, y \in \mathcal{C}$ is denoted by $x \otimes y$.
The tensor product $(x \otimes y) \otimes z$ of three objects $x, y, z \in \mathcal{C}$ is related to $x \otimes (y \otimes z)$ by a natural isomorphism $\alpha_{x, y, z}: (x \otimes y) \otimes z \rightarrow x \otimes (y \otimes z)$ called an associator, which satisfies the following pentagon equation:
\begin{equation}
\begin{tikzcd}
 & (x \otimes y) \otimes (z \otimes w) \arrow[dr, "\alpha_{x, y, z \otimes w}"] & \\
 ((x \otimes y) \otimes z) \otimes w \arrow[ru, "\alpha_{x \otimes y, z, w}"] \arrow[d, "\alpha_{xyz} \otimes \mathrm{id}_w"'] & & x \otimes (y \otimes (z \otimes w))\\
 (x \otimes (y \otimes z)) \otimes w \arrow[rr, "\alpha_{x, y \otimes z, w}"'] & & x \otimes ((y \otimes z) \otimes w) \arrow[u, "\mathrm{id}_x \otimes \alpha_{yzw}"']
\end{tikzcd}
\label{eq: pentagon}
\end{equation}
There is a unit object $1 \in \mathcal{C}$ that behaves as a unit of the tensor product, i.e. $1 \otimes x \cong x \otimes 1 \cong x$.
The isomorphisms $l_x: 1 \otimes x \rightarrow x$ and $r_x: x \otimes 1 \rightarrow x$ are called a left unit morphism and a right unit morphism respectively.
These isomorphisms satisfy the following commutative diagram:
\begin{equation}
\begin{tikzcd}
(x \otimes 1) \otimes y \arrow[r, "\alpha_{x, 1, y}"] \arrow[dr, "r_x \otimes \mathrm{id}_y"'] & x \otimes (1 \otimes y) \arrow[d, "\mathrm{id}_x \otimes l_y"] \\
& x \otimes y
\end{tikzcd}
\end{equation}
We can always take $l_x$ and $r_x$ as the identity morphism $\mathrm{id}_x$ by identifying $1 \otimes x$ and $x \otimes 1$ with $x$.
In sections \ref{sec: Pullback of fusion category TQFTs by tensor functors} and \ref{sec: State sum TQFTs and commuting projector Hamiltonians}, we assume $l_x = r_x = \mathrm{id}_x$.

A unitary fusion category $\mathcal{C}$ also has an additive operation $\oplus: \mathcal{C} \times \mathcal{C} \rightarrow \mathcal{C}$ called a direct sum.
An object $x \in \mathcal{C}$ is called a simple object when it cannot be decomposed into a direct sum of other objects.
In particular, the unit object $1 \in \mathcal{C}$ is simple.
The number of (isomorphism classes of) simple objects is finite, and every object is isomorphic to a direct sum of finitely many simple objects.
Namely, for any object $x \in \mathcal{C}$, we have an isomorphism $x \cong \bigoplus_i N_i a_i$ where $\{a_i\}$ is a set of simple objects and $N_i$ is a non-negative integer.

The Hom space $\mathrm{Hom}(x, y)$ for any objects $x, y \in \mathcal{C}$ is a finite dimensional $\mathbb{C}$-vector space equipped with an adjoint $\dagger: f \in \mathrm{Hom}(x, y) \mapsto f^{\dagger} \in \mathrm{Hom}(y, x)$.
The associators, the left unit morphisms, and the right unit morphisms are unitary with respect to this adjoint, i.e. $\alpha_{x, y, z}^{\dagger} = \alpha_{x, y, z}^{-1}$, $l_x^{\dagger} = l_x^{-1}$, and $r_x^{\dagger} = r_x^{-1}$.
We note that the endomorphism space of a simple object $a_i$ is one-dimensional, i.e. $\mathrm{End}(a_i) := \mathrm{Hom}(a_i, a_i) \cong \mathbb{C}$.

For every object $x \in \mathcal{C}$, we have a dual object $x^{*} \in \mathcal{C}$ and a pair of morphisms $\mathrm{ev}_x^L: x^* \otimes x \rightarrow 1$ and $\mathrm{coev}_x^L: 1 \rightarrow x \otimes x^*$ that satisfy the following relations:
\begin{align}
r_x \circ (\mathrm{id}_x \otimes \mathrm{ev}_x^L) \circ \alpha_{x, x^*, x} \circ (\mathrm{coev}_x^L \otimes \mathrm{id}_x) \circ l_x^{-1} & = \mathrm{id}_x, \label{eq: zigzag1} \\
l_{x^*} \circ (\mathrm{ev}_x^L \otimes \mathrm{id}_{x^*}) \circ \alpha_{x^*, x, x^*}^{-1} \circ (\mathrm{id}_{x^*} \otimes \mathrm{coev}_x^L) \circ r_{x^*}^{-1} & = \mathrm{id}_{x^*}.
\label{eq: zigzag2}
\end{align}
These morphisms are called left evaluation and left coevaluation morphisms respectively.
The adjoints of these morphisms $\mathrm{ev}_x^R := (\mathrm{coev}_x^L)^{\dagger}$ and $\mathrm{coev}_x^R := (\mathrm{ev}_x^L)^{\dagger}$ are called right evaluation and right coevaluation morphisms, which satisfy similar relations to eqs. \eqref{eq: zigzag1} and \eqref{eq: zigzag2}.

A tensor functor $F: \mathcal{C} \rightarrow \mathcal{C}^{\prime}$ between fusion categories $\mathcal{C}$ and $\mathcal{C}^{\prime}$ is a functor equipped with a natural isomorphism $J_{x, y}: F(x) \otimes F(y) \rightarrow F(x \otimes y)$ and an isomorphism $\phi: 1^{\prime} \rightarrow F(1)$ that satisfy the following commutative diagrams:
\begin{equation}
\begin{tikzcd}
& F(x \otimes y) \otimes F(z) \arrow[dr, "J_{x \otimes y, z}"] & \\
(F(x) \otimes F(y)) \otimes F(z) \arrow[ur, "J_{x, y} \otimes \mathrm{id}_{F(z)}"] \arrow[d, "\alpha^{\prime}_{F(x), F(y), F(z)}"'] & & F((x \otimes y) \otimes z) \arrow[d, "F(\alpha_{x, y, z})"] \\
F(x) \otimes (F(y) \otimes F(z)) \arrow[dr, "\mathrm{id}_{F(x)} \otimes J_{y, z}"'] & & F(x \otimes (y \otimes z)) \\
& F(x) \otimes F(y \otimes z) \arrow[ur, "J_{x, y \otimes z}"'] &
\end{tikzcd}
\label{eq: monoidal structure}
\end{equation}
\begin{equation}
\begin{tikzcd}
1^{\prime} \otimes F(x) \arrow[r, "l^{\prime}_{F(x)}"] \arrow[d, "\phi \otimes \mathrm{id}_{F(x)}"'] & F(x)\\
F(1) \otimes F(x) \arrow[r, "J_{1, x}"'] & F(1 \otimes x) \arrow[u, "F(l_x)"']
\end{tikzcd}
\quad
\begin{tikzcd}
F(x) \otimes 1^{\prime} \arrow[r, "r^{\prime}_{F(x)}"] \arrow[d, "\mathrm{id}_{F(x)} \otimes \phi"'] & F(x) \\
F(x) \otimes F(1) \arrow[r, "J_{x, 1}"'] & F(x \otimes 1) \arrow[u, "F(r_x)"']
\end{tikzcd}
\label{eq: monoidal structure unit}
\end{equation}
Here, $1$ and $1^{\prime}$ are unit objects of $\mathcal{C}$ and $\mathcal{C}^{\prime}$ respectively.
When $\mathcal{C}$ and $\mathcal{C}^{\prime}$ are unitary fusion categories, we require that $J_{x, y}$ and $\phi$ are unitary in the sense that $J_{x, y}^{\dagger} = J_{x, y}^{-1}$ and $\phi^{\dagger} = \phi^{-1}$.
The isomorphism $\phi$ can always be chosen as the identity morphism by the identification $1^{\prime} = F(1)$.

A module category $\mathcal{M}$ over a fusion category $\mathcal{C}$ is a category equipped with a bifunctor $\overline{\otimes}: \mathcal{C} \times \mathcal{M} \rightarrow \mathcal{M}$, which represents the action of $\mathcal{C}$ on $\mathcal{M}$.
For any objects $x, y \in \mathcal{C}$ and $M \in \mathcal{M}$, we have a natural isomorphism $m_{x, y, M}: (x \otimes y) \overline{\otimes} M \rightarrow x \overline{\otimes} (y \overline{\otimes} M)$ called a module associativity constraint that satisfies the following commutative diagram:
\begin{equation}
\begin{tikzcd}
 & (x \otimes y) \overline{\otimes} (z \overline{\otimes} M) \arrow[dr, "m_{x, y, z \overline{\otimes} M}"] & \\
((x \otimes y) \otimes z) \overline{\otimes} M \arrow[ur, "m_{x \otimes y, z, M}"] \arrow[d, "\alpha_{x, y, z} \overline{\otimes} \mathrm{id}_M"'] & & x \overline{\otimes} (y \overline{\otimes} (z \overline{\otimes} M)) \\
(x \otimes (y \otimes z)) \overline{\otimes} M \arrow[rr, "m_{x, y \otimes z, M}"'] & & x \overline{\otimes} ((y \otimes z) \overline{\otimes} M) \arrow[u, "\mathrm{id}_x \overline{\otimes} m_{y, z, M}"']
\end{tikzcd}
\end{equation}
The action of the unit object $1 \in \mathcal{C}$ gives an isomorphism $l_M: 1 \overline{\otimes} M \rightarrow M$ called a unit constraint such that the following diagram commutes:
\begin{equation}
\begin{tikzcd}
(x \otimes 1) \overline{\otimes} M \arrow[r, "m_{x, 1, M}"] \arrow[dr, "r_x \overline{\otimes} \mathrm{id}_M"'] & x \overline{\otimes} (1 \overline{\otimes} M) \arrow[d, "\mathrm{id}_x \overline{\otimes} l_M"] \\
& x \overline{\otimes} M
\end{tikzcd}
\end{equation}
A $\mathcal{C}$-module category structure on $\mathcal{M}$ can also be represented by a tensor functor from $C$ to the category of endofunctors of $M$, i.e. $F: \mathcal{C} \rightarrow \mathrm{End}(\mathcal{M})$, which is analogous to an action of an algebra on a module.
A module category $\mathcal{M}$ is said to be indecomposable if it cannot be decomposed into a direct sum of two non-trivial module categories.

When we have a tensor functor $(F, J, \phi): \mathcal{C} \rightarrow \mathcal{C}^{\prime}$, we can regard a $\mathcal{C}^{\prime}$-module category $M$ as a $\mathcal{C}$-module category by defining the action of $\mathcal{C}$ on $\mathcal{M}$ as $x \overline{\otimes} M := F(x) \overline{\otimes}^{\prime} M$ for $x \in \mathcal{C}$ and $M \in \mathcal{M}$, where $\overline{\otimes}^{\prime}$ is the action of $\mathcal{C}^{\prime}$ on $\mathcal{M}$.
The natural isomorphisms $m_{x, y, M}$ and $l_M$ are given by 
\begin{equation}
m_{x, y, M} = m^{\prime}_{F(x), F(y), M} \circ (J_{x, y}^{-1} \overline{\otimes}^{\prime} \mathrm{id}_M), \quad l_M = l^{\prime}_M \circ (\phi^{-1} \overline{\otimes}^{\prime} \mathrm{id}_M),
\label{eq: pullback of mod-cat}
\end{equation}
where $m^{\prime}$ and $l^{\prime}$ are the module associativity constraint and the unit constraint for the $\mathcal{C}^{\prime}$-module category structure on $\mathcal{M}$.

An important example of a unitary fusion category is the category ${}_K \mathcal{M}_K$ of $K$-$K$ bimodules where $K$ is a finite dimensional semisimple algebra.
We review this category in some detail for later convenience.
The objects and morphisms of ${}_K \mathcal{M}_K$ are $K$-$K$ bimodules and $K$-$K$ bimodule maps respectively.
The monoidal structure on ${}_K \mathcal{M}_K$ is given by the tensor product over $K$, which is usually denoted by $\otimes_K$.
To describe the tensor product $Y_1 \otimes_K Y_2$ of $K$-$K$ bimodules $Y_1, Y_2 \in {}_K \mathcal{M}_K$, we first recall that a finite dimensional semisimple algebra $K$ is a Frobenius algebra.
Here, an algebra $K$ equipped with multiplication $m_K: K \otimes K \rightarrow K$ and a unit $\eta_K: \mathbb{C} \rightarrow K$ is called a Frobenius algebra if it is also a coalgebra equipped with comultiplication $\Delta_K: K \rightarrow K \otimes K$ and a counit $\epsilon_K: K \rightarrow \mathbb{C}$ such that the following Frobenius relation is satisfied:
\begin{equation}
(m_K \otimes \mathrm{id}_K) \circ (\mathrm{id}_K \otimes \Delta_K) = (\mathrm{id}_K \otimes m_K) \circ (\Delta_K \otimes \mathrm{id}_K) = \Delta_K \circ m_K.
\end{equation}
In the string diagram notation, the above relation is represented as
\begin{equation}
\adjincludegraphics[valign = c, width = 2.5cm]{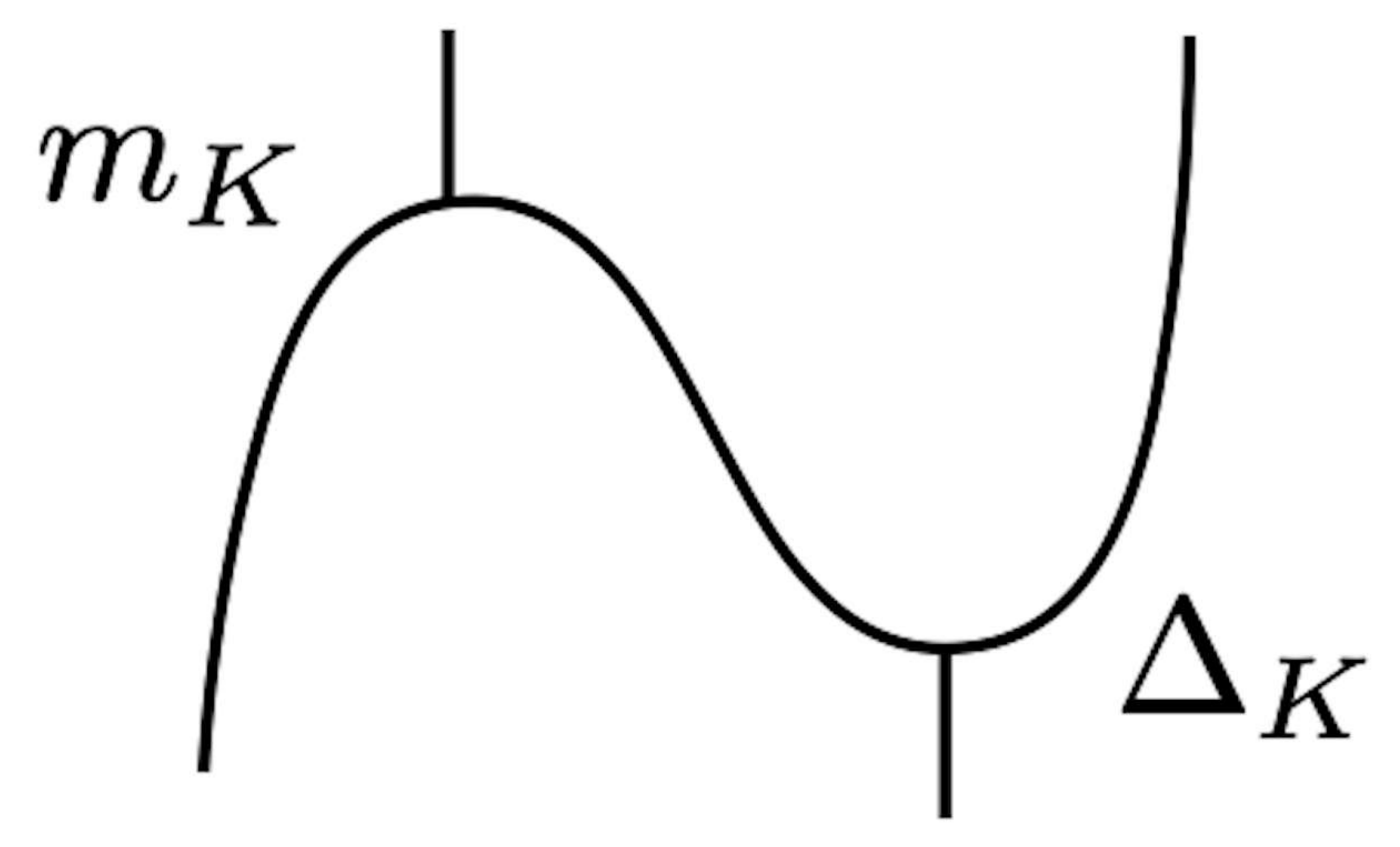} ~ = ~ \adjincludegraphics[valign = c, width = 2cm]{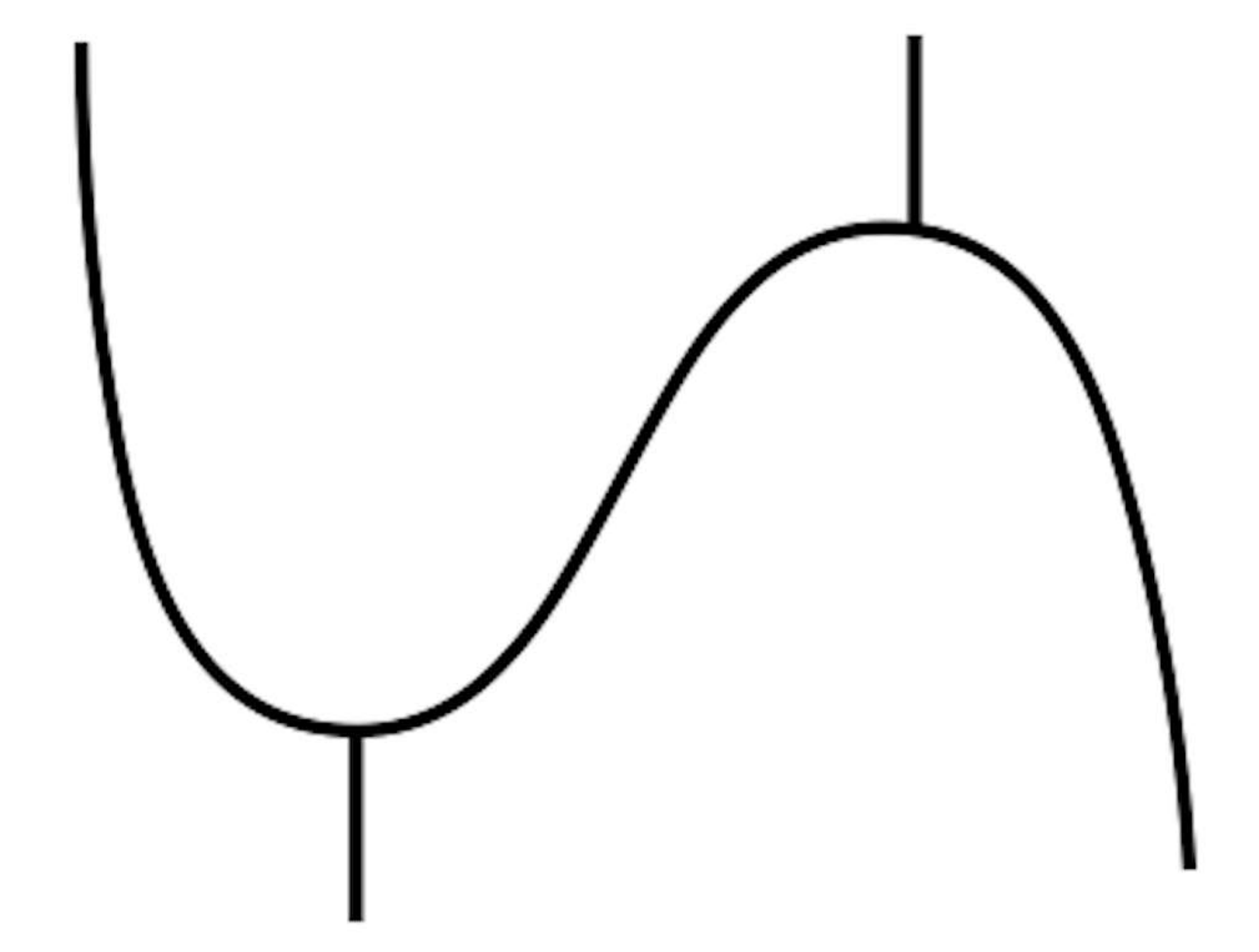} ~ = ~ \adjincludegraphics[valign = c, width = 1.2cm]{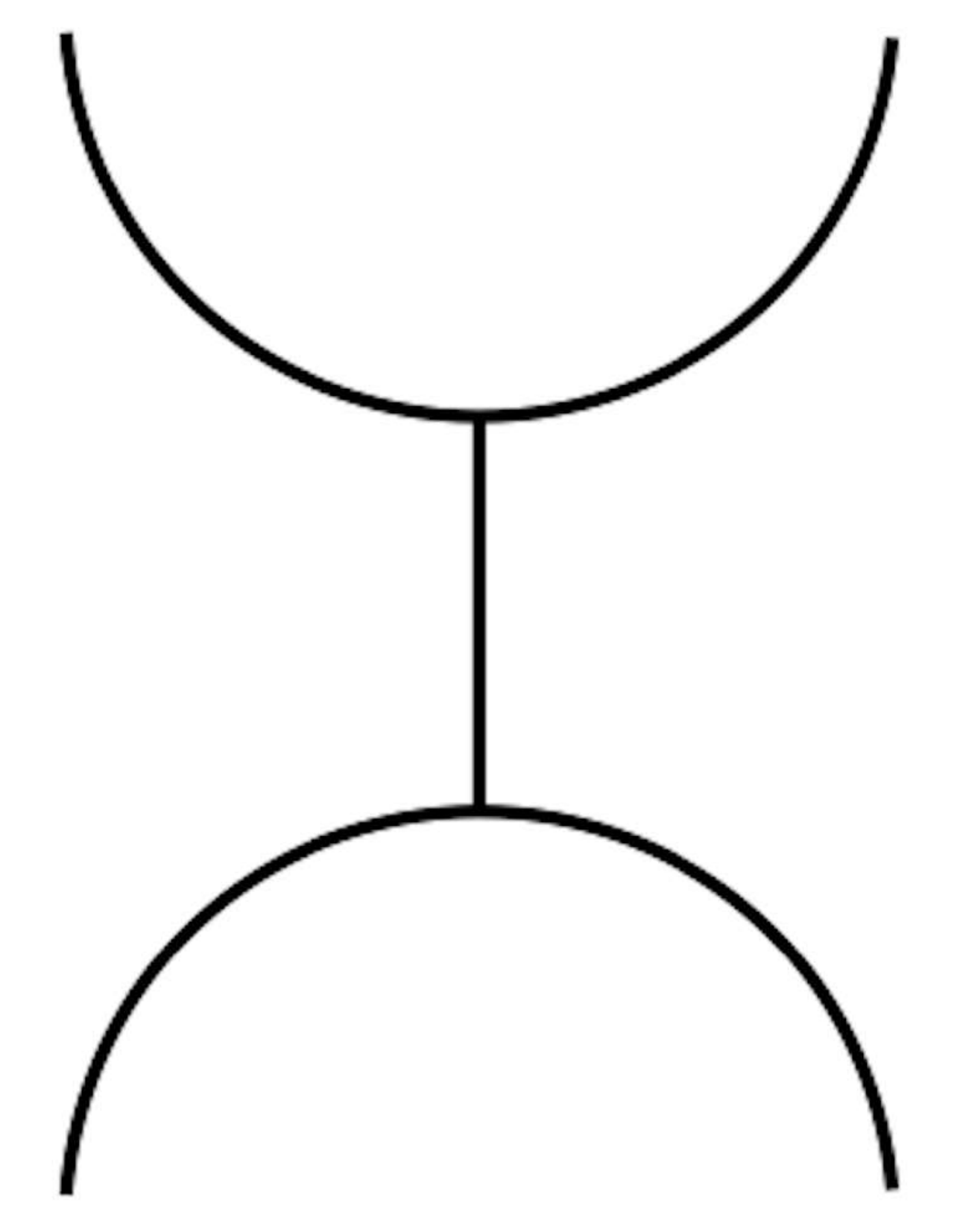}, 
\label{eq: Frobenius relation}
\end{equation}
where each string and junction represent the algebra $K$ and the (co)multiplication respectively.
In our convention, we read these diagrams from bottom to top.
The comultiplication $\Delta_K$ and the counit $\epsilon_K$ can be written in terms of the multiplication $m_K$ and the unit $\eta_K$ as follows \cite{FS2008}: 
\begin{equation}
\adjincludegraphics[valign = c, width = 1.4cm]{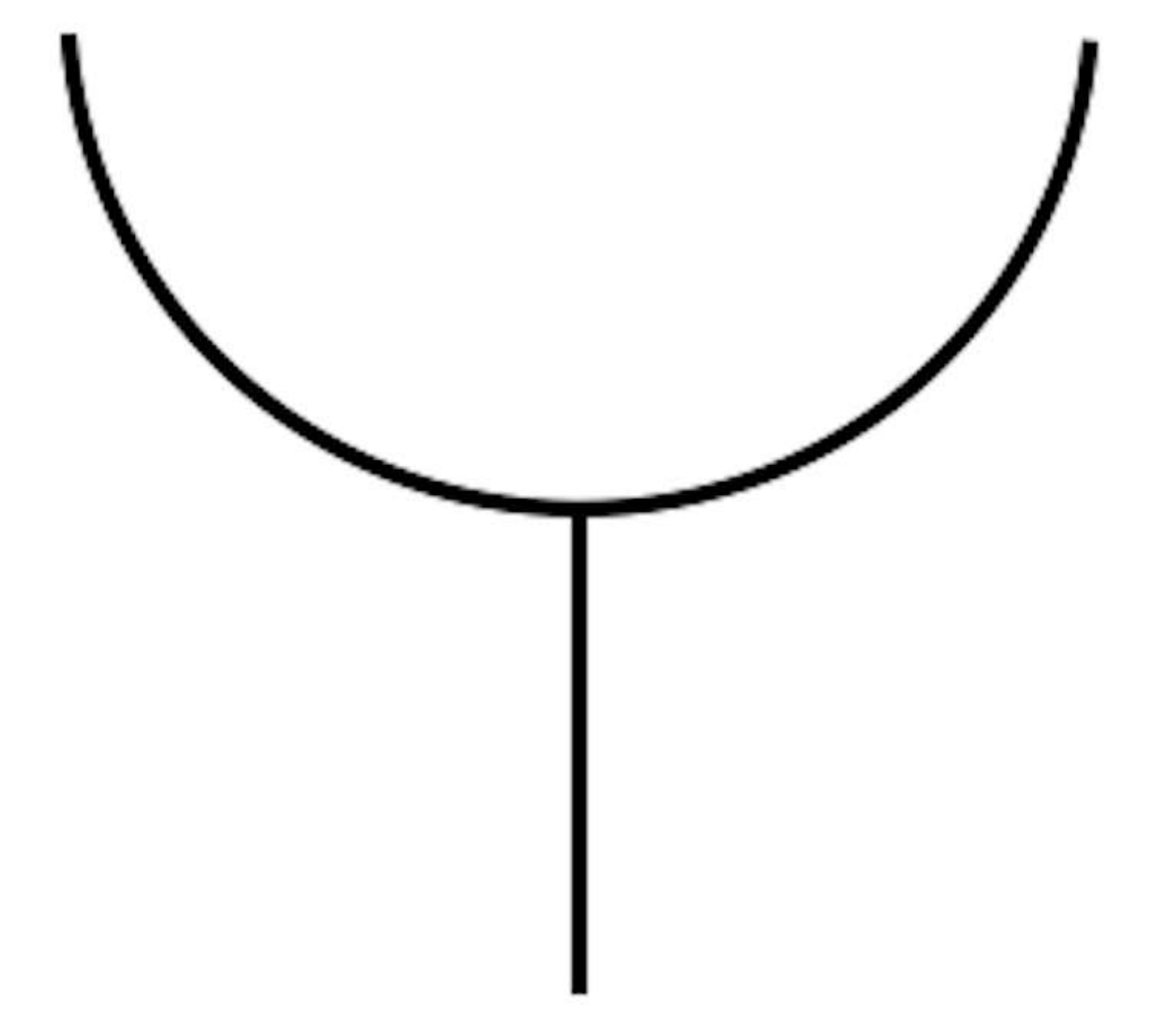} ~ = ~ 
\adjincludegraphics[valign = c, width = 2.2cm]{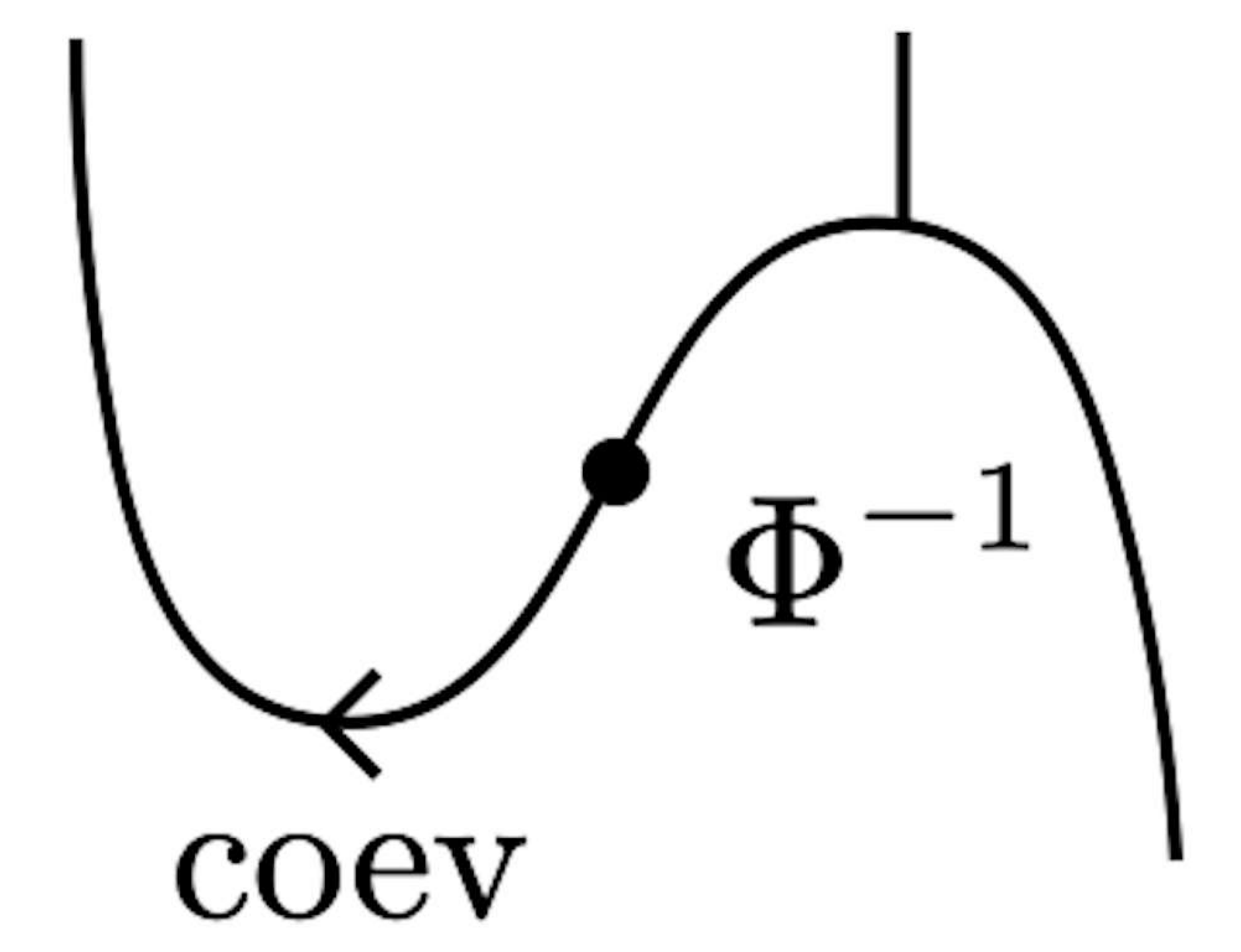}, \quad
\adjincludegraphics[valign = c, width = 0.9cm]{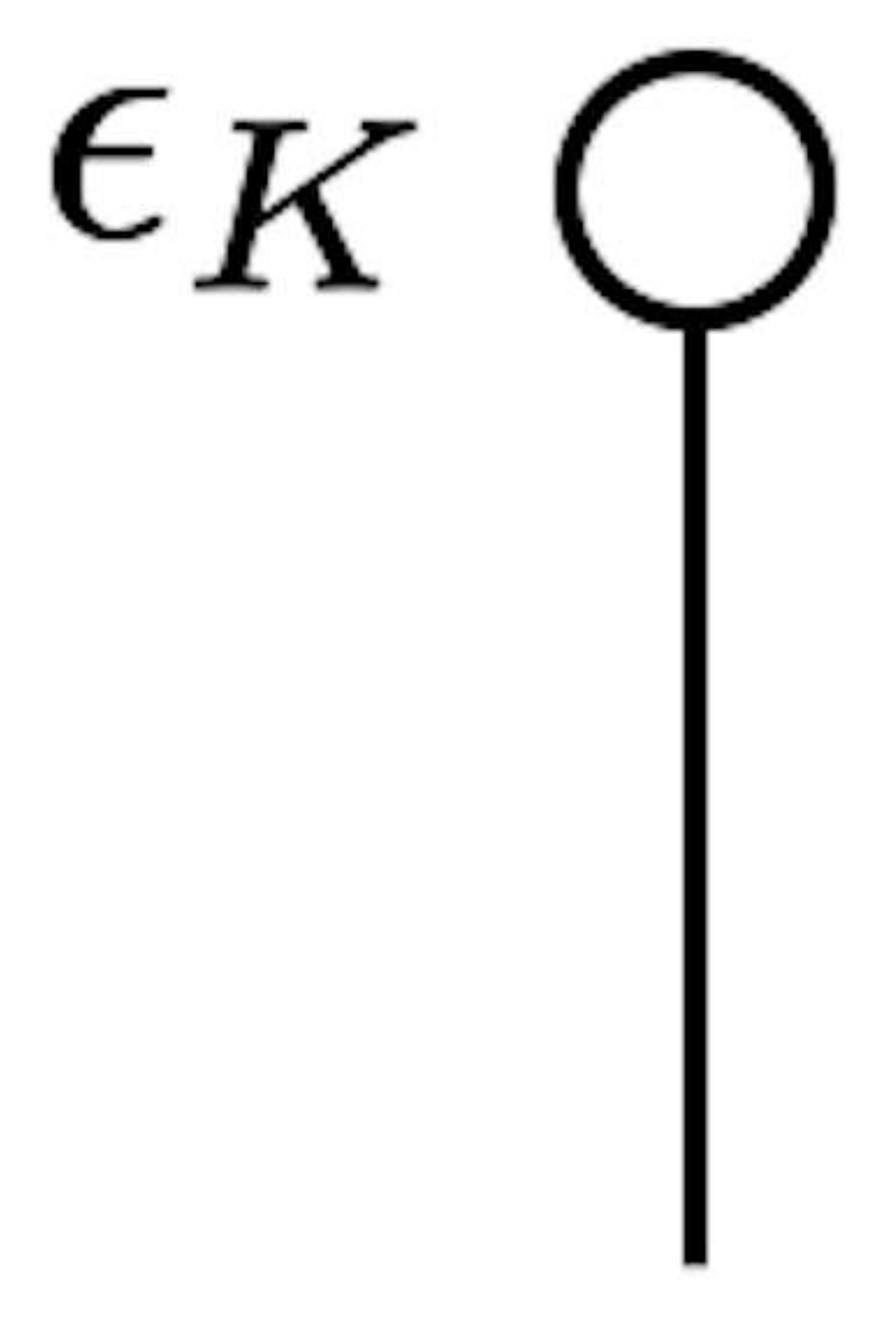} ~ = ~ 
\adjincludegraphics[valign = c, width = 1.35cm]{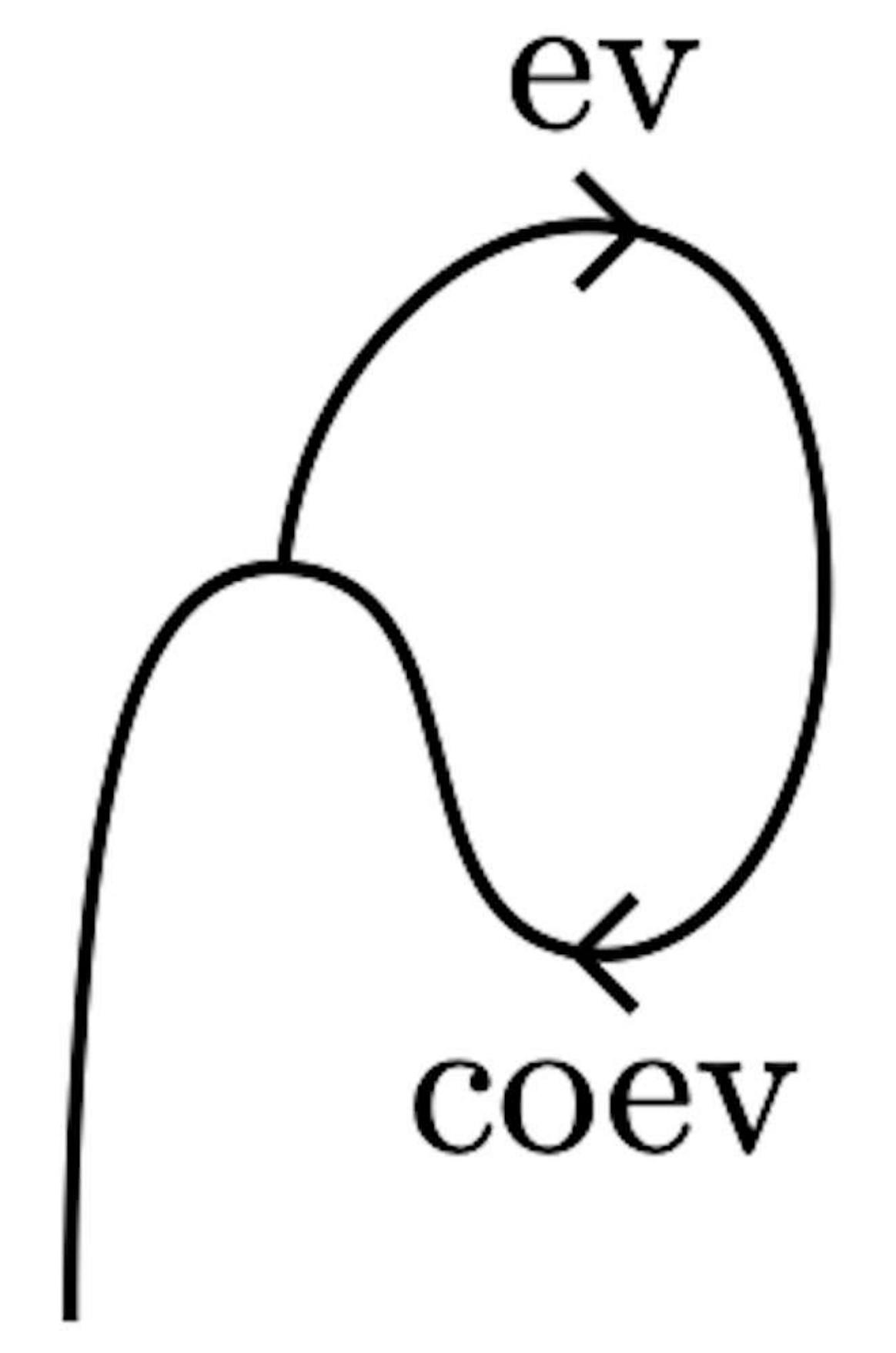}, \quad \text{where }
\adjincludegraphics[valign = c, width = 1cm]{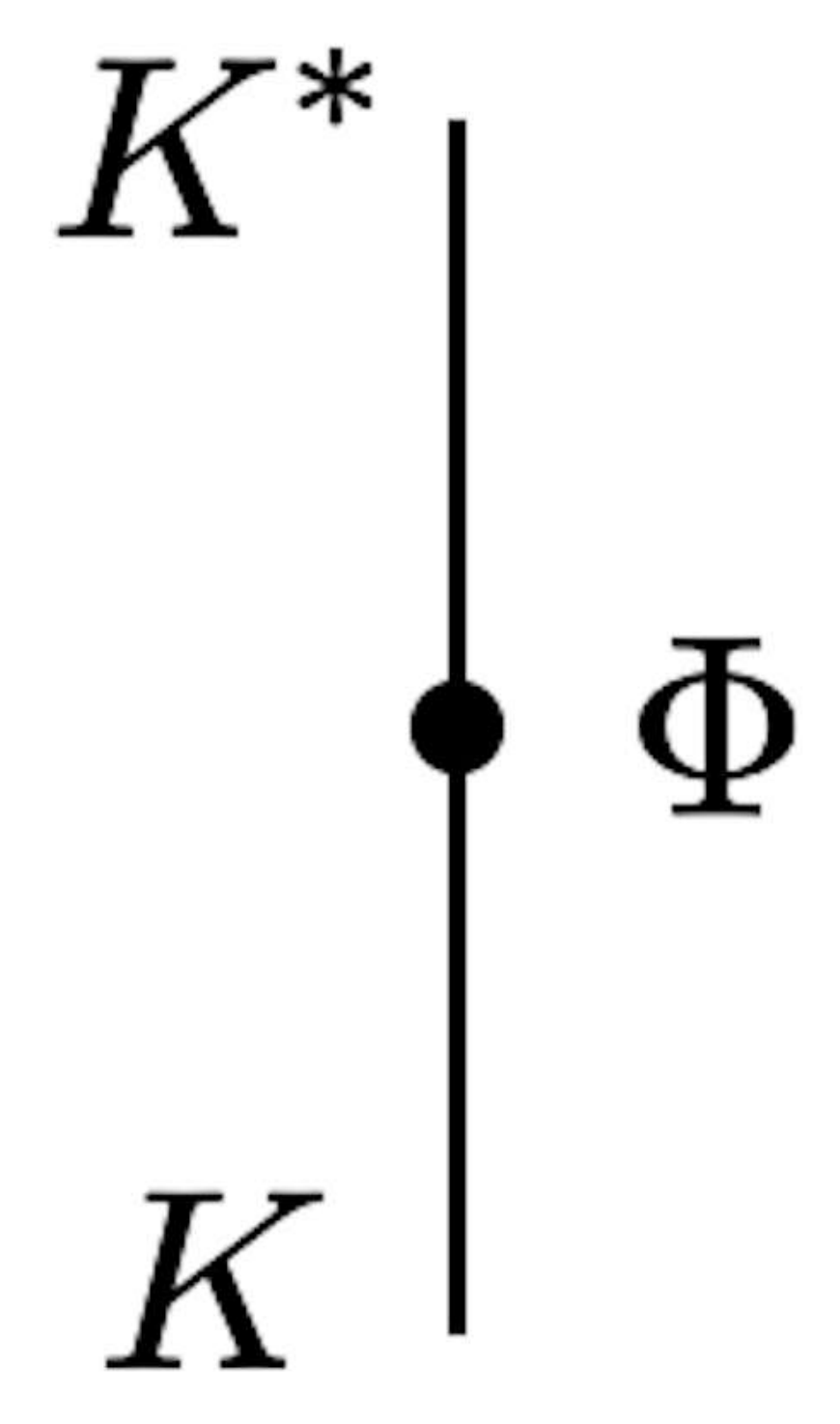} ~ = ~ 
\adjincludegraphics[valign = c, width = 2cm]{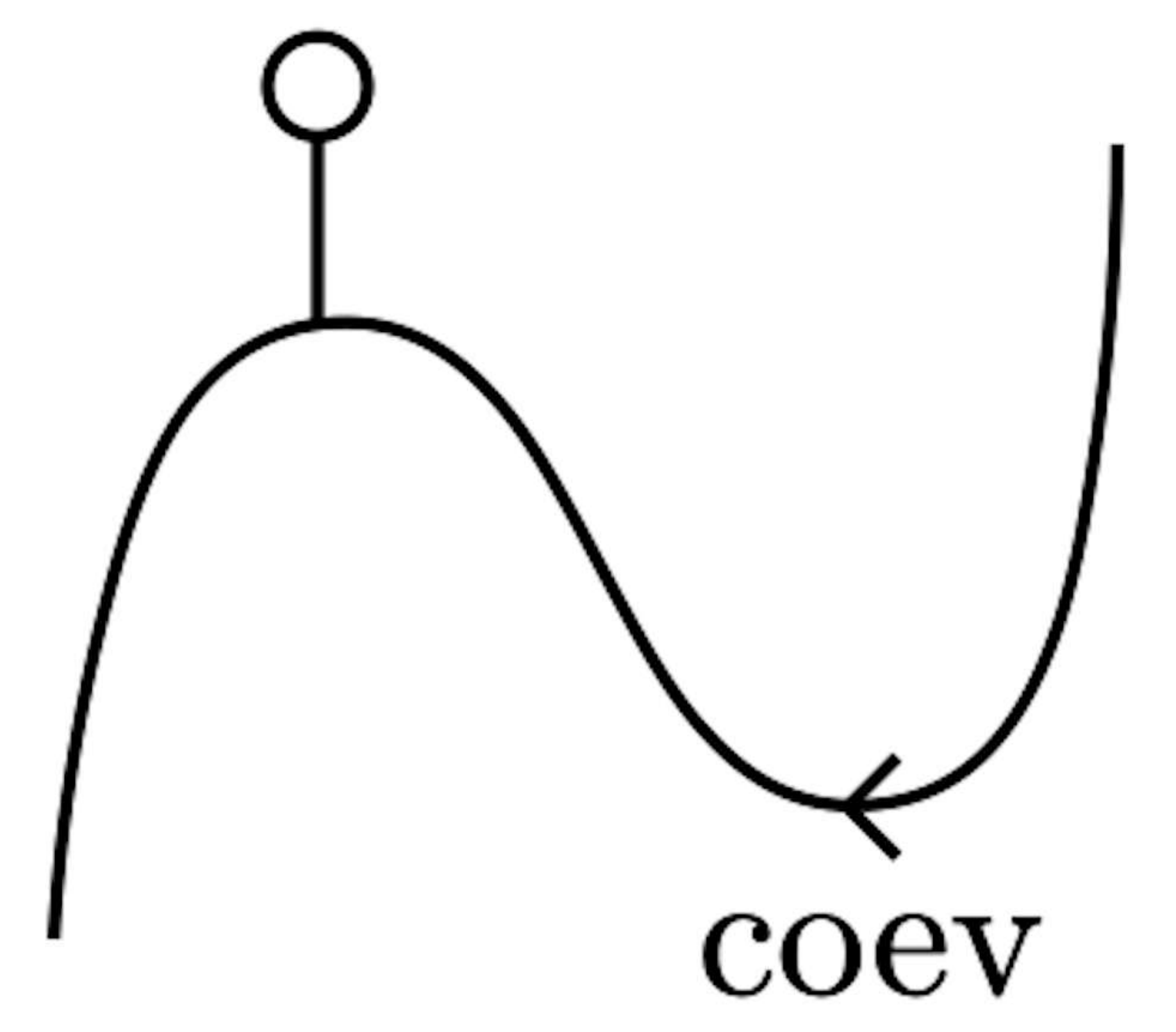}.
\label{eq: Frobenius structure}
\end{equation}
In the above equation, $K^*$ denotes the dual vector space of $K$.
The linear maps $\mathrm{ev}$ and $\mathrm{coev}$ are the evaluation and coevaluation morphisms of the category of vector spaces. 
Specifically, we have
\begin{equation}
\mathrm{ev} (a \otimes f) = f(a), \quad \mathrm{coev}(\lambda) = \lambda \sum_i u_i \otimes u^i, \quad \forall a \in K, \forall f \in K^*, \forall \lambda \in \mathbb{C},
\label{eq: ev/coev}
\end{equation}
where $\{u_i\}$ and $\{u^i\}$ are dual bases of $K$ and $K^*$.
It turns out that the Frobenius algebra structure given by eq. \eqref{eq: Frobenius structure} satisfies the following two properties \cite{FRS2002}:\footnote{In general, given a $\Delta$-separable symmetric Frobenius algebra object $K$ in a fusion category $\mathcal{C}$, the category of $K$-$K$ bimodules in $\mathcal{C}$ also becomes a fusion category. The category ${}_K \mathcal{M}_K$ is a special case where $\mathcal{C}$ is the category of vector spaces.}
\begin{equation}
\text{$\Delta$-separability: } \adjincludegraphics[valign = c, width = 1cm]{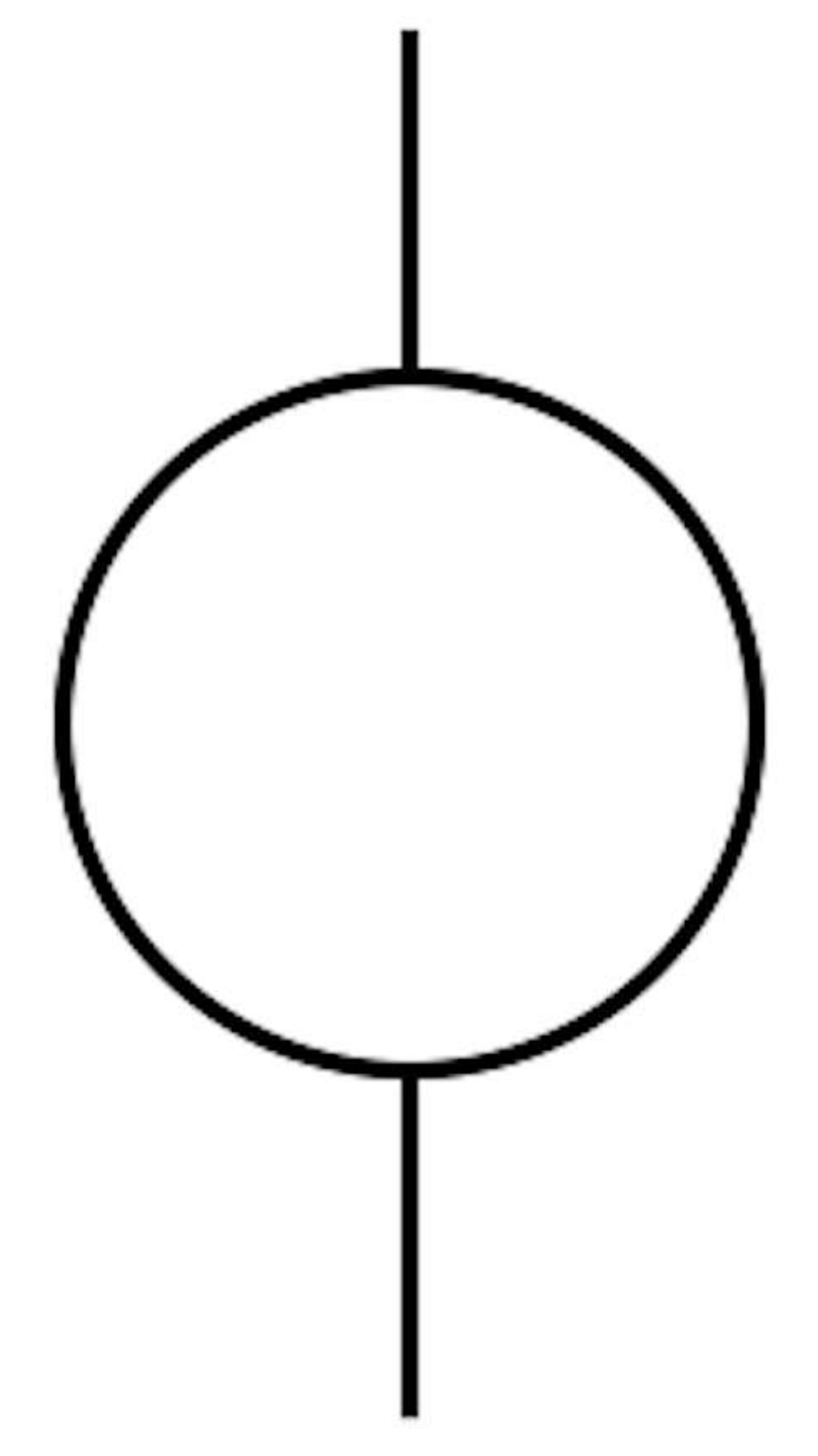} ~ = ~ 
\adjincludegraphics[valign = c, width = 0.25cm]{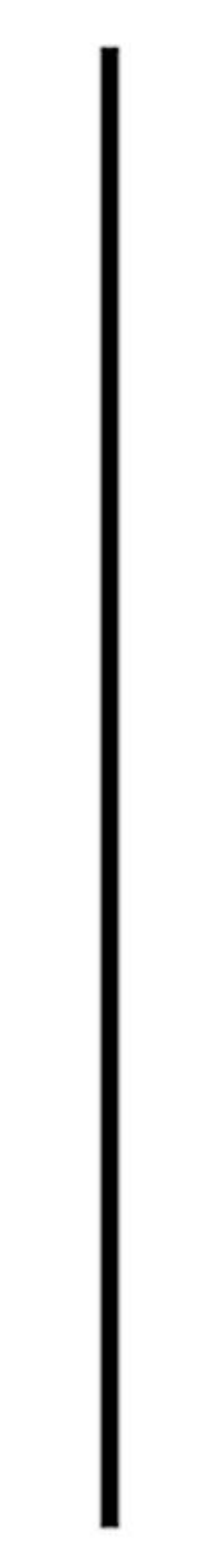}, \quad 
\text{symmetricity: } \adjincludegraphics[valign = c, width = 2cm]{Phi2.pdf} ~ = ~ 
\adjincludegraphics[valign = c, width = 2cm]{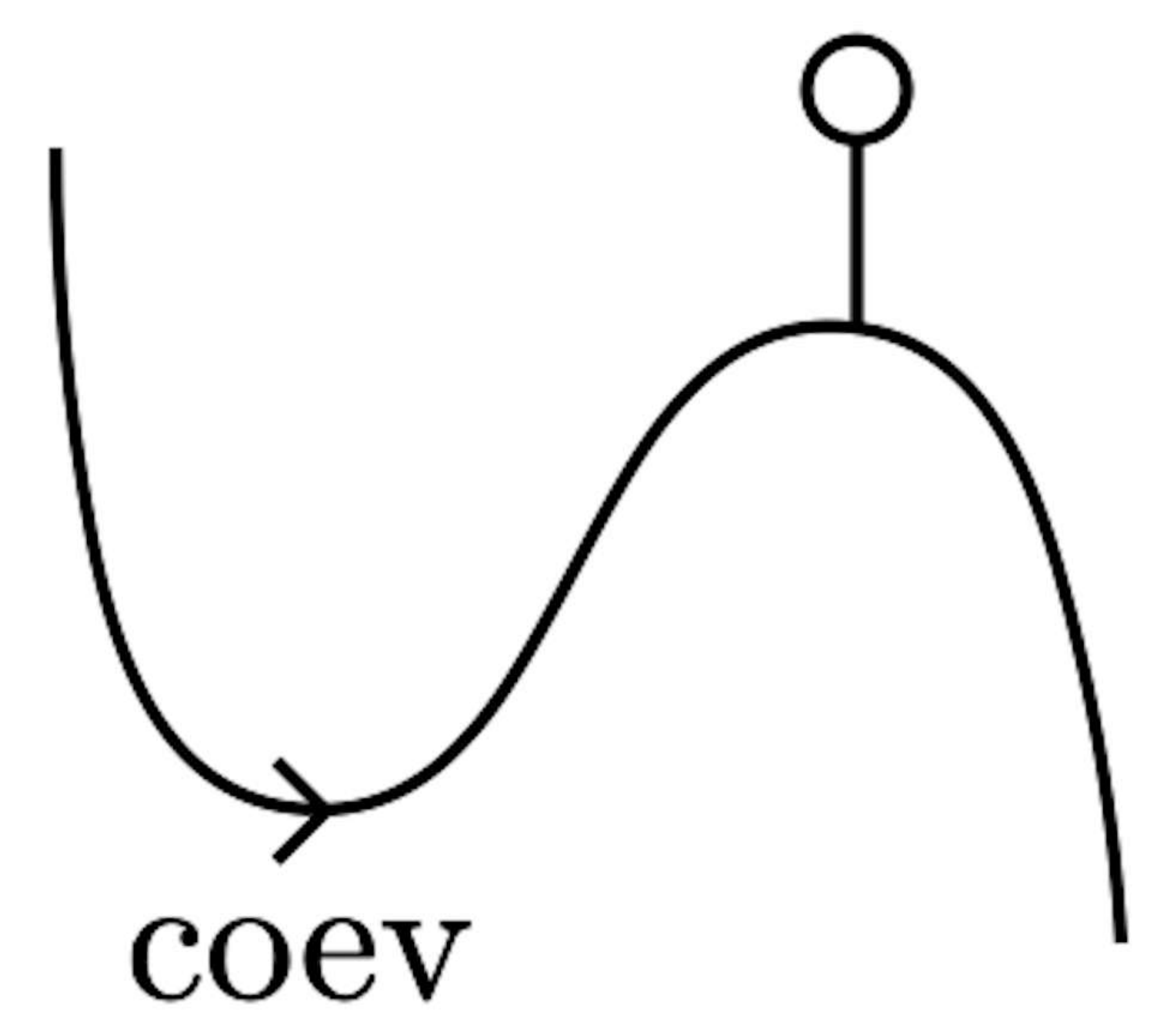}.
\label{eq: Delta-separable symmetric}
\end{equation}
The tensor product $Y_1 \otimes_K Y_2$ is defined as the image of a projector $p_{Y_1, Y_2}: Y_1 \otimes Y_2 \rightarrow Y_1 \otimes Y_2$ that is represented by the following string diagram
\begin{equation}
p_{Y_1, Y_2} = \adjincludegraphics[valign = c, width = 1.8cm]{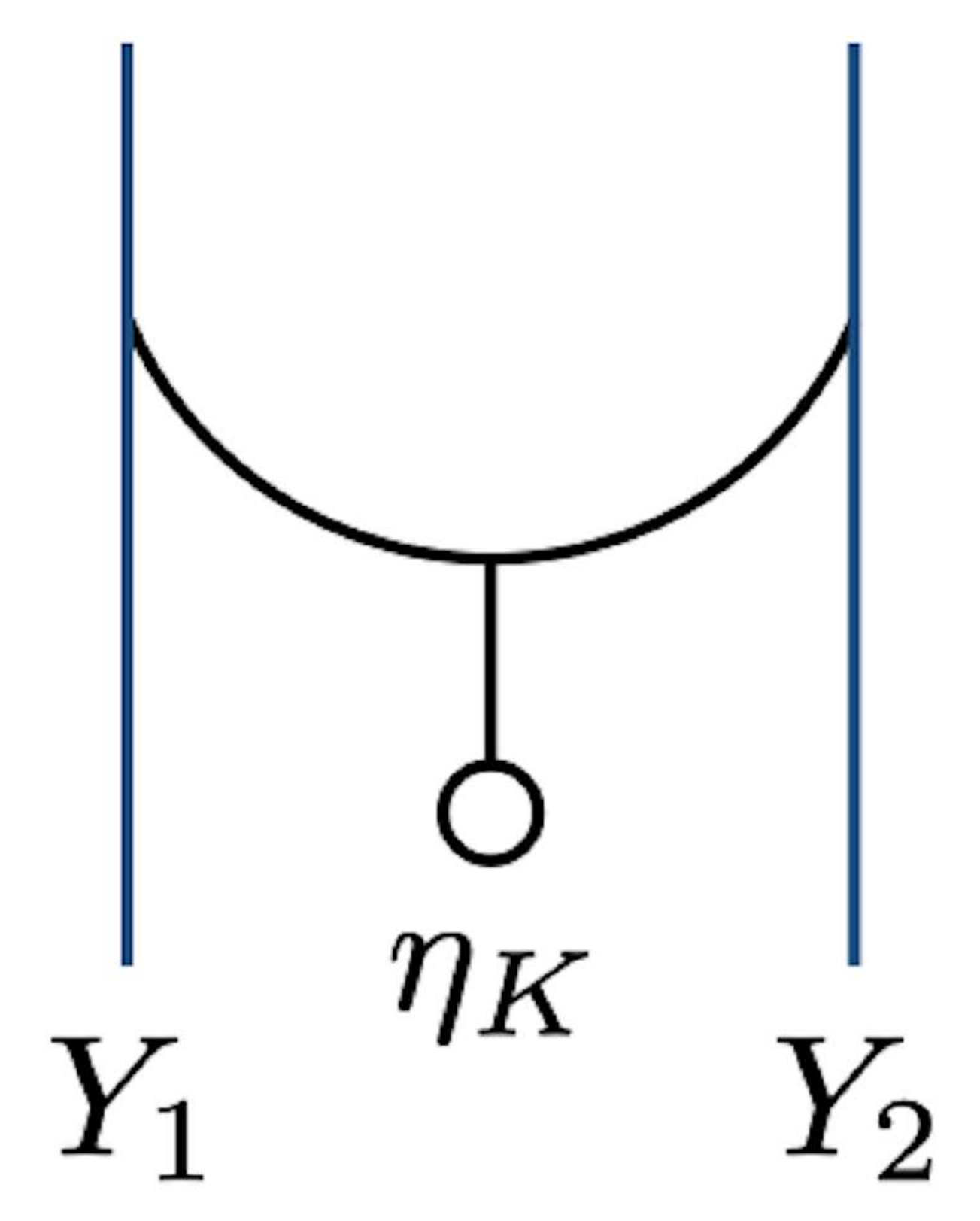},
\label{eq: projector p}
\end{equation}
where the junction of $Y_1$ ($Y_2$) and $K$ represents a right (left) $K$-module action.
We note that the unit object for the tensor product over $K$ is $K$ itself.
The splitting maps of the projector \eqref{eq: projector p} are denoted by $\pi_{Y_1, Y_2}: Y_1 \otimes Y_2 \rightarrow Y_1 \otimes_K Y_2$ and $\iota_{Y_1, Y_2}: Y_1 \otimes_K Y_2 \rightarrow Y_1 \otimes Y_2$, which obey $\iota_{Y_1, Y_2} \circ \pi_{Y_1, Y_2} = p_{Y_1, Y_2}$ and $\pi_{Y_1, Y_2} \circ \iota_{Y_1, Y_2} = \mathrm{id}_{Y_1 \otimes_K Y_2}$.
The associator $\alpha_{Y_1, Y_2, Y_3}: (Y_1 \otimes_K Y_2) \otimes_K Y_3 \rightarrow Y_1 \otimes_K (Y_2 \otimes_K Y_3)$ is given by a composition of these splitting maps as 
\begin{equation}
\alpha_{Y_1, Y_2, Y_3} = \pi_{Y_1, Y_2 \otimes_K Y_3} \circ (\mathrm{id}_{Y_1} \otimes \pi_{Y_2, Y_3}) \circ (\iota_{Y_1, Y_2} \otimes \mathrm{id}_{Y_3}) \circ \iota_{Y_1 \otimes_K Y_2, Y_3}.
\label{eq: associator of KMK}
\end{equation}
The tensor product of morphisms $f \in \mathrm{Hom}_{KK}(Y_1, Y_1^{\prime})$ and $g \in \mathrm{Hom}_{KK}(Y_2, Y_2^{\prime})$ is defined in terms of the splitting maps as $f \otimes_K g := \pi_{Y_1, Y_2} \circ (f \otimes g) \circ \iota_{Y_1, Y_2}$, where $\mathrm{Hom}_{KK}(Y, Y^{\prime})$ denotes the space of $K$-$K$ bimodule maps from $Y$ to $Y^{\prime}$.

We finally notice that the category ${}_K \mathcal{M}$ of left $K$-modules is a ${}_K \mathcal{M}_K$-module category, on which ${}_K \mathcal{M}_K$ acts by the tensor product over $K$.
The module associativity constraint $m_{Y_1, Y_2, M}: (Y_1 \otimes_K Y_2) \otimes_K M \rightarrow Y_1 \otimes_K (Y_2 \otimes_K M)$ for $Y_1, Y_2 \in {}_K \mathcal{M}_K$ and $M \in {}_K \mathcal{M}$ is given by the composition of the splitting maps as the associator \eqref{eq: associator of KMK}:
\begin{equation}
m_{Y_1, Y_2, M} = \pi_{Y_1, Y_2 \otimes_K M} \circ (\mathrm{id}_{Y_1} \otimes \pi_{Y_2, M}) \circ (\iota_{Y_1, Y_2} \otimes \mathrm{id}_{M}) \circ \iota_{Y_1 \otimes_K Y_2, M}.
\label{eq: module associativity KM}
\end{equation}

\subsection{Hopf algebras, (co)module algebras, and smash product}
\label{sec: Hopf algebra, (co)module algebra, and smash product}
In this subsection, we briefly review the definitions and some basic properties of Hopf algebras.
For details, see for example \cite{Montgomery1993, Montgomery2001, Schneider1995}.
We first give the definition.
A $\mathbb{C}$-vector space $H$ is called a Hopf algebra if it is equipped with structure maps $(m, 1, \Delta, \epsilon, S)$ that satisfy the following conditions:
\begin{enumerate}
\item $(H, m, 1)$ is a unital associative algebra where $m: H \otimes H \rightarrow H$ is the multiplication and $1 \in H$ is the unit.
\item $(H, \Delta, \epsilon)$ is a counital coassociative coalgebra where $\Delta: H \rightarrow H \otimes H$ is the comultiplication and $\epsilon: H \rightarrow \mathbb{C}$ is the counit.\footnote{The comultiplication $\Delta$ for the Hopf algebra structure on a semisimple Hopf algebra $H$ is different from the comultiplication $\Delta_{H}$ for the Frobenius algebra structure on $H$. The same comment applies to $\epsilon$ and $\epsilon_H$.}
\item The comultiplication $\Delta$ is a unit-preserving algebra homomorphism
\begin{equation}
\Delta(g h) = \Delta(g) \Delta(h), \quad \Delta(1) = 1 \otimes 1, \quad \forall g, h \in H, 
\end{equation}
where we denote the multiplication of $g$ and $h$ as $g h$.
The multiplication on $H \otimes H$ is induced by that on $H$.
\item The counit $\epsilon$ is a unit-preserving algebra homomorphism\footnote{The right-hand side of the second equation of \eqref{eq: alg-hom epsilon} is just a number $1 \in \mathbb{C}$, which defers from the unit of $H$.}
\begin{equation}
\epsilon (gh) = \epsilon (g) \epsilon (h), \quad \epsilon(1) = 1, \quad \forall g, h \in H.
\label{eq: alg-hom epsilon}
\end{equation}
\item The antipode $S: H \rightarrow H$ satisfies
\begin{equation}
m \circ (\mathrm{id} \otimes S) \circ \Delta = m \circ (S \otimes \mathrm{id}) \circ \Delta = 1 \epsilon.
\end{equation}
\end{enumerate}
In particular, the antipode $S$ squares to the identity when $H$ is semisimple, i.e. $S^2 = \mathrm{id}$.
In the rest of this paper, we only consider finite dimensional semisimple Hopf algebras and do not distinguish between $S$ and $S^{-1}$.

When $H$ is a Hopf algebra, the opposite algebra $H^{\mathrm{op}}$ is also a Hopf algebra, whose underlying vector space is $H$ and whose structure maps are given by $(m^{\mathrm{op}}, 1, \Delta, \epsilon, S^{-1})$.
Here, the opposite multiplication $m^{\mathrm{op}}: H^{\mathrm{op}} \otimes H^{\mathrm{op}} \rightarrow H^{\mathrm{op}}$ is defined by $m^{\mathrm{op}}(g \otimes h) = hg$ for all $g, h \in H$.
Similarly, the coopposite coalgebra $H^{\mathrm{cop}}$ also becomes a Hopf algebra, whose underlying vector space is $H$ and whose structure maps are given by $(m, 1, \Delta^{\mathrm{cop}}, \epsilon, S^{-1})$.
Here, the coopposite comultiplication $\Delta^{\mathrm{cop}}: H^{\mathrm{cop}} \rightarrow H^{\mathrm{cop}} \otimes H^{\mathrm{cop}}$ is defined by $\Delta^{\mathrm{cop}}(h) = h_{(2)} \otimes h_{(1)}$ for all $h \in H$.\footnote{We use Sweedler's notation for the comultiplication $\Delta(h) = h_{(1)} \otimes h_{(2)}$.}

In the subsequent sections, we will use the string diagram notation where the above conditions 1--5 are represented as follows:
\begin{enumerate}
\item
\begin{equation}
\adjincludegraphics[valign = c, width = 1.8cm]{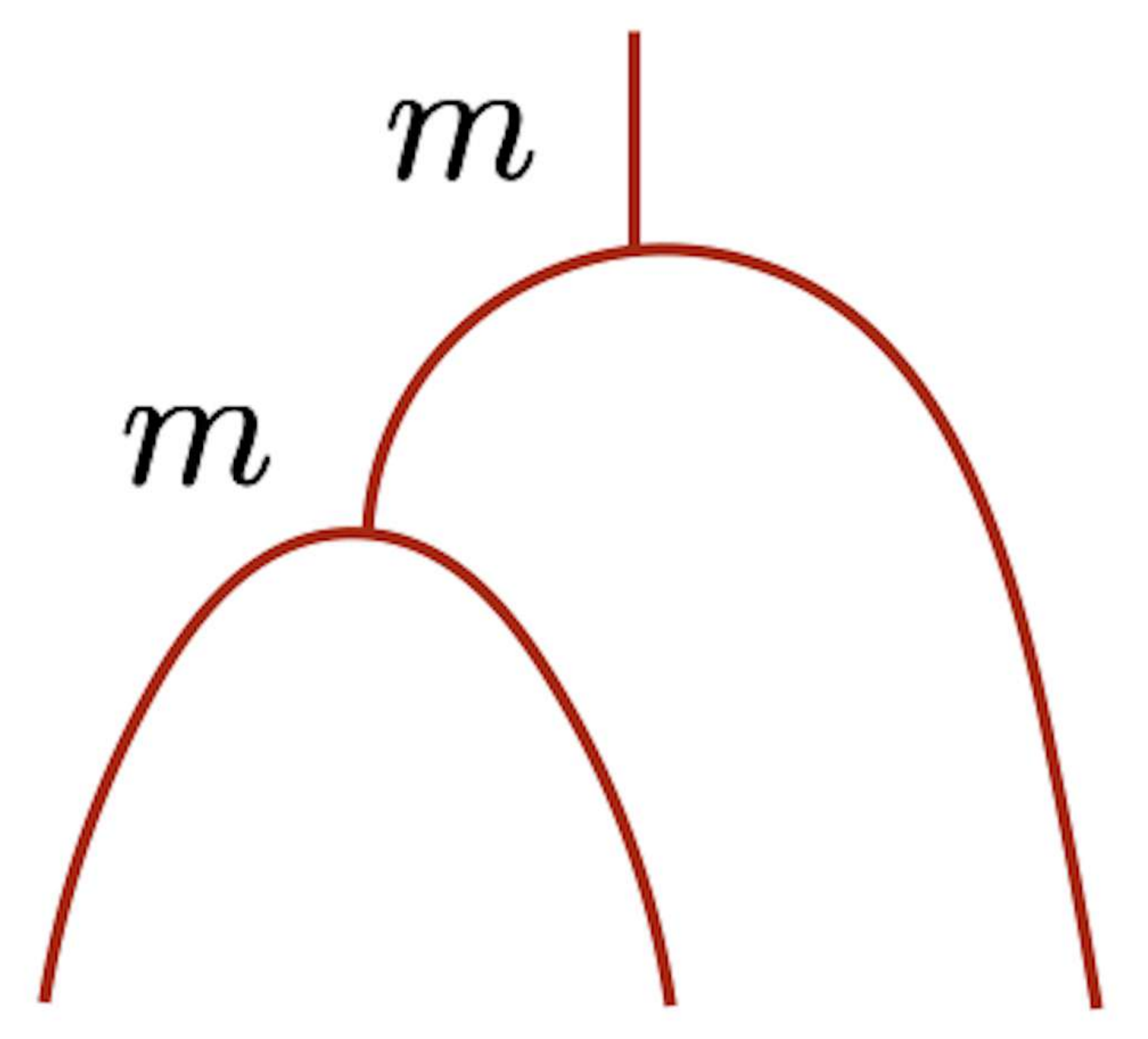} ~ = ~ 
\adjincludegraphics[valign = c, width = 1.8cm]{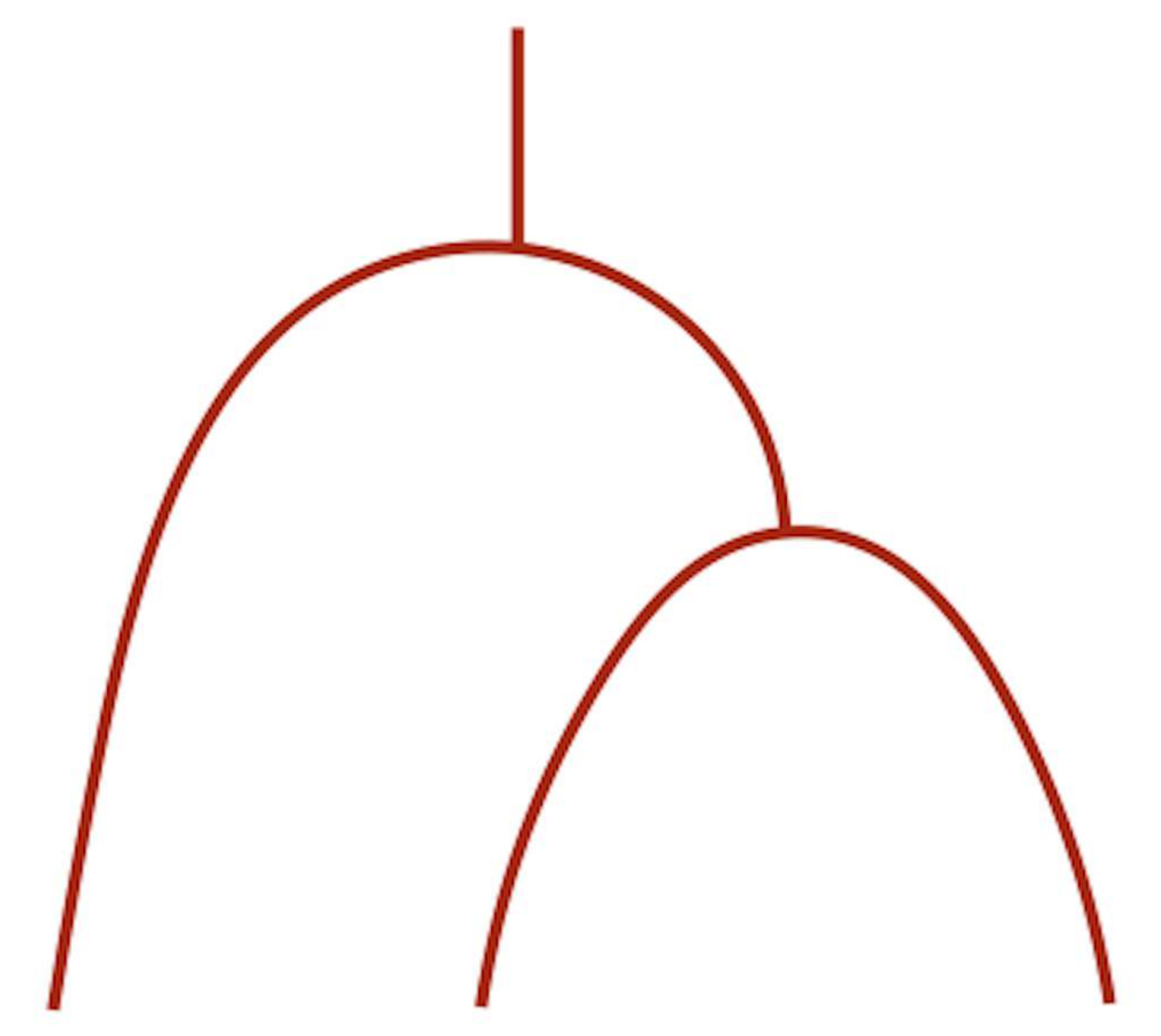}, \quad
\adjincludegraphics[valign = c, width = 1.35cm]{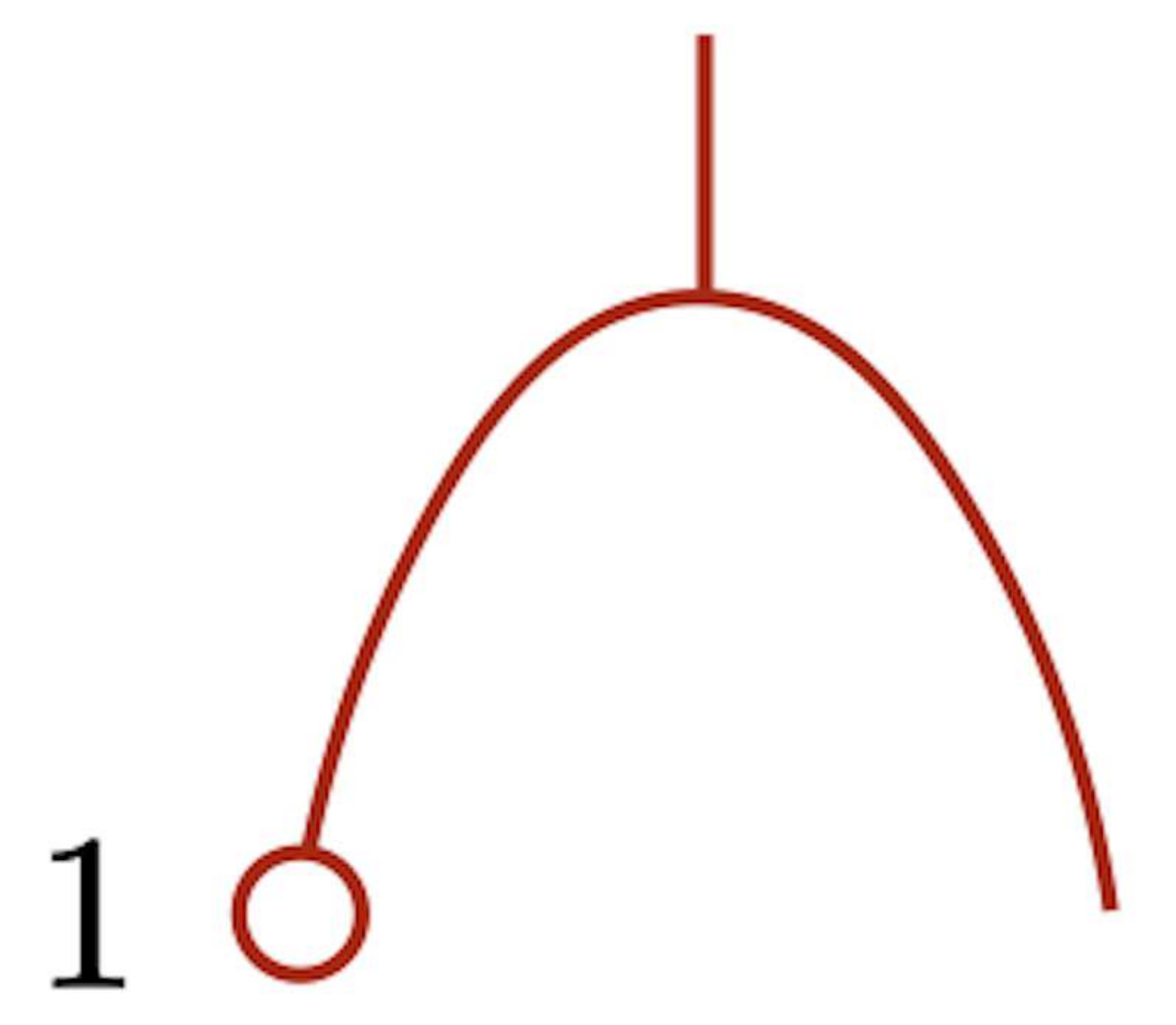} ~ = ~ 
\adjincludegraphics[valign = c, width = 1.3cm]{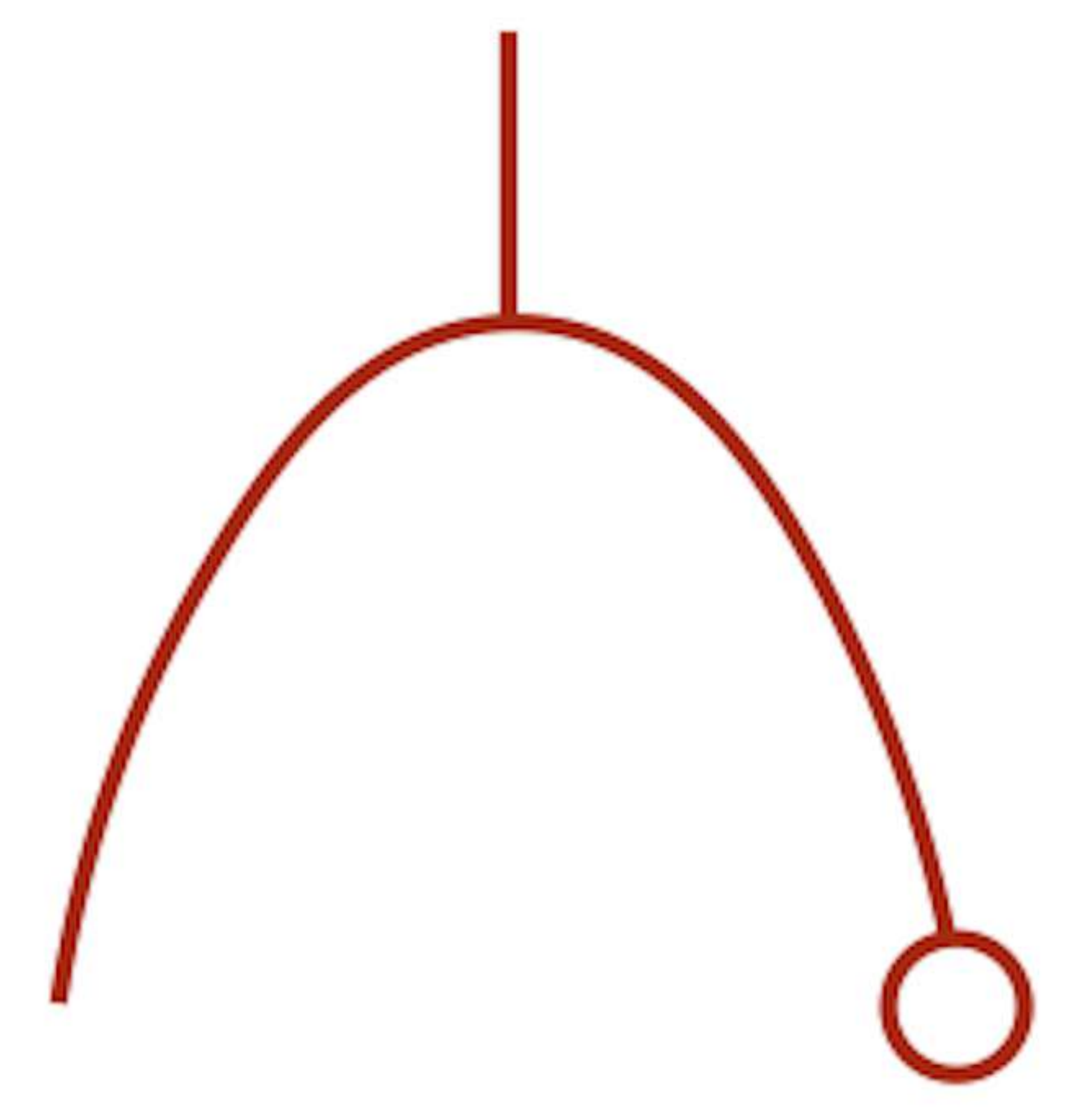} ~ = ~ 
\adjincludegraphics[valign = c, width = 0.15cm]{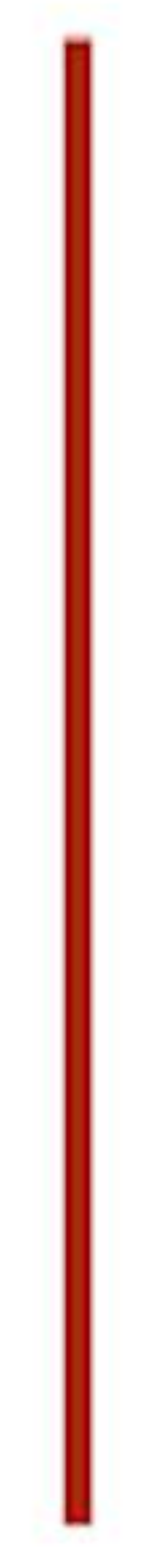}.
\end{equation}
\item
\begin{equation}
\adjincludegraphics[valign = c, width = 1.8cm]{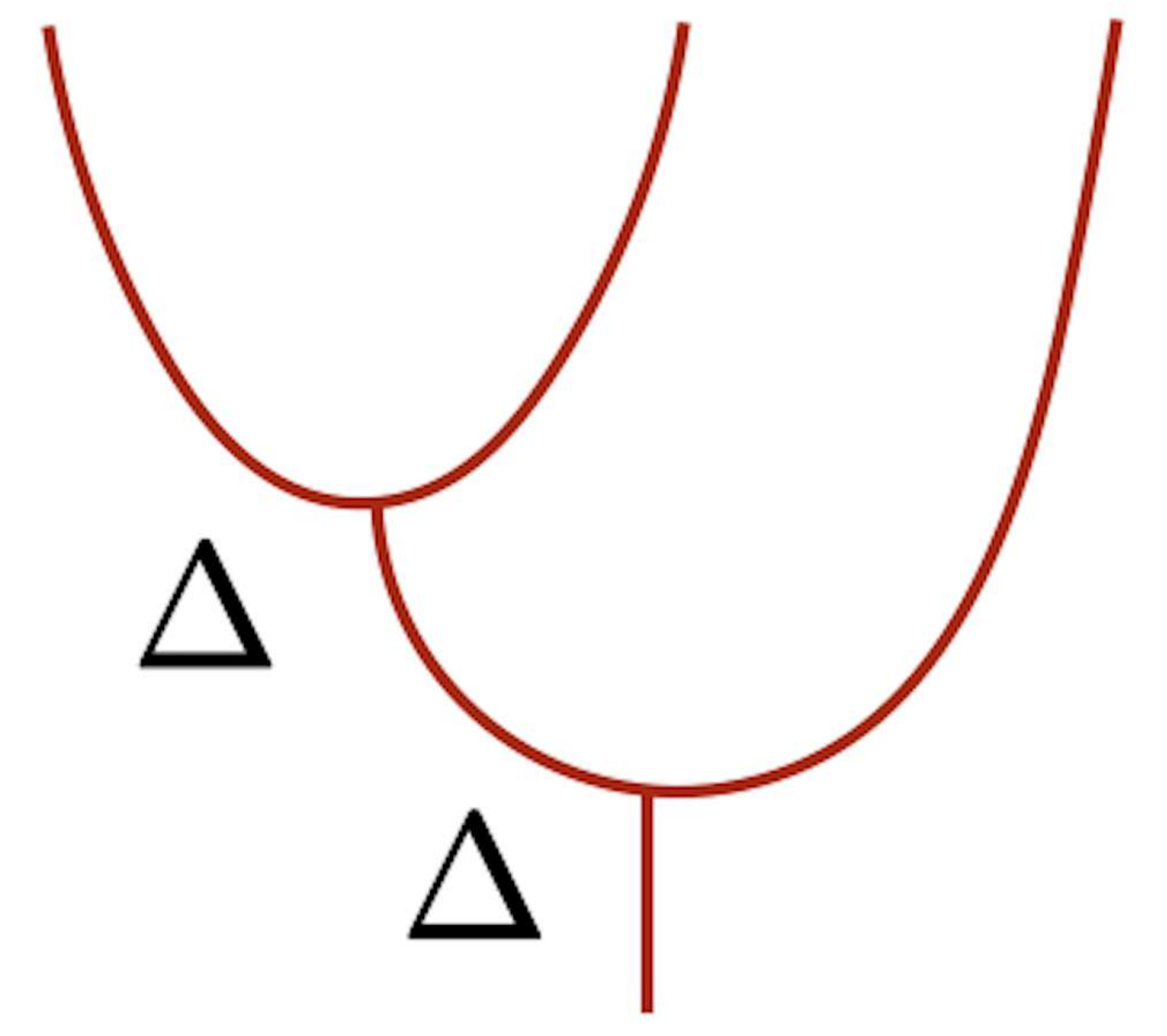} ~ = ~ 
\adjincludegraphics[valign = c, width = 1.8cm]{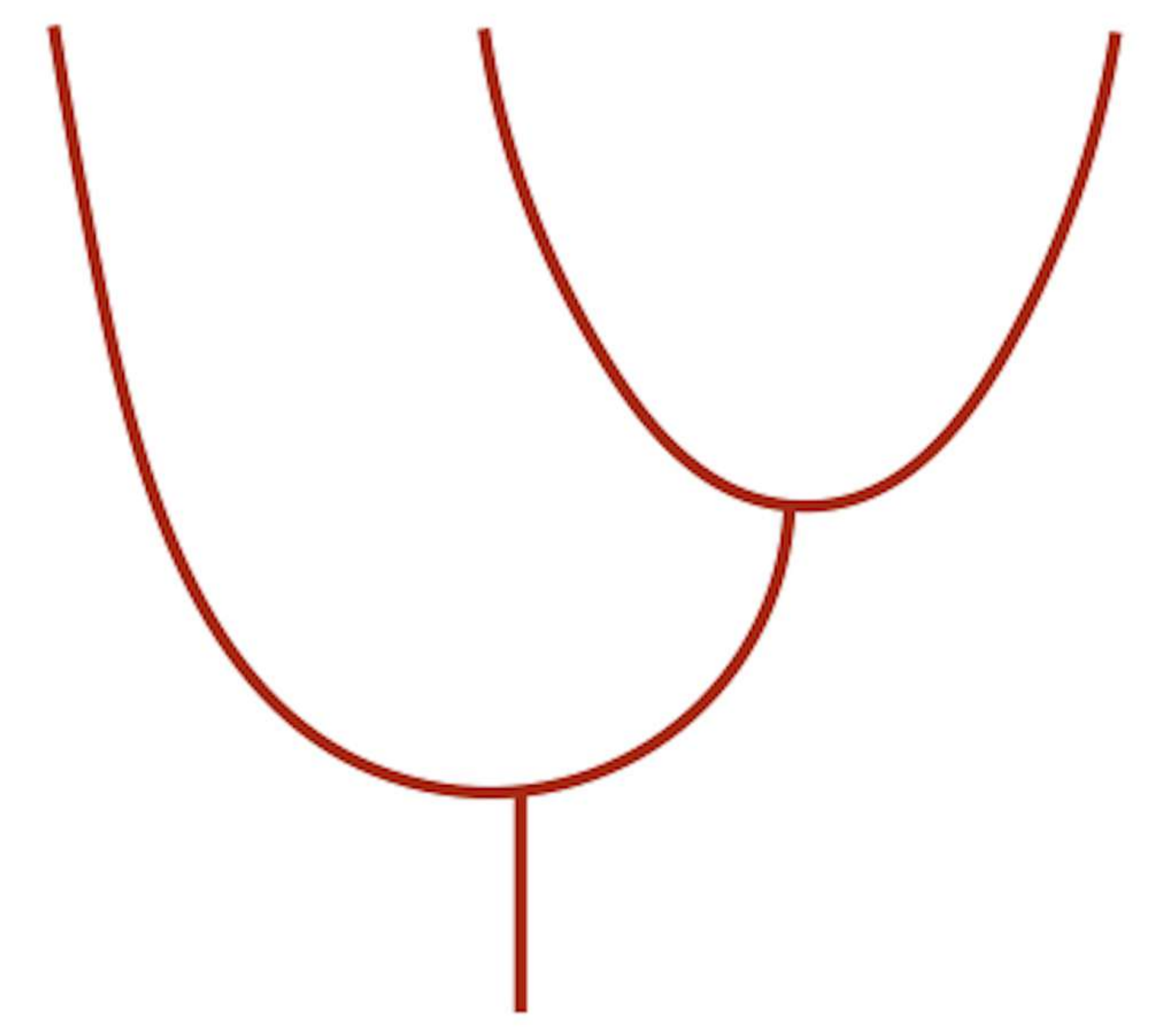}, \quad
\adjincludegraphics[valign = c, width = 1.35cm]{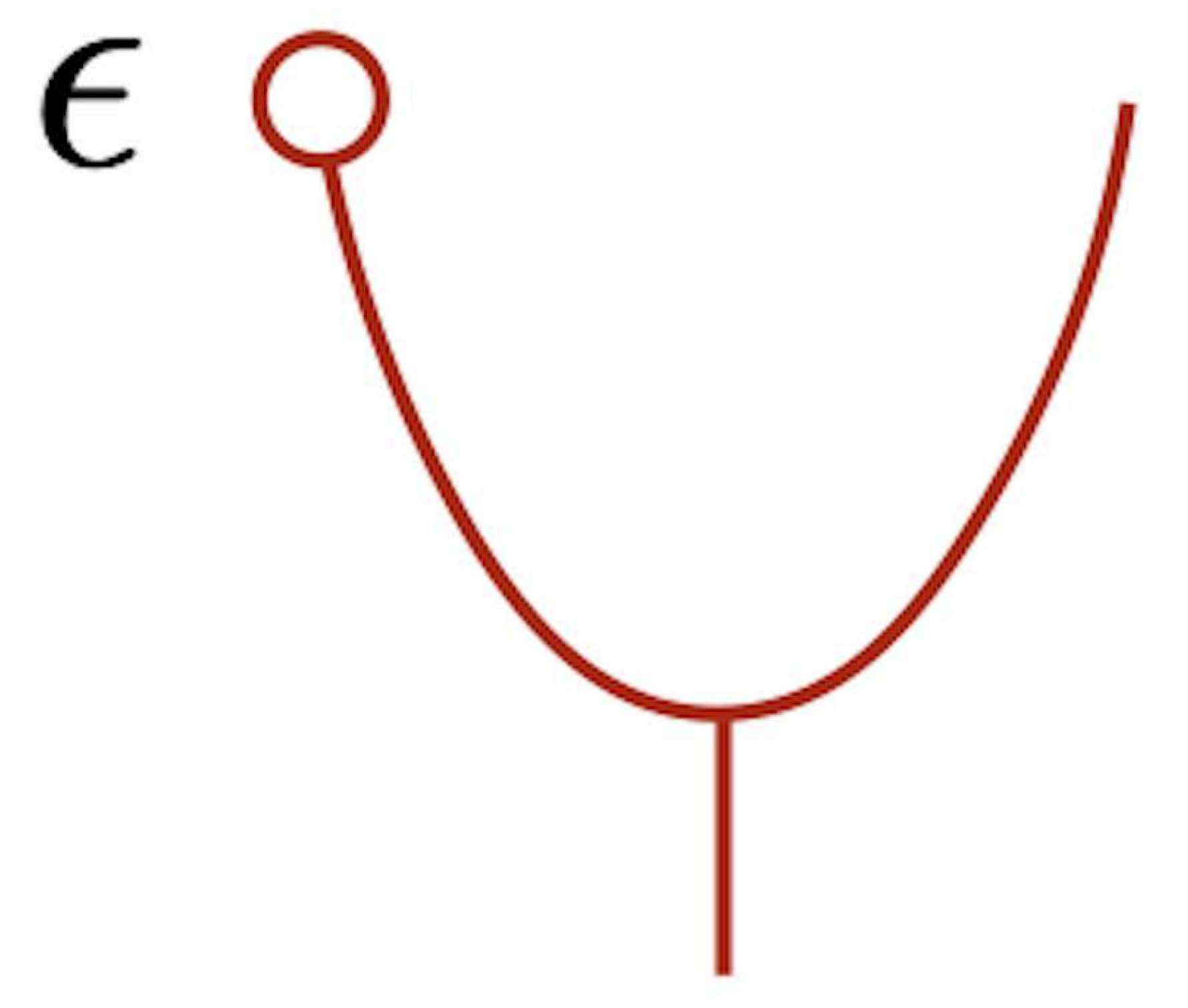} ~ = ~ 
\adjincludegraphics[valign = c, width = 1.3cm]{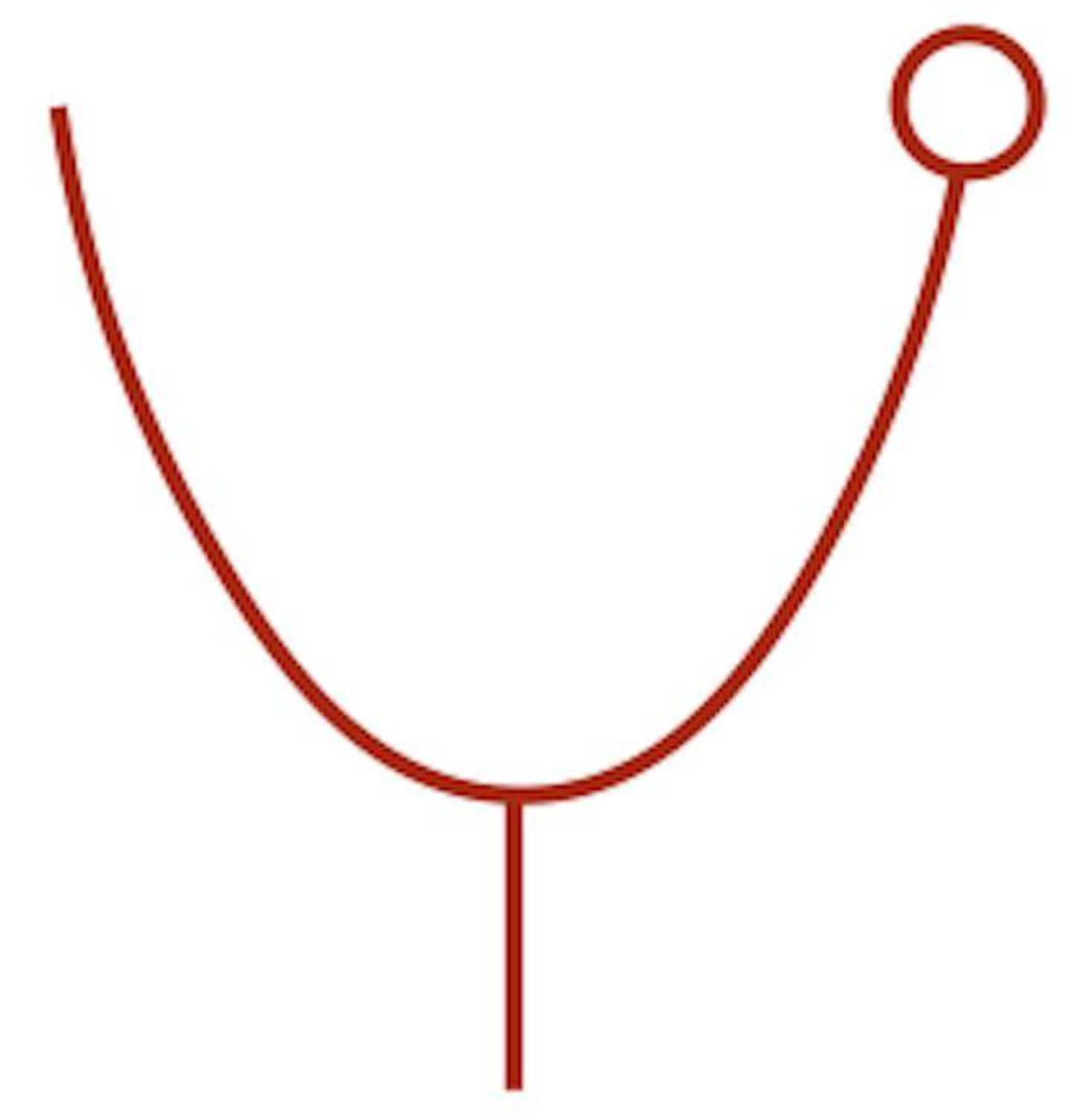} ~ = ~ 
\adjincludegraphics[valign = c, width = 0.15cm]{Hopf_unit3.pdf}.
\end{equation}
\item
\begin{equation}
\adjincludegraphics[valign = c, width = 1.2cm]{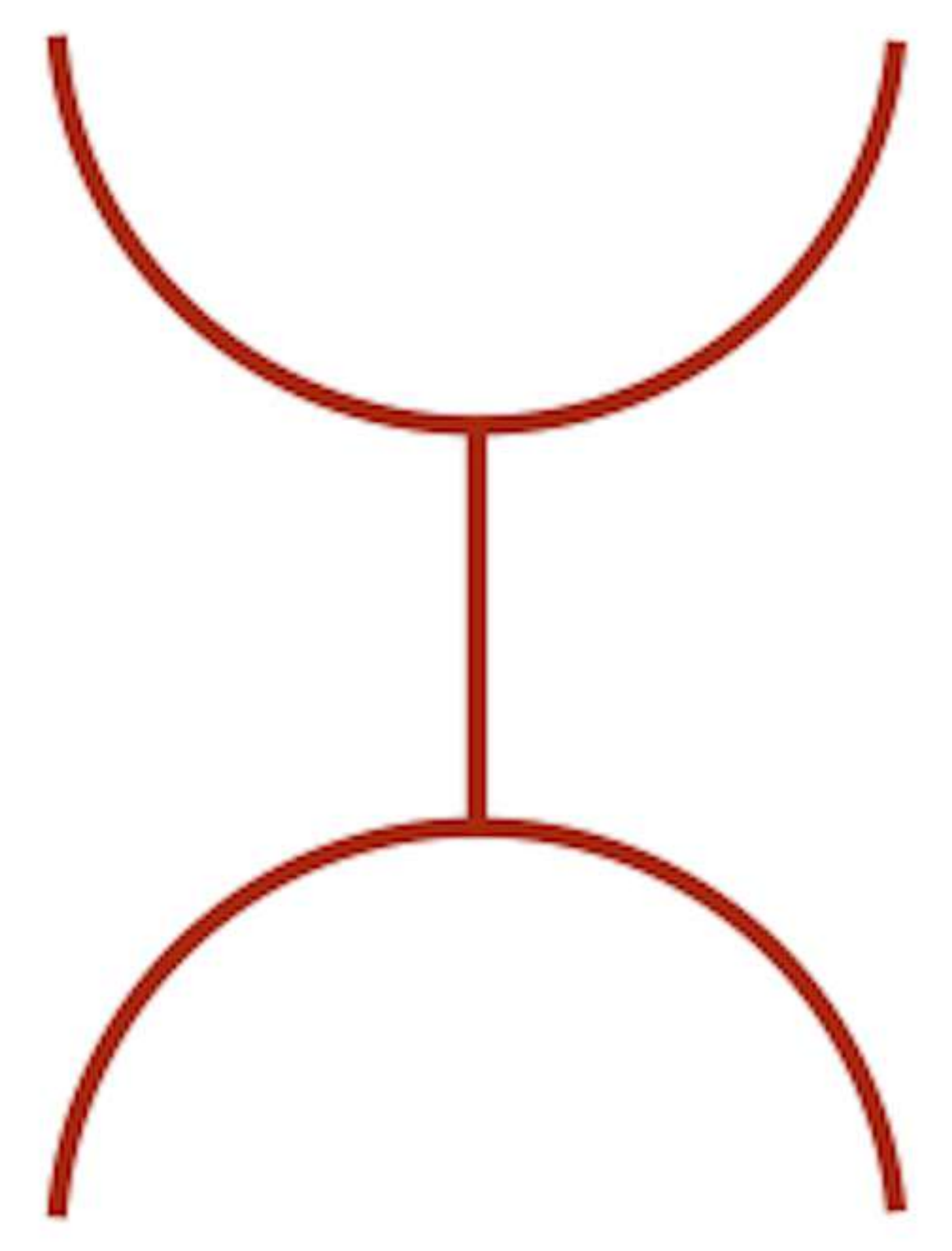} ~ = ~ 
\adjincludegraphics[valign = c, width = 2cm]{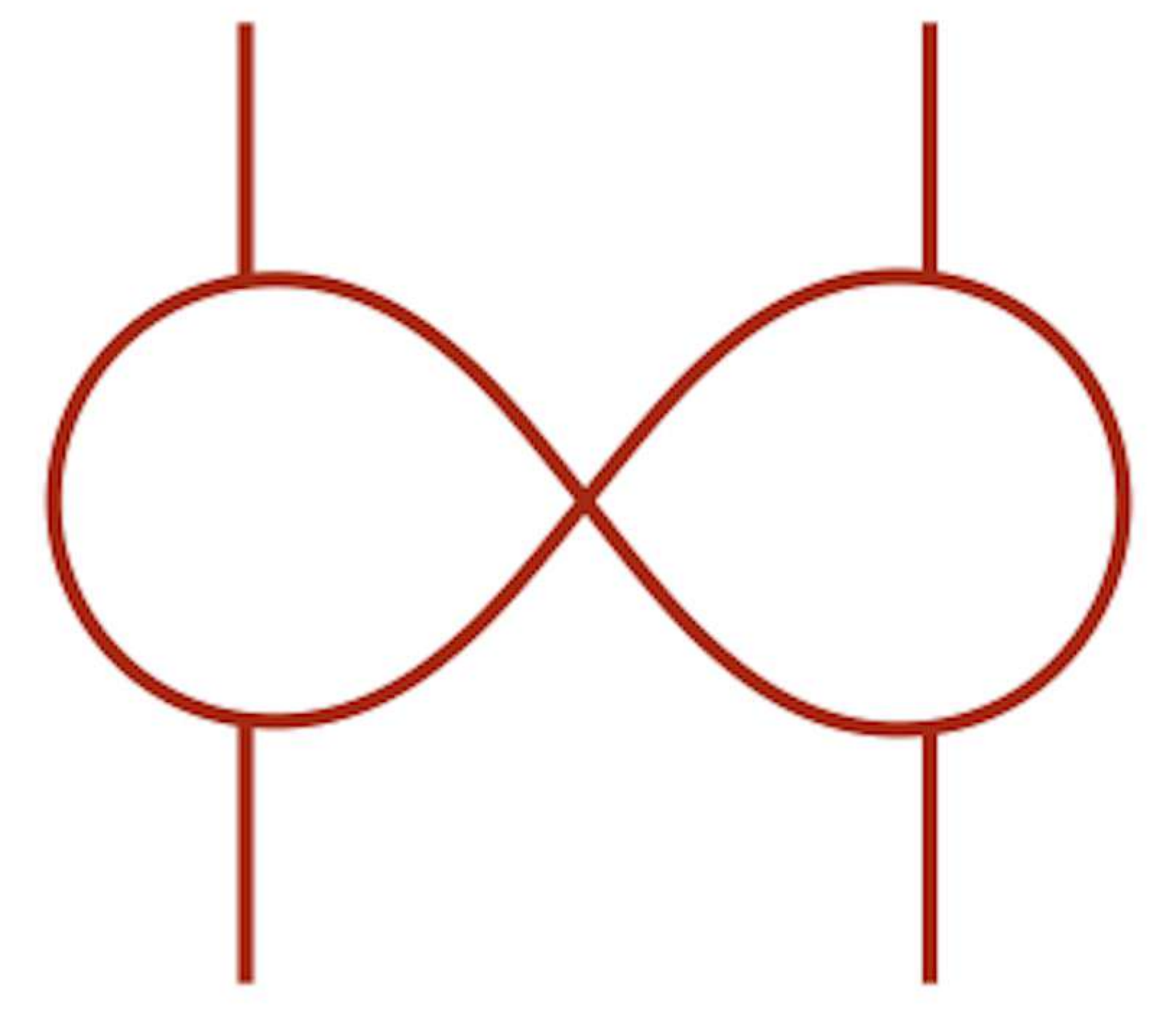}, \quad
\adjincludegraphics[valign = c, width = 1.2cm]{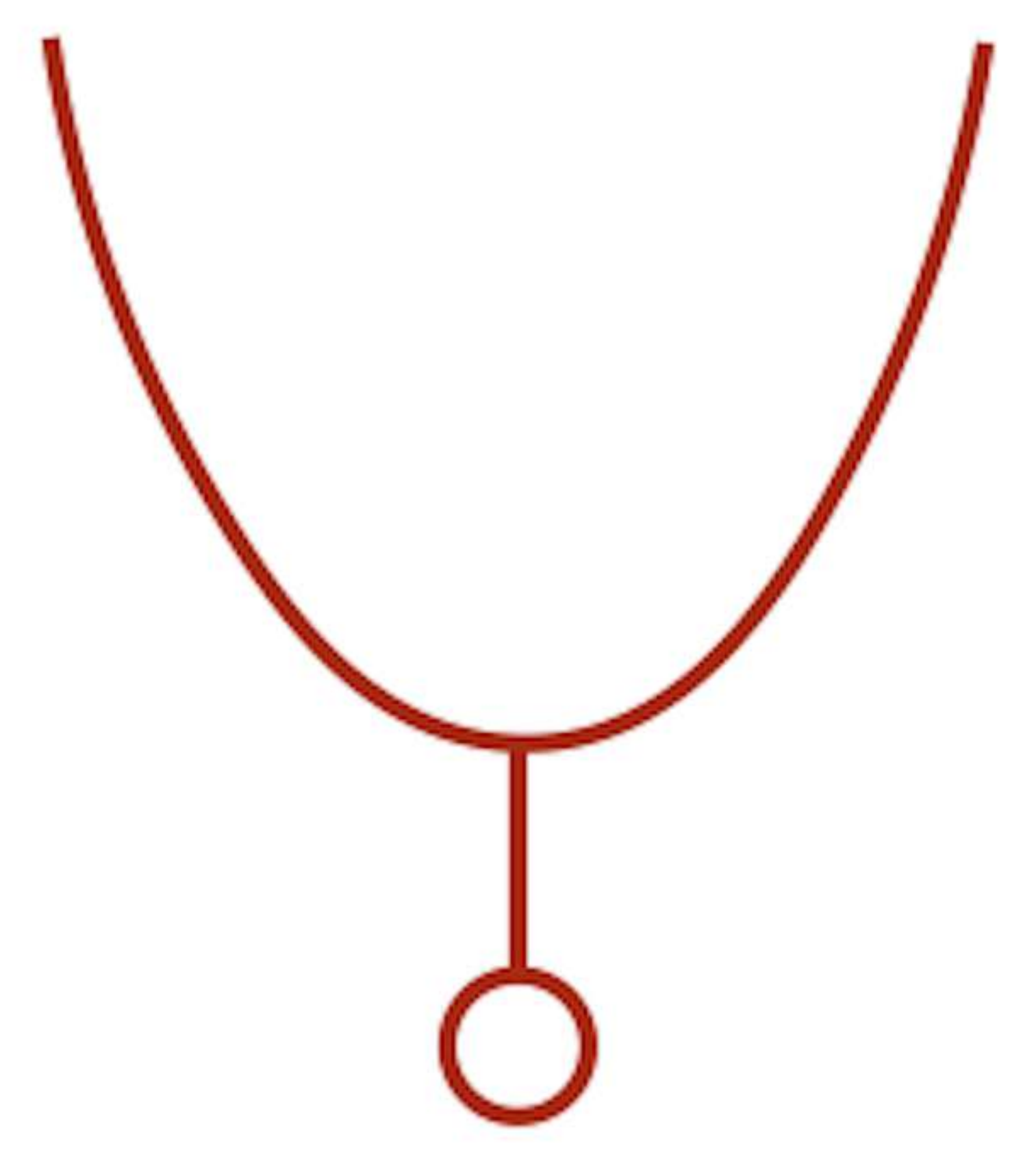} ~ = ~ 
\adjincludegraphics[valign = c, width = 1cm]{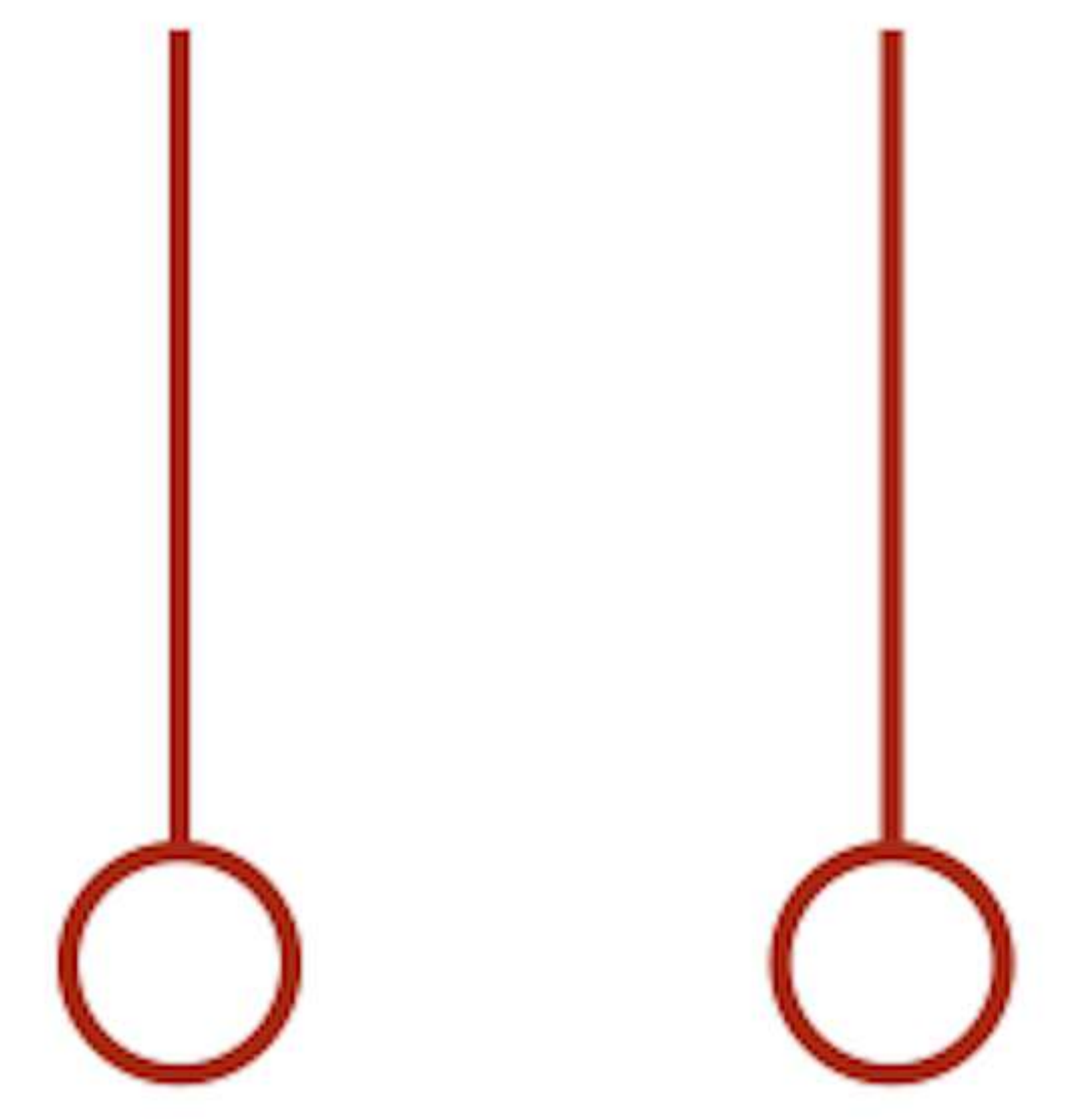}.
\end{equation}
\item
\begin{equation}
\adjincludegraphics[valign = c, width = 1.2cm]{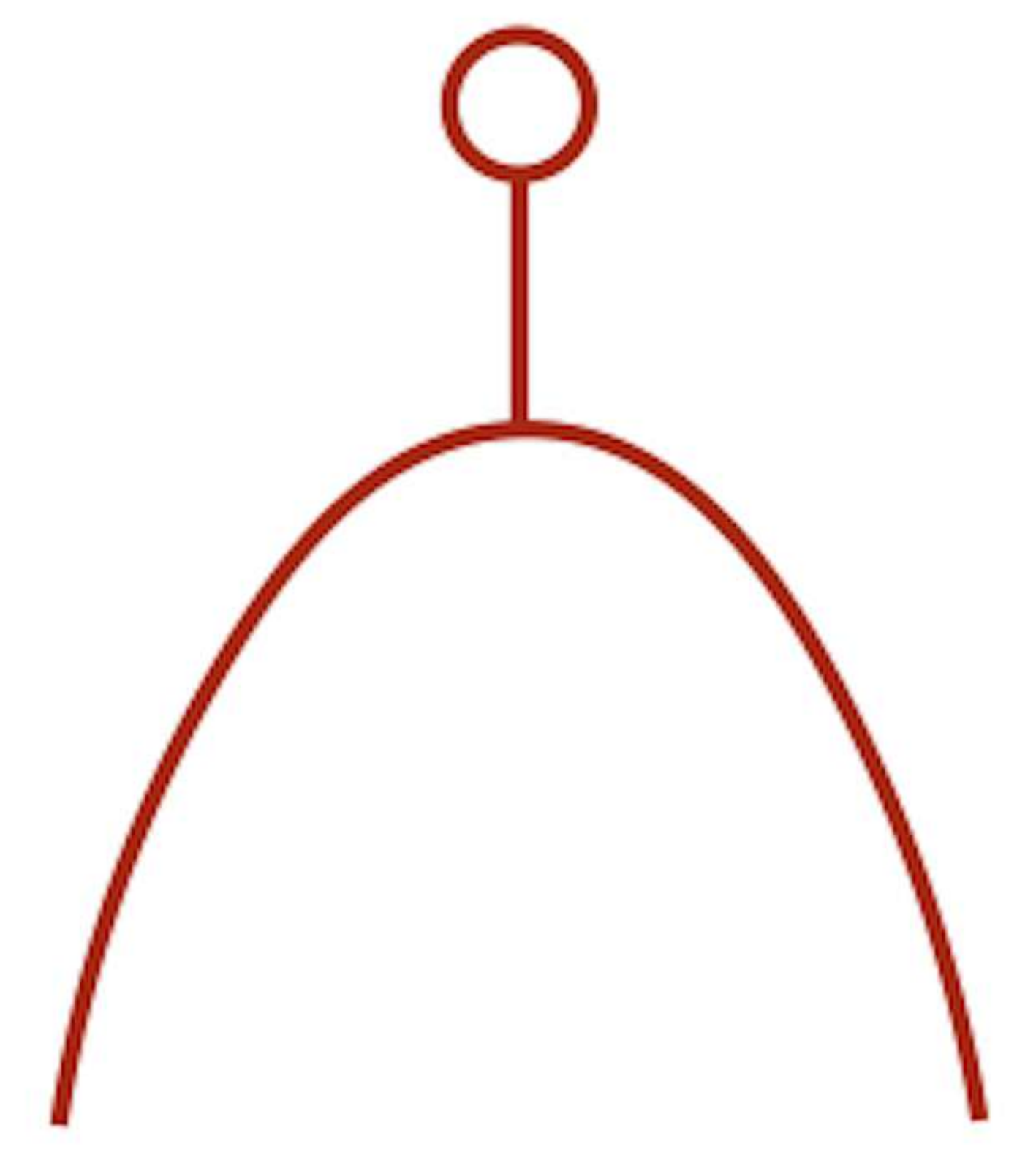} ~ = ~ 
\adjincludegraphics[valign = c, width = 1cm]{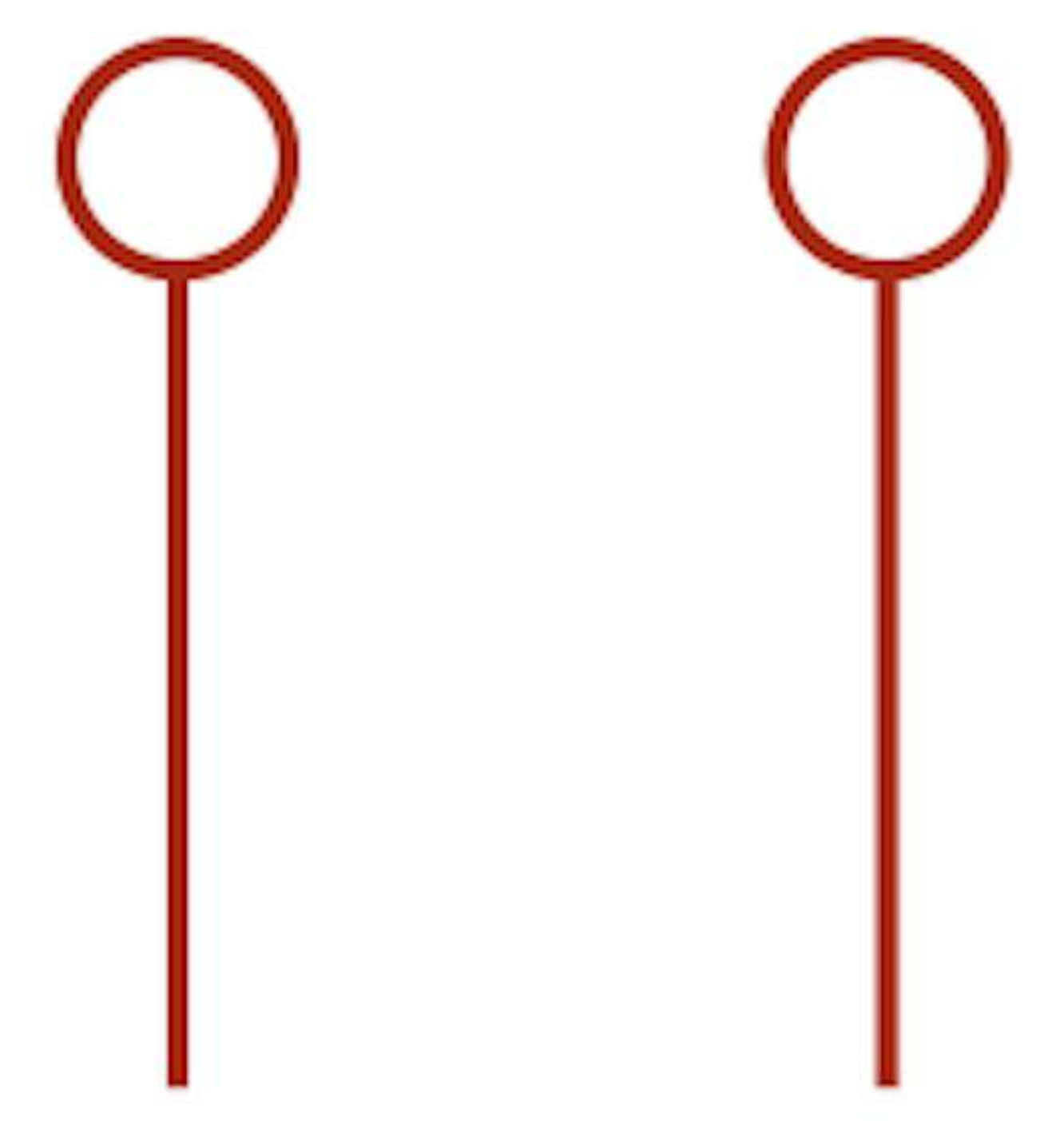}, \quad
\adjincludegraphics[valign = c, width = 0.3cm]{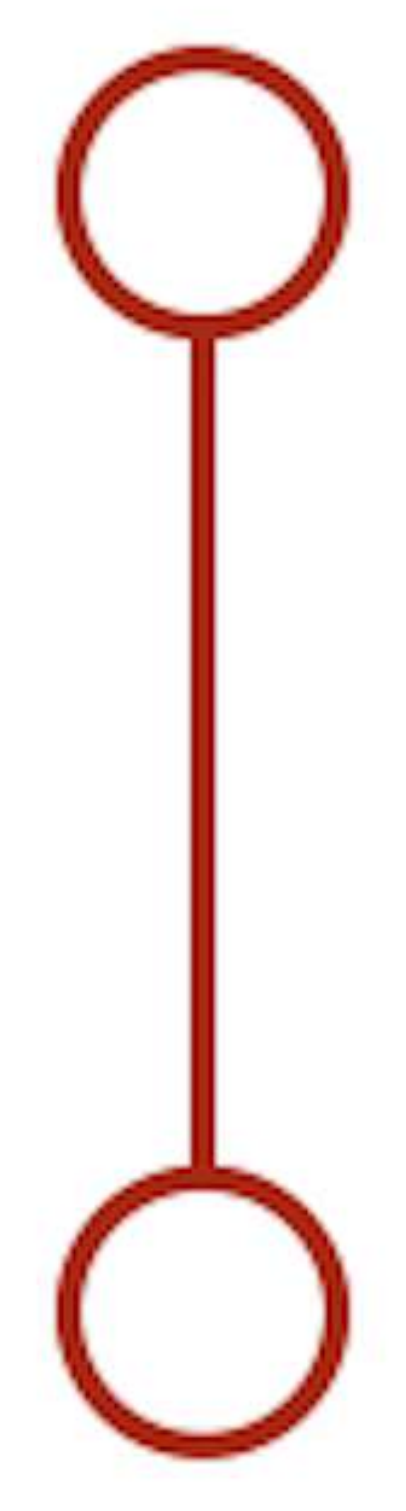} ~ = 1.
\end{equation}
\item
\begin{equation}
\adjincludegraphics[valign = c, width = 1.2cm]{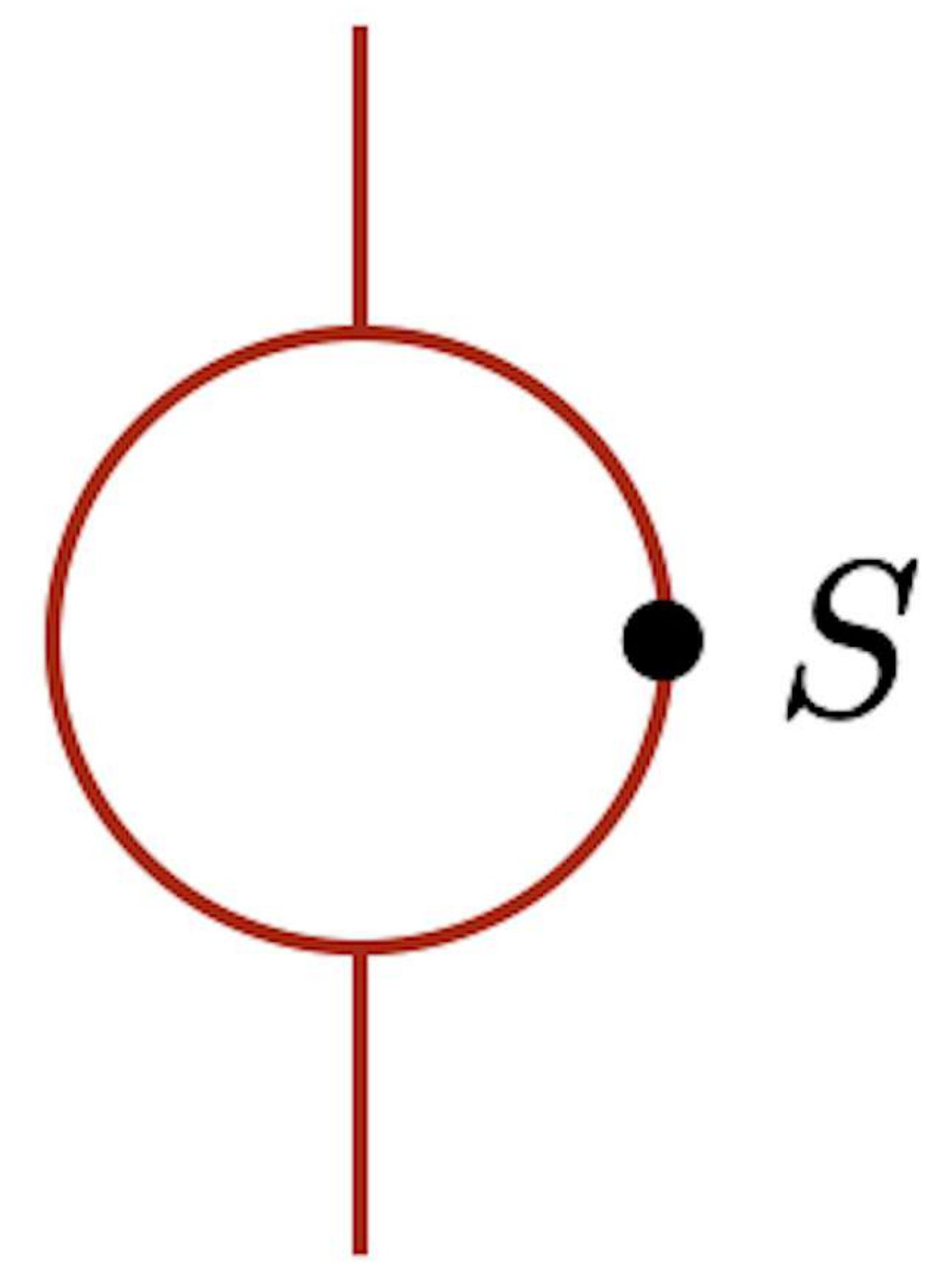} ~ = ~ 
\adjincludegraphics[valign = c, width = 1.2cm]{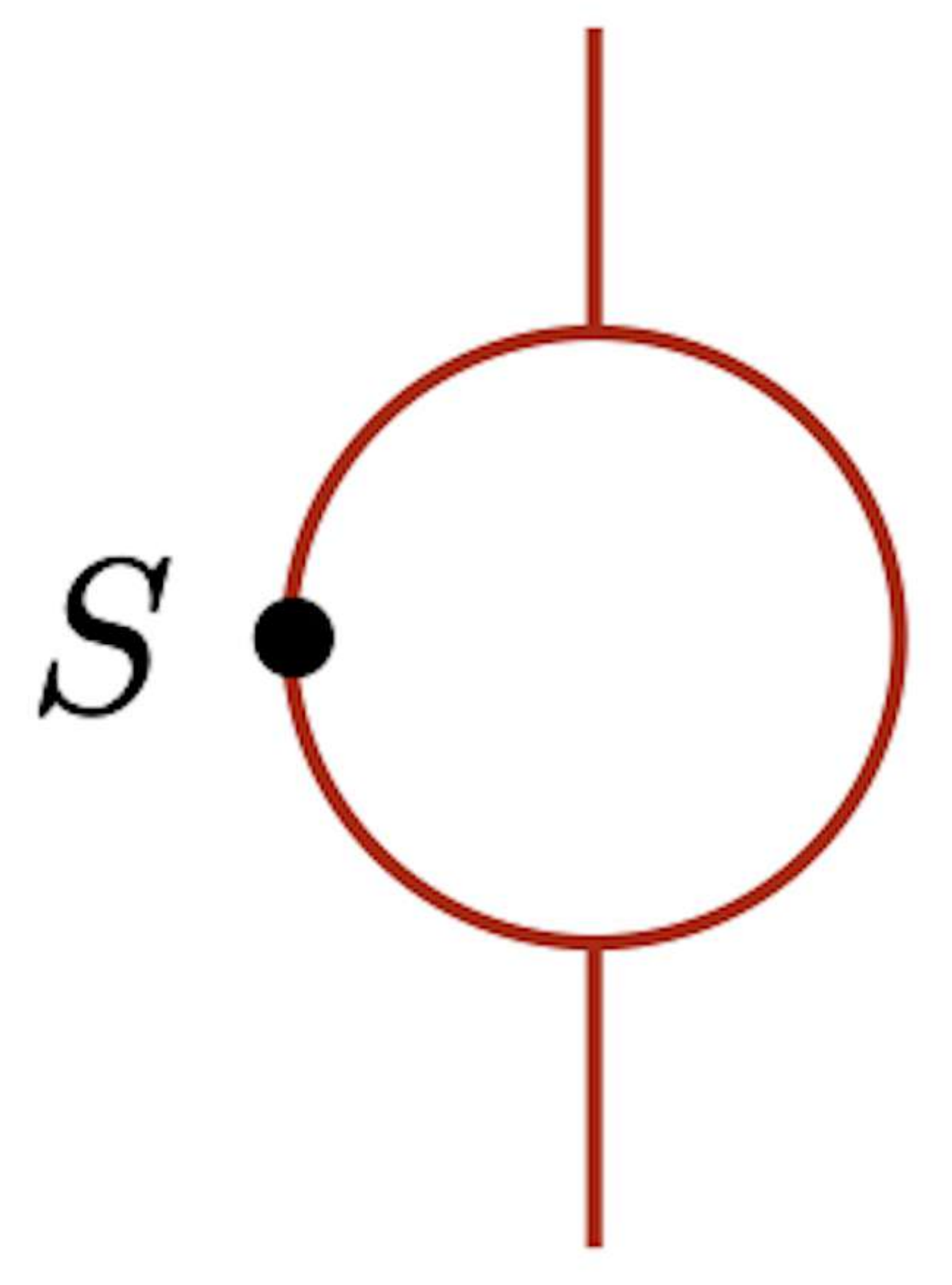} ~ = ~
\adjincludegraphics[valign = c, width = 0.25cm]{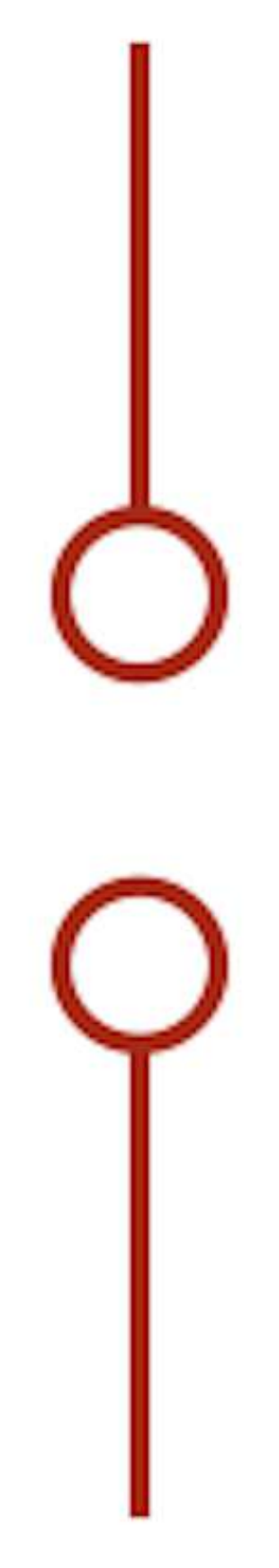}.
\end{equation}
\end{enumerate}

A left $H$-module $A$ is a vector space on which $H$ acts from the left.
The $H$-module action $\rho_A: H \otimes A \rightarrow A$ is a linear map that satisfies
\begin{equation}
\adjincludegraphics[valign = c, width = 2cm]{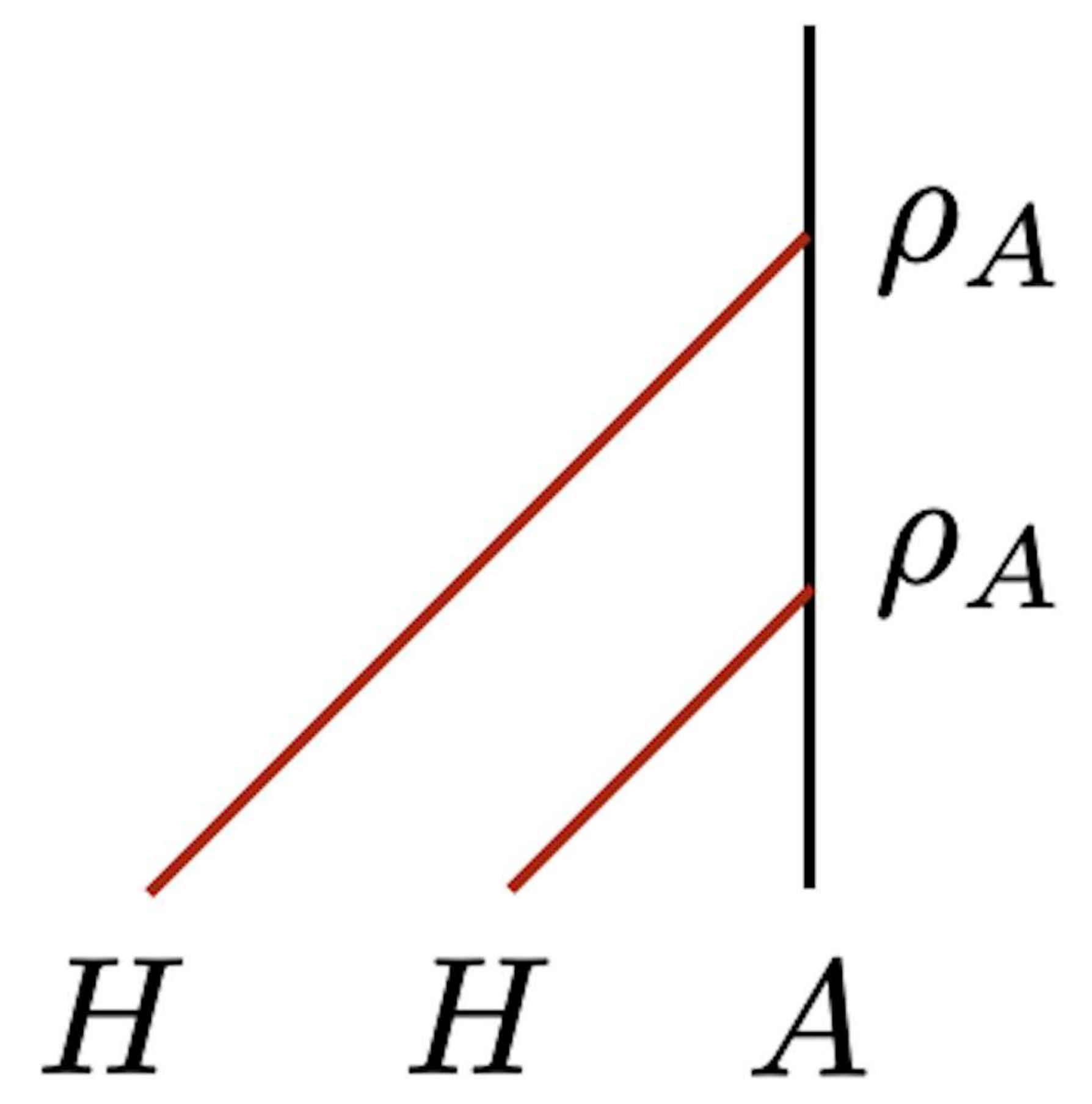} ~ = ~ 
\adjincludegraphics[valign = c, width = 1.8cm]{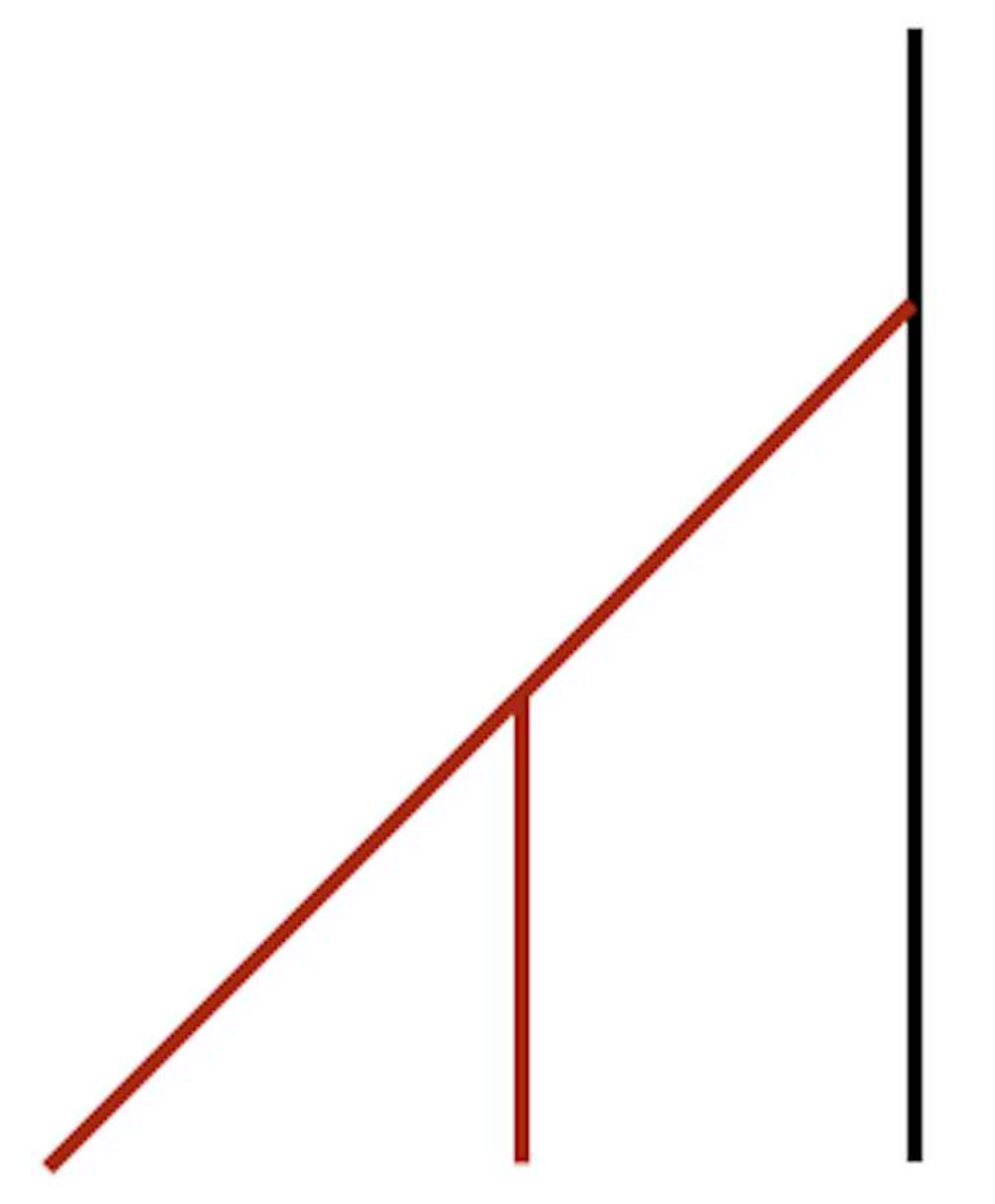}, \quad
\adjincludegraphics[valign = c, width = 1.4cm]{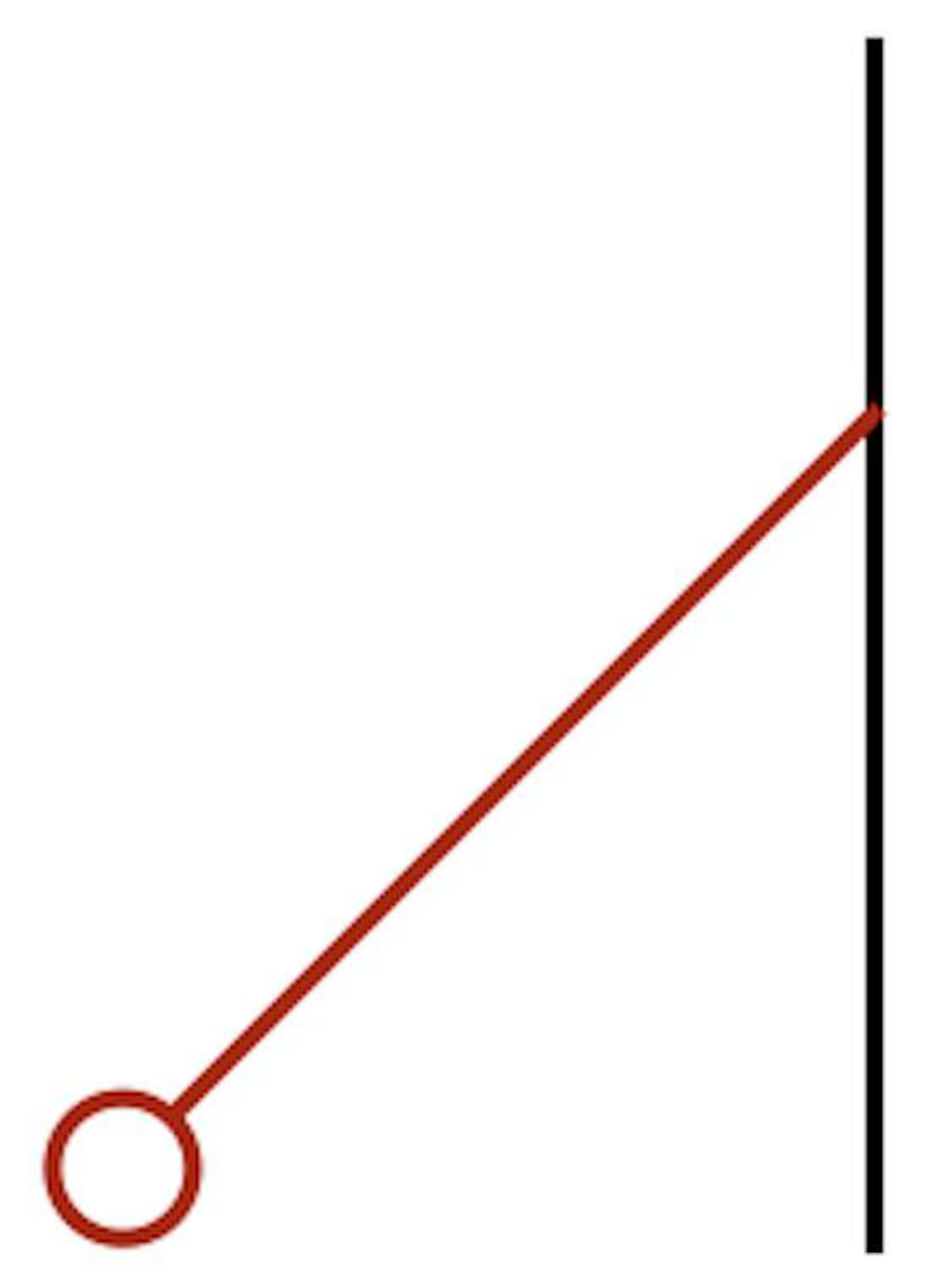} ~ = ~
\adjincludegraphics[valign = c, width = 0.28cm]{separable2.pdf}.
\end{equation}
When a left $H$-module $A$ has an algebra structure that is compatible with the $H$-module structure, $A$ is called a left $H$-module algebra.
More precisely, a left $H$-module $A$ with a module action $\rho_A: H \otimes A \rightarrow A$ is a left $H$-module algebra if $(A, m_A, \eta_A)$ is a unital associative algebra such that
\begin{equation}
\adjincludegraphics[valign = c, width = 2.2cm]{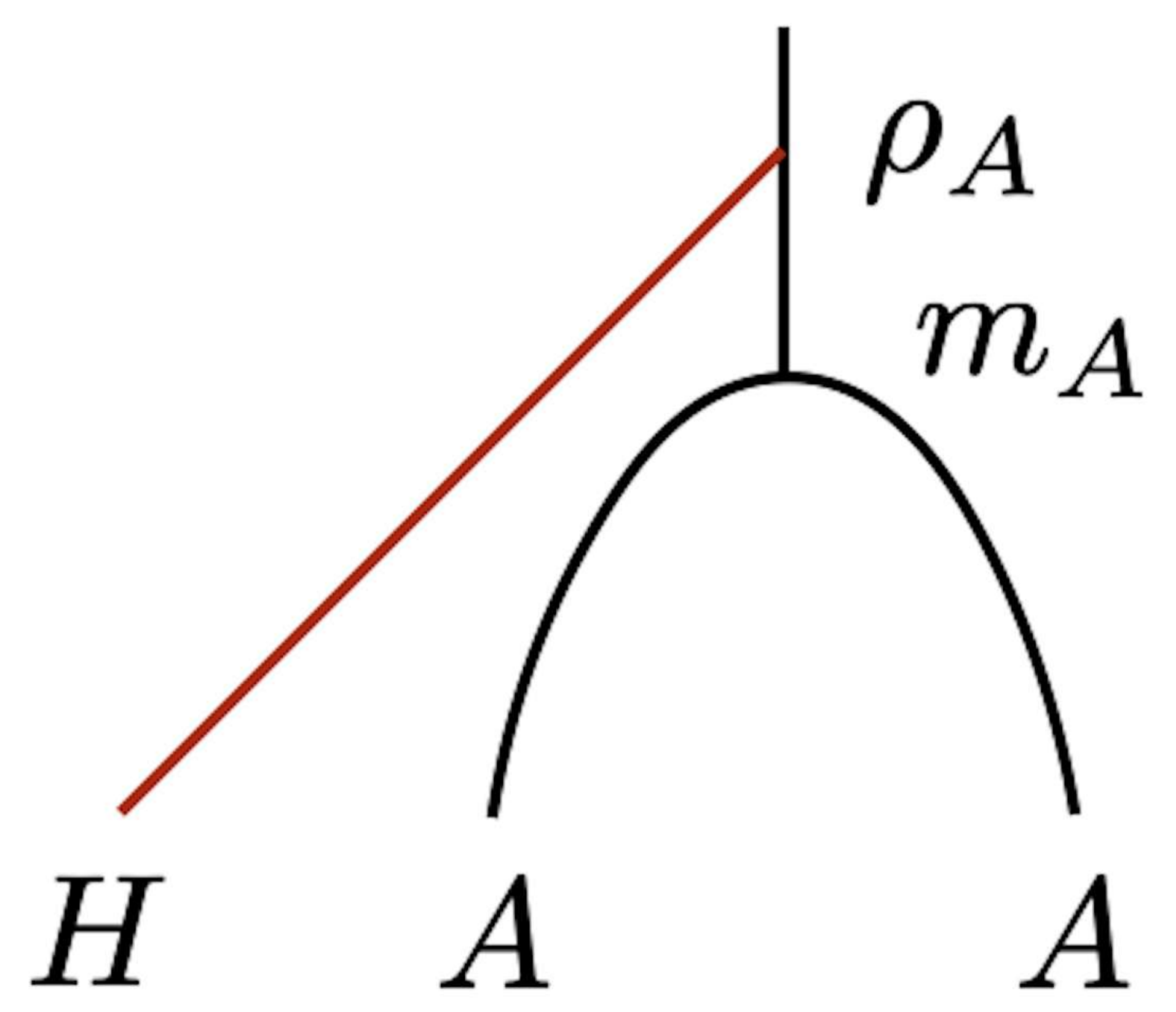} ~ = ~ 
\adjincludegraphics[valign = c, width = 1.8cm]{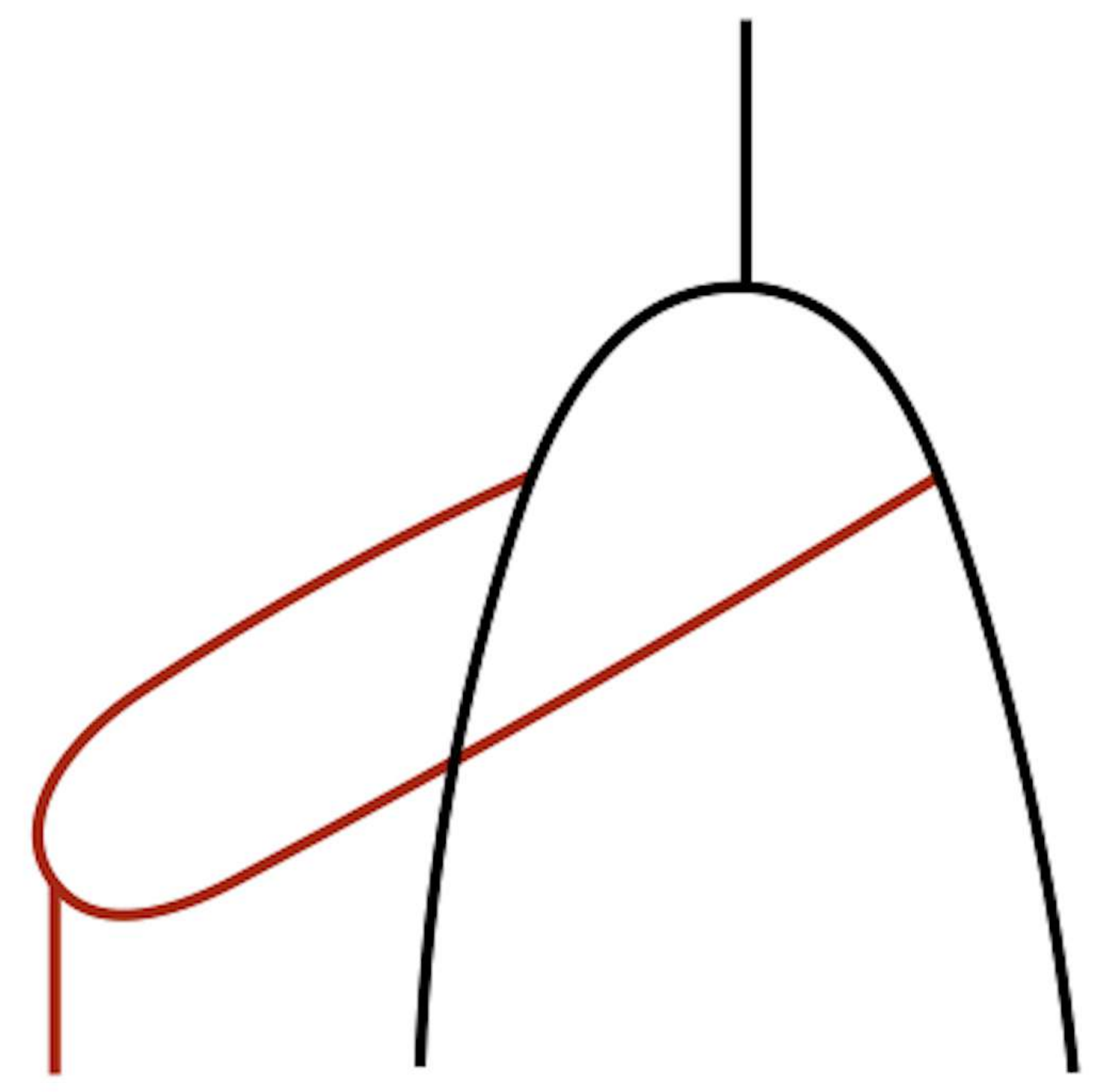}, \quad
\adjincludegraphics[valign = c, width = 1.5cm]{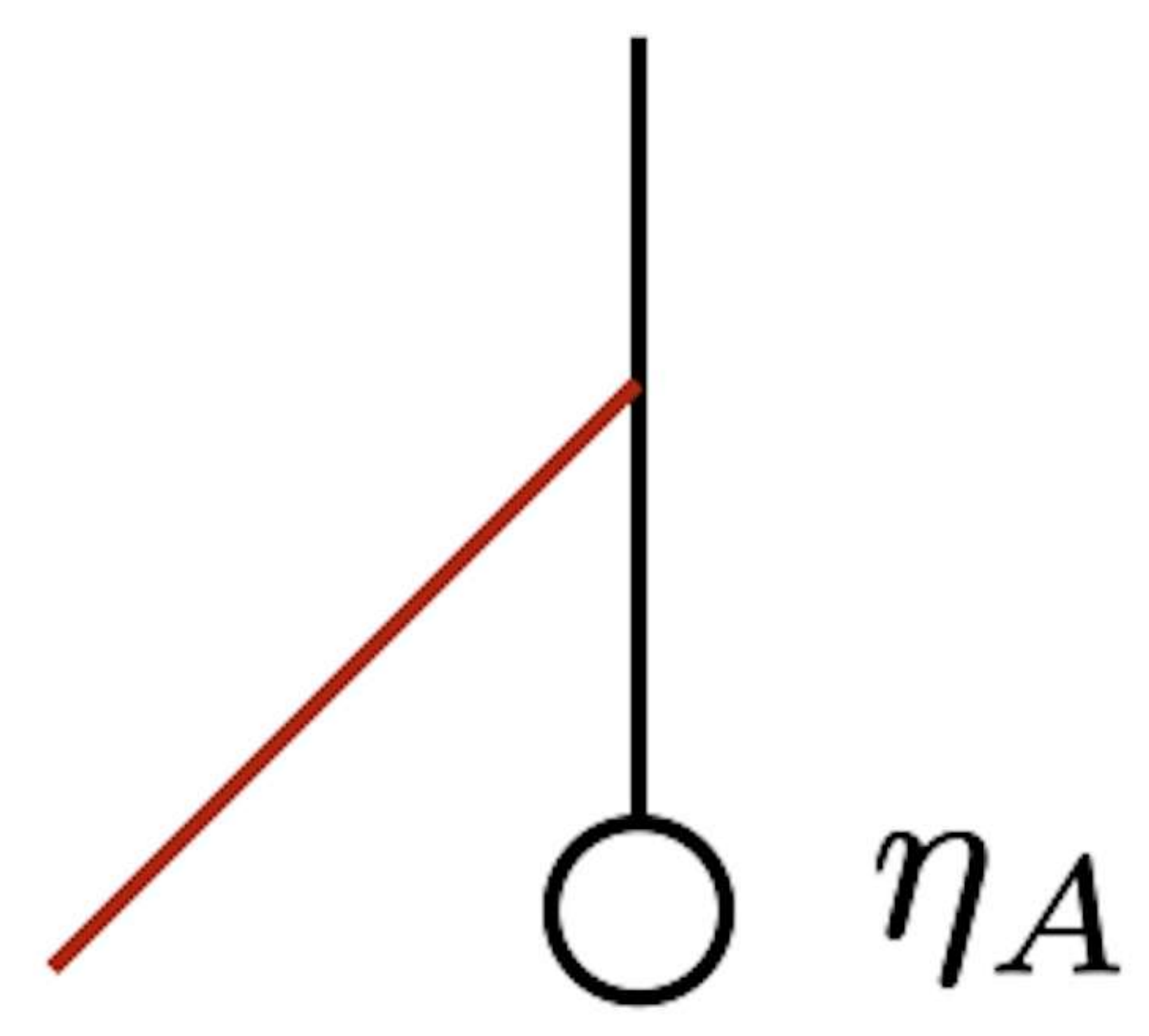} ~ = ~ 
\adjincludegraphics[valign = c, width = 1cm]{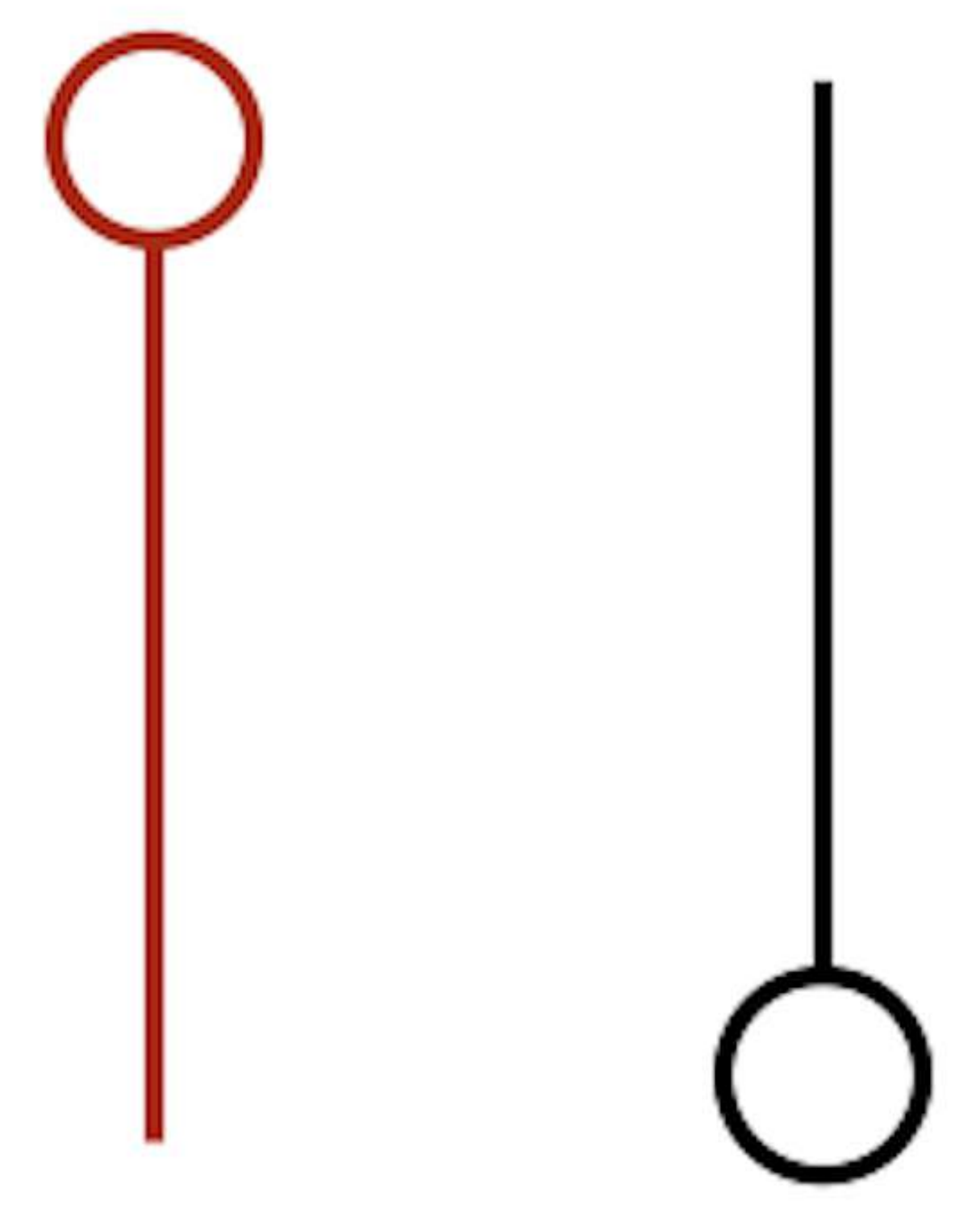}.
\end{equation}
A left $H$-module algebra $A$ is said to be $H$-simple if $A$ does not have any proper non-zero ideal $I$ such that $\rho_A(H \otimes I) \subset I$.

We can also define a left $H$-comodule algebra similarly. 
A left $H$-comodule algebra $K$ is a unital associative algebra whose algebra structure $(K, m_K, \eta_K)$ is compatible with the $H$-comodule action $\lambda_K: K \rightarrow H \otimes K$ in the following sense:
\begin{equation}
\adjincludegraphics[valign = c, width = 1.8cm]{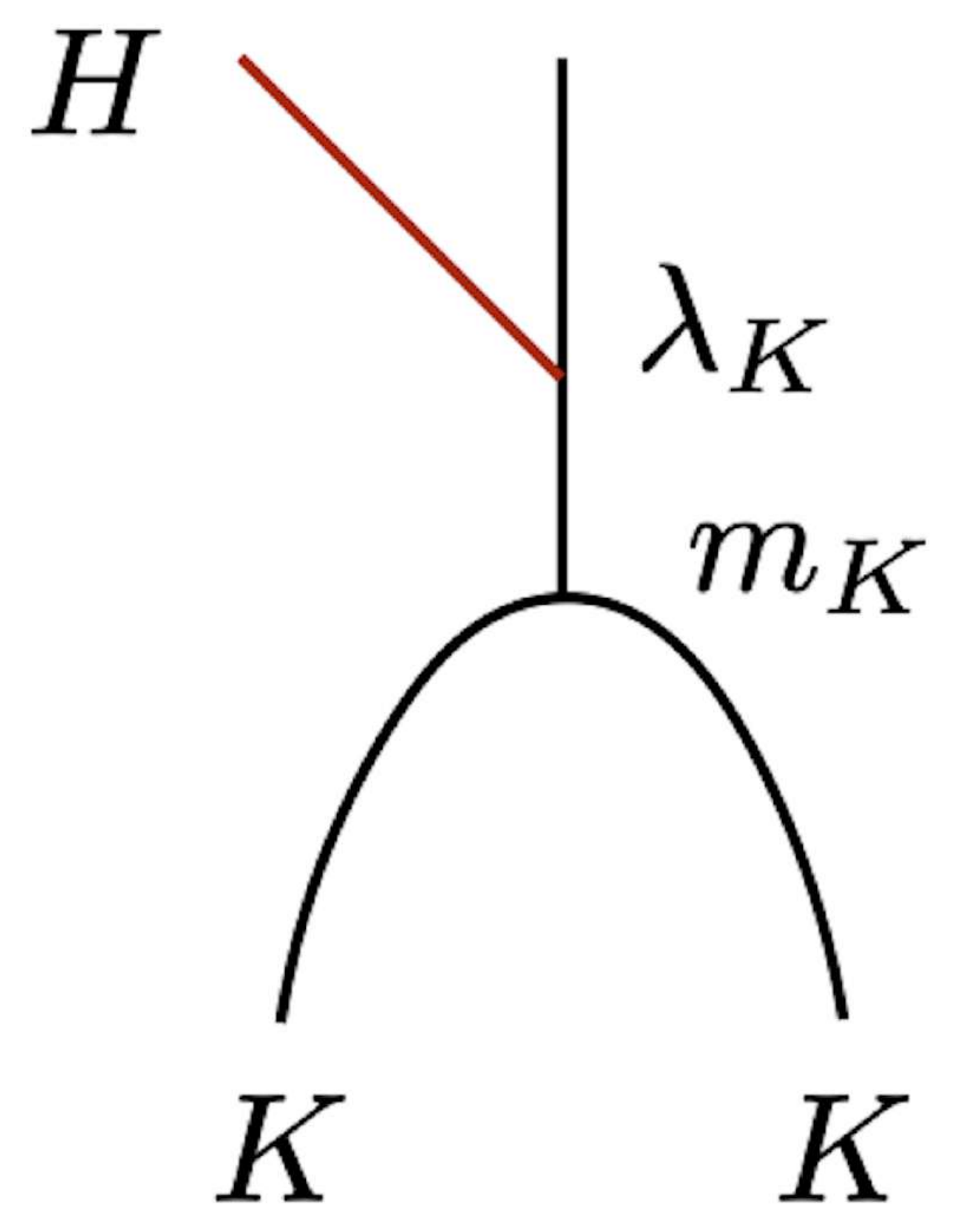} ~ = ~ 
\adjincludegraphics[valign = c, width = 1.8cm]{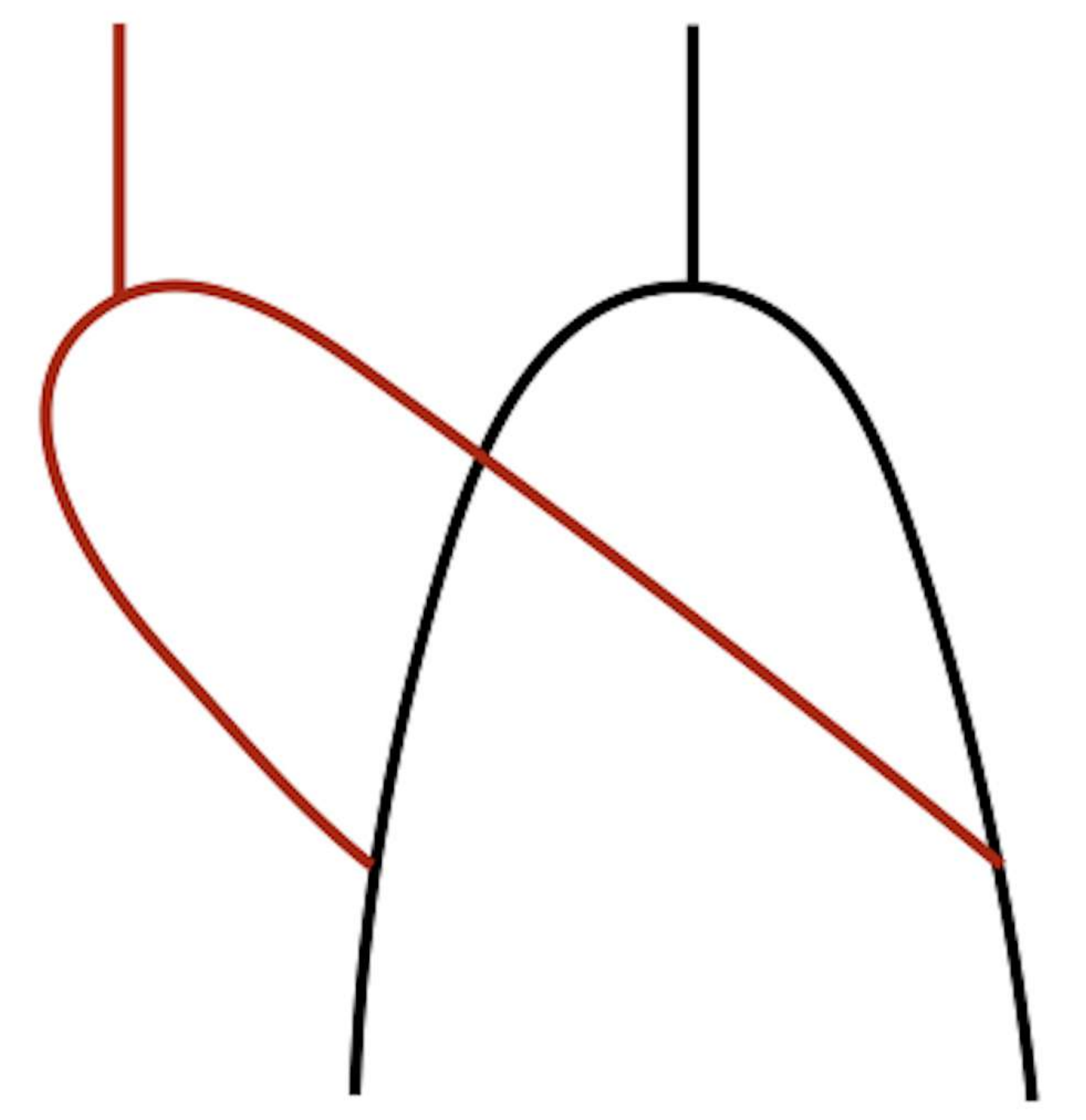}, \quad
\adjincludegraphics[valign = c, width = 1.5cm]{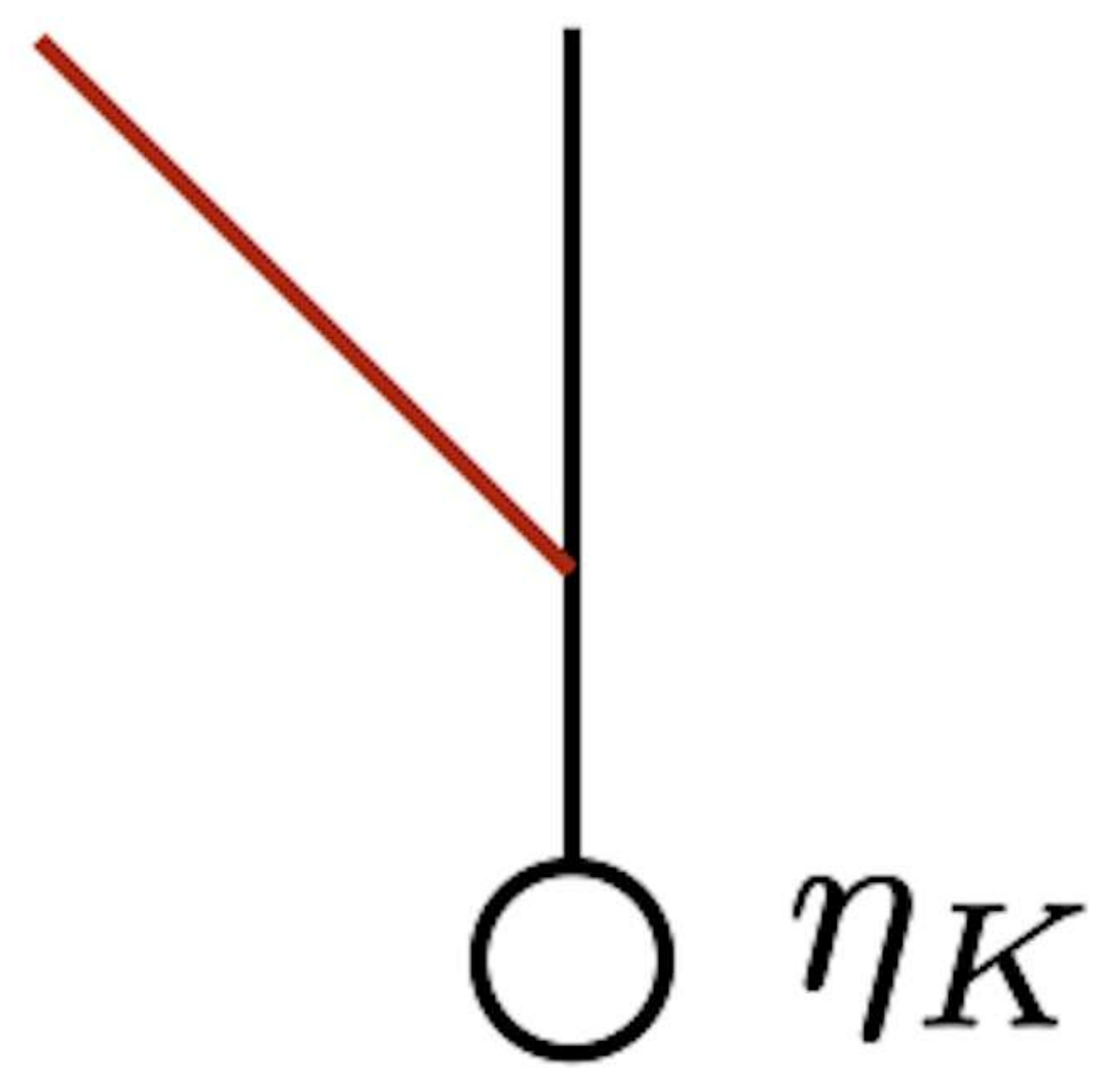} ~ = ~ 
\adjincludegraphics[valign = c, width = 1.2cm]{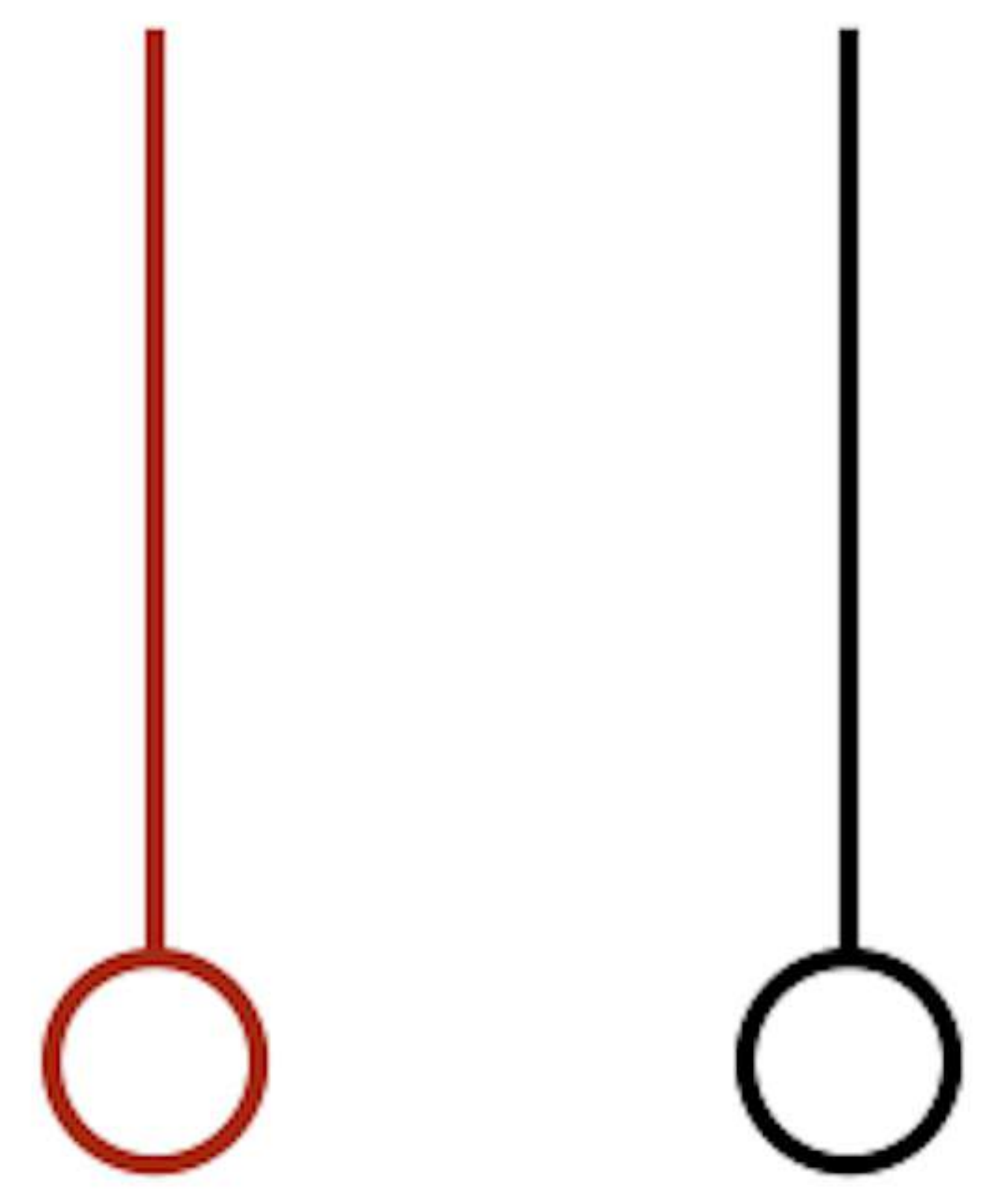}.
\end{equation}
A left $H$-comodule algebra $K$ is said to be $H$-simple if $K$ does not have any proper non-zero ideal $I$ such that $\lambda_K (I) \subset H \otimes I$.
In particular, an $H$-simple left $H$-comodule algebra $K$ is semisimple \cite{Skr2007}.
The left $H$-comodule action on $K$ is said to be inner-faithful if there is no proper Hopf subalgebra $H^{\prime} \subset H$ such that $\lambda_K (K) \subset H^{\prime} \otimes K$ \cite{BB2010}.

Given a left $H$-module algebra $A$, we can construct a left $H^{\mathrm{cop}}$-comodule algebra $A \# H$ called the smash product of $A$ and $H$.
As a vector space, $A \# H$ is the same as the tensor product $A \otimes H$.
The left $H^{\mathrm{cop}}$-comodule action on $A \# H$ is defined via the coopposite comultiplication $\Delta^{\mathrm{cop}}$ as
\begin{equation}
\adjincludegraphics[valign = c, width = 2.4cm]{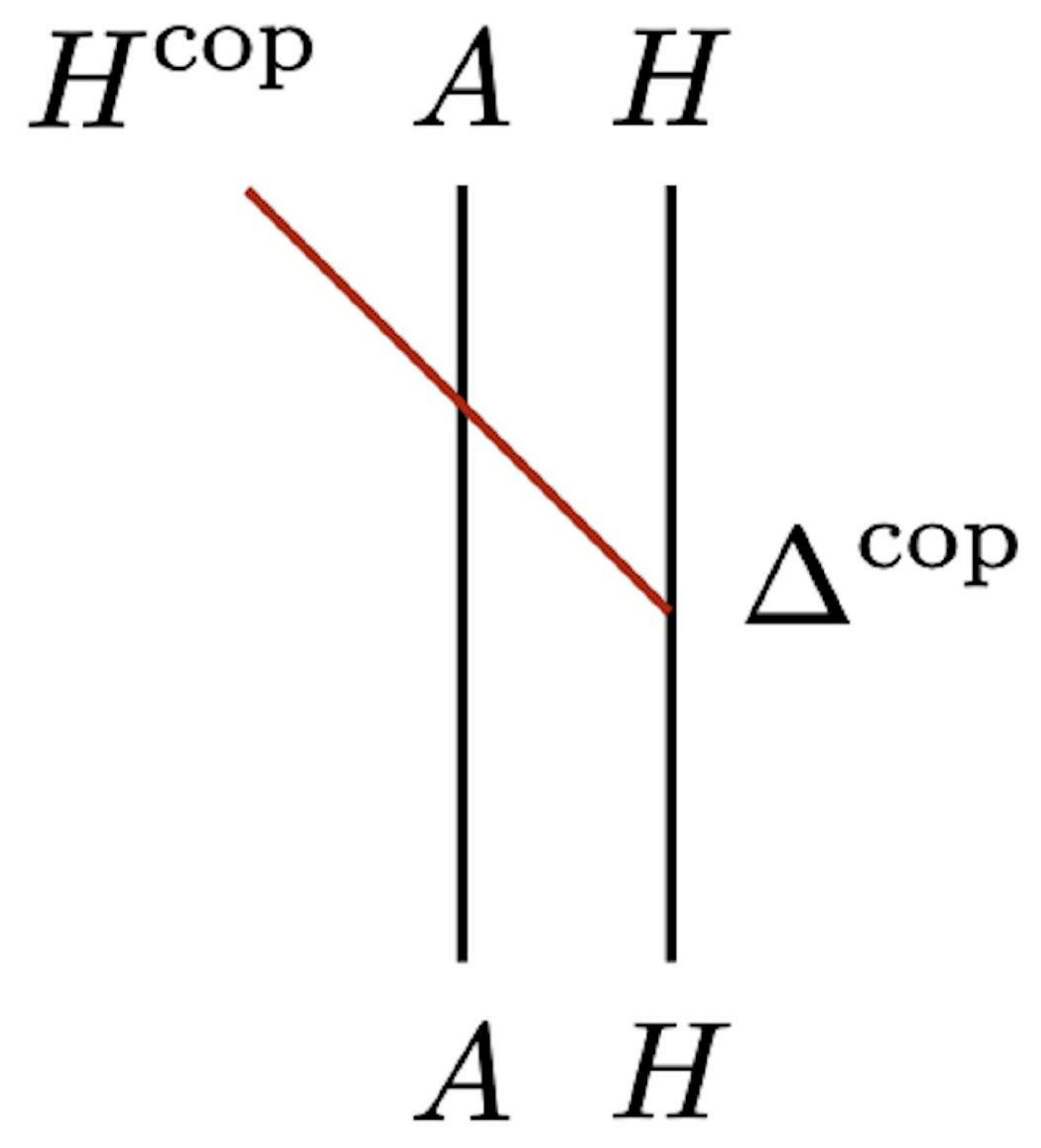}.
\end{equation}
The algebra structure on $A \# H$ is given by
\begin{equation}
(a \# h) \cdot (a^{\prime} \# h^{\prime}) = a(h_{(1)} \cdot a^{\prime}) \# h_{(2)} h^{\prime}, \quad \forall a, a^{\prime} \in A, \forall h, h^{\prime} \in H.
\label{eq: smash multiplication}
\end{equation}

\subsection{Representation categories of Hopf algebras}
\label{sec: Representation categories and module categories}
Every non-anomalous fusion category symmetry is equivalent to the representation category of a Hopf algebra.\footnote{We recall that fusion category symmetries are said to be non-anomalous if and only if they admit SPT phases, i.e. gapped phases with unique ground states \cite{TW2019}. }
In this subsection, we describe the representation category of a Hopf algebra and module categories over it following \cite{AM2007}. 

The representation category $\mathrm{Rep}(H)$ of a Hopf algebra $H$ is a category whose objects are left $H$-modules and whose morphisms are left $H$-module maps.
The tensor product $V \otimes W$ of left $H$-modules $V$ and $W$ is given by the usual tensor product over $\mathbb{C}$.
The left $H$-module structure on the tensor product $V \otimes W$ is defined via the comultiplication $\Delta$.
Specifically, if we denote the left $H$-module action on $V \in \mathrm{Rep}(H)$ as $\rho_V: H \otimes V \rightarrow V$, we have
\begin{equation}
\rho_{V \otimes W} := \adjincludegraphics[valign = c, width = 2cm]{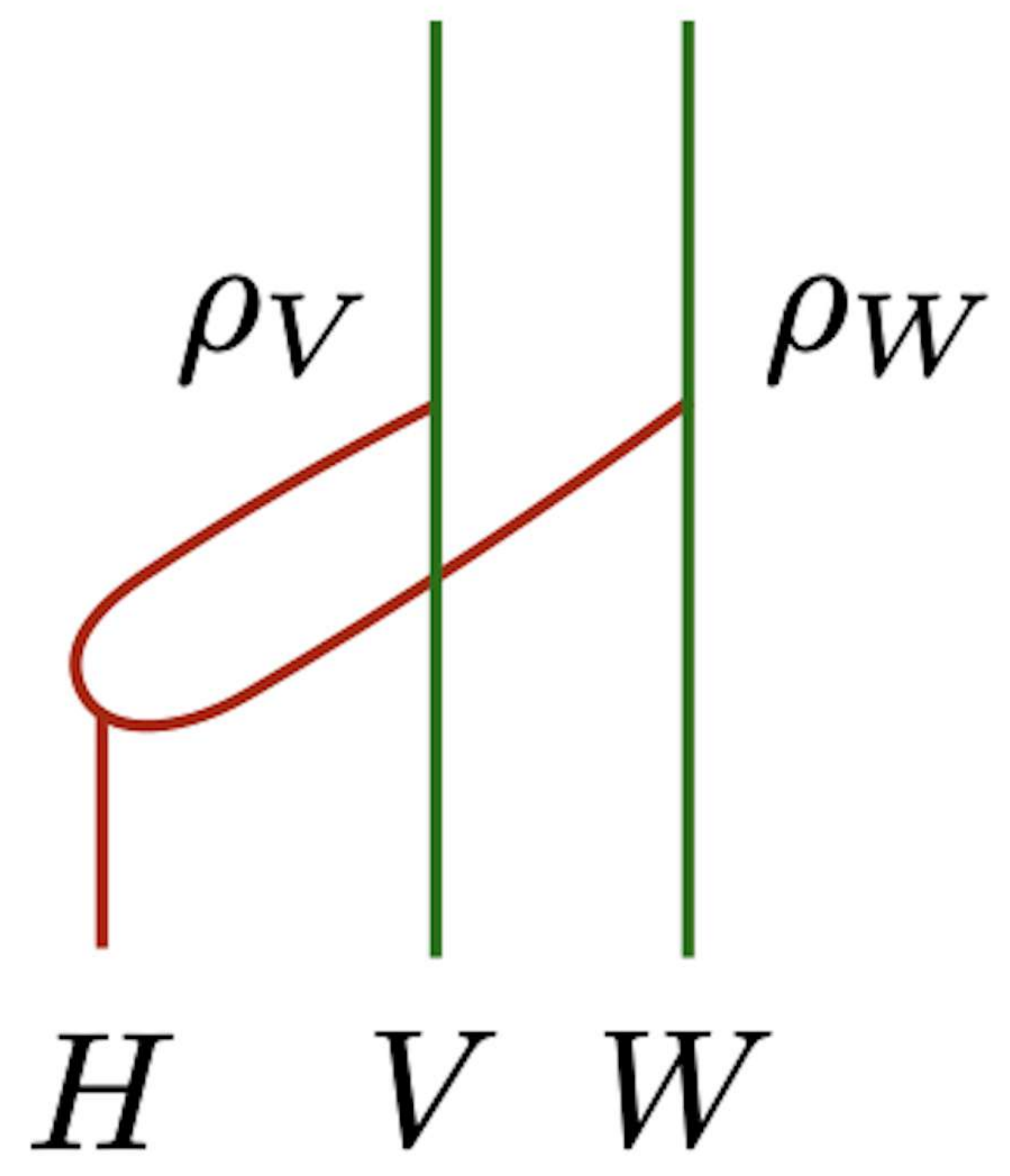}.
\end{equation}
When $V \in \mathrm{Rep}(H)$ is a left $H$-module, the dual vector space $V^*$ is also a left $H$-module with the left $H$-module action given by
\begin{equation}
\adjincludegraphics[valign = c, width = 2.4cm]{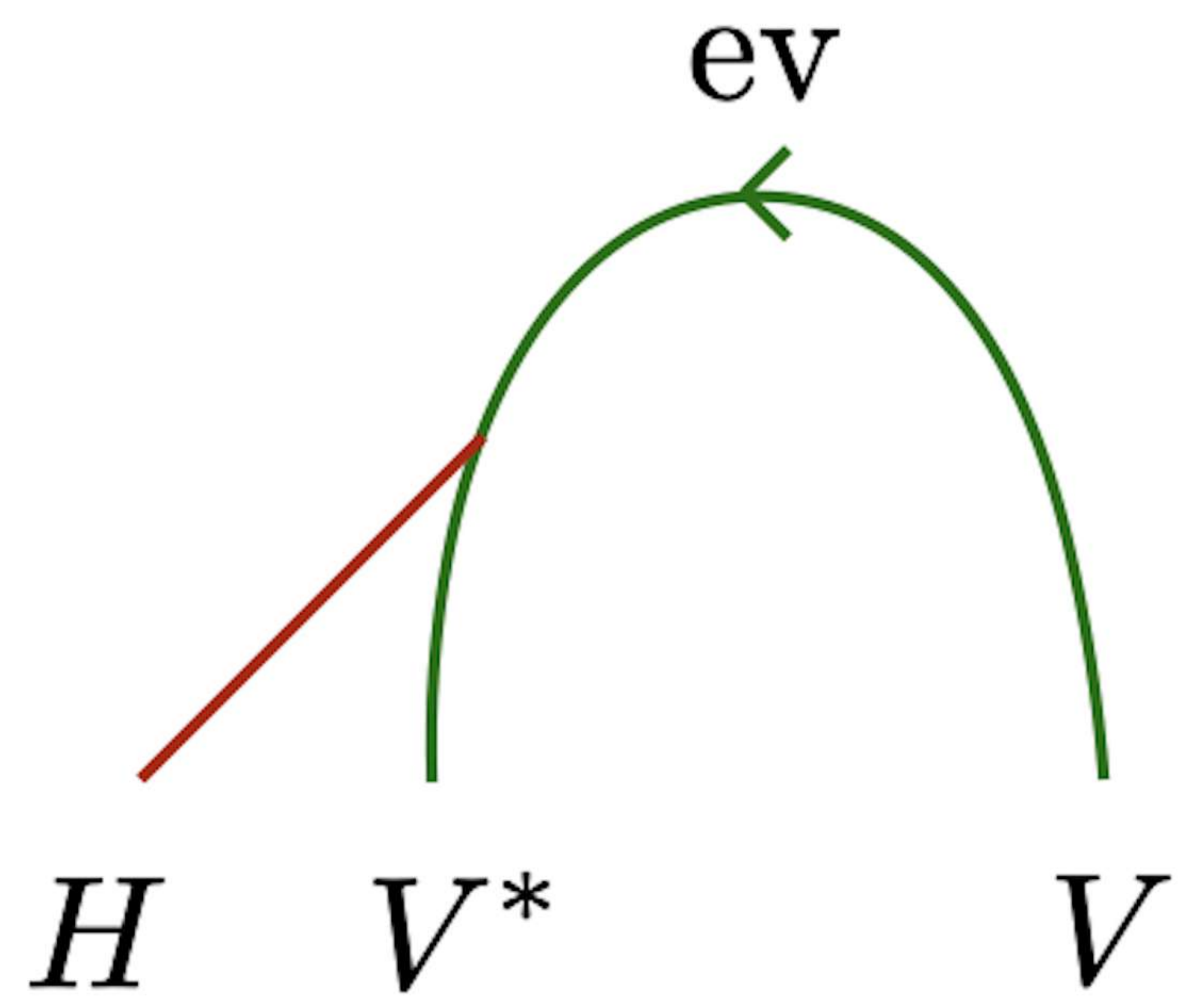} ~ := ~ 
\adjincludegraphics[valign = c, width = 2.4cm]{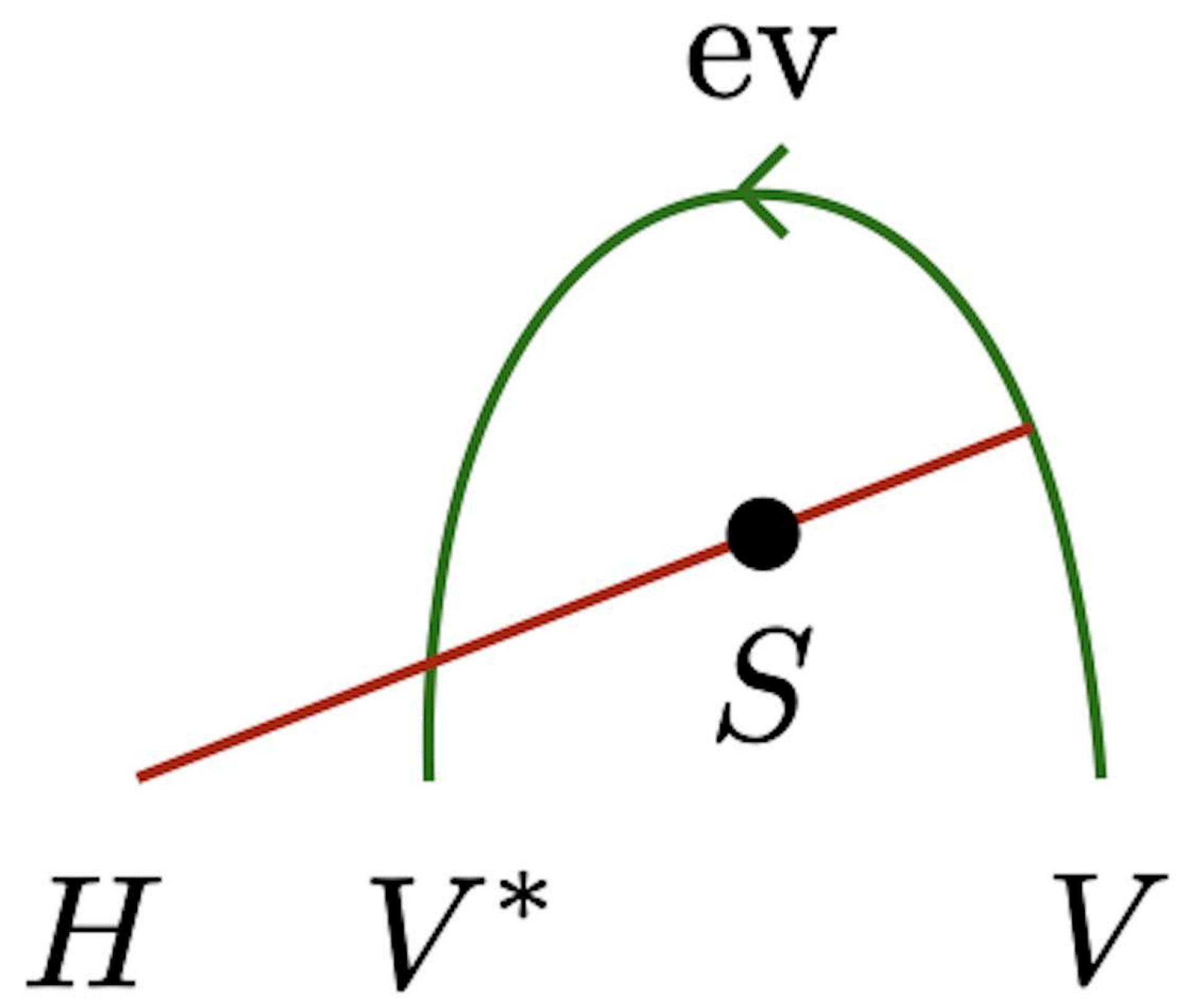},
\label{eq: dual rep}
\end{equation}
where $\mathrm{ev}: V^* \otimes V \rightarrow \mathbb{C}$ is a usual paring defined by $\mathrm{ev}(\varphi \otimes v) := \varphi (v)$ for $\varphi \in V^*, v \in V$.

An indecomposable semisimple module category over $\mathrm{Rep}(H)$ is equivalent to the category of right $A$-modules in $\mathrm{Rep}(H)$ where $A$ is an $H$-simple left $H$-module algebra \cite{EO2004, Ostrik2003}.
We denote this module category as $({}_H \mathcal{M})_A$.
As a module category over $\mathrm{Rep}(H)$, the category $({}_H \mathcal{M})_A$ is equivalent to the category of left $A^{\mathrm{op}} \# H^{\mathrm{cop}}$-modules \cite{AM2007}, which we denote by ${}_{A^{\mathrm{op}} \# H^{\mathrm{cop}}} \mathcal{M}$:
\begin{equation}
({}_H \mathcal{M})_A \cong {}_{A^{\mathrm{op}} \# H^{\mathrm{cop}}} \mathcal{M}.
\end{equation}
We note that $A^{\mathrm{op}}$ is a left $H^{\mathrm{cop}}$-module algebra when $A$ is a left $H$-module algebra, and hence $A^{\mathrm{op}} \# H^{\mathrm{cop}}$ is a left $H$-comodule algebra.
Moreover, when $A$ is $H$-simple as a module algebra, $A^{\mathrm{op}} \# H^{\mathrm{cop}}$ is also $H$-simple as a comodule algebra.
Therefore, every indecomposable semisimple module category over $\mathrm{Rep}(H)$ is equivalent to the category of left modules over an $H$-simple left $H$-comodule algebra $A^{\mathrm{op}} \# H^{\mathrm{cop}}$.
Conversely, for any $H$-simple left $H$-comodule algebra $K$, the category ${}_K \mathcal{M}$ of left $K$-modules becomes an indecomposable semisimple module category over $\mathrm{Rep}(H)$ \cite{AM2007}.
The action of $\mathrm{Rep}(H)$ on ${}_K \mathcal{M}$ is given by the usual tensor product, i.e. $V \overline{\otimes} M := V \otimes M$ for $V \in \mathrm{Rep}(H)$ and $M \in {}_K \mathcal{M}$.
The left $K$-module structure on $V \overline{\otimes} M$ is defined via the $H$-comodule structure on $K$ as
\begin{equation}
\rho_{V \overline{\otimes} M} := \adjincludegraphics[valign = c, width = 2cm]{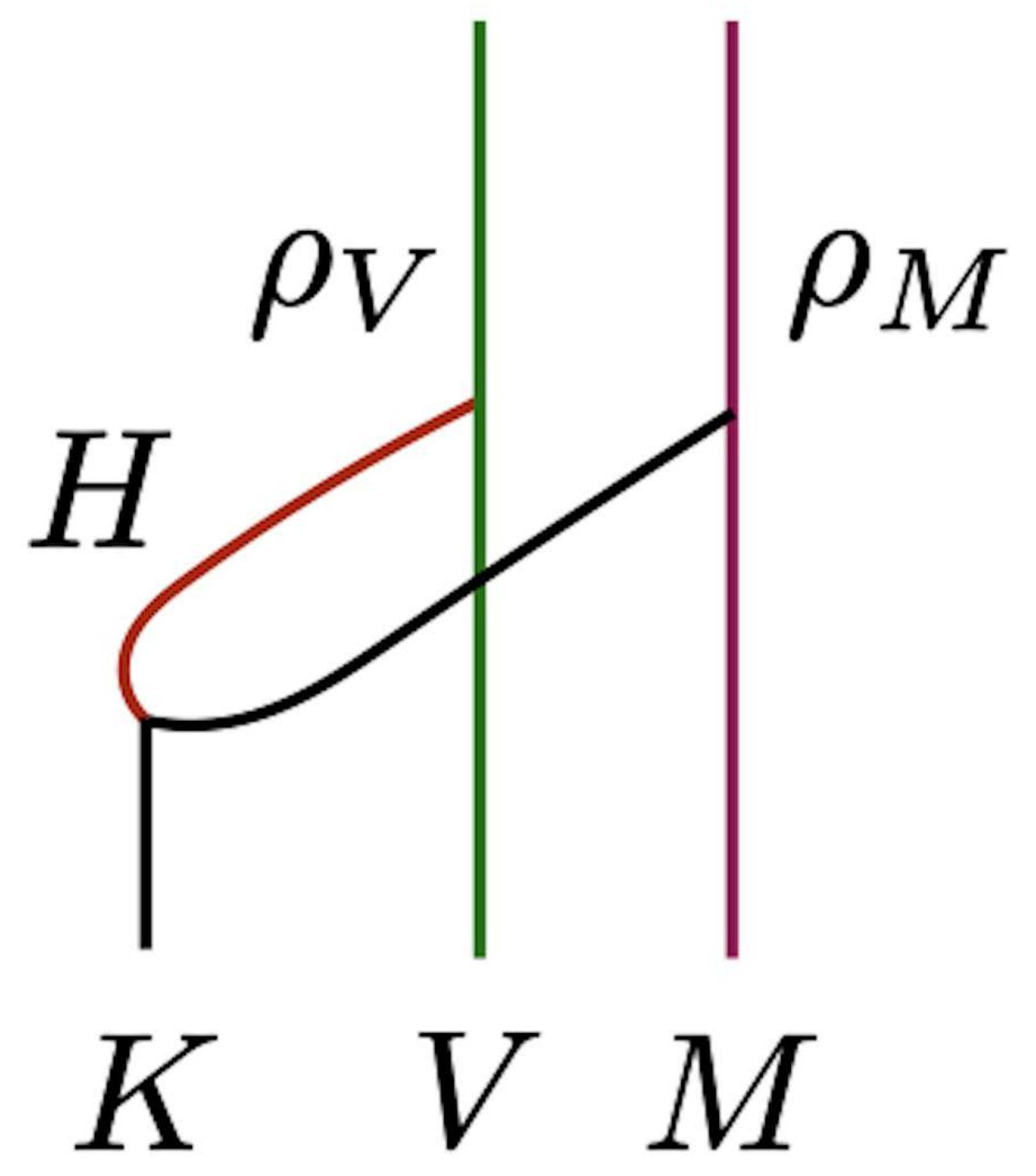},
\end{equation}
where $\rho_M: K \otimes M \rightarrow M$ is the left $K$-module action on $M$.
We note that the $\mathrm{Rep}(H)$-module category structure on ${}_K \mathcal{M}$ is represented by a tensor functor from $\mathrm{Rep}(H)$ to the category $\mathrm{End}({}_K \mathcal{M})$ of endofunctors of ${}_K \mathcal{M}$.
Since the category $\mathrm{End}({}_K \mathcal{M})$ is equivalent to the category of $K$-$K$ bimodules, we have a tensor functor $F_K: \mathrm{Rep}(H) \rightarrow {}_K \mathcal{M}_K$, which maps an object $V \in \mathrm{Rep}(H)$ to a $K$-$K$ bimodule $F_K(V) = V \otimes K$.
This tensor functor induces a $\mathrm{Rep}(H)$-module category structure on a ${}_K \mathcal{M}_K$-module category ${}_K \mathcal{M}$ via eq. \eqref{eq: pullback of mod-cat}.

\section{Pullback of fusion category TQFTs by tensor functors}
\label{sec: Pullback of fusion category TQFTs by tensor functors}
In this section, we show that given a 2d TQFT $Q^{\prime}$ with symmetry $\mathcal{C}^{\prime}$ and a tensor functor $F: \mathcal{C} \rightarrow \mathcal{C}^{\prime}$, we can construct a 2d TQFT $Q$ with symmetry $\mathcal{C}$ by pulling back the TQFT $Q^{\prime}$ by the tensor functor $F$.
In particular, we can construct any 2d TQFT with non-anomalous fusion category symmetry $\mathrm{Rep}(H)$ by pulling back a specific ${}_K \mathcal{M}_K$ symmetric TQFT by a tensor functor $F_{K}: \mathrm{Rep}(H) \rightarrow {}_K \mathcal{M}_K$.
We note that the content of this section can also be applied to anomalous fusion category symmetries as well as non-anomalous ones.

\subsection{TQFTs with fusion category symmetries}
\label{sec: TQFTs with fusion category symmetries}
\begin{figure}
 \centering
 \includegraphics[width = 2.5cm]{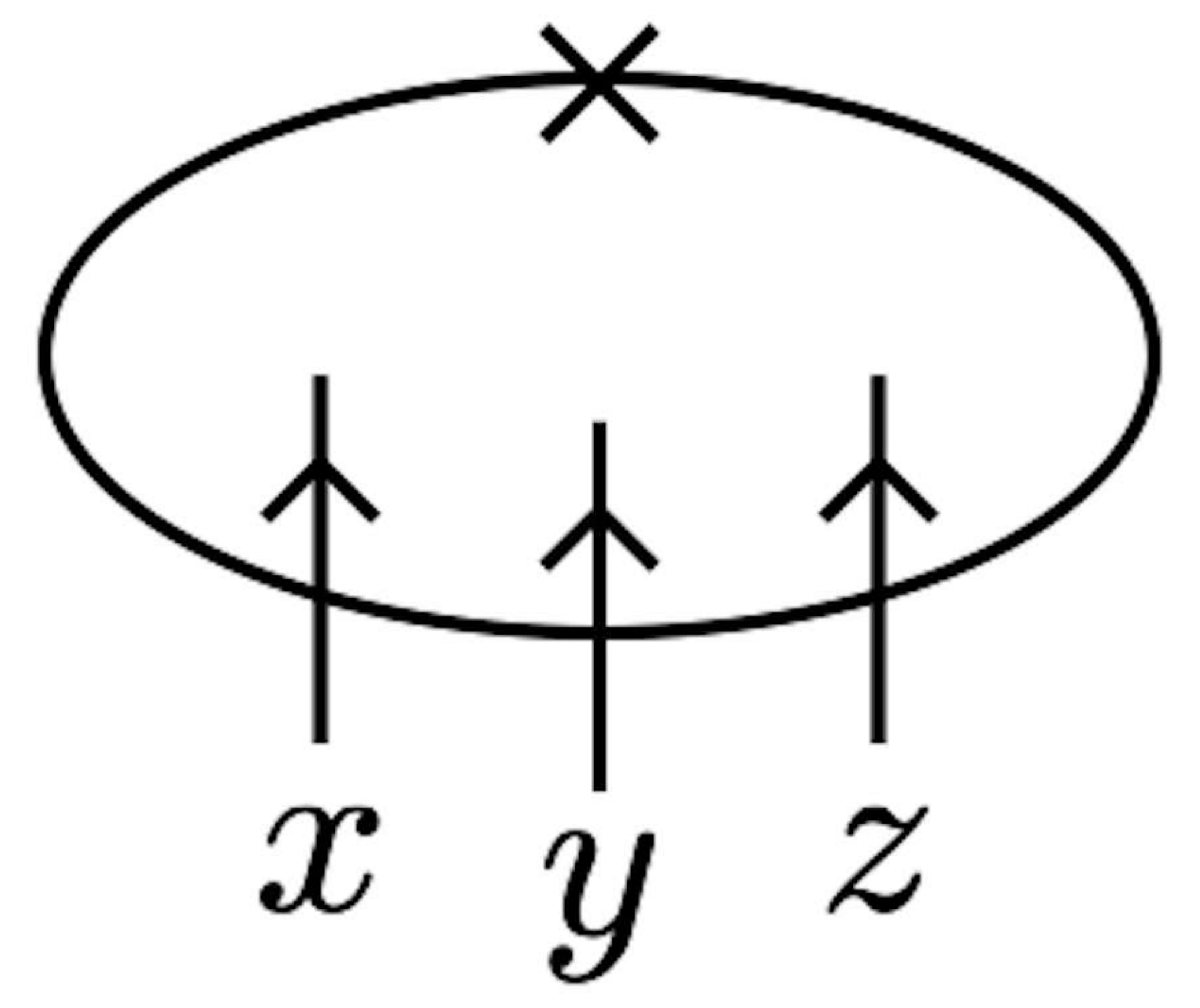}
 \caption{The Hilbert space on the above spatial circle is given by $Z((x \otimes y) \otimes z)$, where the base point is represented by the cross mark in the above figure. We can also assign a Hilbert space to a circle with an arbitrary number of topological defects in a similar way.}
 \label{fig: Hilbert space}
 \end{figure}

We first review the axiomatic formulation of 2d unitary TQFT with fusion category symmetry $\mathcal{C}$ following \cite{BT2018}.
 A 2d TQFT assigns a Hilbert space $Z(x)$ to a spatial circle that has a topological defect $x \in \mathcal{C}$ running along the time direction.
 When the spatial circle has multiple topological defects $x, y, z, \cdots$, the Hilbert space is given by $Z(((x \otimes y) \otimes z) \otimes \cdots)$, where the order of the tensor product is determined by the position of the base point on the circle, see figure \ref{fig: Hilbert space}.
 
 A 2d TQFT also assigns a linear map to a two-dimensional surface decorated by a network of topological defects. 
 The linear map assigned to an arbitrary surface is composed of the following building blocks, see also figure \ref{fig: basic elements}:
 \begin{figure}
 \centering
 \includegraphics[width = 1.8cm]{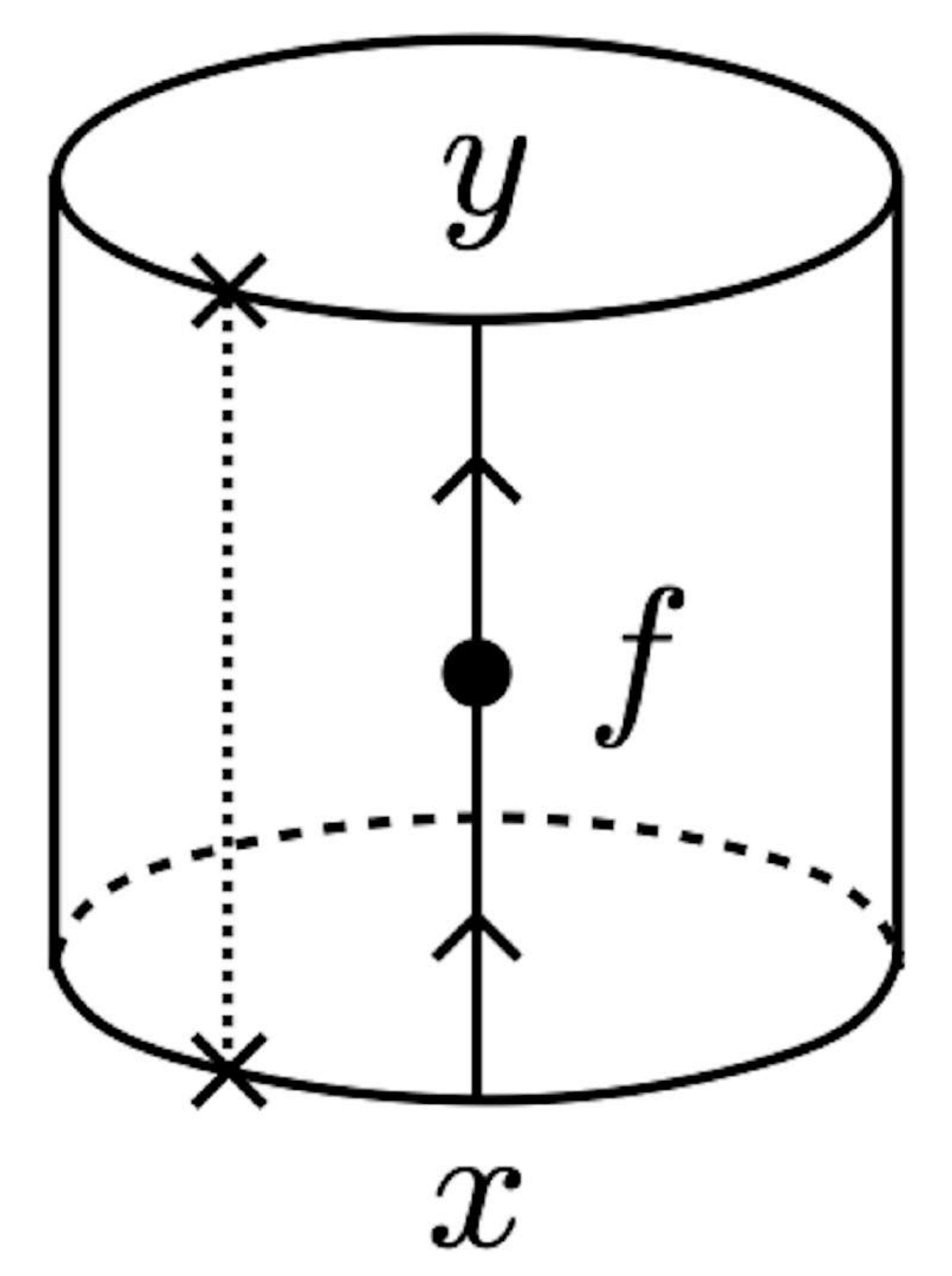} ~
 \includegraphics[width = 1.8cm]{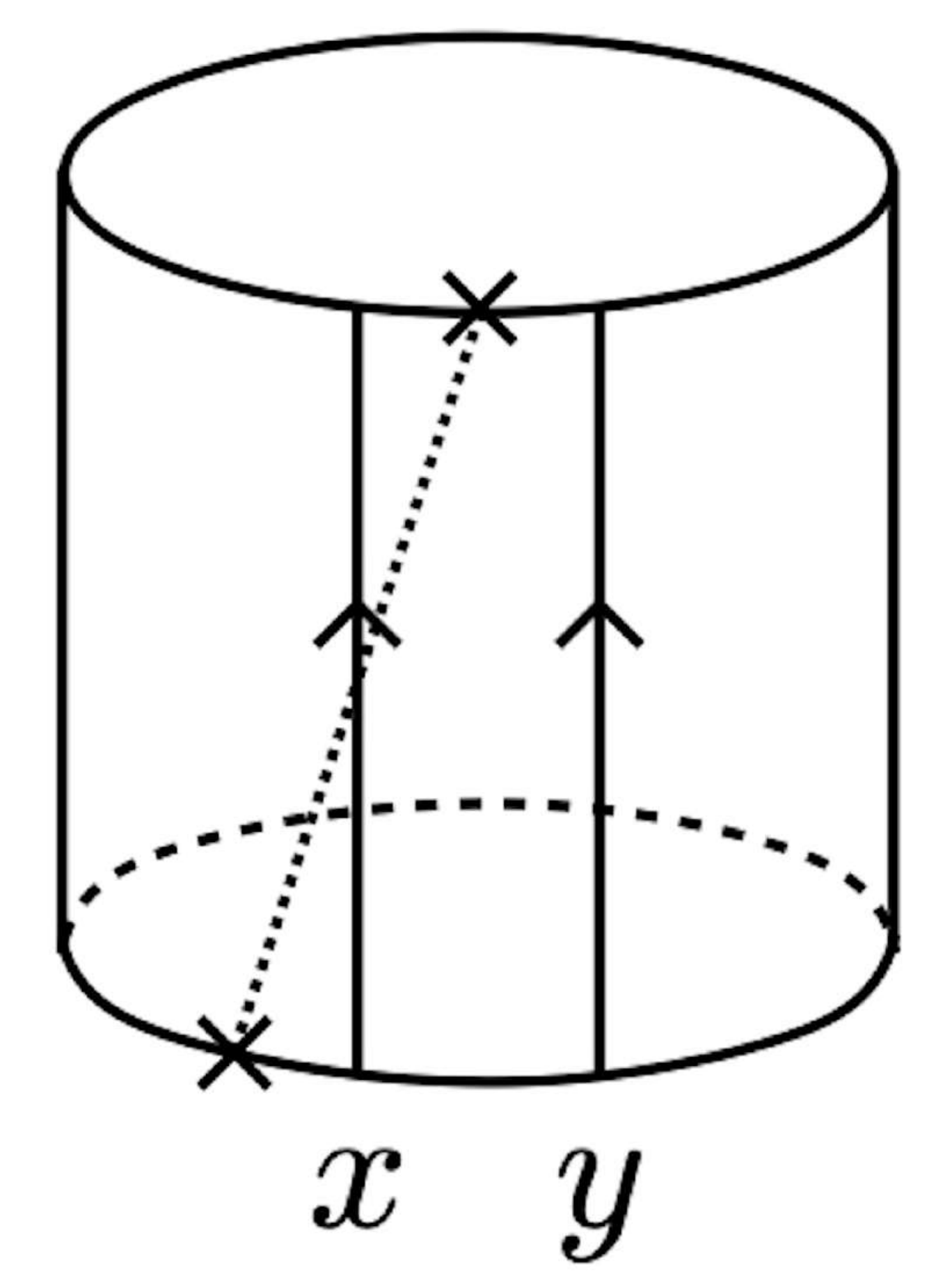} ~
  \includegraphics[ width = 1.6cm]{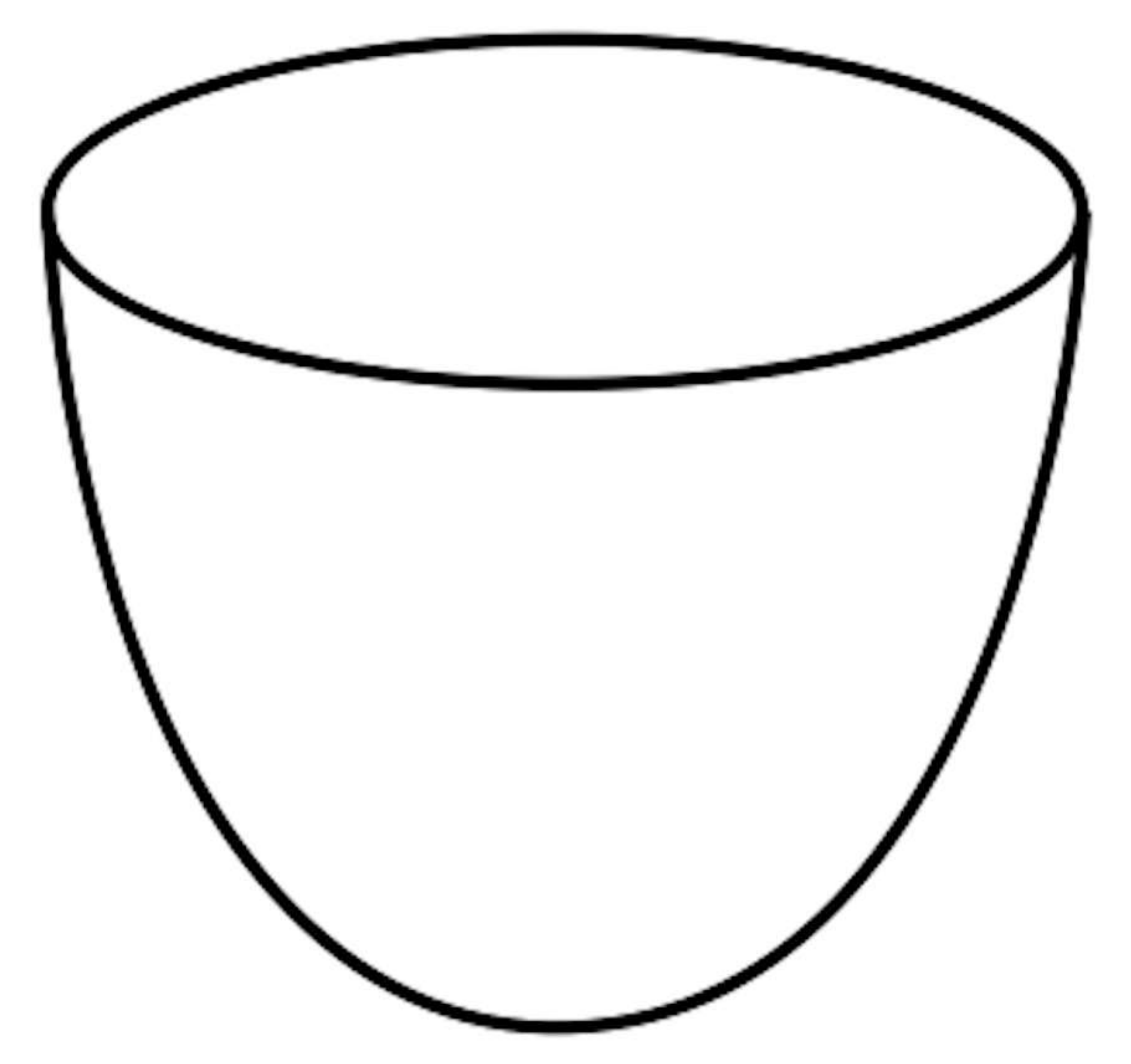} ~
 \includegraphics[width = 2.7cm]{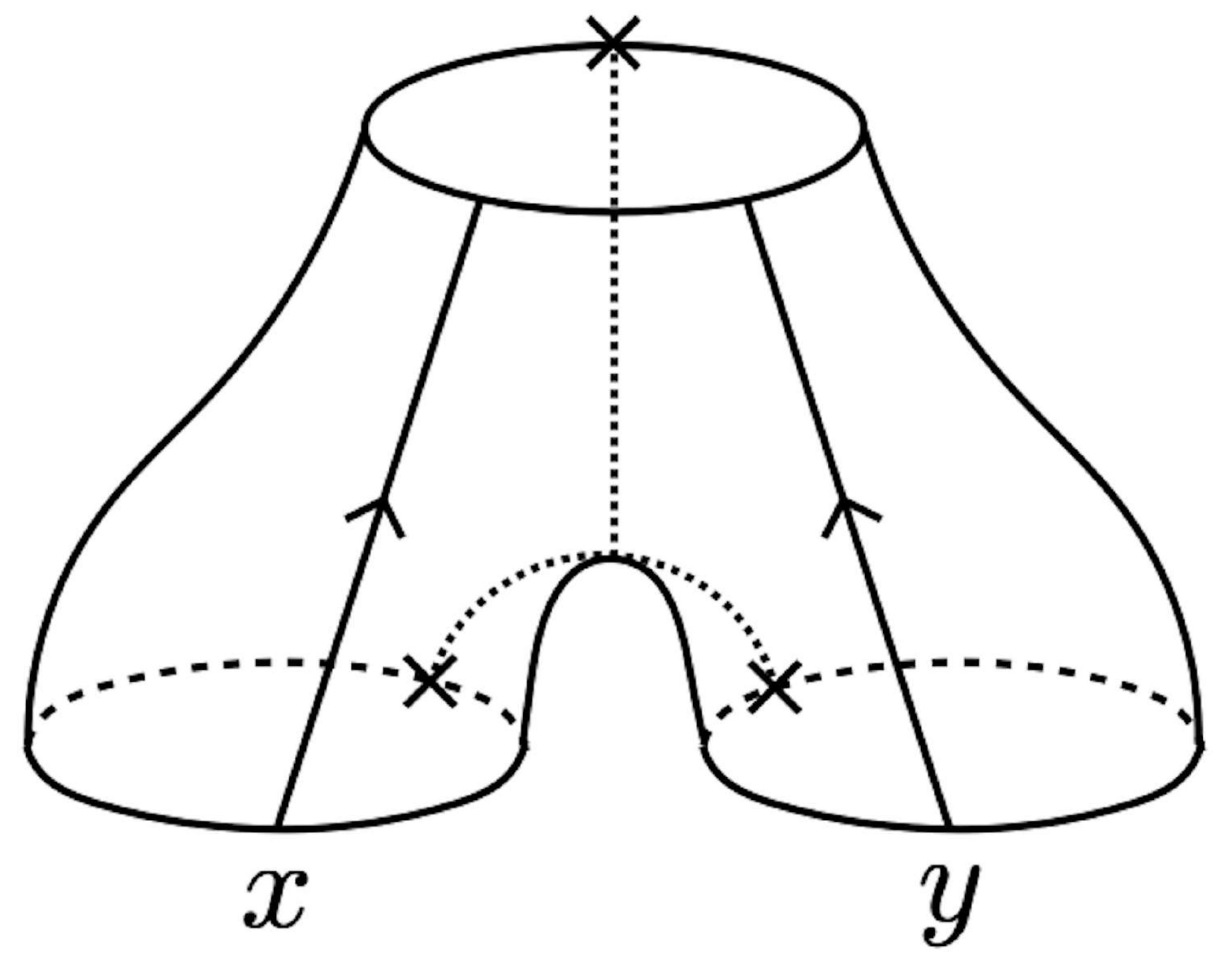} ~
  \includegraphics[width = 1.6cm]{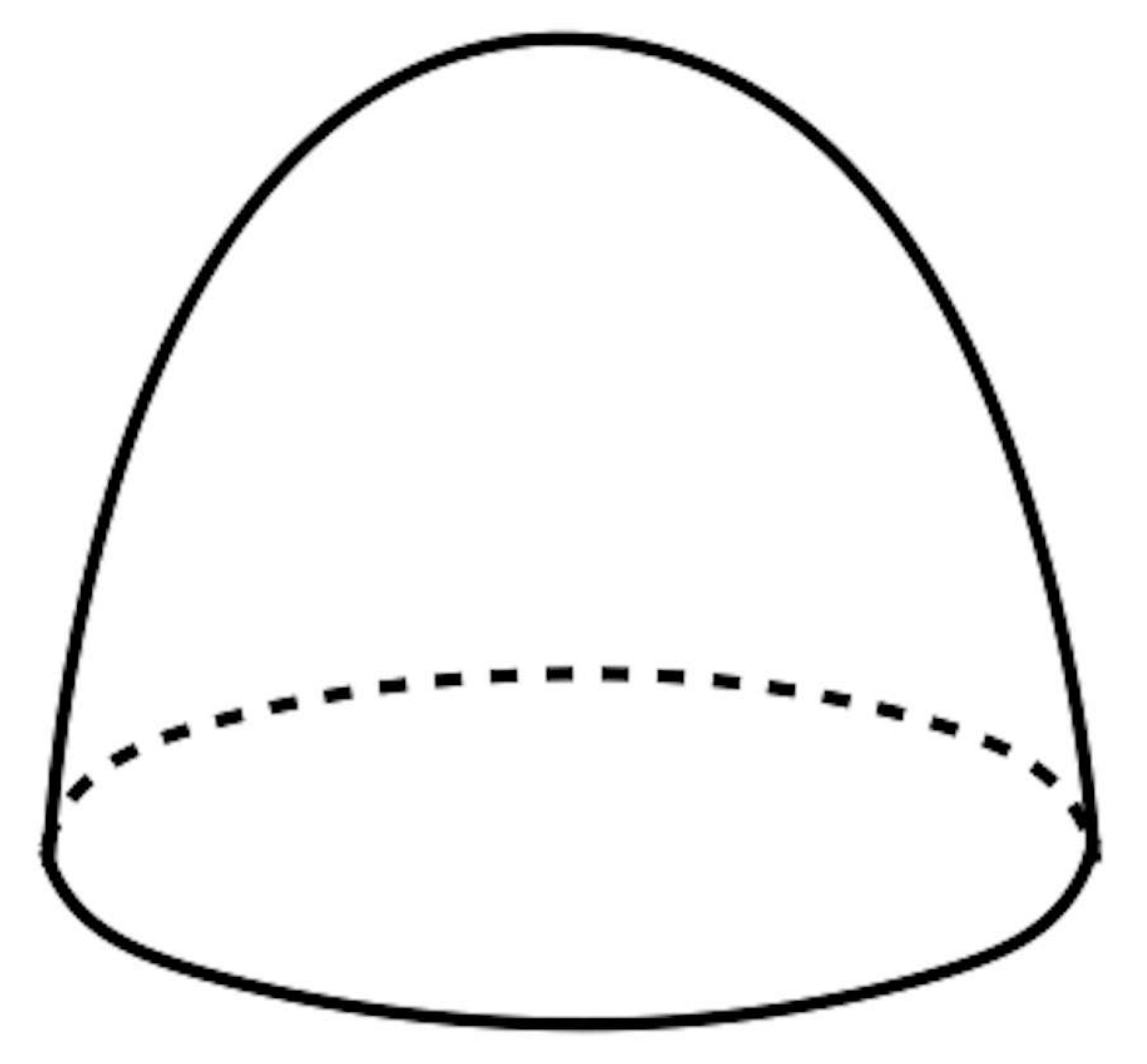} ~
 \includegraphics[width = 2.7cm]{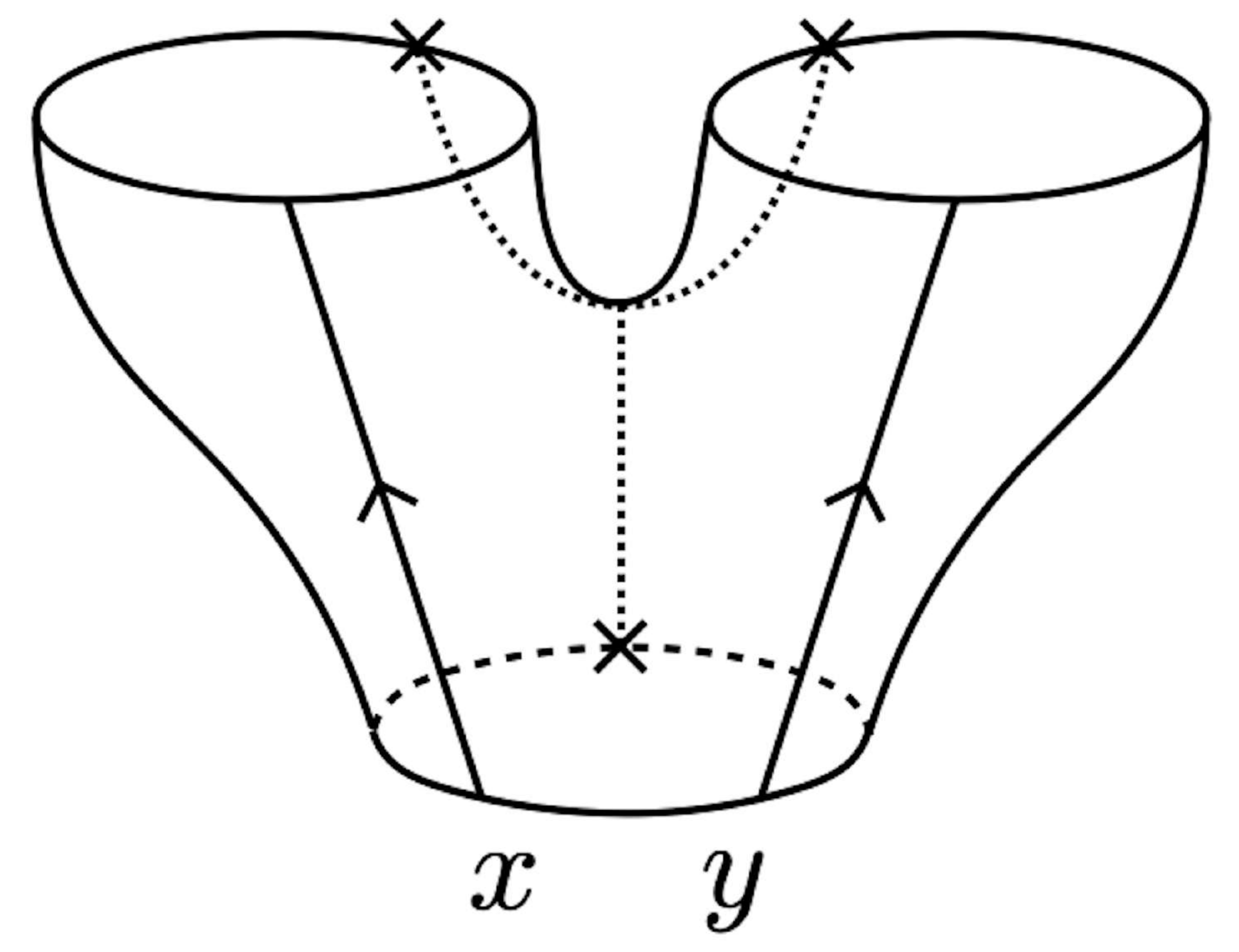} 
 \caption{The building blocks of linear maps: a cylinder amplitude $Z(f)$, a change of the case point $X_{x, y}$, a unit $\eta$, multiplication $M_{x, y}$, a counit $\epsilon$, and comultiplication $\Delta_{x, y}$ (from left to right). Each diagram represents a linear map from the Hilbert space assigned to the bottom to the Hilbert space assigned to the top.}
 \label{fig: basic elements}
 \end{figure}
 \begin{enumerate}
 \item a cylinder amplitude $Z(f): Z(x) \rightarrow Z(y), \quad \forall f \in \mathrm{Hom}(x, y)$,
 \item a change of the base point $X_{x, y}: Z(x \otimes y) \rightarrow Z(y \otimes x)$,
 \item a unit $\eta: \mathbb{C} \rightarrow Z(1)$,
 \item multiplication $M_{x, y}: Z(x) \otimes Z(y) \rightarrow Z(x \otimes y)$,
 \item a counit $\epsilon: Z(1) \rightarrow \mathbb{C}$,
 \item comultiplication $\Delta_{x, y}: Z(x \otimes y) \rightarrow Z(x) \otimes Z(y)$.
 \end{enumerate}
For unitary TQFTs, the counit $\epsilon$ and the comultiplication $\Delta_{x, y}$ are the adjoints of the unit $\eta$ and the multiplication $M_{x, y}$ respectively, i.e. $\epsilon = \eta^{\dagger}$ and $\Delta_{x, y} = M_{x, y}^{\dagger}$.
In particular, the counit $\epsilon$ and the comultiplication $\Delta_{x, y}$ are no longer independent data of a TQFT.

For the well-definedness of the cylinder amplitude, we require that $Z(f)$ is $\mathbb{C}$-linear in morphisms and preserves the composition of morphisms:
\begin{align}
Z(\lambda f + \lambda^{\prime} f^{\prime}) & = \lambda Z(f) + \lambda^{\prime} Z(f^{\prime}), \quad \forall \lambda, \lambda^{\prime} \in \mathbb{C}, \forall f, f^{\prime} \in \mathrm{Hom}(x, y), \\
Z(g \circ f) & = Z(g) \circ Z(f) \quad \forall f \in \mathrm{Hom}(x, y), \forall g \in \mathrm{Hom}(y, z).
\end{align}
Thus, a 2d TQFT with fusion category symmetry $\mathcal{C}$ gives a functor $Z: \mathcal{C} \rightarrow \mathrm{Vec}$ from $\mathcal{C}$ to the category of vector spaces.
This functor obeys various consistency conditions so that the assignment of Hilbert spaces and linear maps are well-defined.
Specifically, a TQFT with fusion category symmetry $\mathcal{C}$ is a functor $Z: \mathcal{C} \rightarrow \mathrm{Vec}$ equipped with a set of linear maps $(X, \eta, M)$ that satisfies the following consistency conditions \cite{BT2018}:
\begin{enumerate}
\item Well-definedness of the change of the base point:
\begin{equation}
X_{x, y} = X_{y, x}^{-1}.
\label{eq: Independence of the auxiliary line}
\end{equation}
\item Naturality of the change of the base point:
\begin{equation}
\begin{aligned}
Z(\mathrm{id}_y \otimes f) \circ X_{x, y} & = X_{x^{\prime}, y} \circ Z(f \otimes \mathrm{id}_y) \quad \forall f \in \mathrm{Hom}(x, x^{\prime}), \\
Z(g \otimes \mathrm{id}_x) \circ X_{x, y} & = X_{x, y^{\prime}} \circ Z(\mathrm{id}_x \otimes g) \quad \forall g \in \mathrm{Hom}(y, y^{\prime}).
\end{aligned}
\label{eq: Transparency of the auxiliary line}
\end{equation}
\item Associativity of the change of the base point:
\begin{equation}
X_{y, z \otimes x} \circ Z(\alpha_{yzx}) \circ X_{x, y \otimes z} \circ Z(\alpha_{xyz}) = Z(\alpha_{zxy}^{-1}) \circ X_{x \otimes y, z}.
\label{eq: Associativity of the change of the base point}
\end{equation}
\item Non-degeneracy of the pairing:
\begin{equation}
\text{The pairing } \eta^{\dagger} \circ Z(\mathrm{ev}^{L}_{x}) \circ M_{x^{*}, x}: Z(x^*) \otimes Z(x) \rightarrow \mathbb{C} \text{ is non-degenerate.}
\label{eq: Non-degeneracy of the pairing}
\end{equation}
\item Unit constraint:
\begin{equation}
M_{1, x} \circ (\eta \otimes \mathrm{id}_{Z(x)}) = \mathrm{id}_{Z(x)} = M_{x, 1} \circ (\mathrm{id}_{Z(x)} \otimes \eta).
\label{eq: Unit constraint}
\end{equation}
\item Associativity of the multiplication:
\begin{equation}
Z(\alpha_{xyz}) \circ M_{x \otimes y, z} \circ (M_{x, y} \otimes \mathrm{id}_{Z(z)}) = M_{x, y \otimes z} \circ (\mathrm{id}_{Z(x)} \otimes M_{y, z}).
\label{eq: Associativity of the multiplication}
\end{equation}
\item Twisted commutativity:
\begin{equation}
M_{y, x} (\psi_y \otimes \psi_x) = X_{x, y} \circ M_{x, y} (\psi_x \otimes \psi_y) \quad \forall \psi_x \in Z(x), \forall \psi_y \in Z(y).
\label{eq: Twisted commutativity}
\end{equation}
\item Naturality of the multiplication:
\begin{equation}
\begin{aligned}
M_{x^{\prime}, y} \circ (Z(f) \otimes \mathrm{id}_{Z(y)}) & = Z(f \otimes \mathrm{id}_y) \circ M_{x, y}, \\
M_{x, y^{\prime}} \circ (\mathrm{id}_{Z(x)} \otimes Z(g)) & = Z(\mathrm{id}_x \otimes g) \circ M_{x, y}.
\end{aligned}
\label{eq: Naturality of the multiplication}
\end{equation}
\item Uniqueness of the multiplication:
\begin{equation}
\begin{aligned}
& Z((\mathrm{id}_x \otimes \mathrm{ev}^L_z) \otimes \mathrm{id}_{y^{*}}) \circ \mathcal{A}_{(x \otimes z^{*}) \otimes (z \otimes y^{*}) \rightarrow (x \otimes (z^{*} \otimes z)) \otimes y^{*}} \circ M_{x \otimes z^{*}, z \otimes y^{*}} \\
& = Z(\mathrm{id}_{x \otimes y^{*}} \otimes \mathrm{ev}^R_z) \circ \mathcal{A}_{(z^* \otimes x) \otimes (y^* \otimes z) \rightarrow (x \otimes y^*) \otimes (z \otimes z^*)} \circ M_{z^* \otimes x, y^* \otimes z} \circ (X_{x, z^*} \otimes X_{z, y^*}),
\end{aligned}
\label{eq: Uniqueness of the multiplication}
\end{equation}
where $\mathcal{A}$ is a generalized associator that we will define below.
\item Consistency on the torus:
\begin{equation}
\begin{aligned}
& \mathcal{A}_{(x \otimes w) \otimes (z \otimes y) \rightarrow (y \otimes x) \otimes (w \otimes z)} \circ M_{x \otimes w, z \otimes y} \circ (X_{w, x} \otimes X_{y, z}) \circ M_{w \otimes x, y \otimes z}^{\dagger} \\
& = M_{y \otimes x, w \otimes z} \circ (X_{x, y} \otimes X_{z, w}) \circ M_{x \otimes y, z \otimes w}^{\dagger} \circ \mathcal{A}_{(w \otimes x) \otimes (y \otimes z) \rightarrow (x \otimes y) \otimes (z \otimes w)}.
\end{aligned}
\label{eq: Consistency on the torus}
\end{equation}
\end{enumerate}
In the last two equations, the generalized associator $\mathcal{A}_{p \rightarrow q}: Z(p) \rightarrow Z(q)$ is defined as a composition of the change of the base point $X$ and the associator $Z(\alpha)$.
We note that the isomorphism $\mathcal{A}_{p \rightarrow q}$ is uniquely determined by $p$ and $q$ \cite{BT2018}. 

In summary, a 2d unitary TQFT with fusion category symmetry $\mathcal{C}$ is a functor $Z: \mathcal{C} \rightarrow \mathrm{Vec}$ equipped with a triple $(X, \eta, M)$ that satisfies the consistency conditions \eqref{eq: Independence of the auxiliary line}--\eqref{eq: Consistency on the torus}.
It is shown in \cite{TW2019, KORS2020} that 2d unitary TQFTs with fusion category symmetry $\mathcal{C}$ are classified by semisimple module categories over $\mathcal{C}$.
Namely, each 2d unitary TQFT with symmetry $\mathcal{C}$ is labeled by a semisimple $\mathcal{C}$-module category.
The TQFT labeled by a $\mathcal{C}$-module category $\mathcal{M}$ has the category of boundary conditions described by $\mathcal{M}$ \cite{MS2006, KORS2020}, whose semisimplicity follows from the unitarity of the TQFT \cite{KTY2017, MS2006}.

\subsection{Pullback of TQFTs by tensor functors}
\label{sec: Pullback of TQFTs by tensor functors and the classification of TQFTs with fusion category symmetry}
Let $(Z^{\prime}, X^{\prime}, \eta^{\prime}, M^{\prime})$ be a 2d TQFT with symmetry $\mathcal{C}^{\prime}$.
Given a tensor functor $(F, J, \phi): \mathcal{C} \rightarrow \mathcal{C}^{\prime}$, we can construct a 2d TQFT $(Z, X, \eta, M)$ with symmetry $\mathcal{C}$ as follows: the functor $Z: \mathcal{C} \rightarrow \mathrm{Vec}$ is given by the composition $Z := Z^{\prime} \circ F$, and the linear maps $(X, \eta, M)$ are defined as
\begin{align}
X_{x, y} & := Z^{\prime}(J_{y, x}) \circ X^{\prime}_{F(x), F(y)} \circ Z^{\prime}(J_{x, y}^{-1}), \\
\eta & := Z^{\prime}(\phi) \circ \eta^{\prime}, \\
M_{x, y} & := Z^{\prime}(J_{x, y}) \circ M^{\prime}_{F(x), F(y)}.
\end{align}
We can show that the quadruple $(Z, X, \eta, M)$ defined as above becomes a 2d TQFT, provided that $(Z^{\prime}, X^{\prime}, \eta^{\prime}, M^{\prime})$ satisfies the consistency conditions \eqref{eq: Independence of the auxiliary line}--\eqref{eq: Consistency on the torus}.
We will explicitly check some of the consistency conditions for $(Z, X, \eta, M)$ below.
The other equations can also be checked similarly.

Let us begin with eq. \eqref{eq: Independence of the auxiliary line}.
This equation holds because the right-hand side can be written as
\begin{equation}
(Z^{\prime}(J_{x, y}) \circ X^{\prime}_{F(y), F(x)} \circ Z^{\prime}(J_{y, x}^{-1}))^{-1} = Z^{\prime}(J_{y, x}) \circ X^{\prime}_{F(x), F(y)} \circ Z^{\prime}(J_{x, y}^{-1}) = (\mathrm{LHS}), 
\end{equation}
where we used the fact that $X^{\prime}$ satisfies eq. \eqref{eq: Independence of the auxiliary line}.
Equation \eqref{eq: Transparency of the auxiliary line} follows from the naturality of $J$:
\begin{equation}
F(g \otimes f) \circ J_{y, x} = J_{y^{\prime}, x^{\prime}} \circ (F(g) \otimes F(f)), \quad \forall g \in \mathrm{Hom}(y, y^{\prime}), \forall f \in \mathrm{Hom}(x, x^{\prime}).
\end{equation}
Indeed, if we choose either $g$ or $f$ as the identity morphism and use eq. \eqref{eq: Transparency of the auxiliary line} for $X^{\prime}$, we obtain eq. \eqref{eq: Transparency of the auxiliary line} for $X$.
To show eq. \eqref{eq: Associativity of the change of the base point}, we note that $F(\alpha_{xyz})$ can be written in terms of the associators $\alpha^{\prime}_{F(x), F(y), F(z)}$ of $\mathcal{C}^{\prime}$ due to the commutative diagram \eqref{eq: monoidal structure} as follows:
\begin{equation}
F(\alpha_{xyz}) = J_{x, y \otimes z} \circ (\mathrm{id}_{F(x)} \otimes J_{y, z}) \circ \alpha^{\prime}_{F(x), F(y), F(z)} \circ (J_{x, y}^{-1} \otimes \mathrm{id}_{F(z)}) \circ J_{x \otimes y, z}^{-1}.
\label{eq: F(alpha)}
\end{equation}
We also notice that the naturality \eqref{eq: Transparency of the auxiliary line} of $X^{\prime}$ implies
\begin{equation}
\begin{aligned}
X^{\prime}_{F(x \otimes y), F(z)} & = Z^{\prime}(\mathrm{id}_{F(z)} \otimes J_{x, y}) \circ X^{\prime}_{F(x) \otimes F(y), F(z)} \circ Z^{\prime}(J_{x, y}^{-1} \otimes \mathrm{id}_{F(z)}), \\
X^{\prime}_{F(x), F(y \otimes z)} & = Z^{\prime}(J_{y, z} \otimes \mathrm{id}_{F(x)}) \circ X^{\prime}_{F(x), F(y) \otimes F(z)} \circ Z^{\prime}(\mathrm{id}_{F(x)} \otimes J_{y, z}^{-1}).
\end{aligned}
\label{eq: X naturality}
\end{equation}
By plugging eqs. \eqref{eq: F(alpha)} and \eqref{eq: X naturality} into the left-hand side of eq. \eqref{eq: Associativity of the change of the base point}, we find
\begin{equation}
\begin{aligned}
(\mathrm{LHS}) & = Z^{\prime}(J_{z \otimes x, y}) \circ Z^{\prime}(J_{z, x} \otimes \mathrm{id}_{F(y)}) \circ X^{\prime}_{F(y), F(z) \otimes F(x)} \circ Z^{\prime}(\alpha_{F(y), F(z), F(x)}^{\prime})\\
& \quad \circ X^{\prime}_{F(x), F(y) \otimes F(z)} \circ Z^{\prime}(\alpha_{F(x), F(y), F(z)}^{\prime}) \circ Z^{\prime}(J_{x, y}^{-1} \otimes \mathrm{id}_{F(z)}) \circ Z^{\prime}(J_{x \otimes y, z}^{-1})\\
& = Z^{\prime}(J_{z \otimes x, y}) \circ Z^{\prime} (J_{z, x} \otimes \mathrm{id}_{F(y)}) \circ Z^{\prime}((\alpha_{F(z), F(x), F(y)}^{\prime})^{-1})\\
& \quad \circ X^{\prime}_{F(x) \otimes F(y), F(z)} \circ Z^{\prime}(J_{x, y}^{-1} \otimes \mathrm{id}_{F(z)}) \circ Z^{\prime}(J_{x \otimes y, z}^{-1})\\
& = (\mathrm{RHS}).
\end{aligned}
\end{equation}
The non-degeneracy condition \eqref{eq: Non-degeneracy of the pairing} for an object $x \in \mathcal{C}$ follows from that for $F(x) \in \mathcal{C}^{\prime}$ because
\begin{equation}
\eta^{\dagger} \circ Z(\mathrm{ev}_x^L) \circ M_{x^*, x} = (\eta^{\prime})^{\dagger} \circ Z^{\prime}(\mathrm{ev}_{F(x)}^L) \circ M^{\prime}_{F(x)^*, F(x)},
\end{equation}
where we used $F(x)^* = F(x^*)$ and $F(\mathrm{ev}_x^L) = \phi \circ \mathrm{ev}_{F(x)}^L \circ J_{x^*, x}^{-1}$, cf. Exercise 2.10.6. in \cite{EGNO2015}.
The unit constraint \eqref{eq: Unit constraint} is an immediate consequence of the commutative diagram \eqref{eq: monoidal structure unit} and eqs. \eqref{eq: Unit constraint} and \eqref{eq: Naturality of the multiplication} for $(\eta^{\prime}, M^{\prime})$.

We can also check the remaining equations similarly.
Thus, we find that the quadruple $(Z, X, \eta, M)$ becomes a 2d TQFT with symmetry $\mathcal{C}$.
We call a TQFT $(Z, X, \eta, M)$ the pullback of a TQFT $(Z^{\prime}, X^{\prime}, \eta^{\prime}, M^{\prime})$ by a tensor functor $(F, J, \phi)$.

By using the pullback, we can construct all the TQFTs with non-anomalous fusion category symmetry $\mathcal{C}$.\footnote{More generally, we can construct all the TQFTs with arbitrary fusion category symmetries including anomalous ones just by replacing a Hopf algebra $H$ with a (semisimple pseudo-unitary connected) weak Hopf algebra in the following discussion, see also section \ref{sec: A comment on a generalization to anomalous symmetries}.}
To see this, we first recall that every non-anomalous fusion category symmetry $\mathcal{C}$ is equivalent to the representation category $\mathrm{Rep}(H)$ of a Hopf algebra $H$.
Indecomposable semisimple module categories over $\mathrm{Rep}(H)$ are given by the categories ${}_K \mathcal{M}$ of left $K$-modules where $K$ is an $H$-simple left $H$-comodule algebra.
Accordingly, we have a tensor functor $F_K: \mathrm{Rep}(H) \rightarrow {}_K \mathcal{M}_K$ that represents the $\mathrm{Rep}(H)$-module category structure on ${}_K \mathcal{M}$.
Therefore, we can pull back a ${}_K \mathcal{M}_K$ symmetric TQFT by $F_K$ to obtain a $\mathrm{Rep}(H)$ symmetric TQFT.
Here, we notice that there is a canonical ${}_K \mathcal{M}_K$ symmetric TQFT labeled by a ${}_K \mathcal{M}_K$-module category ${}_K \mathcal{M}$, whose module category structure was discussed in section \ref{sec: Fusion categories and tensor functors}.
Thus, by pulling back this canonical ${}_K \mathcal{M}_K$ symmetric TQFT ${}_K \mathcal{M}$ by the tensor functor $F_K: \mathrm{Rep}(H) \rightarrow {}_K \mathcal{M}_K$, we obtain a $\mathrm{Rep}(H)$ symmetric TQFT canonically from the data of a $\mathrm{Rep}(H)$-module category ${}_K \mathcal{M}$.
This suggests that the TQFT obtained in this way is a $\mathrm{Rep}(H)$ symmetric TQFT labeled by a module category ${}_K \mathcal{M}$, or equivalently, this is a $\mathrm{Rep}(H)$ symmetric TQFT whose category of boundary conditions is given by ${}_K \mathcal{M}$.
In the next section, we will see that this is the case by showing that the action of the $\mathrm{Rep}(H)$ symmetry on the boundary conditions of this TQFT is described by the $\mathrm{Rep}(H)$-module action on ${}_K \mathcal{M}$.

\section{State sum TQFTs and commuting projector Hamiltonians}
\label{sec: State sum TQFTs and commuting projector Hamiltonians}
The canonical ${}_K \mathcal{M}_K$ symmetric TQFT ${}_K \mathcal{M}$ is obtained by state sum construction \cite{FHK94} whose input datum is a semisimple algebra $K$.
The ${}_K \mathcal{M}_K$ symmetry of this TQFT was first discussed in \cite{DKR2011}.
This symmetry can also be understood from a viewpoint of generalized gauging \cite{FFRS2010, BCP2014a, BCP2014b, BCP2015, CR2016, BT2018, CRS2019}.
In this section, we show that this state sum TQFT actually has $\mathrm{Rep}(H)$ symmetry when the input algebra $K$ is a left $H$-comodule algebra.
Specifically, this TQFT is regarded as the pullback of a ${}_K \mathcal{M}_K$ symmetric TQFT ${}_K \mathcal{M}$ by a tensor functor $F_K: \mathrm{Rep}(H) \rightarrow {}_K \mathcal{M}_K$.
We also construct $\mathrm{Rep}(H)$ symmetric commuting projector Hamiltonians whose ground states are described by the above state sum TQFTs.
These commuting projector Hamiltonians realize all the gapped phases with non-anomalous fusion category symmetries.

\subsection{State sum TQFTs with defects}
\label{sec: State sum TQFTs with defects}
We begin with reviewing state sum TQFTs with defects following \cite{DKR2011}.
We slightly modify the description of topological junctions in \cite{DKR2011} so that it fits into the context of TQFTs with fusion category symmetries discussed in section \ref{sec: Pullback of fusion category TQFTs by tensor functors}.

Let $\Sigma$ be a two-dimensional surface with in-boundary $\partial_{\mathrm{in}} \Sigma$ and out-boundary $\partial_{\mathrm{out}} \Sigma$.
The surface $\Sigma$ is decorated by a network of topological defects that are labeled by objects of the category ${}_K \mathcal{M}_K$.
We assume that the junctions of these topological defects are trivalent and labeled by morphisms of ${}_K \mathcal{M}_K$.
We further assume, as in section \ref{sec: TQFTs with fusion category symmetries}, that the topological defects intersecting the in-boundary (out-boundary) are oriented so that they go into (out of) $\Sigma$

To assign a linear map to $\Sigma$, we first give a triangulation $T(\Sigma)$ of $\Sigma$ such that every face $p$ contains at most one trivalent junction and every edge $e$ intersects at most one topological defect.
The possible configurations of topological defects on a face $p$ are as follows:
\begin{equation}
\text{(i) } \adjincludegraphics[valign = c, width = 1.5cm]{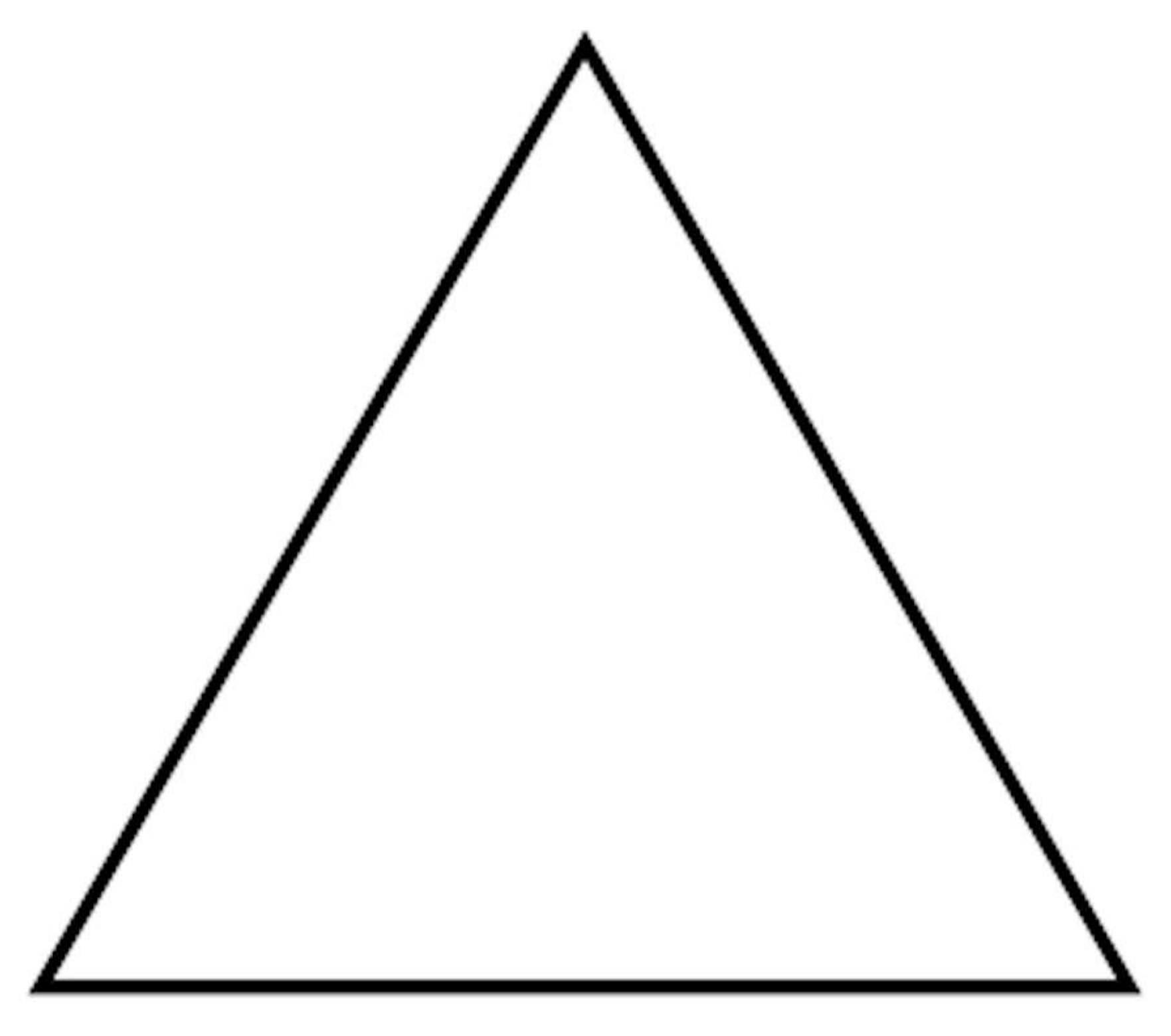}, \quad
\text{(ii) } \adjincludegraphics[valign = c, width = 1.6cm]{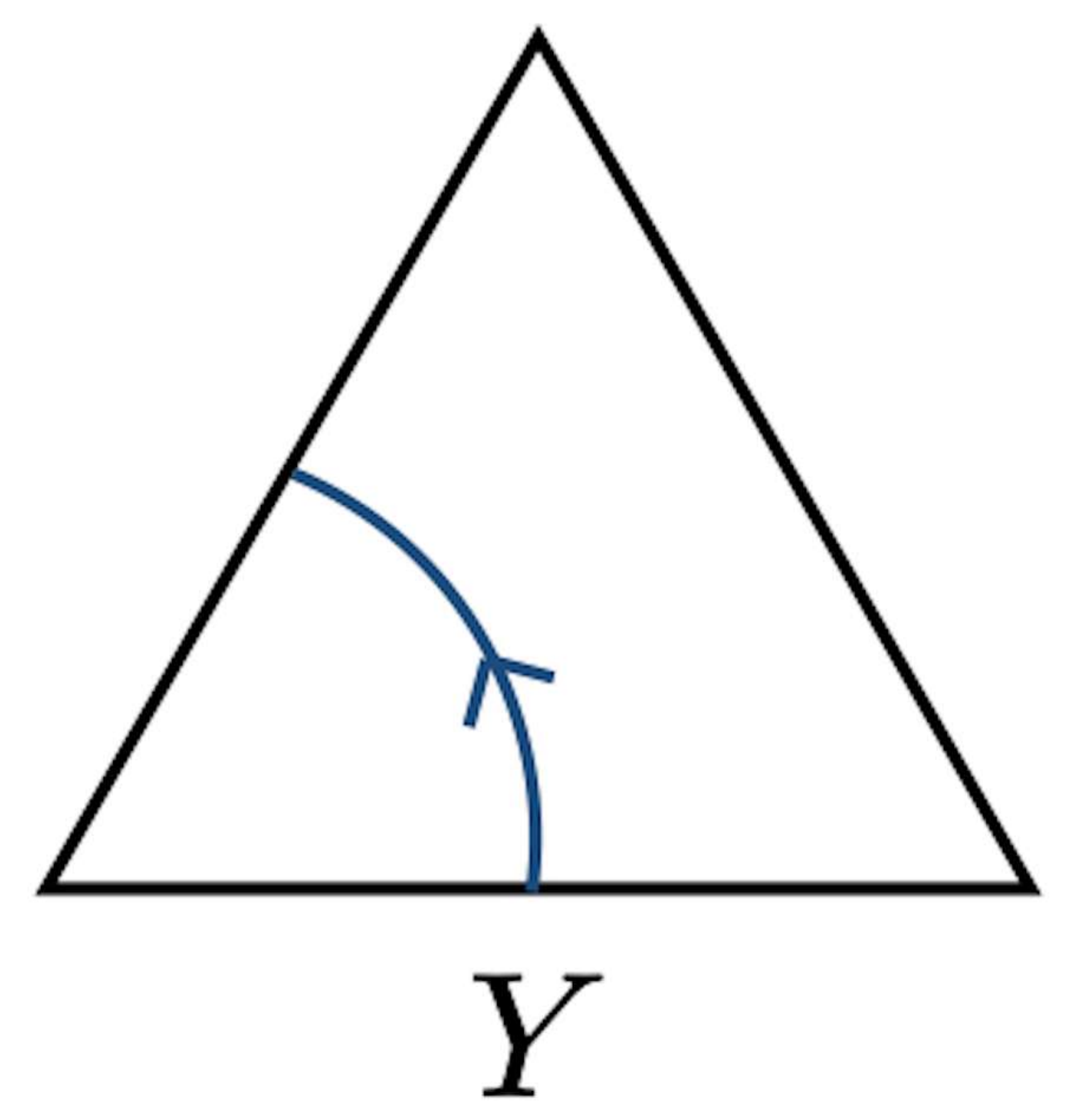}, \quad
\text{(iii) } \adjincludegraphics[valign = c, width = 1.65cm]{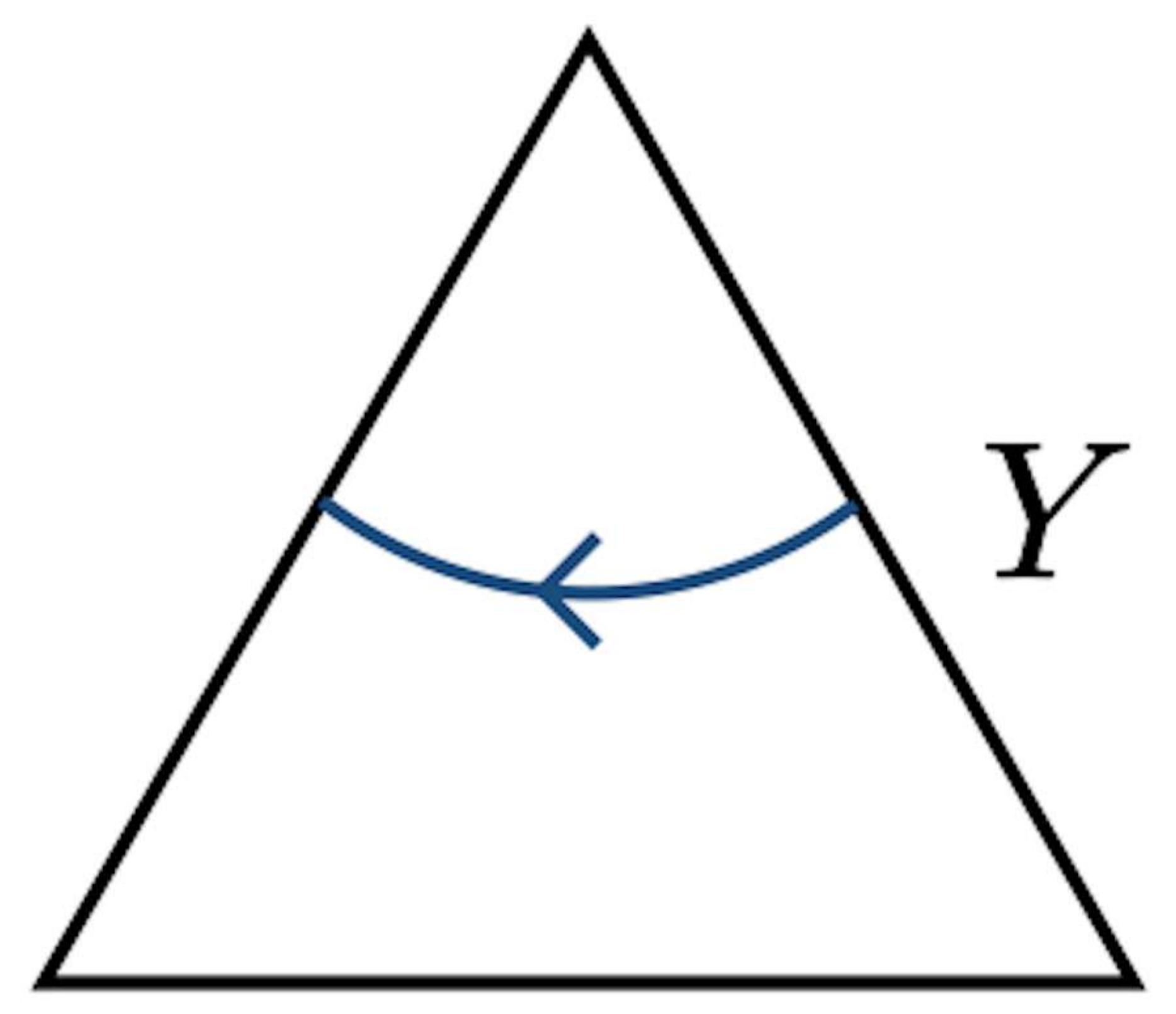}, \quad
\text{(iv) } \adjincludegraphics[valign = c, width = 1.75cm]{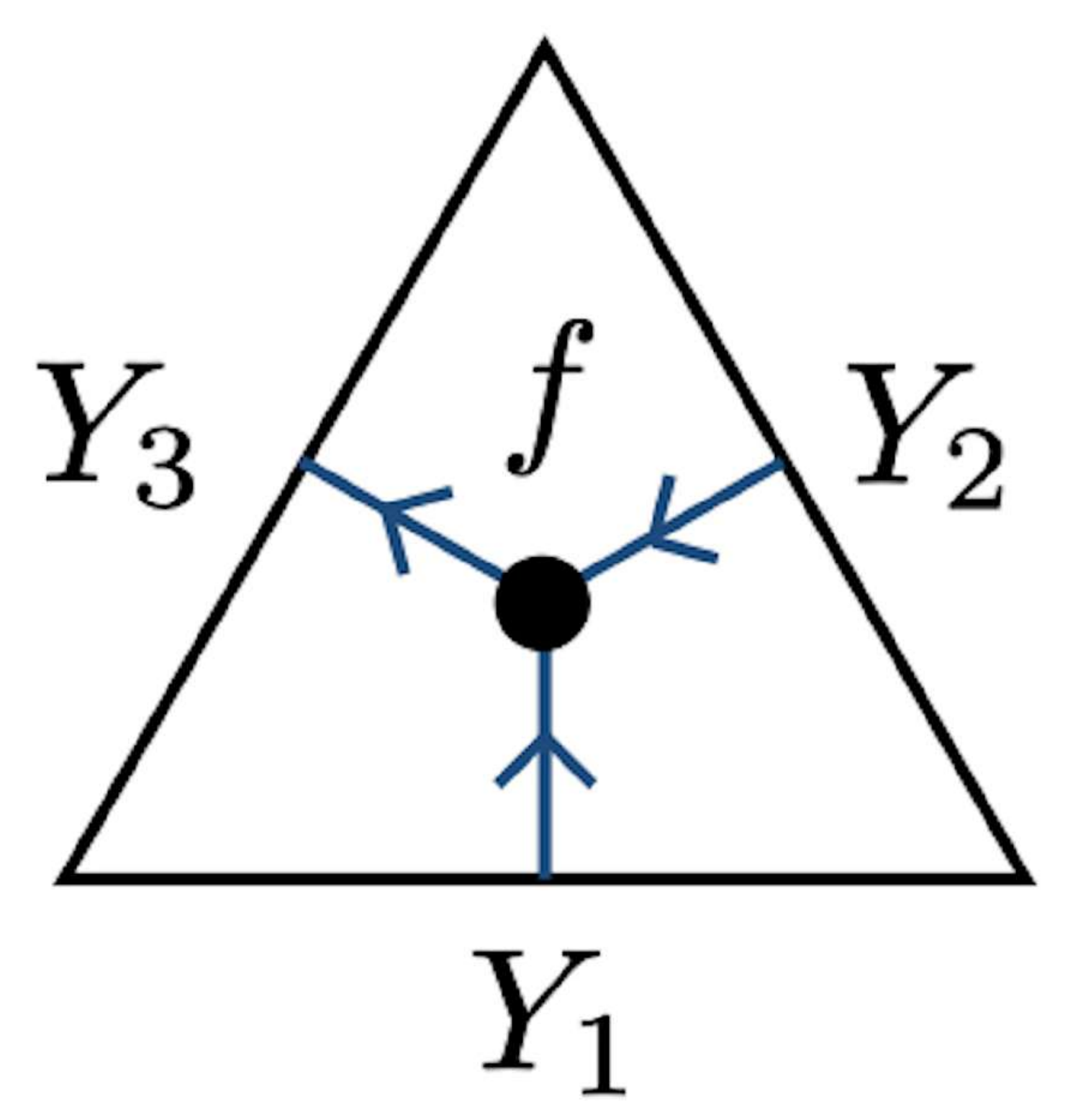}, \quad
\text{(v) } \adjincludegraphics[valign = c, width = 1.75cm]{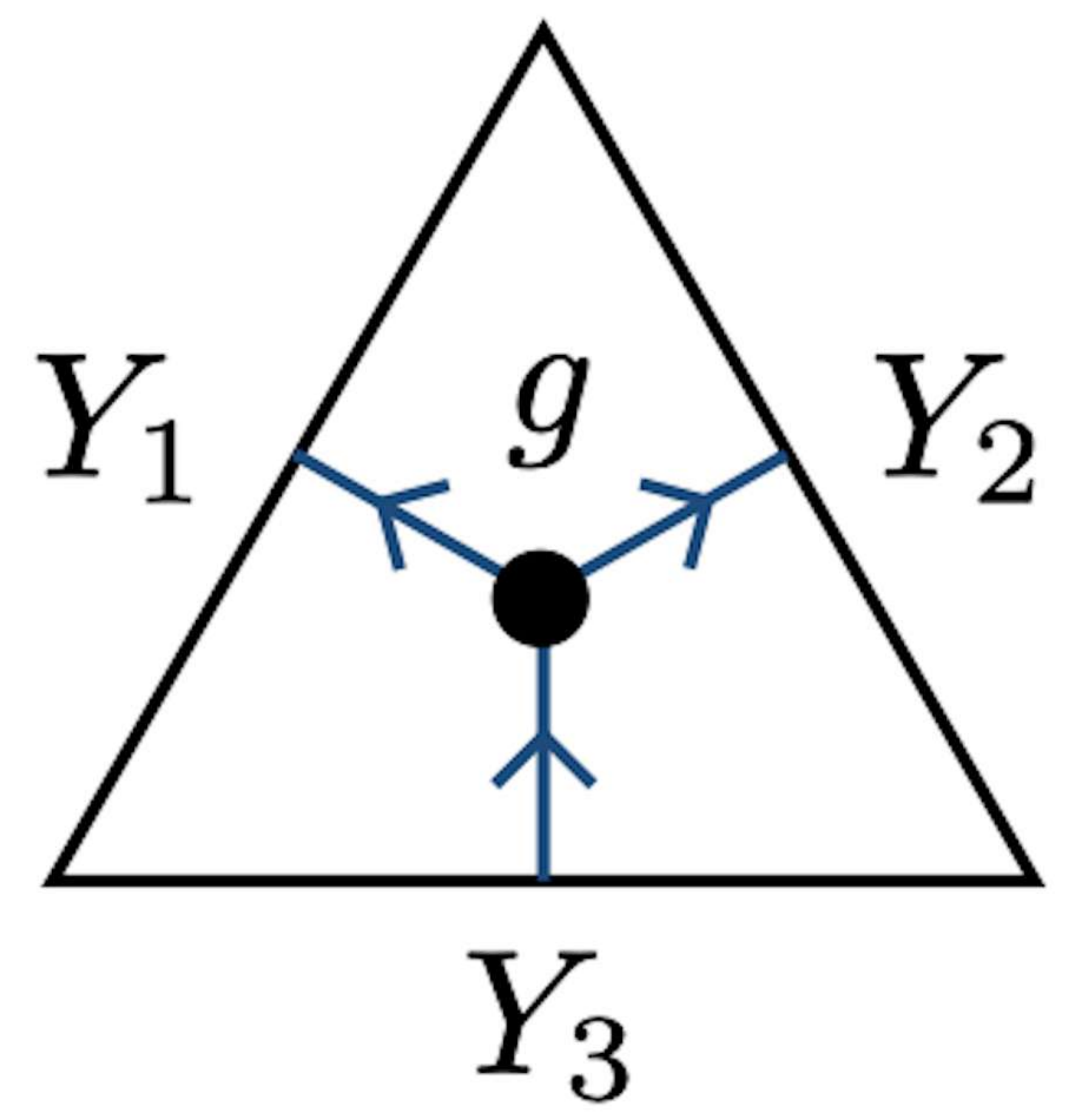}.
\label{eq: configurations}
\end{equation}
Here, topological defects are labeled by $K$-$K$ bimodules $Y, Y_1, Y_2, Y_3 \in {}_K \mathcal{M}_K$, and trivalent junctions are labeled by bimodule maps $f \in \mathrm{Hom}_{KK}(Y_1 \otimes_K Y_2, Y_3), g \in \mathrm{Hom}_{KK}(Y_3, Y_1 \otimes_K Y_2)$.
We note that all of the above configurations are obtained from configuration (iv) by choosing some of the topological defects as trivial defects or replacing some of the topological defects with their duals. Nevertheless, we distinguish these configurations for convenience.

For the triangulated surface $T(\Sigma)$, we define a linear map $Z_T(\Sigma)$ as \cite{DKR2011}
\begin{equation}
Z_T(\Sigma): Z_T(\partial_{\mathrm{in}} \Sigma) \xrightarrow{\mathrm{id} \otimes P(\Sigma)} Z_T(\partial_{\mathrm{in}} \Sigma) \otimes Q(\Sigma) \otimes Z_T(\partial_{\mathrm{out}} \Sigma) \xrightarrow{E(\Sigma) \otimes \mathrm{id}} Z_T(\partial_{\mathrm{out}} \Sigma).
\label{eq: linear map ZT}
\end{equation}
The constituents of this linear map are described below.
\begin{description}
\item[The vector spaces $Z_T(\partial_{\mathrm{in}} \Sigma)$ and $Z_T(\partial_{\mathrm{out}} \Sigma)$] ~\\
The vector space $Z_T(\partial_{a} \Sigma)$ for $a = \mathrm{in}, \mathrm{out}$ is defined as the tensor product of vector spaces $R_e$ assigned to edges $e \in \partial_a \Sigma$, namely
\begin{equation}
Z_T(\partial_{a} \Sigma) := \bigotimes_{e \in \partial_{a} \Sigma} R_e,
\label{eq: vector space ZT}
\end{equation}
where the vector spaces $R_e$ are given as follows: 
\begin{equation}
R_e := 
\begin{cases}
K \quad & \text{when } e \text{ does not intersect a topological defect}, \\
Y \quad & \text{when } e \text{ intersects a topological defect $Y \in {}_K\mathcal{M}_K$}.
\end{cases}
\label{eq: Re}
\end{equation}
We recall that the orientation of a topological defect $Y$ on a boundary edge $e \in \partial_a \Sigma$ is uniquely determined by assumption.

\item[The vector space $Q(\Sigma)$] ~\\
Similarly, we define the vector space $Q(\Sigma)$ as the tensor product of the vector spaces $Q_{(p, e)}$ assigned to flags $(p, e) \in \Sigma$ except for those whose edge $e$ is contained in the in-boundary $\partial_{\mathrm{in}} \Sigma$:
\begin{equation}
Q(\Sigma) := \bigotimes_{(p, e) \in \Sigma \text{ s.t. } e \notin \partial_{\mathrm{in}} \Sigma} Q_{(p. e)}.
\label{eq: vector space Q}
\end{equation}
Here, a flag $(p, e)$ is a pair of a face $p$ and an edge $e$ on the boundary of $p$.
The vector space $Q_{(p, e)}$ depends on both the label and the orientation of a topological defect that intersects the edge $e$.
Concretely, we define
\begin{equation}
Q_{(p, e)} :=
\begin{cases}
K \quad & \text{when } e \text{ does not intersect a topological defect}, \\
Y \quad & \text{when a topological defect } Y \text{ goes into }  p \text{ across } e, \\
Y^{*} \quad & \text{when a topological defect } Y \text{ goes out of }  p \text{ across } e.
\end{cases}
\label{eq: Q(p, e)}
\end{equation}

\item[The linear map $P(\Sigma)$] ~\\
The linear map $P(\Sigma): \mathbb{C} \rightarrow Q(\Sigma) \otimes Z_T(\partial_{\mathrm{out}} \Sigma)$ is also defined in the form of the tensor product
\begin{equation}
P(\Sigma) := \bigotimes_{e \in \Sigma \setminus \partial_{\mathrm{in}} \Sigma} P_e,
\label{eq: linear map P}
\end{equation}
where the tensor product is taken over all edges $e$ of $\Sigma$ except for those on the in-boundary.
The linear map $P_e$ for each edge $e \in \Sigma \setminus \partial_{\mathrm{in}} \Sigma$ is given by
\begin{equation}
P_e := 
\begin{cases}
\Delta_K \circ \eta_K \quad & \text{when } e \text{ does not intersect a topological defect}, \\
\mathrm{coev}_Y \quad & \text{when } e \text{ intersects a topological defect } Y,
\end{cases}
\label{eq: Pe}
\end{equation}
where $\Delta_{K}: K \rightarrow K \otimes K$ and $\eta_{K}: \mathbb{C} \rightarrow K$ are the comultiplication and the unit of the Frobenius algebra $K$, see section \ref{sec: Fusion categories and tensor functors}.
The coevaluation map $\mathrm{coev}_Y: \mathbb{C} \rightarrow Y \otimes Y^*$ is given by the usual embedding analogous to eq. \eqref{eq: ev/coev}.

\item[The linear map $E(\Sigma)$] ~\\
Finally, the linear map $E(\Sigma): Z_T(\partial_{\mathrm{in}} \Sigma) \otimes Q(\Sigma) \rightarrow \mathbb{C}$ is again given by the tensor product
\begin{equation}
E(\Sigma) := \bigotimes_{p \in \Sigma} E_p,
\label{eq: linear map E}
\end{equation}
where the linear map $E_p$ for each face $p \in \Sigma$ depends on a configuration of topological defects on $p$.
We have five different configurations (i)--(v) as shown in eq. \eqref{eq: configurations}, and define the linear map $E_p$ for each of them as follows:
\begin{equation}
E_p := 
\begin{cases}
\text{(i)} & \epsilon_K \circ m_K \circ (m_K \otimes \mathrm{id}_K): K \otimes K \otimes K \rightarrow \mathbb{C}, \\
\text{(ii)} & \mathrm{ev}_Y \circ (\mathrm{id}_{Y^*} \otimes \rho_Y^R): Y^{*} \otimes Y \otimes K \rightarrow \mathbb{C}, \\
\text{(iii)} & \mathrm{ev}_Y \circ (\mathrm{id}_{Y^*} \otimes \rho_Y^L): Y^* \otimes K \otimes Y \rightarrow \mathbb{C}, \\
\text{(iv)} & \mathrm{ev}_{Y_3} \circ (\mathrm{id}_{Y_3^*} \otimes (f \circ \pi_{Y_1, Y_2})): Y_3^* \otimes Y_1 \otimes Y_2 \rightarrow \mathbb{C}, \\
\text{(v)} & \mathrm{ev}_{Y_1 \otimes Y_2} \circ (\mathrm{id}_{(Y_1 \otimes Y_2)^*} \otimes (\iota_{Y_1, Y_2} \circ g)): (Y_1 \otimes Y_2)^* \otimes Y_3 \rightarrow \mathbb{C}.
\end{cases}
\label{eq: Ep}
\end{equation}
Here, $\rho_Y^L: K \otimes Y \rightarrow Y$ and $\rho_Y^{R}: Y \otimes K \rightarrow Y$ denote the left and right $K$-module actions on $Y$ respectively, and $\pi_{Y_1, Y_2}$ and $\iota_{Y_1, Y_2}$ are the splitting maps defined in section \ref{sec: Fusion categories and tensor functors}.
The linear maps $f \in \mathrm{Hom}_{KK} (Y_1 \otimes_K Y_2, Y_3)$ and $g \in \mathrm{Hom}_{KK}(Y_3, Y_1 \otimes_K Y_2)$ are morphisms in the category of $K$-$K$ bimodules.
As we mentioned before, the linear maps for (i)--(iii) and (v) are obtained from that for (iv) with an appropriate choice of $Y_1, Y_2, Y_3$, and $f$.
\end{description}

Combining the above definitions \eqref{eq: vector space ZT}--\eqref{eq: Ep}, we obtain the linear map $Z_T(\Sigma): Z_T(\partial_{\mathrm{in}} \Sigma) \rightarrow Z_T(\partial_{\mathrm{out}} \Sigma)$ via eq. \eqref{eq: linear map ZT}.
However, $Z_T(\Sigma)$ is not yet an appropriate transition amplitude of a TQFT because the linear map $Z_T(\partial_{\mathrm{in}} \Sigma \times [0, 1])$ assigned to a cylinder $\partial_{\mathrm{in}} \Sigma \times [0, 1]$ is not the identity map.
In particular, the linear map $Z_T(\partial_{\mathrm{in}} \Sigma \times [0, 1])$ is an idempotent on $Z_T(\partial_{\mathrm{in}} \Sigma)$, whose image will be denoted by $Z(\partial_{\mathrm{in}} \Sigma)$.
It turns out that  $Z(\partial_{\mathrm{in}} \Sigma)$ is mapped to $Z(\partial_{\mathrm{out}} \Sigma)$ by $Z_T(\Sigma)$.
Hence, we obtain a linear map $Z(\Sigma): Z(\partial_{\mathrm{in}} \Sigma) \rightarrow Z(\partial_{\mathrm{out}} \Sigma)$ by restricting the domain of the linear map \eqref{eq: linear map ZT} to $Z(\partial_{\mathrm{in}} \Sigma)$.
We note that the linear map assigned to a cylinder is now the identity map.
It is shown in \cite{DKR2011} that the assignment of the vector spaces $Z(\partial_{\text{in/out}} \Sigma)$ and the linear map $Z(\Sigma)$ gives a TQFT with defects.\footnote{The proof of the topological invariance in \cite{DKR2011} can still be applied even though the description of topological junctions is slightly changed.}
Based on the above definition, we find that the two possible ways to resolve a quadrivalent junction into two trivalent junctions are related by the associator $\alpha_{Y_1, Y_2, Y_3}$ defined by eq. \eqref{eq: associator of KMK} as follows:
\begin{equation}
\adjincludegraphics[valign = c, width = 3cm]{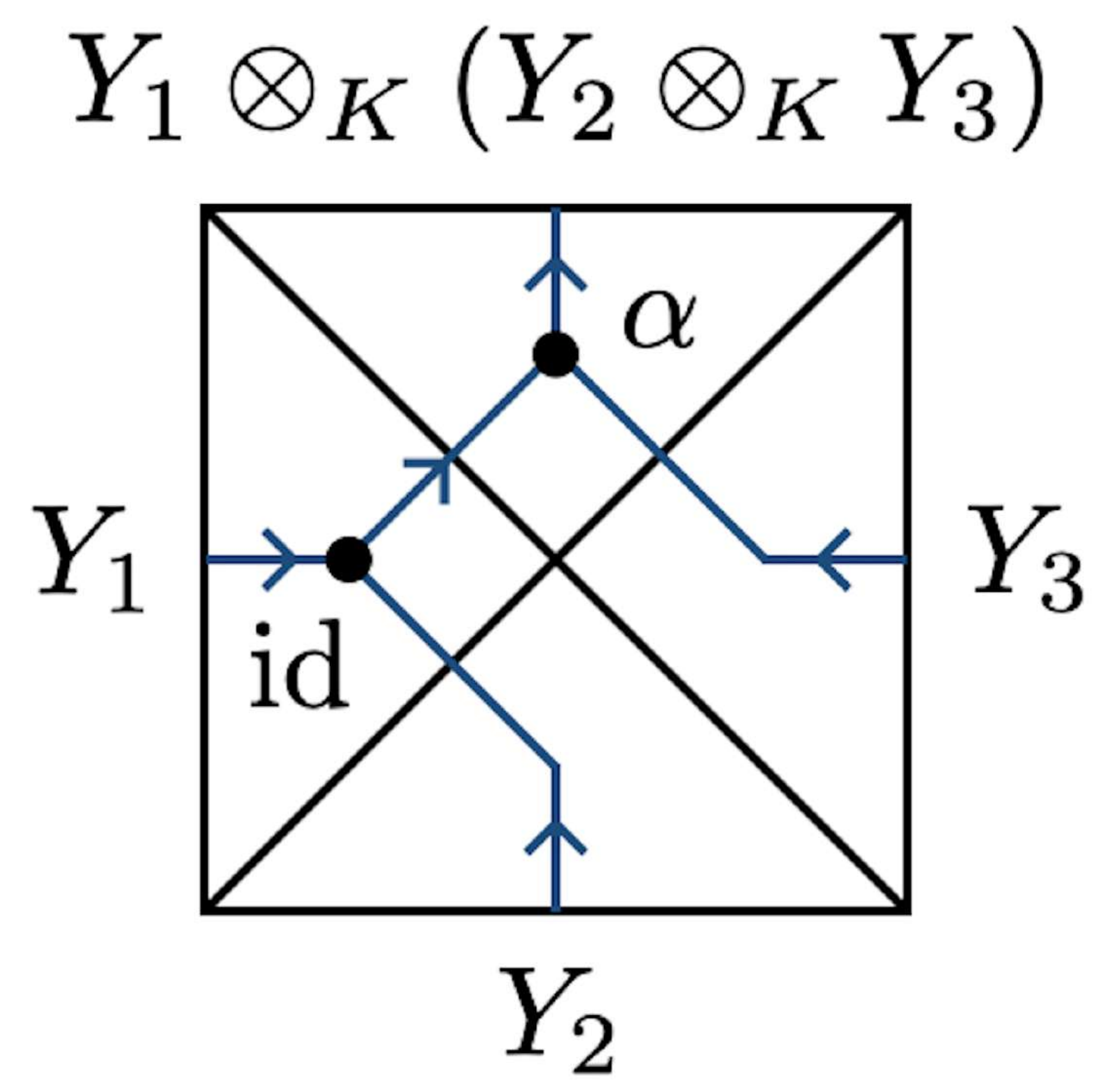} ~ = ~ \adjincludegraphics[valign = c, width = 3cm]{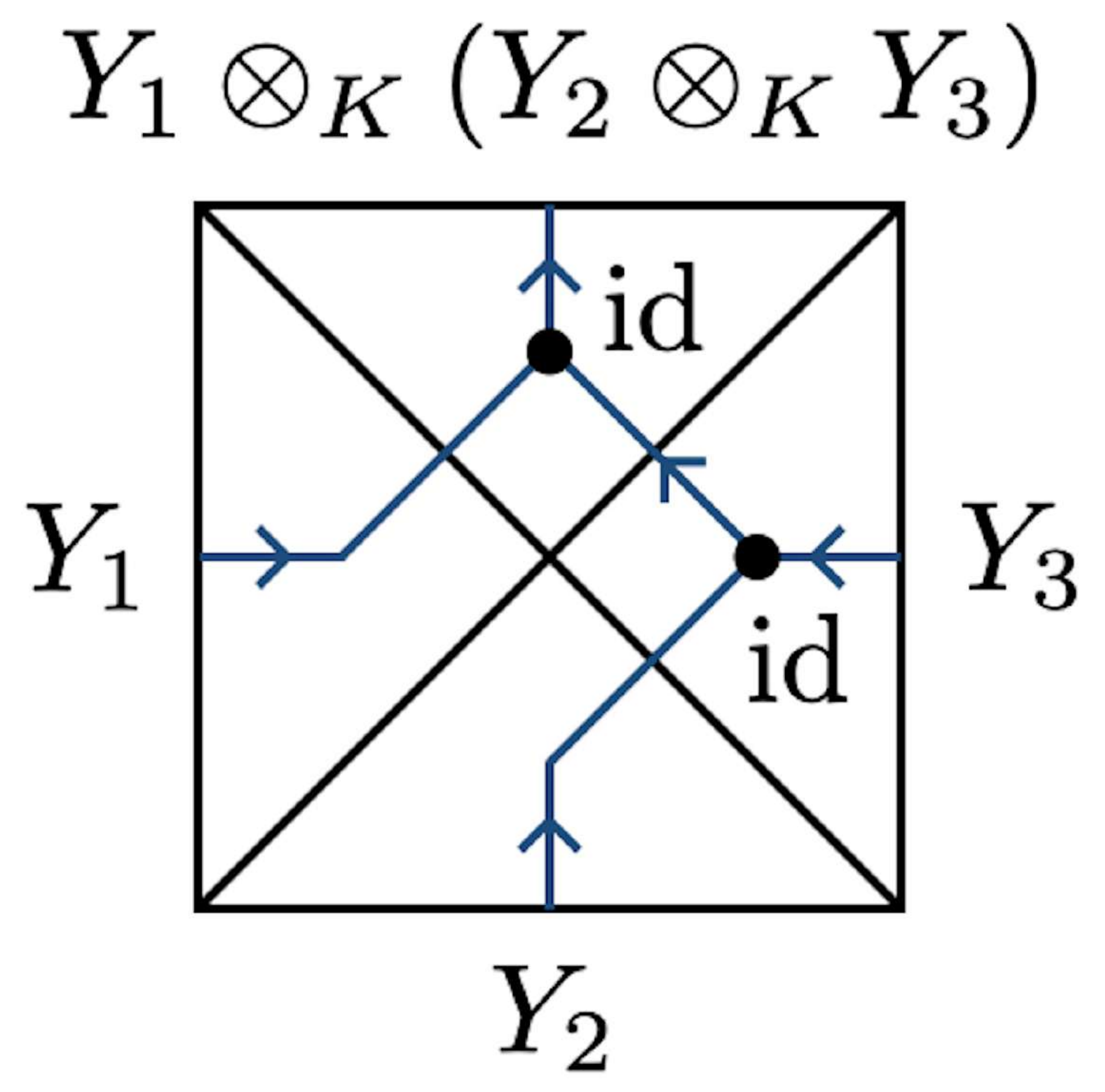}.
\label{eq: KMK associator}
\end{equation}
The square in the above equation represents a local patch of an arbitrary triangulated surface.
This equation \eqref{eq: KMK associator} implies that the symmetry of the state sum TQFT is precisely described by ${}_K \mathcal{M}_K$.

To argue that the state sum TQFT obtained above is the canonical ${}_K \mathcal{M}_K$ symmetric TQFT ${}_K \mathcal{M}$, we first notice that the state sum construction can be viewed as a generalized gauging of the trivial TQFT \cite{CRS2019}.
Here, the generalized gauging of a TQFT $Q$ with fusion category symmetry $\mathcal{C}$ is the procedure to condense a $\Delta$-separable symmetric Frobenius algebra object $A \in \mathcal{C}$ on a two-dimensional surface.
This procedure gives rise to a new TQFT $Q/A$ whose symmetry is given by the category ${}_A \mathcal{C}_A$ of $A$-$A$ bimodules in $\mathcal{C}$ \cite{CR2016, BT2018}.
To examine the relation between $Q$ and $Q/A$ in more detail, we consider the categories of boundary conditions of these TQFTs.
Let $\mathcal{B}$ be the category of boundary conditions of the original TQFT $Q$.
We note that $\mathcal{B}$ is the category of right $B$-modules in $\mathcal{C}$ for some $\Delta$-separable symmetric Frobenius algebra object $B \in \mathcal{C}$ because $\mathcal{B}$ is a left $\mathcal{C}$-module category \cite{EO2004, Ostrik2003}.
Then, the category of boundary conditions of the gauged TQFT $Q/A$ should be the category of left $A$-modules in $\mathcal{B}$ \cite{KORS2020}, which is a left ${}_A \mathcal{C}_A$-module category.
This is because the algebra object $A$ is condensed in the gauged theory and hence a boundary condition in $\mathcal{B}$ survives after gauging only when it is a left $A$-module.\footnote{The reason why we use left $A$-modules instead of right $A$-modules is that the category $\mathcal{B}$ of boundary conditions is supposed to be a left module category over $\mathcal{C}$.}
For example, gauging the algebra object $H^{*}$ in a $\mathrm{Rep}(H)$ symmetric TQFT ${}_K \mathcal{M}$ would result in a $\mathrm{Rep}(H^*)$ symmetric TQFT $({}_{H^*} \mathcal{M})_{K^{\mathrm{op}}} \cong {}_{\mathrm{Stab}_{K^{\mathrm{op}}}(V)} \mathcal{M}$ due to Theorem 3.10. of \cite{AM2007},\footnote{This generalized gauging is relevant for relating the state sum models to the anyon chain models as we mentioned in section \ref{sec: Introduction and summary}.\label{foot: mapping}} where $K$ is an $H$-simple left $H$-comodule algebra, $V$ is a left $K^{\mathrm{op}}$-module, and $\mathrm{Stab}_{K^{\mathrm{op}}}(V)$ is the Yan-Zhu stabilizer of $K^{\mathrm{op}}$ with respect to $V$ \cite{YZ1998}.
The $\mathrm{Rep}(H^*)$-module category ${}_{\mathrm{Stab}_{K^{\mathrm{op}}}(V)} \mathcal{M}$ does not depend on a choice of $V$ \cite{AM2007}.
In the case of the state sum TQFT with the input $K$, the condensed algebra object is $K \in \mathrm{Vec}$ and the category of boundary conditions of the original TQFT is $\mathrm{Vec}$.
Therefore, the category of boundary conditions of the state sum TQFT would be the category ${}_K \mathcal{M}$ of left $K$-modules.

\begin{figure}
\centering
\includegraphics[width = 3cm]{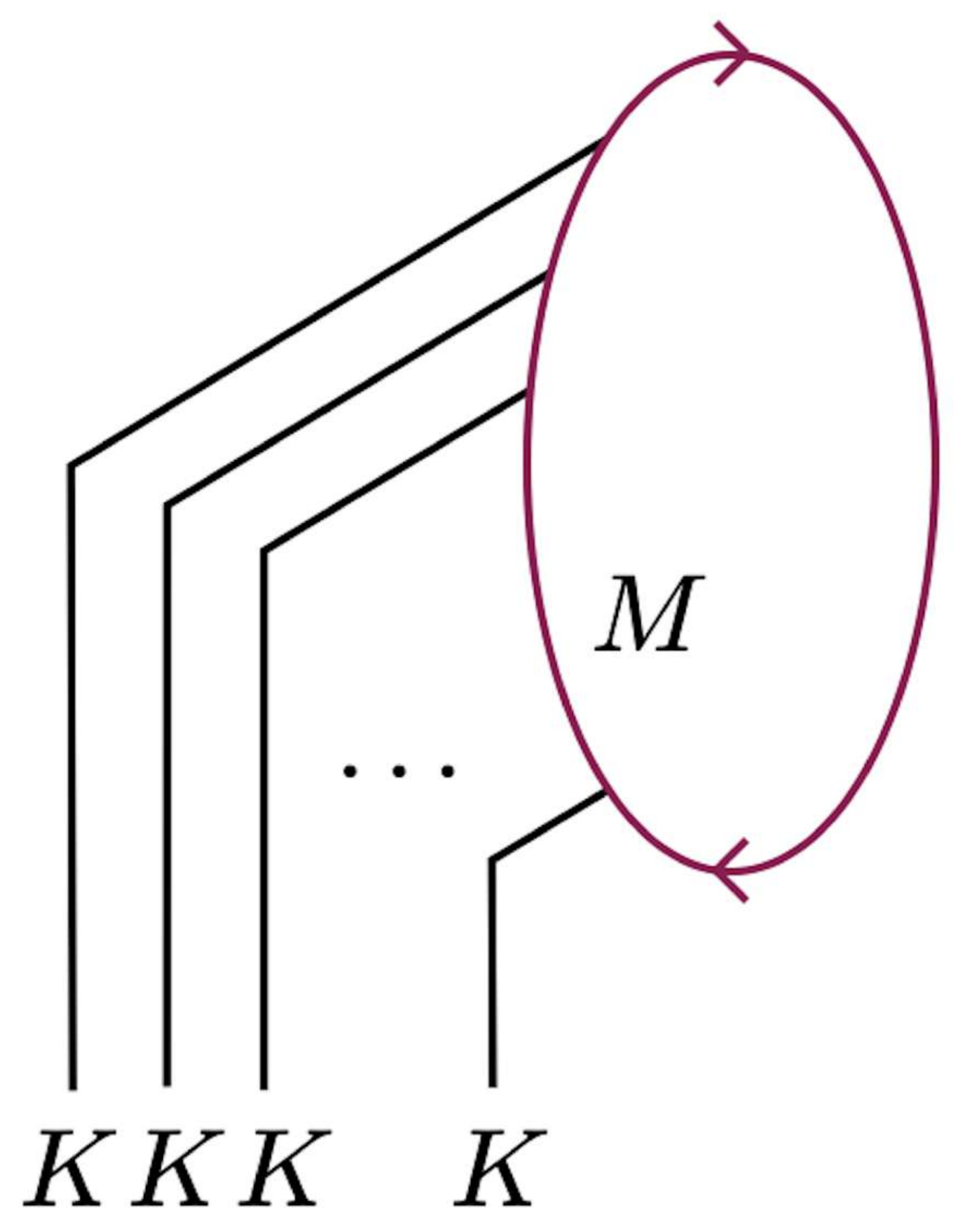}
\caption{The boundary state \eqref{eq: MPS} is represented by the above string diagram where the circle labeled by $M$ corresponds to taking the trace on $M$. This string diagram is invariant under the composition of the cylinder amplitude $Z_T(S^1 \times [0, 1])$ at the bottom.}
\label{fig: boundary state}
\end{figure}

We can also see this more explicitly by computing the action of the ${}_K \mathcal{M}_K$ symmetry on the boundary states of the state sum TQFT.
For this purpose, we first notice that a boundary of the state sum TQFT is equivalent to an interface between the state sum TQFT and the trivial TQFT.
Since the trivial TQFT is a state sum TQFT with the trivial input $\mathbb{C}$, interfaces are described by $K$-$\mathbb{C}$ bimodules, or equivalently, left $K$-modules.
The wave function of the boundary state $\bra{M}$ corresponding to the boundary condition $M \in {}_K \mathcal{M}$ is the linear map assigned to a triangulated disk
\begin{equation}
\bra{M} := \adjincludegraphics[valign = c, width = 2cm]{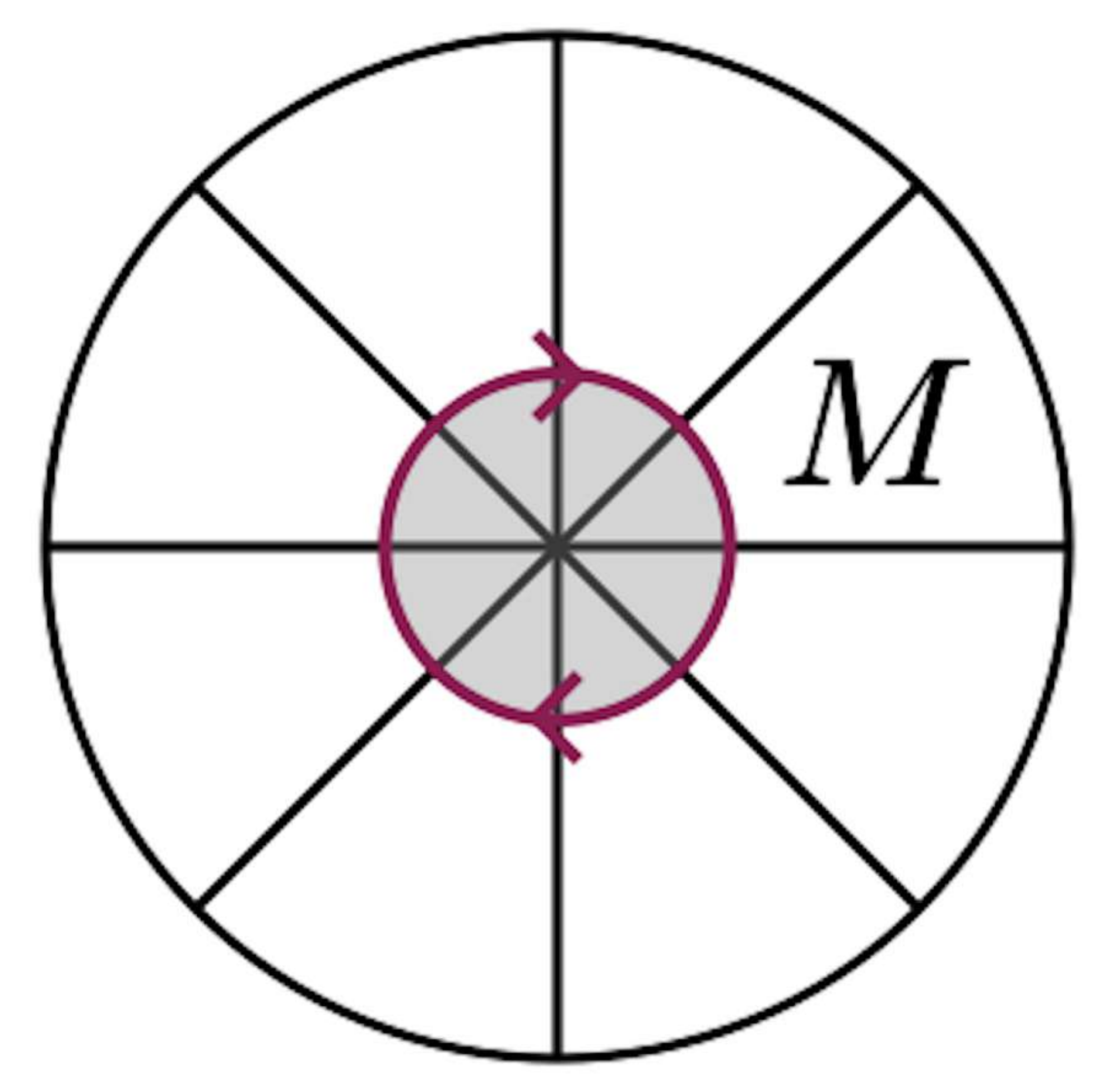}, 
\end{equation}
where the outer circle is an in-boundary and the inner circle labeled by $M$ is the interface between the trivial TQFT (shaded region) and the state sum TQFT with the input $K$ (unshaded region).
We can compute the linear map assigned to the above disk by using a left $K$-module $M$ instead of a $K$-$K$ bimodule $Y$ in eqs. \eqref{eq: Re}, \eqref{eq: Q(p, e)}, \eqref{eq: Pe}, and \eqref{eq: Ep} \cite{DKR2011}, see also appendix \ref{sec: State sum TQFTs on surfaces with interfaces} for more details.
Specifically, we can express the wave function $\bra{M}$ in the form of a matrix product state (MPS) as \cite{KTY2017, SR2017}
\begin{equation}
\bra{M} = \sum_{i_1, \cdots, i_N} \mathrm{tr}[T_M(e_{i_1}) \cdots T_M(e_{i_N})] \bra{e_{i_1}, \cdots, e_{i_N}},
\label{eq: MPS}
\end{equation}
where $\{e_i\}$ is a basis of $K$, $N$ is the number of edges on the boundary, and $T_M: K \rightarrow \mathrm{End}(M)$ is the $K$-module action on $M$.
In the string diagram notation, this MPS can be represented as shown in figure \ref{fig: boundary state}.
We notice that the MPS \eqref{eq: MPS} satisfies the additive property
\begin{equation}
\bra{M_1 \oplus M_2} = \bra{M_1} + \bra{M_2},
\label{eq: additive}
\end{equation}
due to which it suffices to consider simple modules $M_j \in {}_K \mathcal{M}$.

A topological defect $Y \in {}_K \mathcal{M}_K$ acts on a boundary state $\bra{M_j}$ by winding around the spatial circle.
We denote the wave function of the resulting state by $\bra{Y \cdot M_j}$.
By giving a specific triangulation of a disk as follows, we can compute the action of $Y$ on the boundary state $\bra{M_j}$ as
\begin{equation}
\bra{Y \cdot M_j} = \adjincludegraphics[valign = c, width = 3cm]{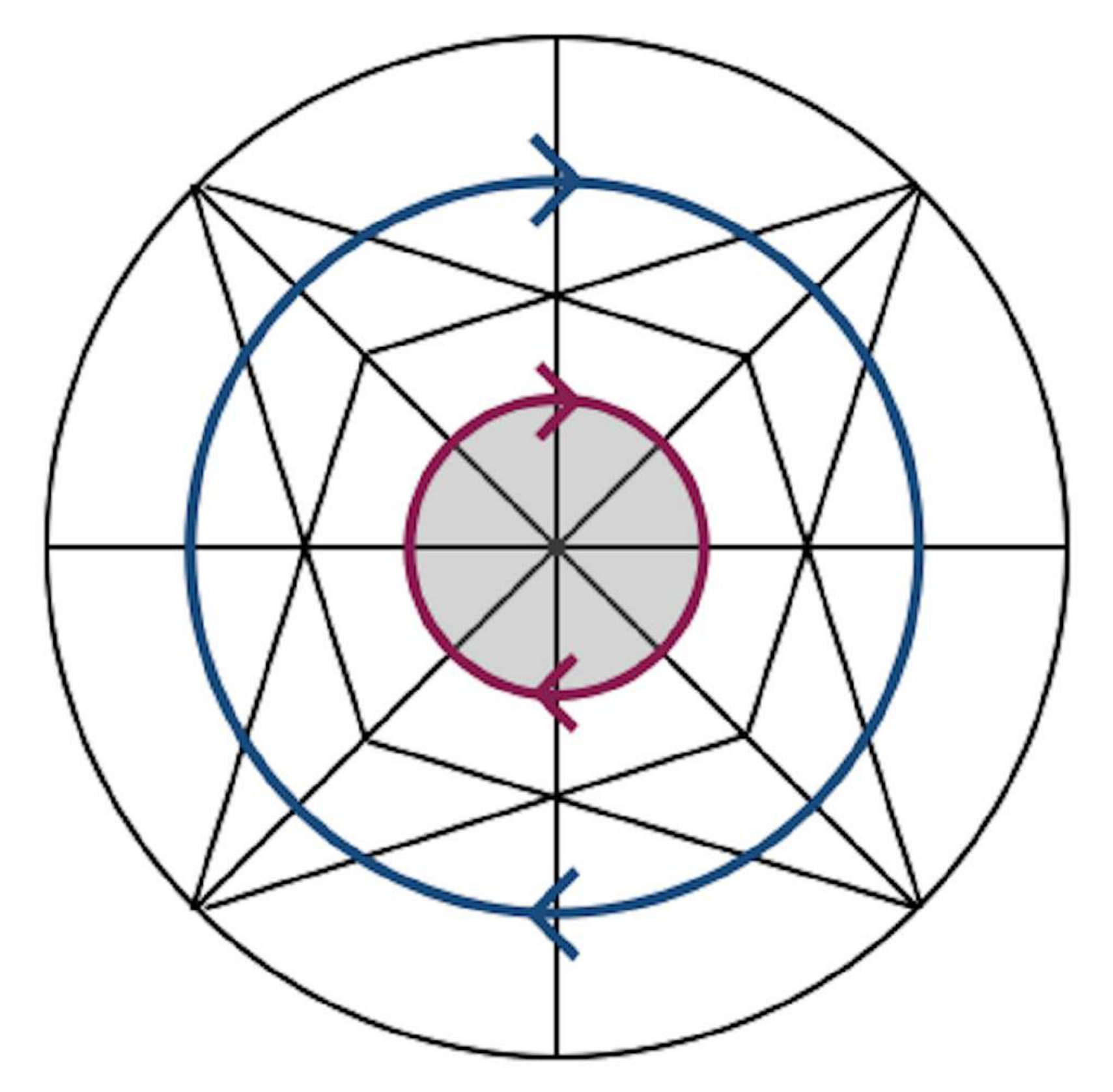} = \bra{Y \otimes_K M_j} = \sum_i N^Y_{ij} \bra{M_i},
\label{eq: KMK action}
\end{equation}
where the blue circle and the purple circle represent a topological defect $Y \in {}_K \mathcal{M}_K$ and a boundary condition $M_j \in {}_K \mathcal{M}$ respectively, and $N^Y_{ij}$ is a non-negative integer that appears in the direct sum decomposition of $Y \otimes_K M_j \cong \bigoplus_i N^Y_{ij} M_i$.
We note that the boundary states form a non-negative integer matrix representation (NIM rep) of the fusion ring of ${}_K \mathcal{M}_K$.
Equation \eqref{eq: KMK action} implies that the action of the ${}_K \mathcal{M}_K$ symmetry on  boundary conditions is described by a module category ${}_K \mathcal{M}$.
The module associativity constraint \eqref{eq: module associativity KM} is also captured in the same way as eq. \eqref{eq: KMK associator}.
Thus, the category of boundary conditions is given by the ${}_K \mathcal{M}_K$-module category ${}_K \mathcal{M}$, which indicates that the state sum TQFT with the input $K$ is a ${}_K \mathcal{M}_K$ symmetric TQFT ${}_K \mathcal{M}$.

\subsection{Pullback of state sum TQFTs}
\label{sec: Rep(H) symmetry of the state sum TQFT}
When $K$ is a left $H$-comodule algebra, the ${}_K \mathcal{M}_K$ symmetric TQFT ${}_K \mathcal{M}$ can be pulled back to a $\mathrm{Rep}(H)$ symmetric TQFT by a tensor functor $F_K: \mathrm{Rep}(H) \rightarrow {}_K \mathcal{M}_K$.
Accordingly, the symmetry of the state sum TQFT with the input $K$ can be regarded as $\mathrm{Rep}(H)$.
Specifically, when a two-dimensional surface $\Sigma$ is decorated by a topological defect network associated with the $\mathrm{Rep}(H)$ symmetry, the assignment of the vector spaces \eqref{eq: Re}, \eqref{eq: Q(p, e)} and linear maps \eqref{eq: Pe}, \eqref{eq: Ep} are modified as follows:
\begin{align}
R_e & := 
\begin{cases}
K \quad & \text{when } e \text{ does not intersect a topological defect}, \\
F_K(V) \quad & \text{when } e \text{ intersects a topological defect } V \in \mathrm{Rep}(H).
\end{cases}
\label{eq: Re Rep(H)}\\
Q_{(p, e)} & := 
\begin{cases}
K \quad & \text{when } e \text{ does not intersect a topological defect}, \\
F_K(V) \quad & \text{when a topological defect } V \text{ goes into } p \text{ across } e, \\
F_K(V)^{*} \quad & \text{when a topological defect } V \text { goes out of } p \text{ across } e. 
\end{cases}
\label{eq: Q(p, e) Rep(H)}\\
P_e & := 
\begin{cases}
\Delta_K \circ \eta_K \quad & \text{when } e \text{ does not intersect a topological defect}, \\
\mathrm{coev}_{F_K(V)} \quad & \text{when } e \text{ intersects a topological defect } V.
\end{cases}
\label{eq: Pe Rep(H)}\\
E_p & := 
\begin{cases}
\text{(i)} & \epsilon_K \circ m_K \circ (m_K \otimes \mathrm{id}_K), \\
\text{(ii)} & \mathrm{ev}_{F_K(V)} \circ (\mathrm{id}_{F_K(V)^*} \otimes \rho_{F_K(V)}^R), \\
\text{(iii)} & \mathrm{ev}_{F_K(V)} \circ (\mathrm{id}_{F_K(V)^*} \otimes \rho_{F_K(V)}^L), \\
\text{(iv)} & \mathrm{ev}_{F_K(V_3)} \circ (\mathrm{id}_{F_K(V_3)^*} \otimes (F_K(f) \circ J_{V_1, V_2} \circ \pi_{F_K(V_1), F_K(V_2)})), \\
\text{(v)} & \mathrm{ev}_{F_K(V_1) \otimes F_K(V_2)} \circ ( \mathrm{id}_{(F_K(V_1) \otimes F_K(V_2))^*} \otimes (\iota_{F_K(V_1), F_K(V_2)} \circ J_{V_1, V_2}^{-1} \circ F_K(g))).
\end{cases}
\label{eq: Ep Rep(H)}
\end{align}
(i)--(v) in eq. \eqref{eq: Ep Rep(H)} refer to the configurations \eqref{eq: configurations} of topological defects on a face $p$, where the topological defects and junctions of ${}_K \mathcal{M}_K$ symmetry are replaced with those of $\mathrm{Rep}(H)$ symmetry, i.e. $V, V_1, V_2, V_3 \in \mathrm{Rep}(H)$, $f \in \mathrm{Hom}_{\mathrm{Rep}(H)}(V_1 \otimes V_2, V_3)$, and $g \in \mathrm{Hom}_{\mathrm{Rep}(H)}(V_3, V_1 \otimes V_2)$.
The linear map $J_{V_1, V_2}: F_K(V_1) \otimes_K F_K(V_2) \rightarrow F_K(V_1 \otimes V_2)$ is a natural isomorphism associated with the tensor functor $F_K: \mathrm{Rep}(H) \rightarrow {}_K \mathcal{M}_K$.
Here, we again point out that (i)--(iii) and (v) are obtained from (iv) by choosing topological defects and junctions appropriately.
By substituting the above definitions to eq. \eqref{eq: linear map ZT}, we assign a linear map $Z_T(\Sigma)$ to a triangulated surface $\Sigma$ decorated by defects of the $\mathrm{Rep}(H)$ symmetry.
Then, as in the previous subsection, we obtain a TQFT with $\mathrm{Rep}(H)$ symmetry by restricting the domain of the linear map $Z_T(\Sigma)$ to the image of the cylinder amplitude $Z_T(\partial_{\mathrm{in}} \Sigma \times [0, 1])$.
The topological invariance of this $\mathrm{Rep}(H)$ symmetric TQFT readily follows from the topological invariance of the ${}_K \mathcal{M}_K$ symmetric state sum TQFT defined in the previous subsection.
It is also straightforward to check that the difference between two possible resolutions of a quadrivalent junction into two trivalent junctions is described by the associator of $\mathrm{Rep}(H)$.

We can compute the action of the $\mathrm{Rep}(H)$ symmetry on the boundary states as follows.
We first recall that the action of the ${}_K \mathcal{M}_K$ symmetry on the boundary states is represented by the ${}_K \mathcal{M}_K$-module action on ${}_K \mathcal{M}$ as eq. \eqref{eq: KMK action}.
This ${}_K \mathcal{M}_K$ action induces a $\mathrm{Rep}(H)$ action on the boundary states because topological defects associated with the $\mathrm{Rep}(H)$ symmetry are mapped to ${}_K \mathcal{M}_K$ symmetry defects by a tensor functor $F_K: \mathrm{Rep}(H) \rightarrow {}_K \mathcal{M}_K$.
Specifically, a topological defect $V \in \mathrm{Rep}(H)$ acts on a boundary state $\bra{M}$ as
\begin{equation}
\bra{V \cdot M} = \bra{F_K(V) \otimes_K M} = \bra{V \overline{\otimes} M},
\label{eq: Rep(H) action on boundary states}
\end{equation}
where $\overline{\otimes}$ represents the $\mathrm{Rep}(H)$-module action on ${}_K \mathcal{M}$, see section \ref{sec: Representation categories and module categories}.
Combined with the additive property \eqref{eq: additive}, this equation implies that the boundary states form a NIM rep of the fusion ring of $\mathrm{Rep}(H)$.
Furthermore, if we extend the definition \eqref{eq: Ep Rep(H)} in the presence of boundaries as we describe in appendix \ref{sec: State sum TQFTs on surfaces with interfaces}, we find that the two possible resolutions of a junction on the boundary are related to each other by the module associativity constraint.
Therefore, the category of boundary conditions of this TQFT is given by a $\mathrm{Rep}(H)$-module category ${}_K \mathcal{M}$.
Since every semisimple indecomposable module category over $\mathrm{Rep}(H)$ is equivalent to the category ${}_K \mathcal{M}$ of left modules over an $H$-simple left $H$-comodule algebra $K$, we conclude that any semisimple indecomposable TQFTs with $\mathrm{Rep}(H)$ symmetry are obtained via the above state sum construction.\footnote{The state sum TQFTs with ordinary finite group symmetry $G$ are originally discussed in \cite{Tur1999, KTY2017, SR2017}. We can reproduce these TQFTs as special cases where the Hopf algebra $H$ is a dual group algebra $\mathbb{C}[G]^*$.
This is because an ordinary finite group symmetry $G$ is described by $\mathrm{Rep}(\mathbb{C}[G]^*)$ as a fusion category symmetry.
We note that a left $\mathbb{C}[G]^*$-comodule algebra $K$ is a $G$-equivariant algebra, which agrees with the input algebra used in \cite{Tur1999, KTY2017, SR2017}.}
Here, we recall that an $H$-simple left $H$-comodule algebra is semisimple \cite{Skr2007} and hence can be used as an input of the state sum construction.

\subsection{Commuting projector Hamiltonians}
\label{sec: Commuting projector Hamiltonians for gapped phases with Rep(H) symmetry}
In this subsection, we write down $\mathrm{Rep}(H)$ symmetric commuting projector Hamiltonians whose ground states are described by the $\mathrm{Rep}(H)$ symmetric TQFTs ${}_K \mathcal{M}$.
For concreteness, we choose $K = A^{\mathrm{op}} \# H^{\mathrm{cop}}$ where $A$ is an $H$-simple left $H$-module algebra.
This is always possible because every semisimple indecomposable module category over $\mathrm{Rep}(H)$ is equivalent to ${}_{A^{\mathrm{op}} \# H^{\mathrm{cop}}} \mathcal{M}$ as we discussed in section \ref{sec: Representation categories and module categories}.
We first define the Hilbert space on a circular lattice $T(S^1)$, which is a triangulation of a circle $S^1$.
The Hilbert space $\mathcal{H}$ on the lattice $T(S^1)$ is the tensor product of local Hilbert spaces $K_i := K$ on edges $i = 1, 2, \cdots, N$, i.e. $\mathcal{H} := \bigotimes_{i} K_i$.
We define a commuting projector Hamiltonian $H$ acting on this Hilbert space $\mathcal{H}$ as \cite{WW2017, Bullivant2018}\footnote{We can also define the Hilbert space and the Hamiltonian of a twisted sector analogously. Specifically, when a topological defect $V \in \mathrm{Rep}(H)$ intersects an edge $i$, the local Hilbert space on $i$ is given by $F_K(V) = V \otimes K$ according to eq. \eqref{eq: Re Rep(H)}. Namely, we attach a vector space $V$ to a topological defect $V \in \mathrm{Rep}(H)$. This generalizes the twisted sectors of finite gauge theories discussed in \cite{HO2021}.}
\begin{equation}
H := \sum_i (1 - h_{i, i+1}), \quad h_{i, i+1} := \Delta_K \circ m_K: K_i \otimes K_{i+1} \rightarrow K_{i} \otimes K_{i+1},
\label{eq: commuting projector Hamiltonian}
\end{equation}
where the comultiplication $\Delta_K$ for the Frobenius algebra structure on $K$ is given by eq. \eqref{eq: Frobenius structure}.
The fact that $K$ is a $\Delta$-separable symmetric Frobenius algebra \eqref{eq: Delta-separable symmetric} guarantees that the linear map $h_{i, i+1}$ becomes a local commuting projector, i.e. $h_{i, i+1} h_{j, j+1} = h_{j, j+1} h_{i, i+1}$ and $h_{i, i+1}^2 = h_{i, i+1}$.
The local commuting projector $h_{i, i+1}$ can also be written in terms of a string diagram as
\begin{equation}
h_{i, i+1} = \adjincludegraphics[valign = c, width = 1.8cm]{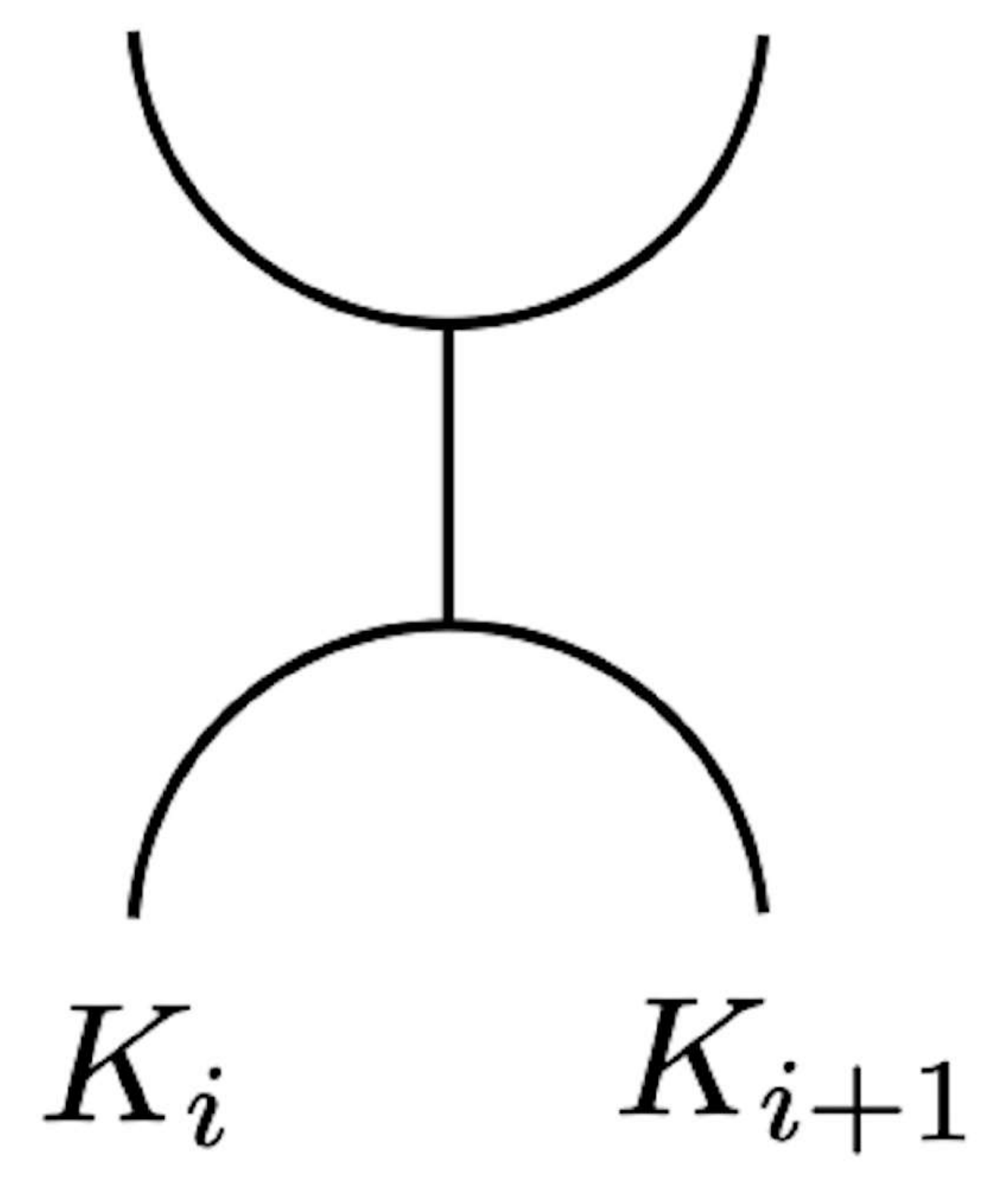} ~ = ~  \adjincludegraphics[valign = c, width = 1.8cm]{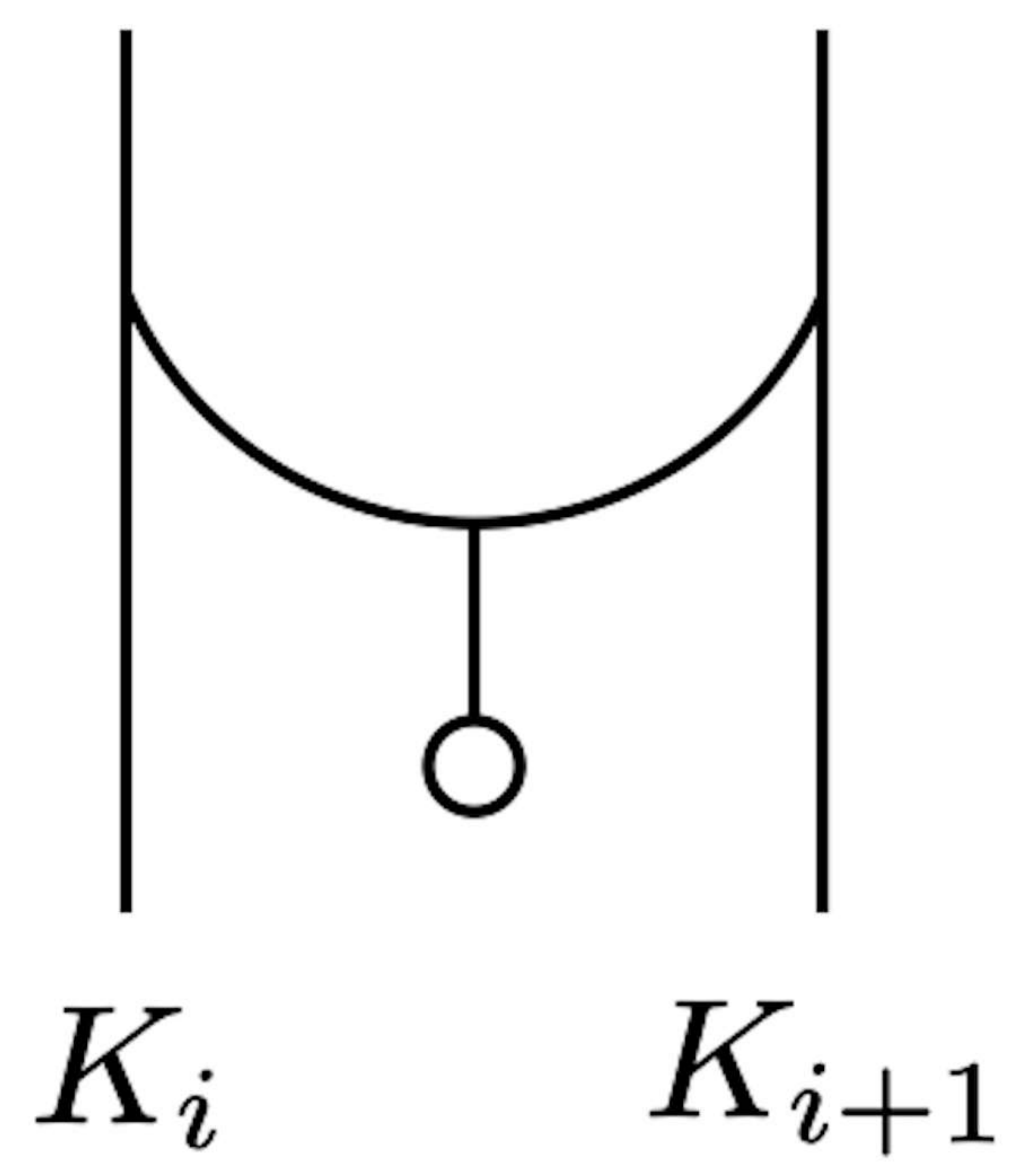},
\label{eq: local CP}
\end{equation}
where we used the Frobenius relation \eqref{eq: Frobenius relation}.
The projector $\Pi$ to the subspace of $\mathcal{H}$ spanned by the ground states of the Hamiltonian \eqref{eq: commuting projector Hamiltonian} is given by the composition of the local commuting projectors $h_{i, i+1}$ for all edges $i = 1, 2, \cdots, N$.
This projector $\Pi: \mathcal{H} \rightarrow \mathcal{H}$ can be represented by the following string diagram:
\begin{equation}
\Pi = \adjincludegraphics[valign = c, width = 6cm]{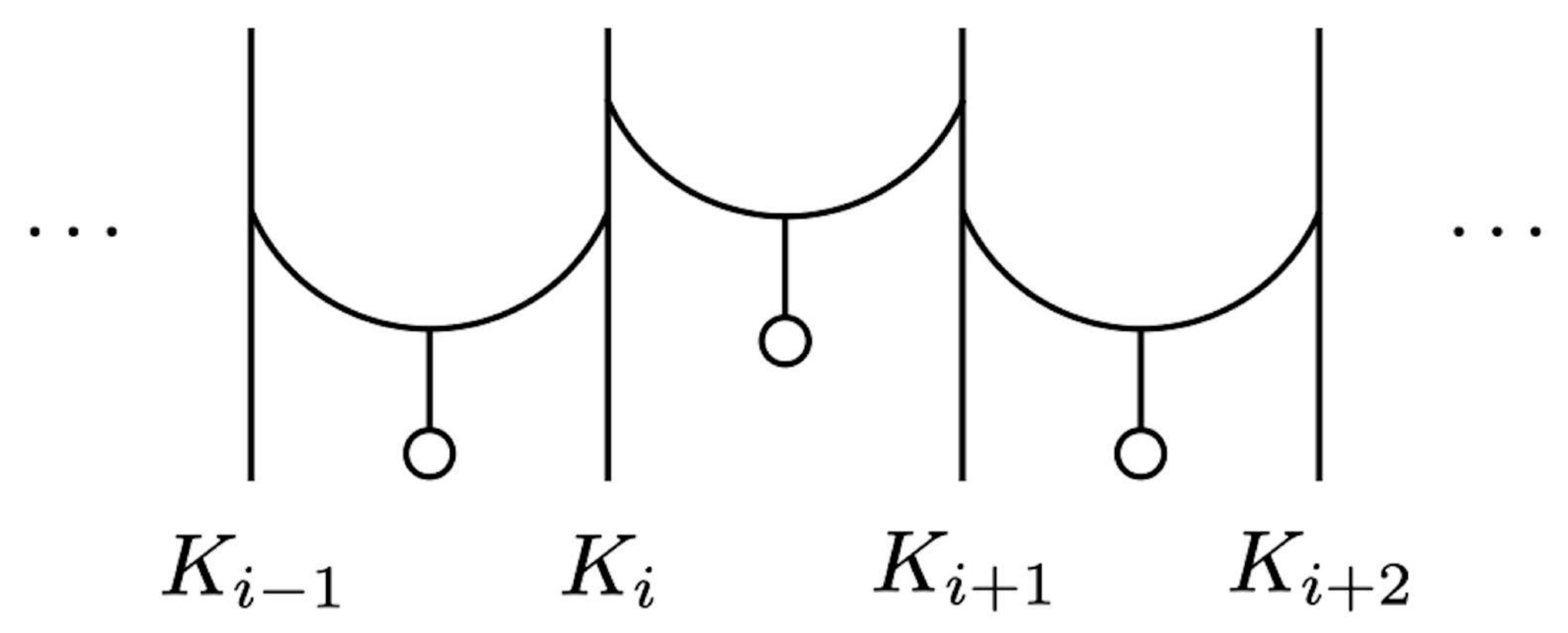}.
\end{equation}
This coincides with the string diagram representation of the linear map $Z_T(S^1 \times [0, 1])$ assigned to a triangulated cylinder $S^1 \times [0, 1]$.
Therefore, the ground states of the commuting projector Hamiltonian \eqref{eq: commuting projector Hamiltonian} agree with the vacua of the state sum TQFT whose input algebra is $K$.

We can define the action of the $\mathrm{Rep}(H)$ symmetry on the lattice Hilbert space $\mathcal{H}$ via the $H$-comodule structure on $K$.
Concretely, the adjoint of the action $U_V: \mathcal{H} \rightarrow \mathcal{H}$ associated with a topological defect $V \in \mathrm{Rep}(H)$ is given by the following string diagram
\begin{equation}
U_V^{\dagger} = \adjincludegraphics[valign = c, width = 4cm]{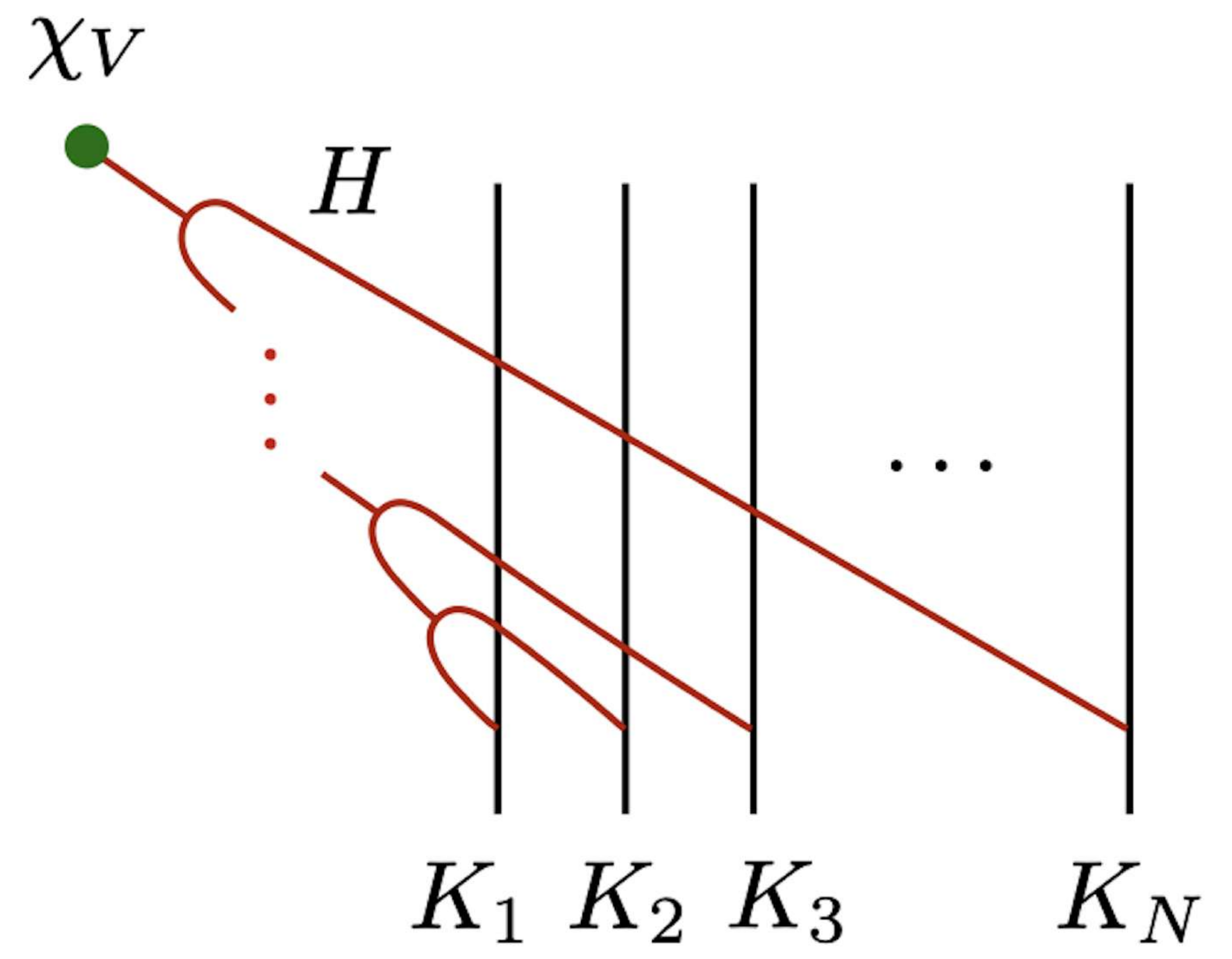} ~ = ~ \adjincludegraphics[valign = c, width = 4cm]{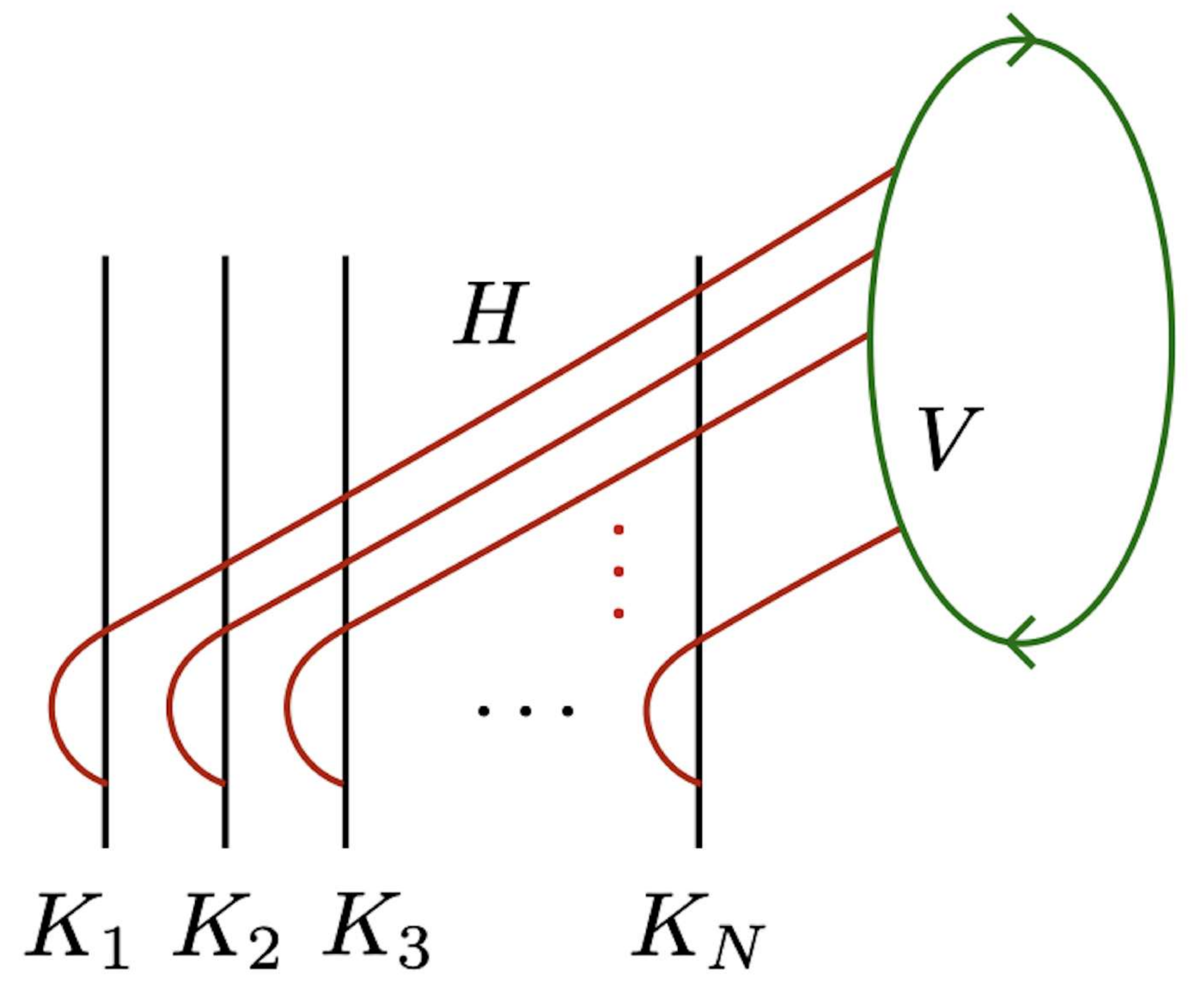},
\label{eq: Rep(H) action}
\end{equation}
where $\chi_V \in H^*$ is the character of the representation $V \in \mathrm{Rep}(H)$, which is defined as the trace of the left $H$-module action on $V$.\footnote{We note that the $\mathrm{Rep}(H)$ action \eqref{eq: Rep(H) action} does not involve the algebra structure on $K$, which means that we can define the $\mathrm{Rep}(H)$ action on the lattice as long as the local Hilbert space is a left $H$-comodule.}
The above $\mathrm{Rep}(H)$ action obeys the fusion rule of $\mathrm{Rep}(H)$, i.e. $U_{V_i} U_{V_j} = \sum_{k} N_{ij}^k U_{V_k}$ for irreducible representations $V_i, V_j \in \mathrm{Rep}(H)$, where $N_{ij}^k$ is a fusion coefficient $V_i \otimes V_j \cong \bigoplus_k N_{ij}^k V_k$.
The cyclic symmetry of the character guarantees that the action \eqref{eq: Rep(H) action} is well-defined on a periodic lattice $T(S^1)$.
Moreover, this action is faithful since the left $H$-comodule action on $K = A^{\mathrm{op}} \# H^{\mathrm{cop}}$ is inner-faithful.\footnote{Another choice of $K$ is also possible as long as the $\mathrm{Rep}(H)$ symmetry acts faithfully on the lattice Hilbert space.}

Let us now show the commutativity of the $\mathrm{Rep}(H)$ action \eqref{eq: Rep(H) action} and the commuting projector Hamiltonian \eqref{eq: commuting projector Hamiltonian}.
It suffices to check that the $\mathrm{Rep}(H)$ action commutes with each local commuting projector $h_{i, i+1}$.
Namely, we need to check
\begin{equation}
\adjincludegraphics[valign = c, width = 2cm]{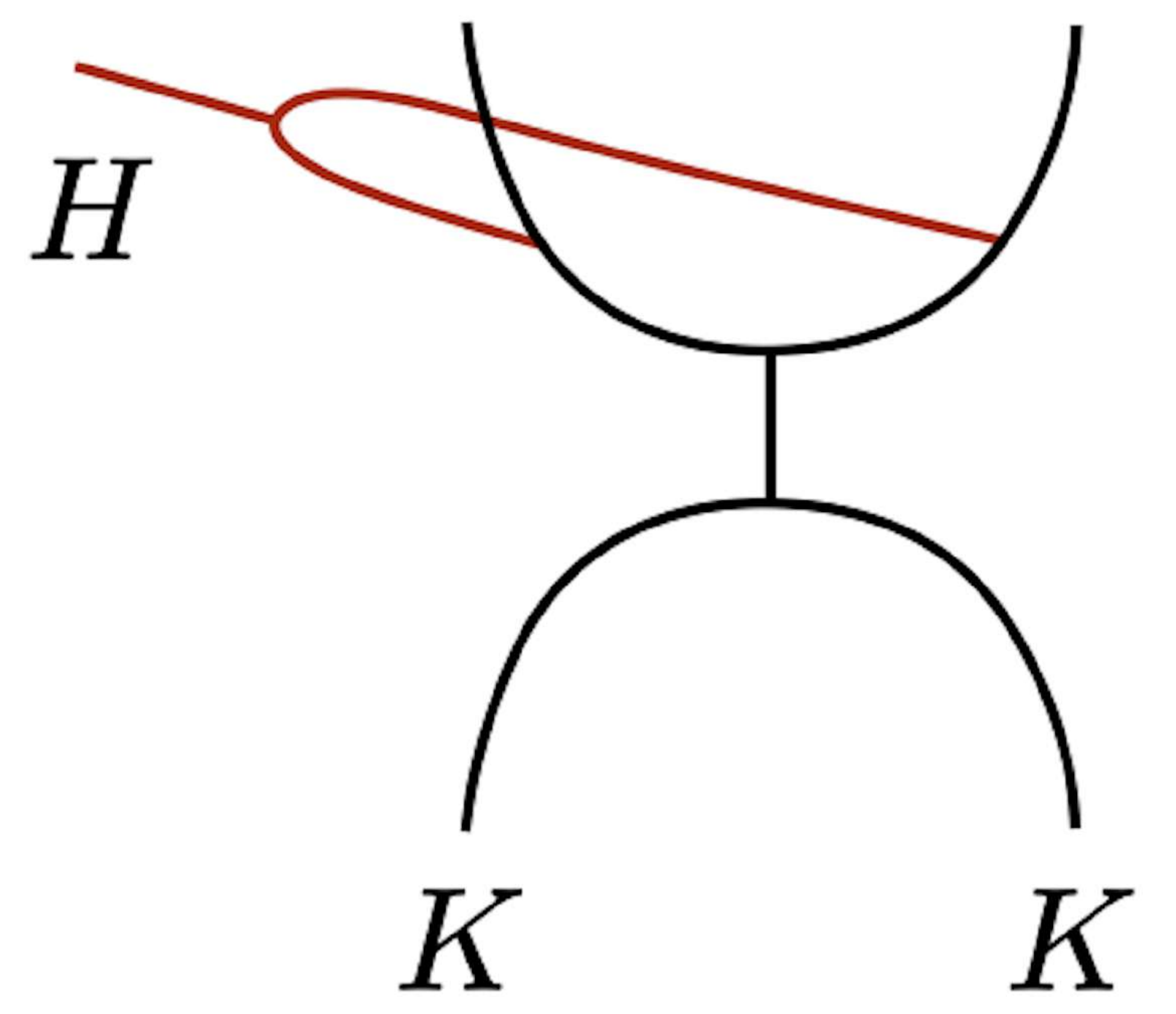} ~ = ~ 
\adjincludegraphics[valign = c, width = 2cm]{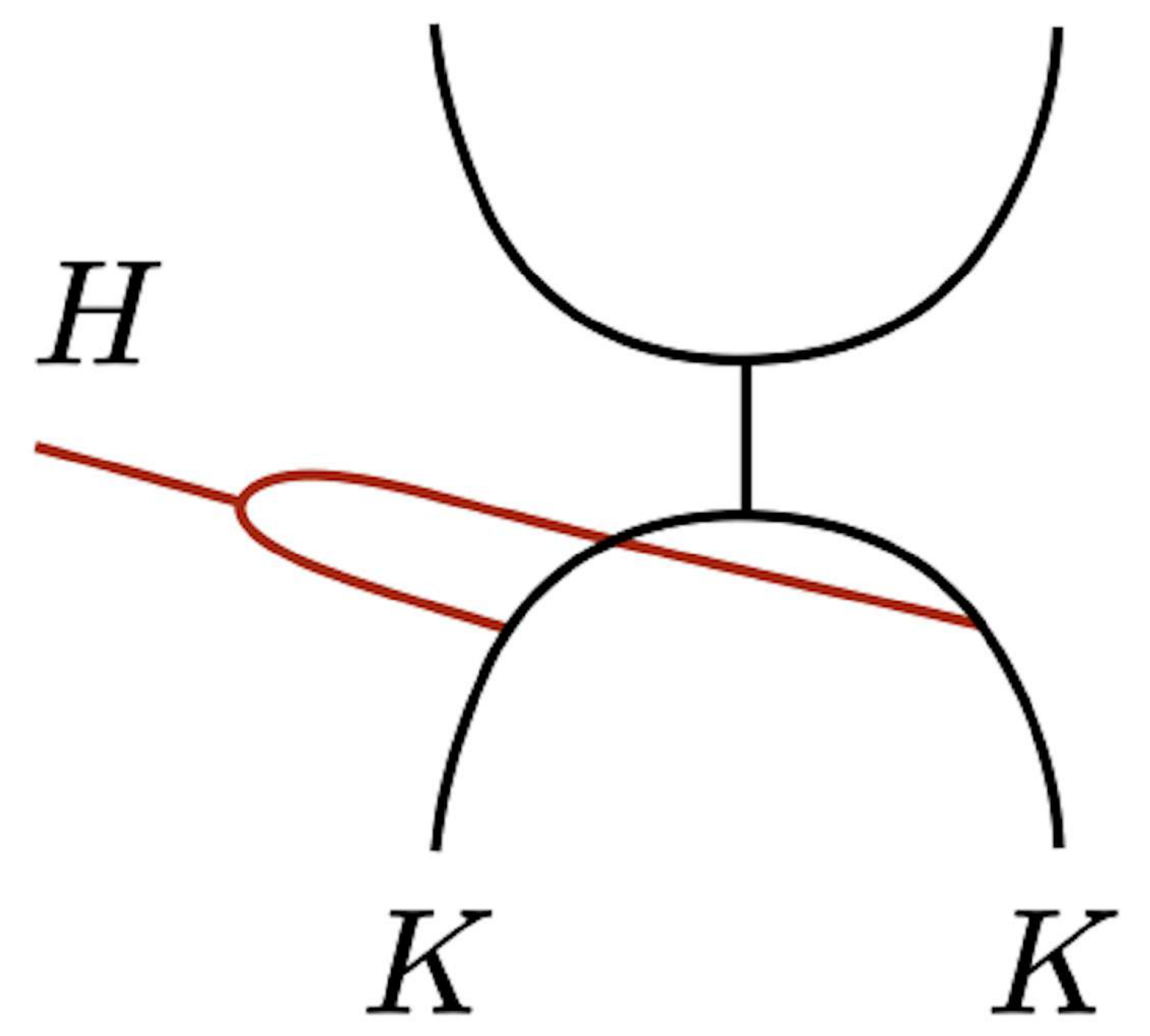}, \text{ or equivalently, }
\adjincludegraphics[valign = c, width = 2cm]{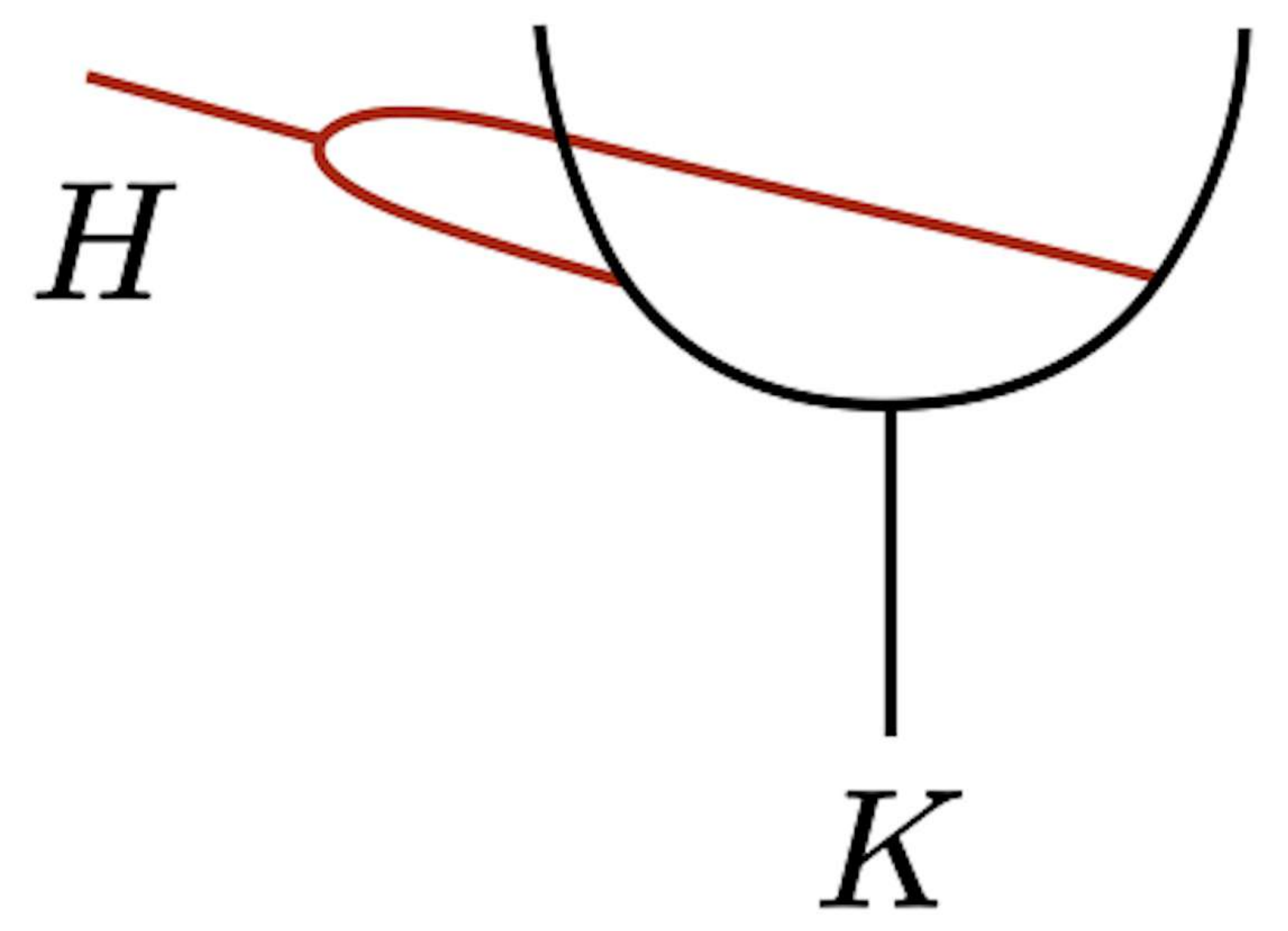} ~ = ~ 
\adjincludegraphics[valign = c, width = 1.6cm]{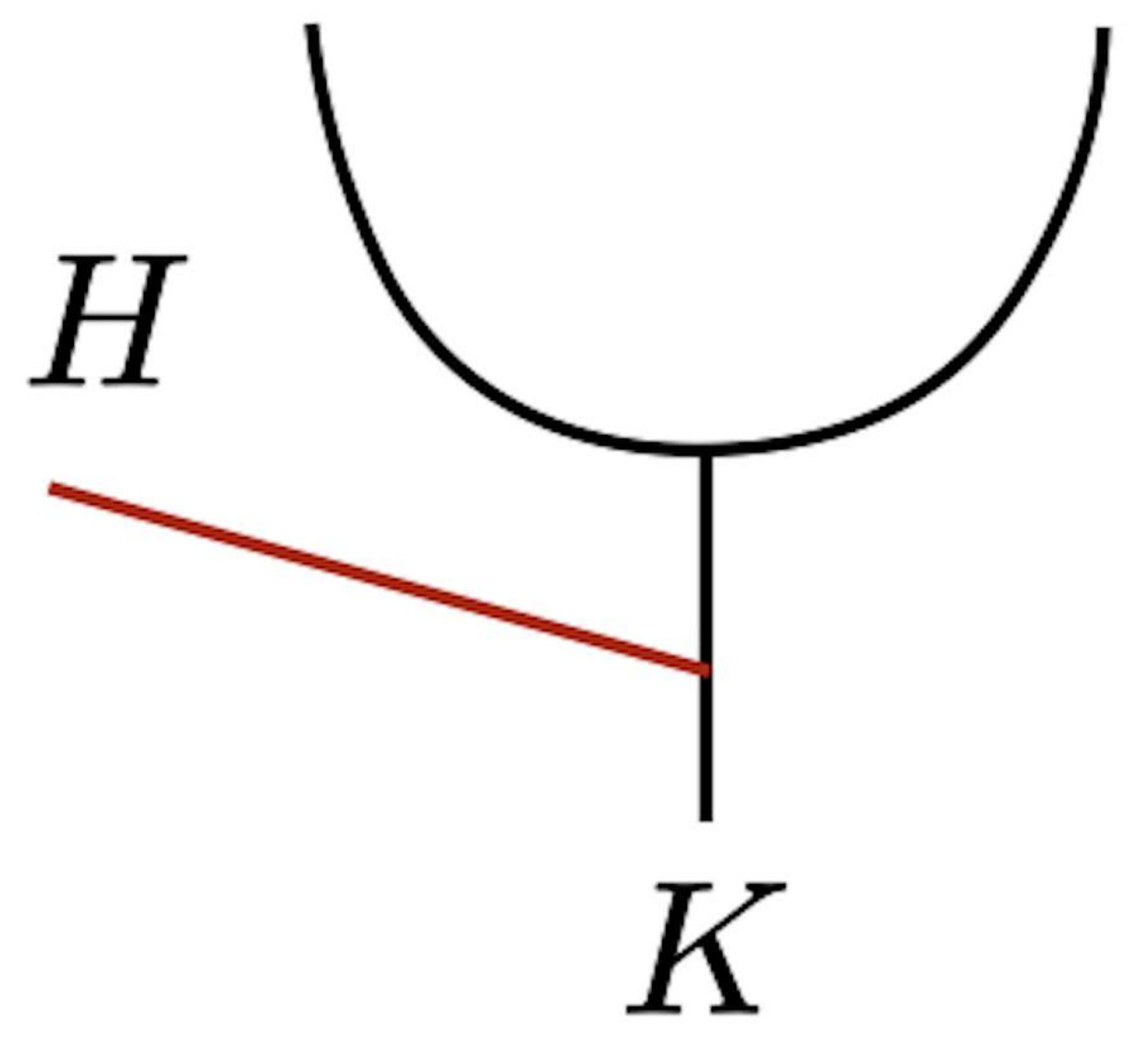}.
\label{eq: commutativity}
\end{equation}
The first equality follows from the second equality because $K$ is a left $H$-comodule algebra.
Conversely, we can derive the second equality from the first equality by composing a unit at the bottom of the diagram.

To show eq. \eqref{eq: commutativity}, we first notice that the counit $\epsilon$ given by eq. \eqref{eq: Frobenius structure} satisfies
\begin{equation}
\adjincludegraphics[valign = c, width = 1cm]{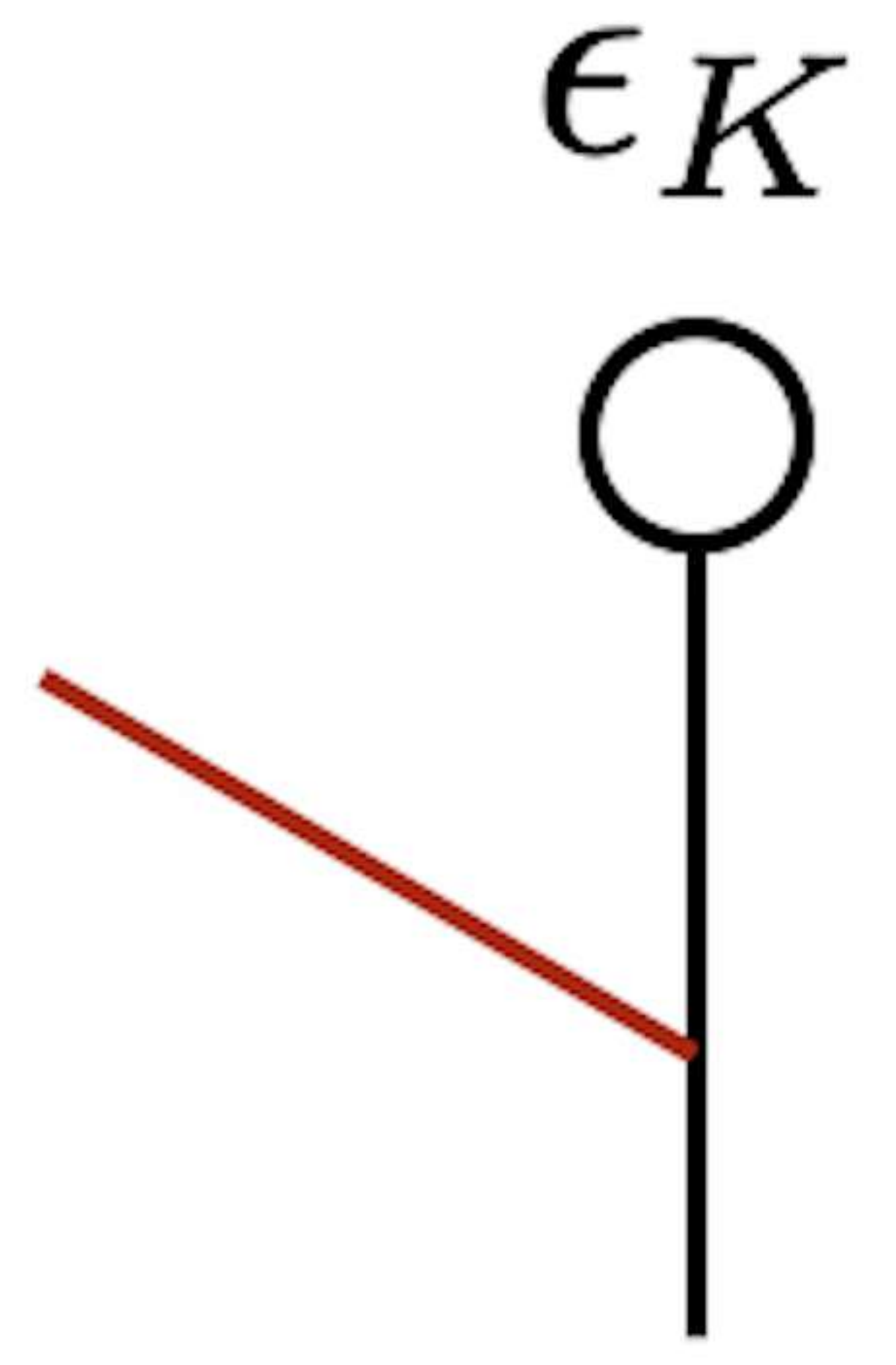} ~ = 
\adjincludegraphics[valign = c, width = 2cm]{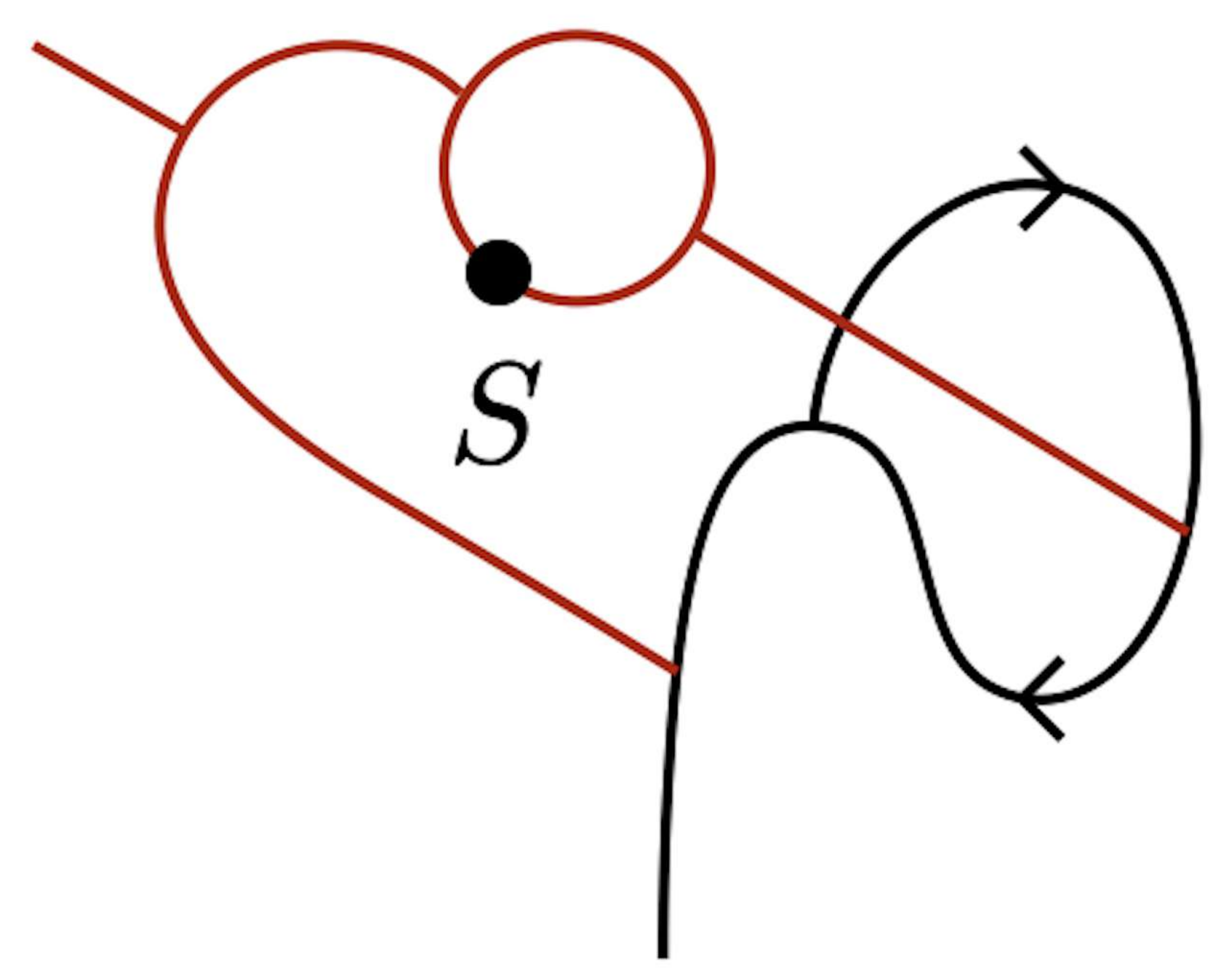} ~ = 
\adjincludegraphics[valign = c, width = 1.8cm]{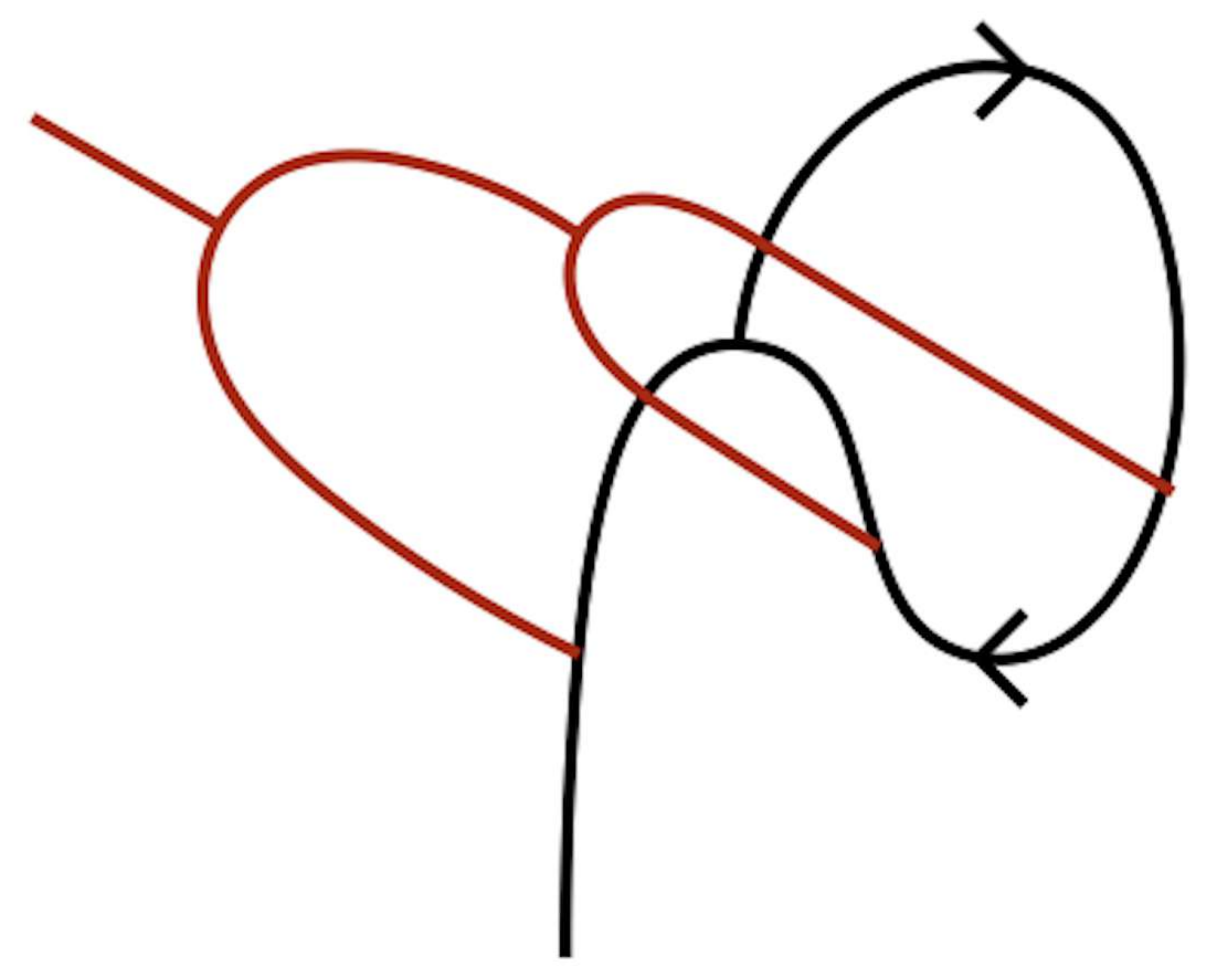} ~ = 
\adjincludegraphics[valign = c, width = 1.8cm]{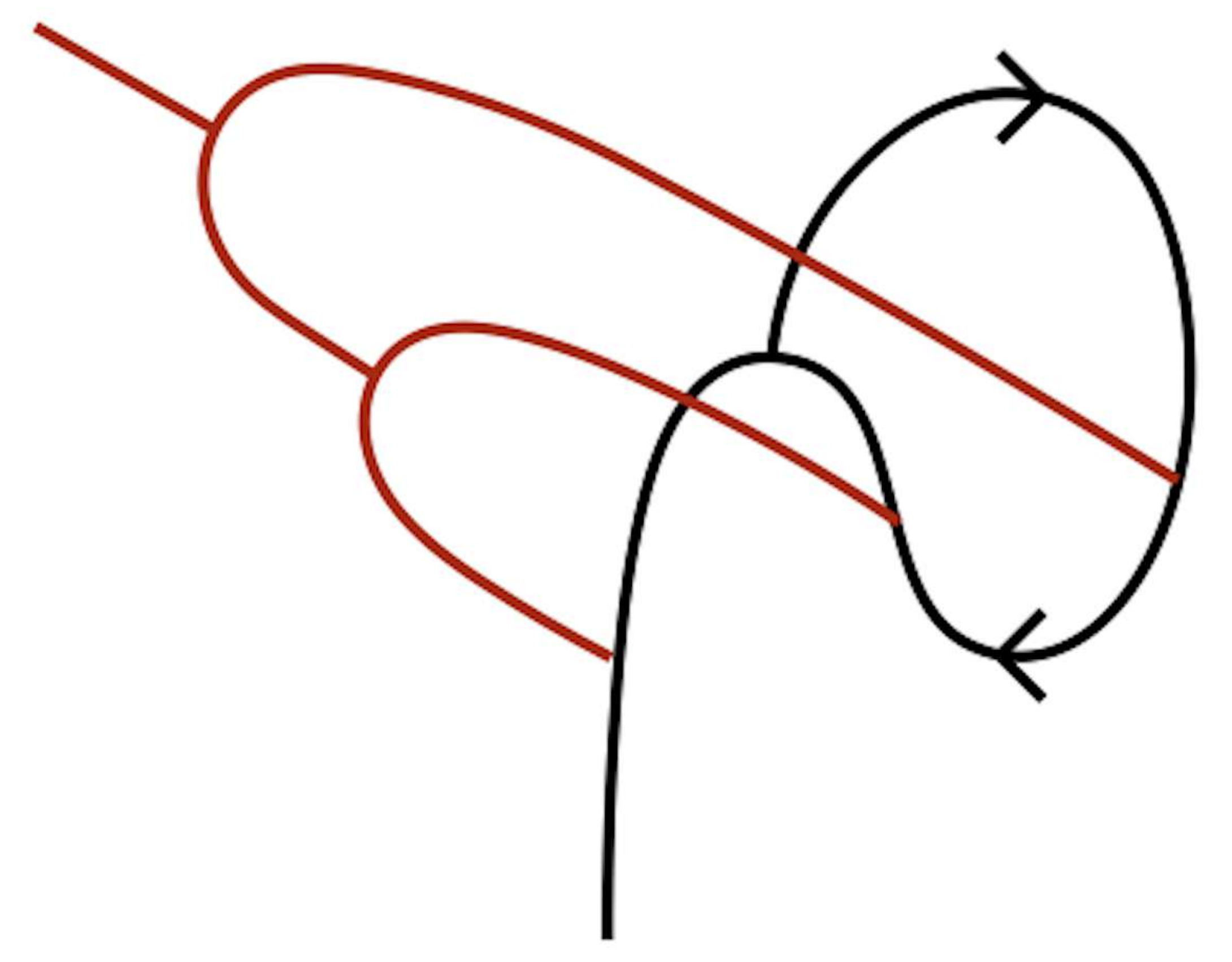} ~ = 
\adjincludegraphics[valign = c, width = 1.5cm]{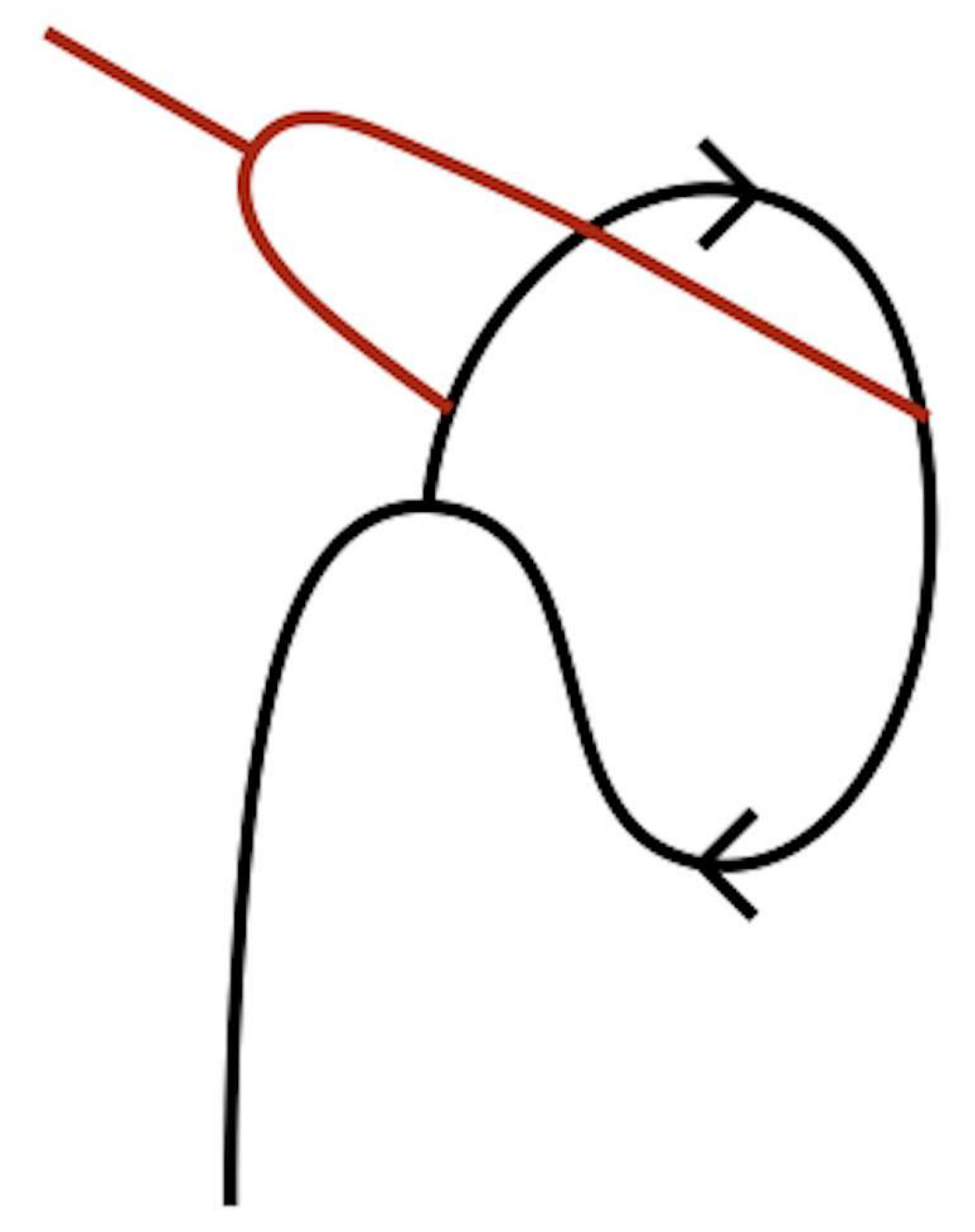} ~ = 
\adjincludegraphics[valign = c, width = 1.8cm]{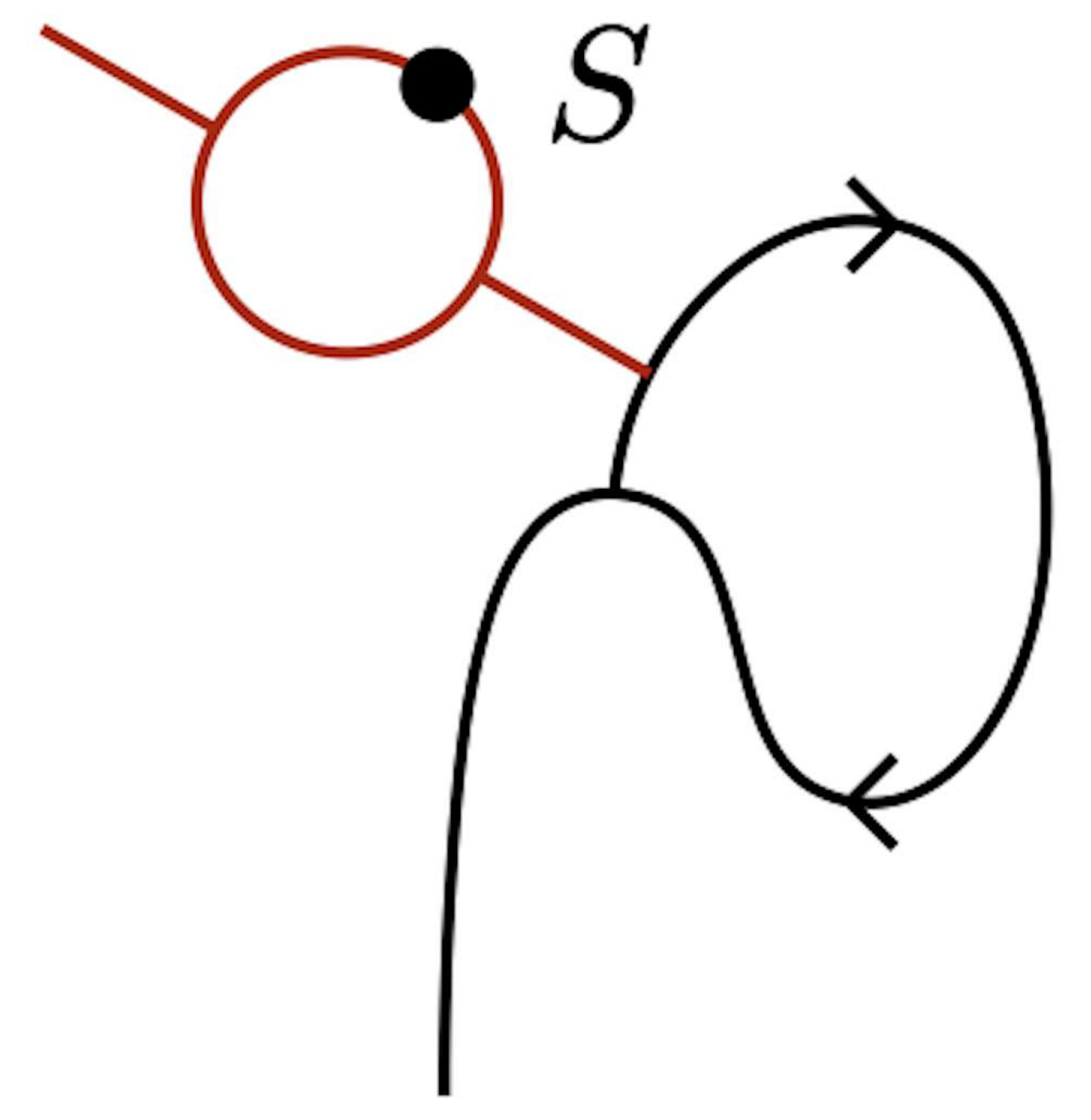} ~ = 
\adjincludegraphics[valign = c, width = 1cm]{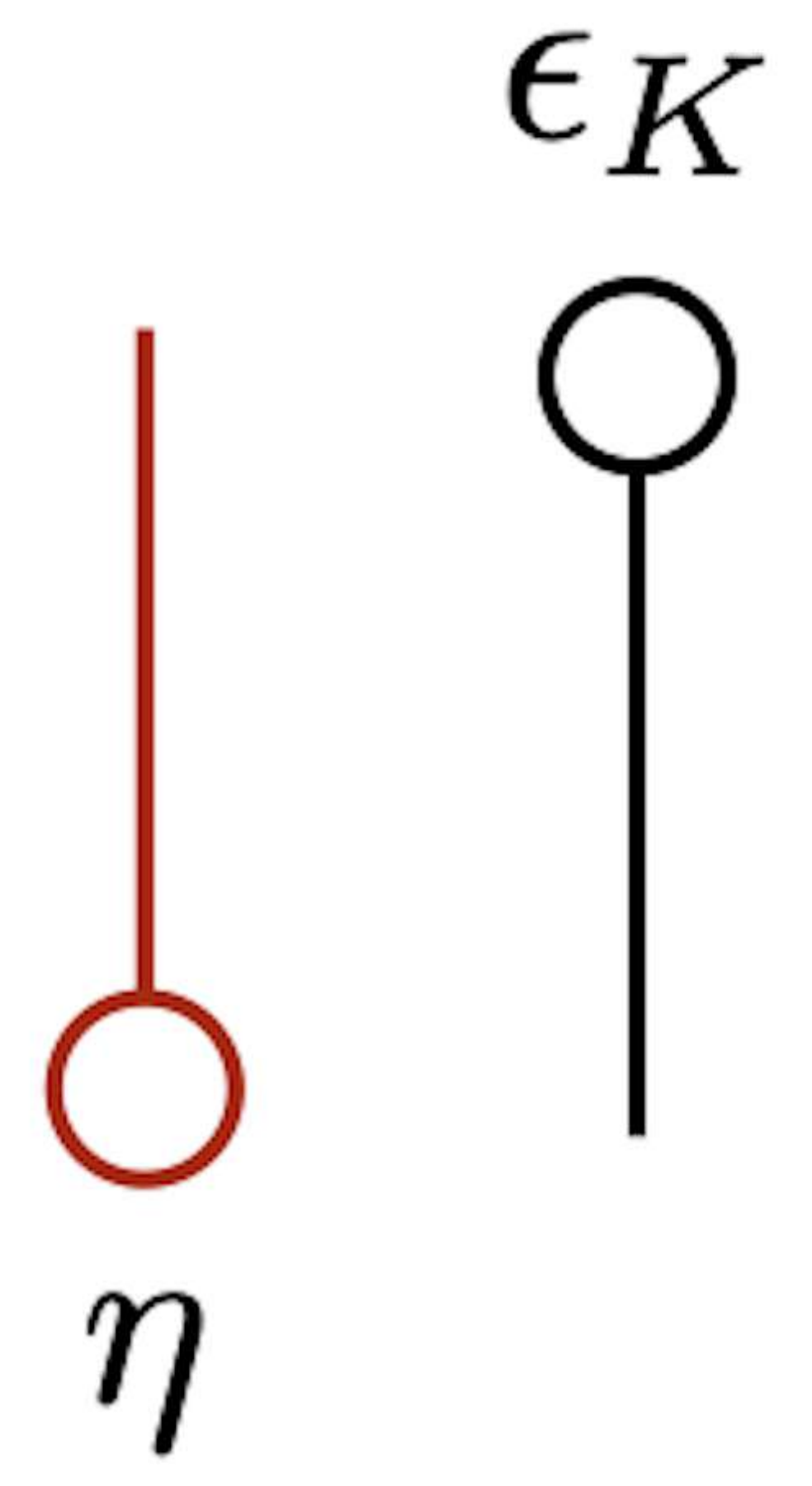},
\label{eq: H-comod action on epsilon}
\end{equation} 
where we used the left $H$-comodule action on $K^*$ defined in a similar way to eq. \eqref{eq: dual rep}.
We note that the above equation relies on the fact that the antipode $S$ of a semisimple Hopf algebra $H$ squares to the identity.
Equation \eqref{eq: H-comod action on epsilon} in turn implies that the isomorphism $\Phi: K \rightarrow K^*$ defined in eq. \eqref{eq: Frobenius structure} is an $H$-comodule map because
\begin{equation}
\adjincludegraphics[valign = c, width = 1.5cm]{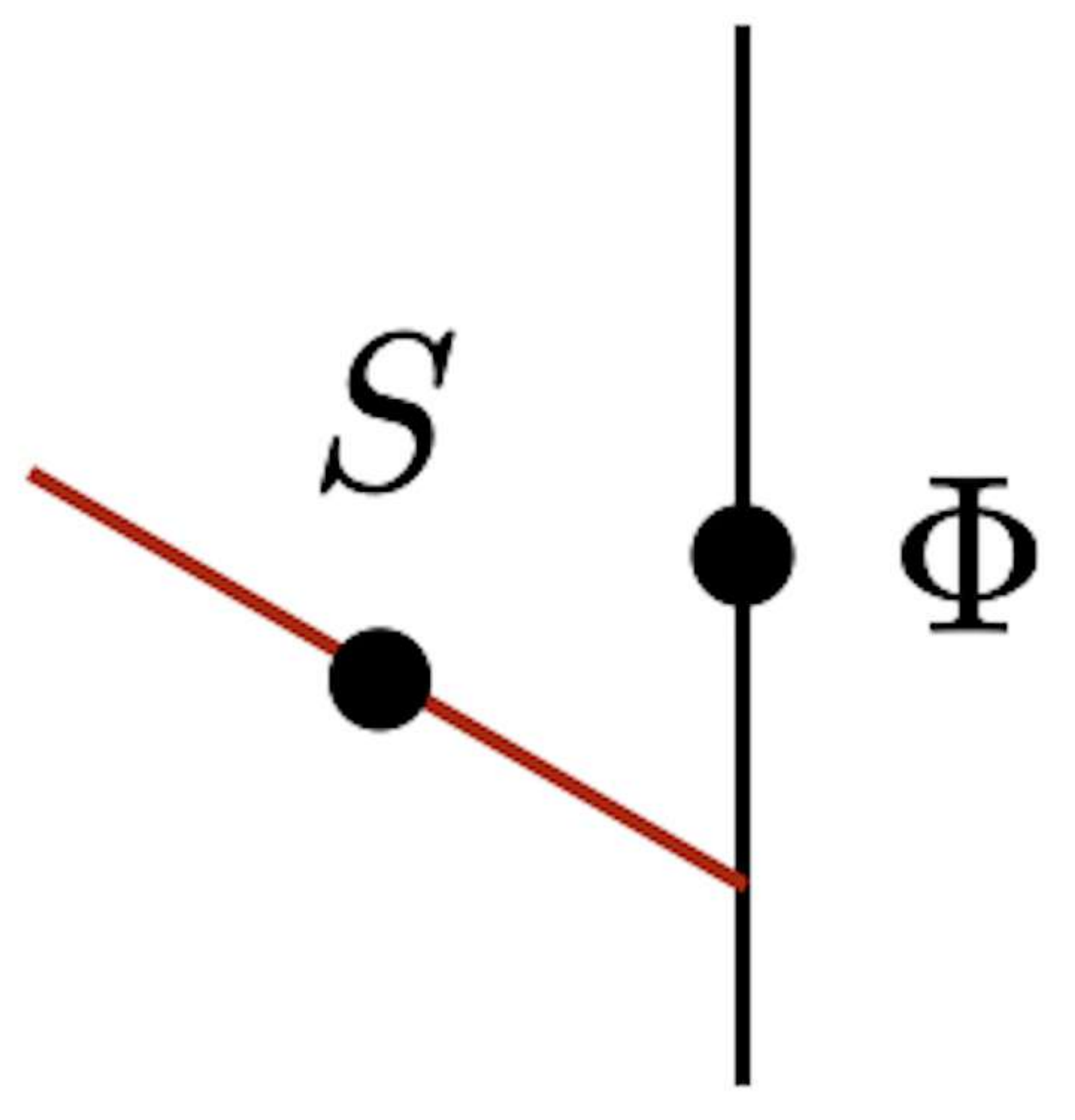} ~ = ~ 
\adjincludegraphics[valign = c, width = 2.6cm]{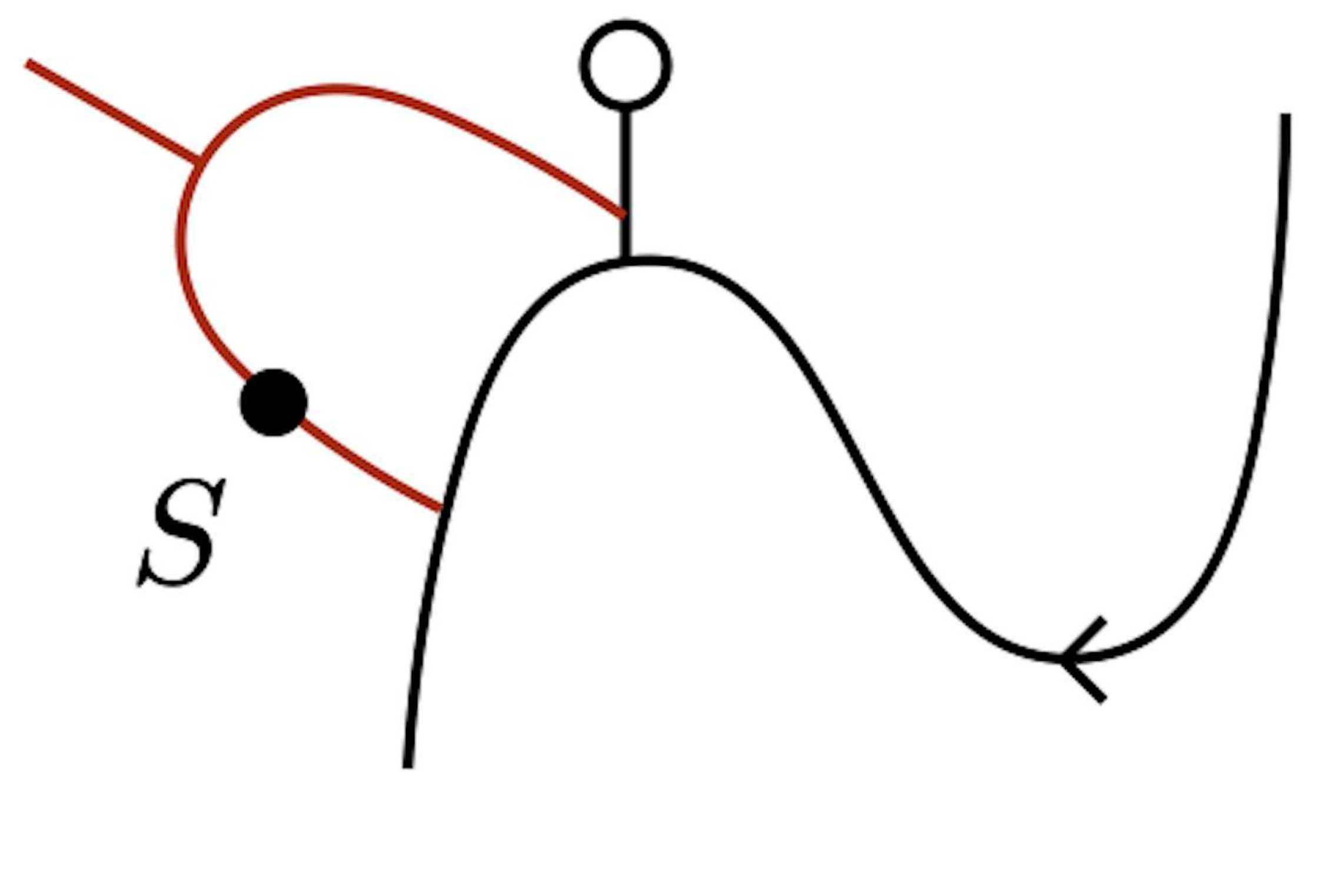} ~ = ~ 
\adjincludegraphics[valign = c, width = 2.8cm]{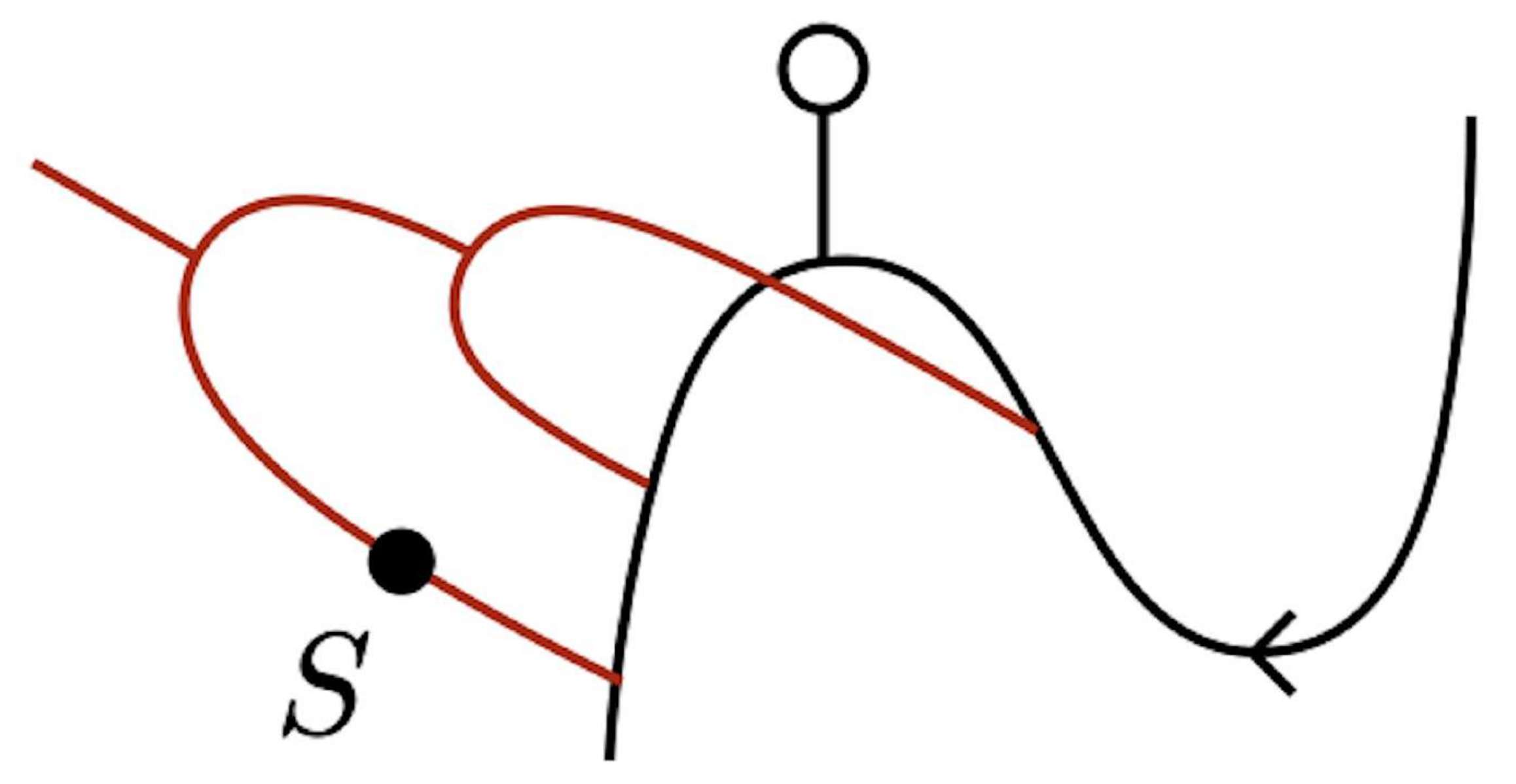} ~ = ~ 
\adjincludegraphics[valign = c, width = 2.8cm]{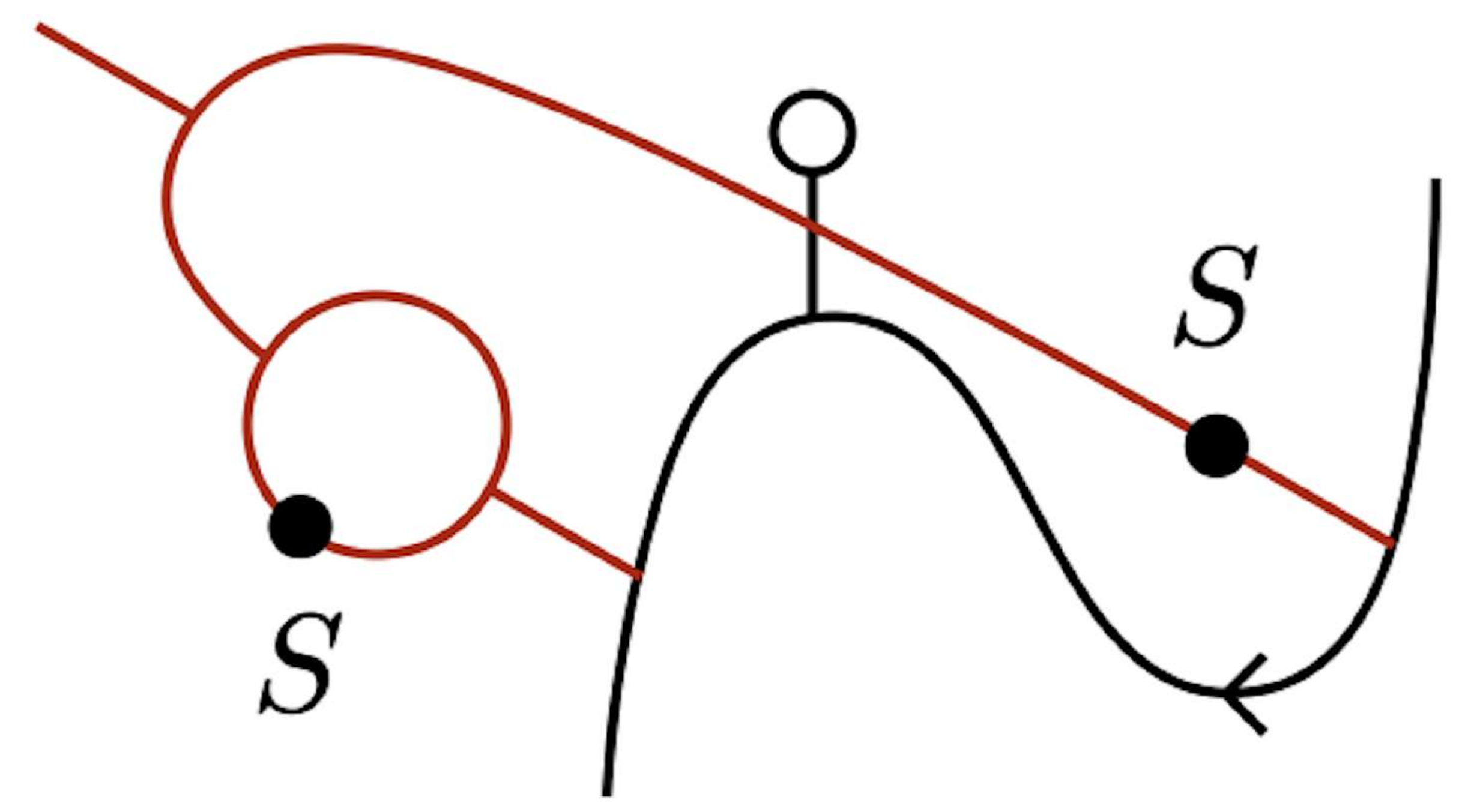} ~ = ~ 
\adjincludegraphics[valign = c, width = 1.4cm]{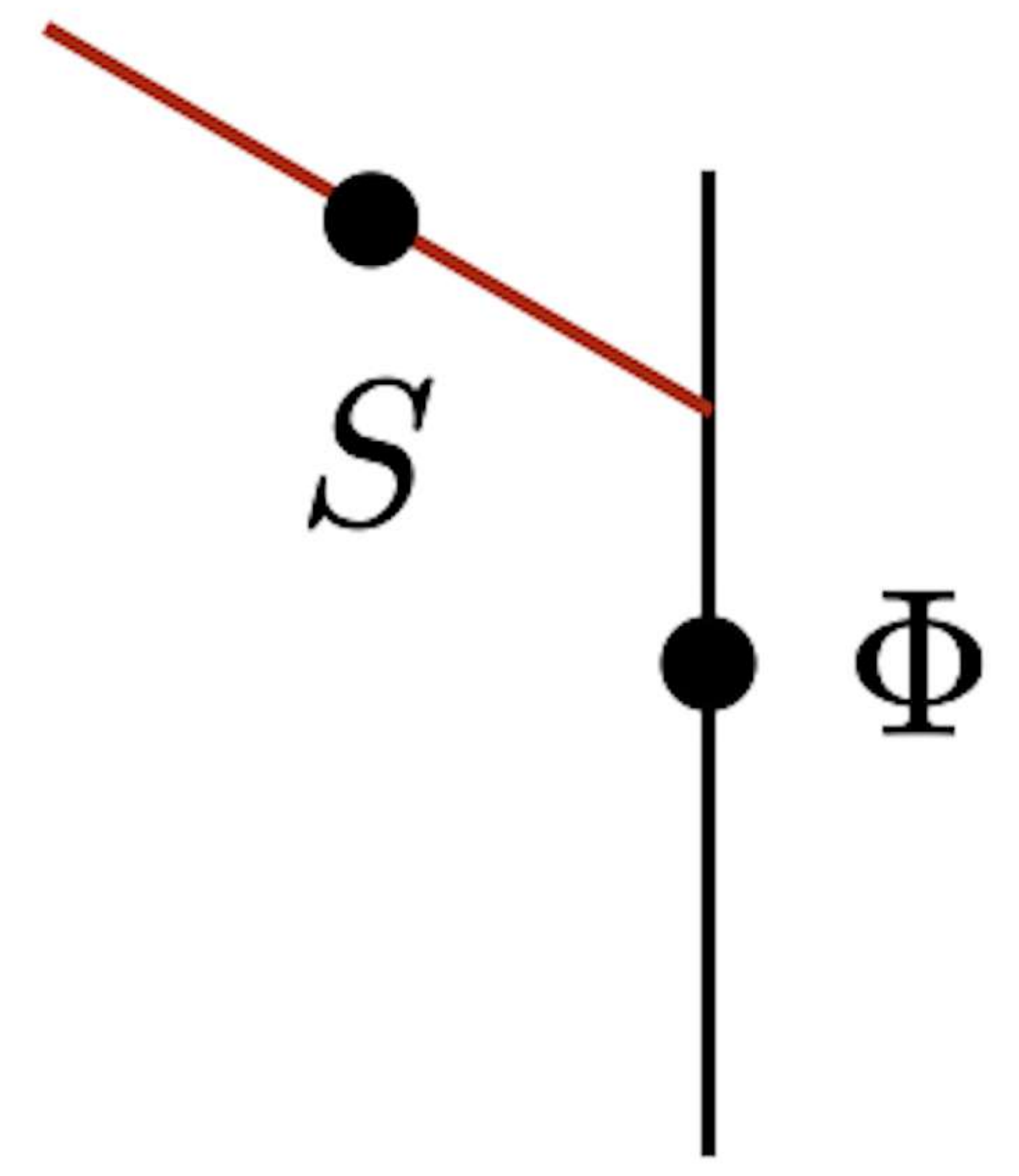}.
\end{equation}
This indicates that $\Phi^{-1}: K^* \rightarrow K$ is an $H$-comodule map as well.
Therefore, we have
\begin{equation}
\adjincludegraphics[valign = c, width = 2cm]{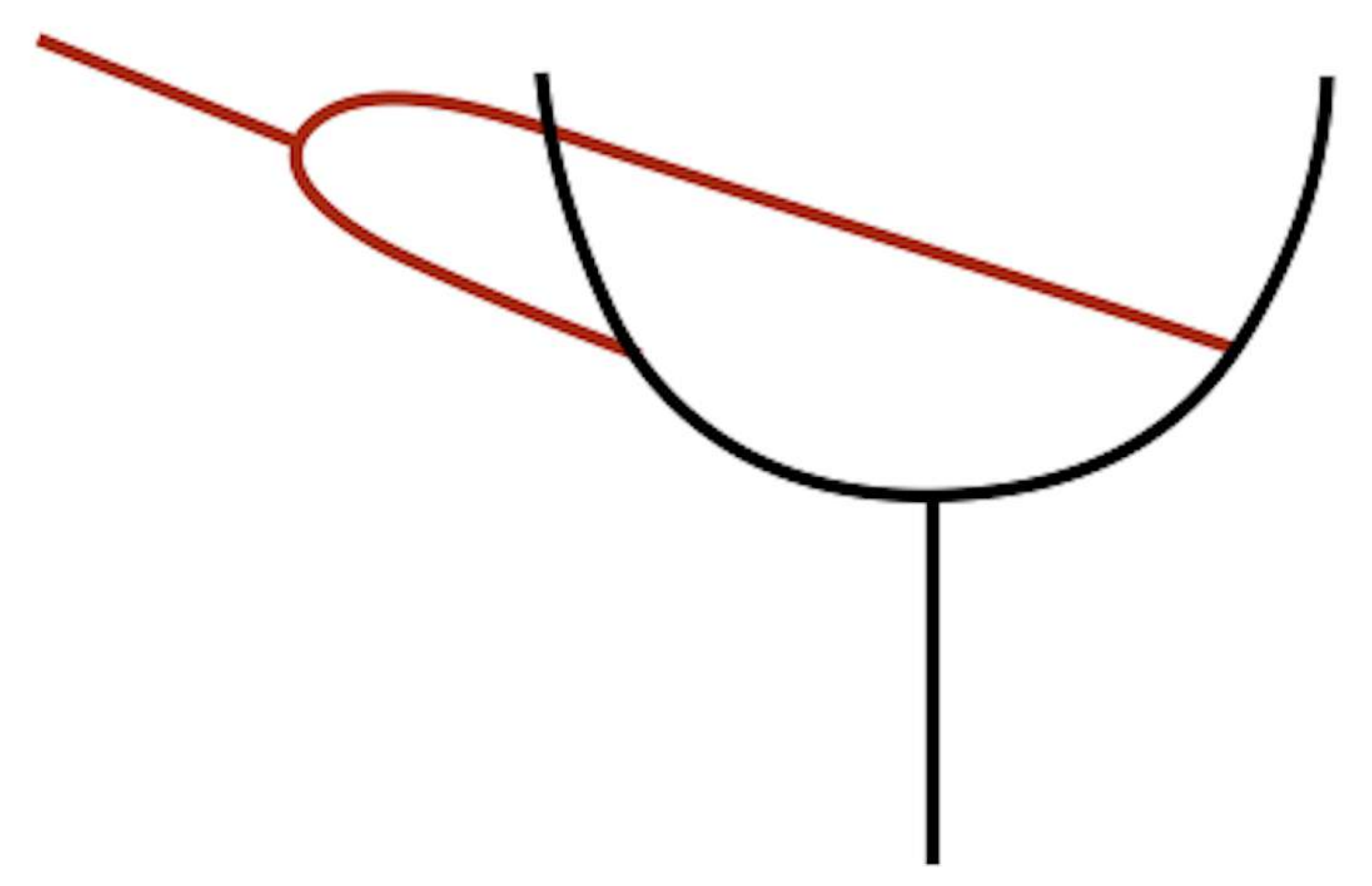} ~ = ~ 
\adjincludegraphics[valign = c, width = 2.4cm]{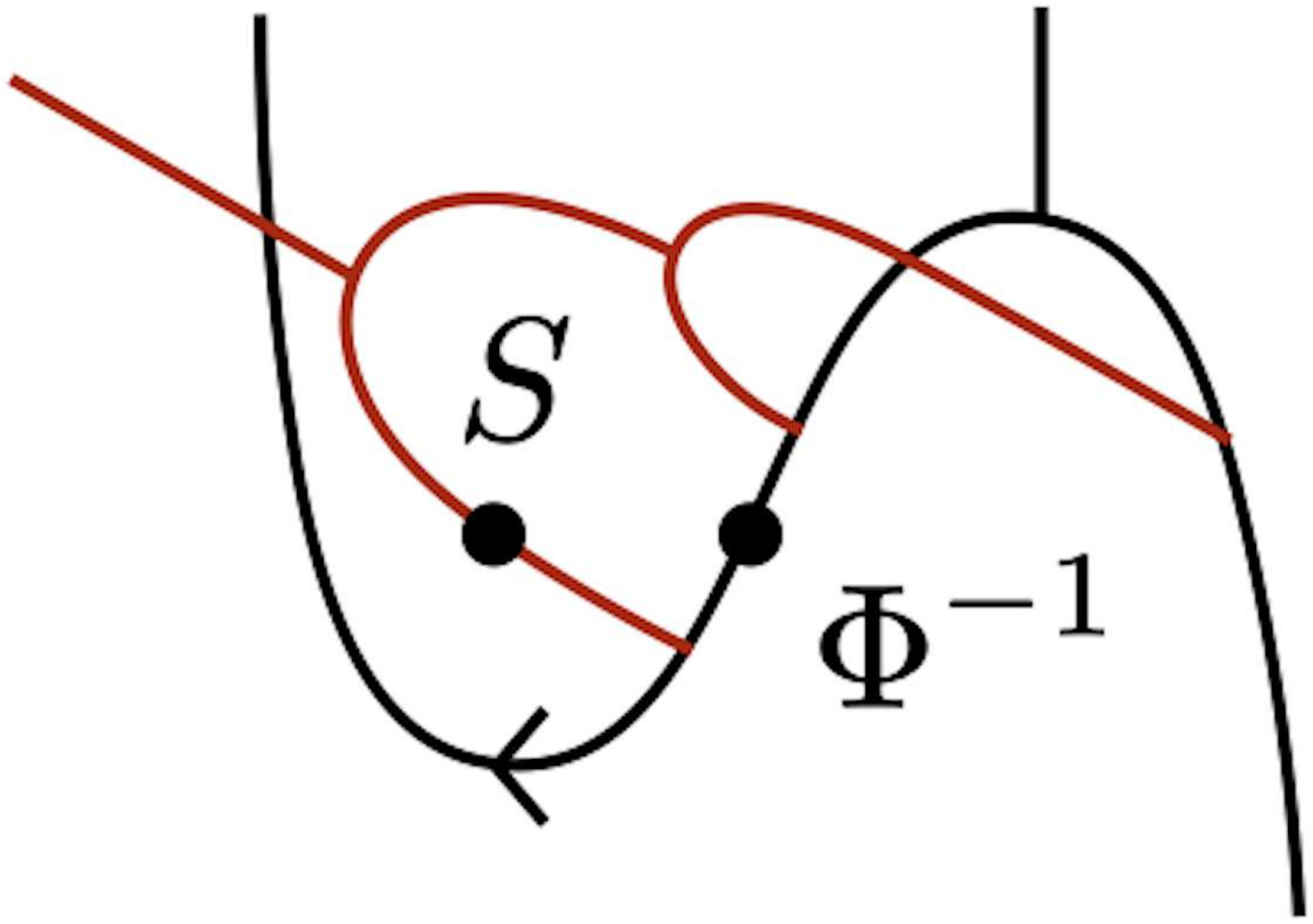} ~ = ~ 
\adjincludegraphics[valign = c, width = 2.8cm]{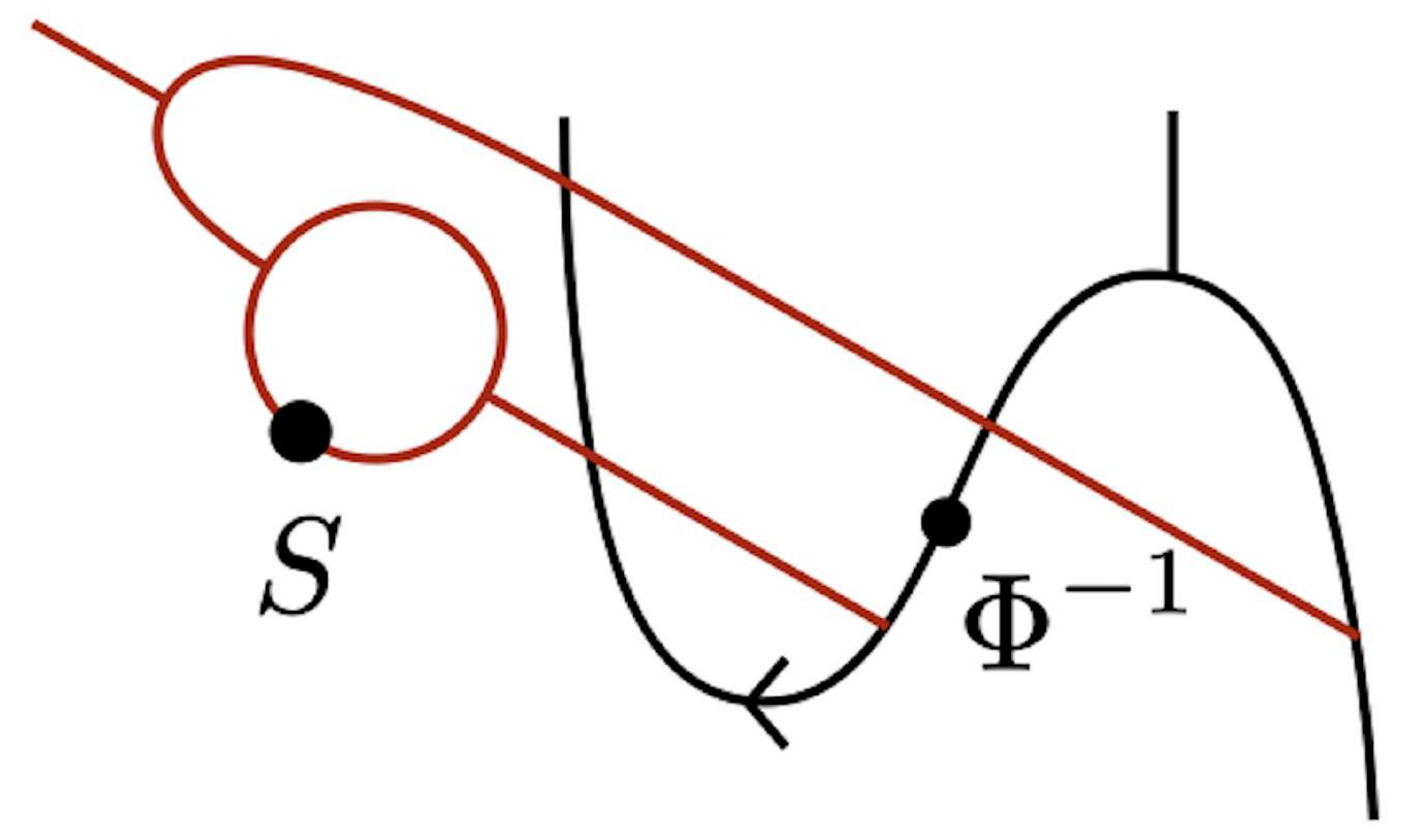} ~ = ~ 
\adjincludegraphics[valign = c, width = 1.6cm]{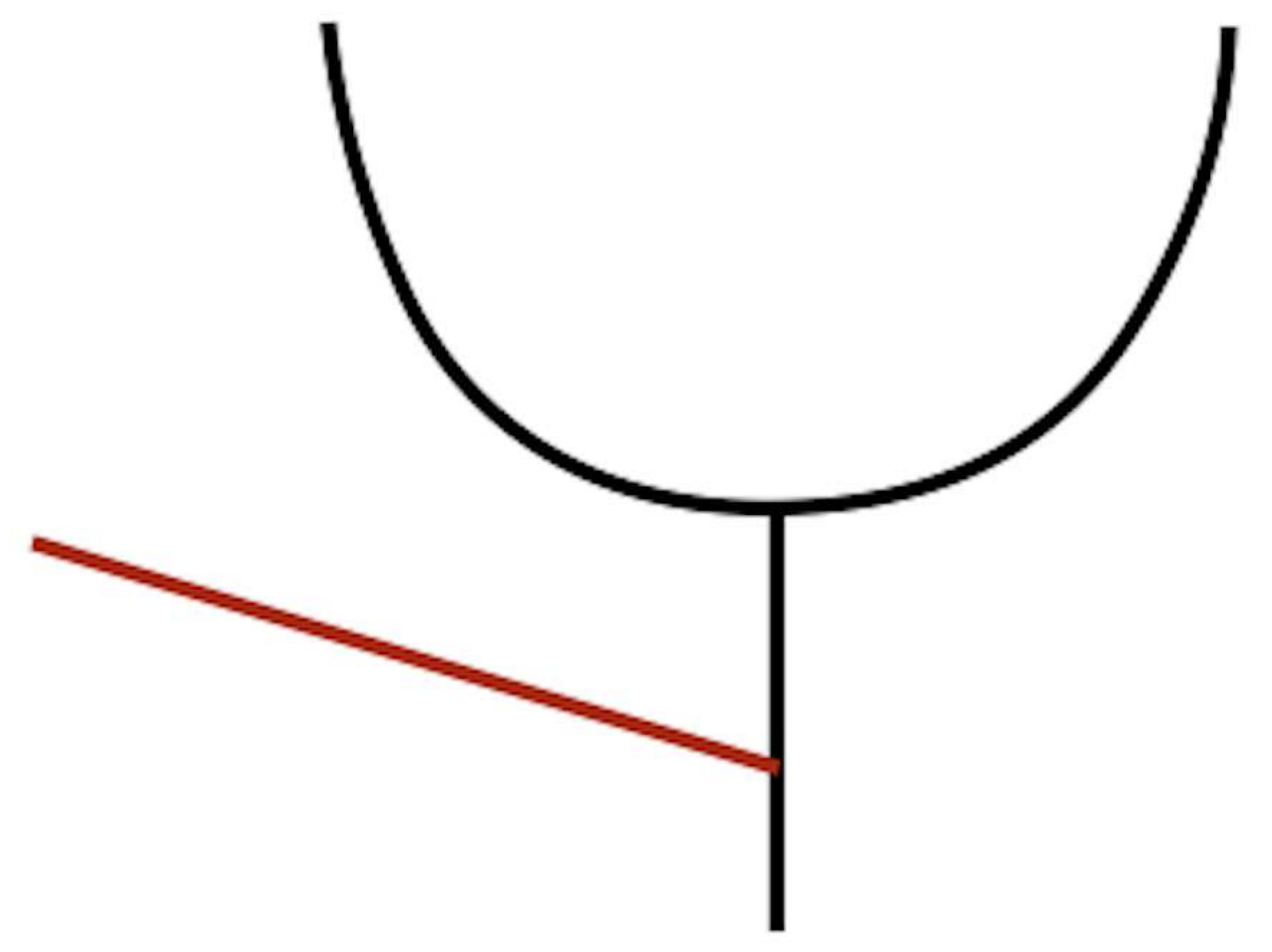},
\end{equation}
which shows eq. \eqref{eq: commutativity}.

We can also compute the action \eqref{eq: Rep(H) action} of the $\mathrm{Rep}(H)$ symmetry on the ground states of the Hamiltonian \eqref{eq: commuting projector Hamiltonian}.
To perform the computation, we recall that the ground states of \eqref{eq: commuting projector Hamiltonian} are in one-to-one correspondence with the vacua of the state sum TQFT, and hence can be written as the boundary states \eqref{eq: MPS} \cite{MS2006}.
The $\mathrm{Rep}(H)$ symmetry action $U_V$ on a boundary state $\bra{M}$ is given by
\begin{equation}
\bra{M} U_V^{\dagger} = 
\adjincludegraphics[valign = c, width = 2.8cm]{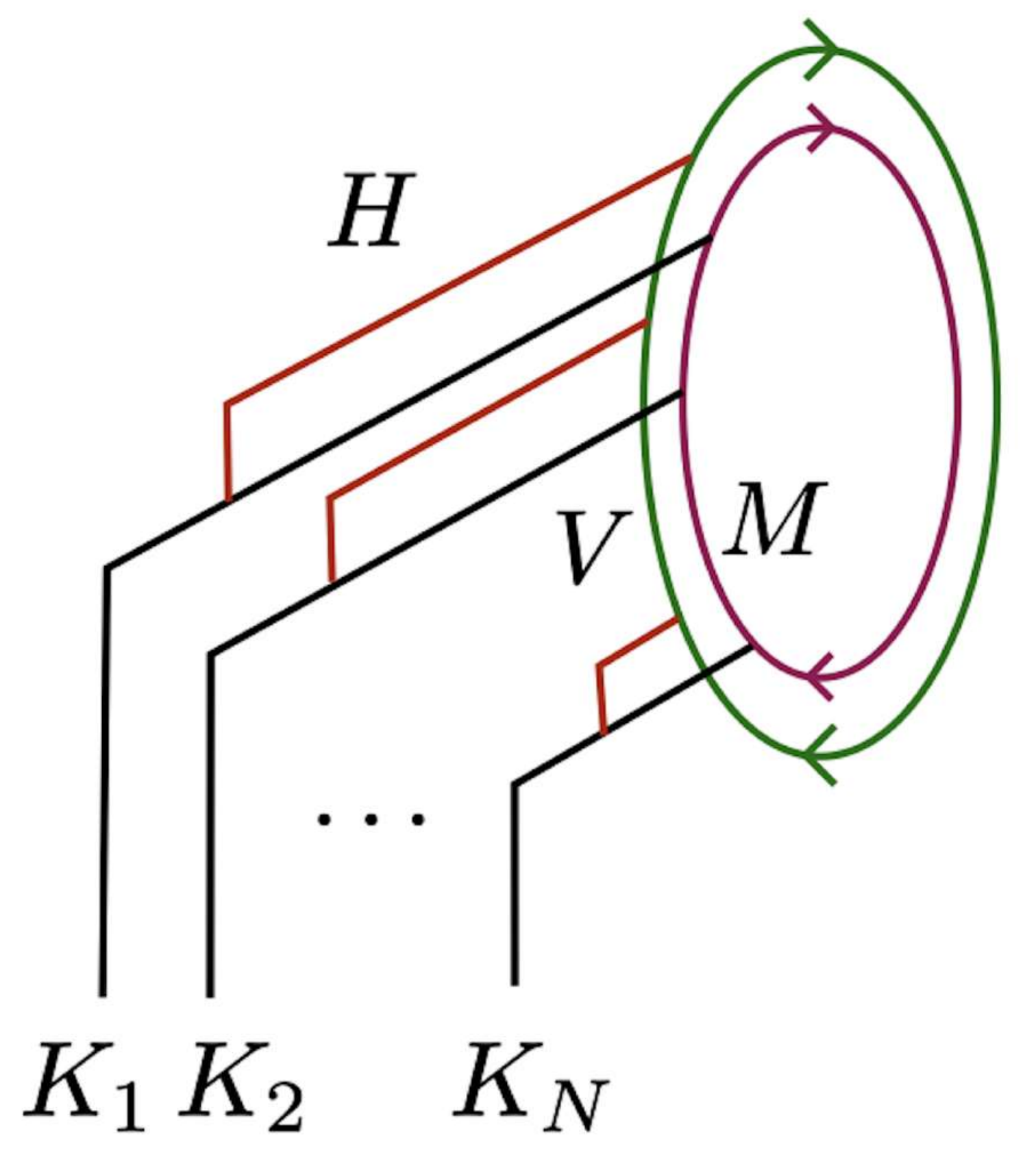} ~ = ~
\adjincludegraphics[valign = c, width = 2.8cm]{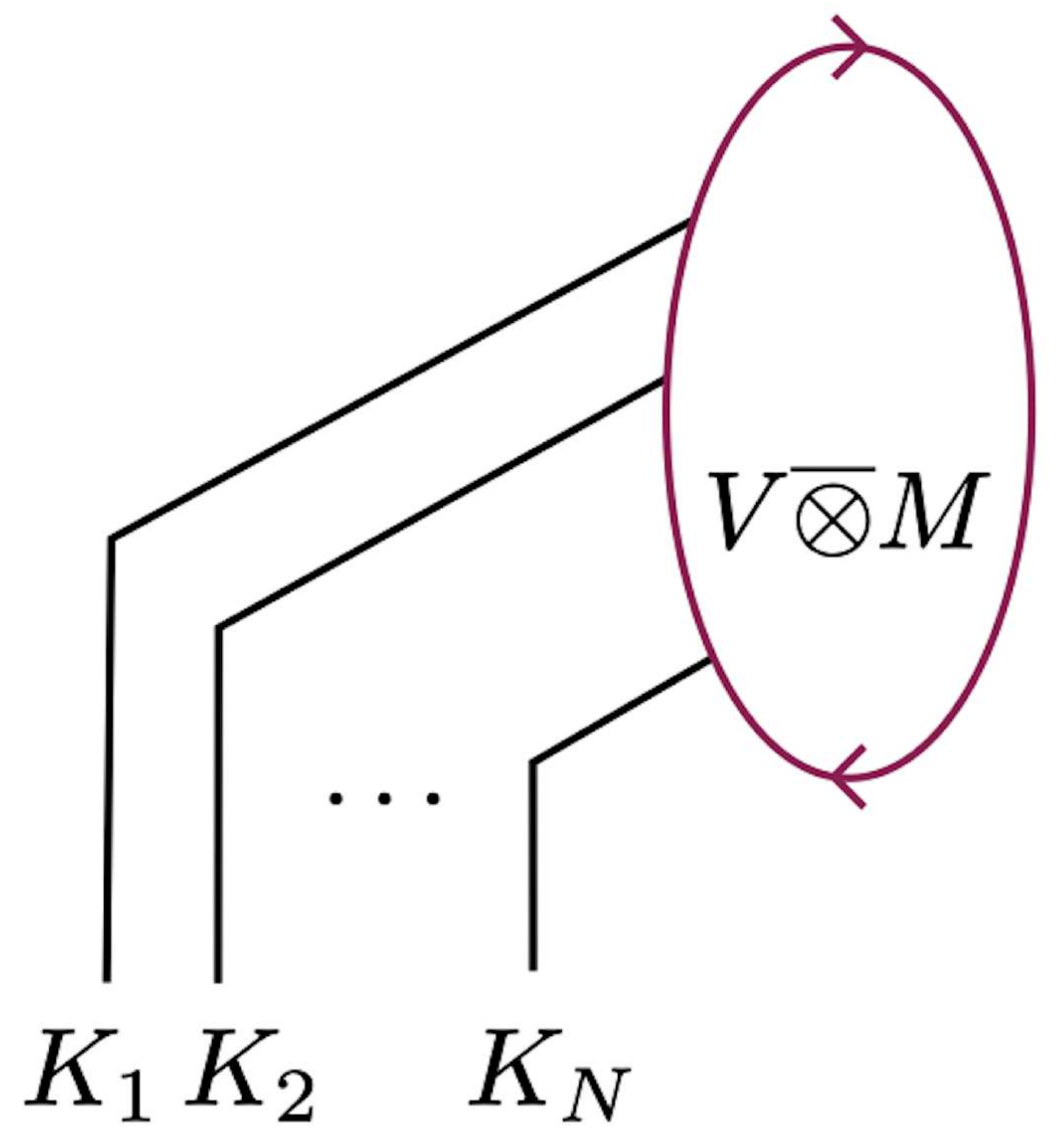} ~ = \bra{V \overline{\otimes} M},
\end{equation}
which coincides with the $\mathrm{Rep}(H)$ symmetry action \eqref{eq: Rep(H) action on boundary states} of the state sum TQFT.
This implies that the commuting projector Hamiltonian \eqref{eq: commuting projector Hamiltonian} is a lattice realization of a $\mathrm{Rep}(H)$ symmetric TQFT ${}_K \mathcal{M}$.
In summary, every semisimple indecomposable TQFT ${}_K \mathcal{M} \cong ({}_H \mathcal{M})_A$ with $\mathrm{Rep}(H)$ symmetry can be realized by a commuting projector Hamiltonian \eqref{eq: commuting projector Hamiltonian} on the lattice where the input datum $K$ is an $H$-simple left $H$-comodule algebra $K = A^{\mathrm{op}} \# H^{\mathrm{cop}}$.

\subsection{Examples: gapped phases of finite gauge theories}
\label{sec: Examples}
Let $G$ be a finite group and $\mathbb{C}[G]$ be a group algebra.
Gapped phases of $G$ gauge theory are labeled by a pair $(H, \omega)$ \cite{TW2019} where $H$ is a subgroup of $G$ to which the gauge group $G$ is Higgsed down and $\omega \in H^2(H, \mathrm{U}(1))$ is a discrete torsion \cite{Vafa1986}.
The symmetry of $G$ gauge theory is described by $\mathrm{Rep}(G) := \mathrm{Rep}(\mathbb{C}[G])$, which is generated by the Wilson lines.
Therefore, we can realize these phases by the commuting projector Hamiltonians \eqref{eq: commuting projector Hamiltonian} where the input algebra $K$ is a left $\mathbb{C}[G]$-comodule algebra.
Specifically, the input algebra $K$ for the gapped phase labeled by $(H, \omega)$ is given by $K = (\mathbb{C}[G] \otimes_{\mathbb{C}[H]} \mathrm{End}(U))^{\mathrm{op}} \# \mathbb{C}[G]$, where $U$ is a projective representation of $H$ characterized by $\omega$ \cite{AM2007}.\footnote{We note that the group algebra $\mathbb{C}[G]$ is cocommutative, i.e. $\mathbb{C}[G]^{\mathrm{cop}} = \mathbb{C}[G]$.}
The action \eqref{eq: Rep(H) action} of a representation $V \in \mathrm{Rep}(G)$ is expressed as
\begin{equation}
U_{V}^{\dagger} \ket{a_1 \# g_1, a_2 \# g_2, \cdots, a_N \# g_N} = \chi_V(g_1 g_2 \cdots g_N) \ket{a_1 \# g_1, a_2 \# g_2, \cdots, a_N \# g_N}
\end{equation}
for $a_i \in (\mathbb{C}[G] \otimes_{\mathbb{C}[H]} \mathrm{End}(U))^{\mathrm{op}}$ and $g_i \in G$.
In the following, we will explicitly describe the actions of the commuting projector Hamiltonians \eqref{eq: commuting projector Hamiltonian} for gapped phases of $G$ gauge theory by choosing a specific basis of $K$.
For simplicity, we will only consider two limiting cases where the gauge group $G$ is not Higgsed at all or completely Higgsed.

When $G$ is not Higgsed, the gapped phases of $G$ gauge theory are described by Dijkgraaf-Witten theories \cite{DW90}.
The input algebras $K$ for these phases are given by $K = \mathrm{End}(U)^{\mathrm{op}} \# \mathbb{C}[G]$.\footnote{The input algebra of a Dijkgraaf-Witten theory is usually chosen as a twisted group algebra $\mathbb{C}[G]^{\omega}$. The algebras $\mathbb{C}[G]^{\omega}$ and $\mathrm{End}(U)^{\mathrm{op}} \# \mathbb{C}[G]$ give rise to the same TQFT because ${}_{\mathbb{C}[G]^{\omega}} \mathcal{M}$ is equivalent to ${}_{\mathrm{End}(U)^{\mathrm{op}} \# \mathbb{C}[G]} \mathcal{M}$ as a module category over $\mathrm{Rep}(G)$ \cite{AM2007}.}
We choose a basis of the algebra $K$ as $\{E_{ij} \# v_g \mid i, j = 1, 2, \cdots, \mathrm{dim}U, g \in G\}$, where $E_{ij}$ is a $\mathrm{dim}U \times \mathrm{dim}U$ matrix whose $(k, l)$ component is $1$ when $(k, l) = (i, j)$ and otherwise $0$.
If we denote the projective action of $G$ on $U$ by $Q: G \rightarrow \mathrm{End}(U)$, the multiplication \eqref{eq: smash multiplication} on the algebra $K$ is written as 
\begin{equation}
(E_{ij} \# v_g) \cdot (E_{kl} \# v_h) = Q(g) E_{kl} Q(g)^{-1} E_{ij} \# v_{gh}.
\end{equation}
The Frobenius algebra structure on $K$ is characterized by a pairing 
\begin{equation}
\epsilon_K ((E_{ij} \# v_g) \cdot (E_{kl} \# v_h)) = |G| \mathrm{dim}U \delta_{g, h^{-1}} [Q(g) E_{kl} Q(g)^{-1}]_{ji},
\end{equation}
where the last term on the right-hand side represents the $(j, i)$ component of $Q(g) E_{kl} Q(g)^{-1}$.
The above equation implies that $Q(g)^{-1} E_{ji} Q(g) \# v_{g^{-1}} / |G| \mathrm{dim}U$ is dual to $E_{ij} \# v_g$ with respect to the pairing $\epsilon_K \circ m_K$, and hence the comultiplication of the unit element $1_K \in K$ is given by
\begin{equation}
\Delta_K (1_K) = \frac{1}{|G| \mathrm{dim}U} \sum_{i, j} \sum_{g} E_{ij} \# v_g \otimes Q(g)^{-1} E_{ji} Q(g) \# v_{g^{-1}}.
\end{equation}
Therefore, we can explicitly write down the action of the local commuting projector $h: K \otimes K \rightarrow K \otimes K$ defined by eq. \eqref{eq: local CP} as
\begin{equation}
h (E_{ij} \# v_g \otimes E_{kl} \# v_h) = \frac{1}{|G| \mathrm{dim}U} \sum_{m} \sum_{f \in G} Q(g) E_{ml} Q(g)^{-1} E_{ij} \# v_{gf} \otimes Q(f)^{-1} E_{km} Q(f) \# v_{f^{-1}h}.
\end{equation} 

On the other hand, when $G$ is completely Higgsed, the input algebra $K$ is given by $K = \mathbb{C}[G]^{*} \# \mathbb{C}[G]$.
We choose a basis of $K$ as $\{v^g \# v_h \mid g, h \in G\}$ where $v^g \in \mathbb{C}[G]^*$ denotes the dual basis of $v_g \in \mathbb{C}[G]$.
The multiplication \eqref{eq: smash multiplication} on the algebra $K$ is written as
\begin{equation}
(v^g \# v_h) \cdot (v^k \# v_l) = \delta_{g, hk} v^g \# v_{hl},
\end{equation}
where we defined a left $\mathbb{C}[G]$-module action on $\mathbb{C}[G]^{*}$ by the left translation $\rho(v_g) v^h := v^{gh}$.
Since the dual of $v^g \# v_h$ with respect to the Frobenius pairing $\epsilon_K \circ m_K$ is given by $v^{h^{-1}g} \# v_{h^{-1}} /|G|$, we have
\begin{equation}
\Delta_K (1_K) = \frac{1}{|G|} \sum_{g, h} v^g \# v_h \otimes v^{h^{-1}g} \# v_{h^{-1}}.
\end{equation}
Thus, the action of the local commuting projector $h$ is expressed as
\begin{equation}
h(v^g \# v_h \otimes v^k \# v_l) = \frac{\delta_{g, hk}}{|G|} \sum_{m} v^g \# v_{hm} \otimes v^{m^{-1}k} \# v_{m^{-1}l}.
\end{equation}

\subsection{Edge modes of SPT phases with fusion category symmetries}
SPT phases with fusion category symmetry $\mathcal{C}$ are uniquely gapped phases preserving the symmetry $\mathcal{C}$.
Since anomalous fusion category symmetries do not admit SPT phases, it suffices to consider non-anomalous symmetries $\mathcal{C} = \mathrm{Rep}(H)$.
SPT phases with $\mathrm{Rep}(H)$ symmetry are realized by the commuting projector Hamiltonians \eqref{eq: commuting projector Hamiltonian} when $K = A^{\mathrm{op}} \# H^{\mathrm{cop}}$ is a simple algebra.\footnote{Generally, the ground states of the Hamiltonian \eqref{eq: commuting projector Hamiltonian} on a circle are given by the center of $K$ \cite{FHK94}. In particular, when $K$ is simple, the ground state is unique because the center of $K$ is one-dimensional. For example, the complete Higgs phase discussed in section \ref{sec: Examples} is an SPT phase with $\mathrm{Rep}(G)$ symmetry.}
These Hamiltonians have degenerate ground states on an interval even though they have unique ground states on a circle.
Specifically, it turns out that the ground states on an interval are given by the algebra $K$ \cite{LP2007, DKR2011}.
Since $K$ is simple, we can write $K \cong \mathrm{End}(M) \cong M^* \otimes M$ where $M$ is a simple left $K$-module, which is unique up to isomorphism. 
We can interpret $M^*$ and $M$ as the edge modes localized to the left and right boundaries because the bulk is a uniquely gapped state represented by an MPS \eqref{eq: MPS}.
Indeed, if we choose a basis of the local Hilbert space on an edge $e$ as $\{ \ket{v^i}_e \otimes \ket{v_j}_e \in M^* \otimes M \}$, we can write the ground states of the commuting projector Hamiltonian \eqref{eq: commuting projector Hamiltonian} on an interval as $\ket{v^i}_1 \otimes \ket{\Omega}_{1, 2} \otimes \ket{\Omega}_{2, 3} \otimes \cdots \otimes \ket{\Omega}_{N-1, N} \otimes \ket{v_j}_N$, where $\ket{\Omega}_{e, e+1} := \sum_k \ket{v_k}_e \otimes \ket{v^k}_{e+1}$ is the maximally entangled state.
This expression indicates that the degrees of freedom of $M^*$ and $M$ remain on the left and right boundaries respectively.
Therefore, the edge modes of the Hamiltonian \eqref{eq: commuting projector Hamiltonian} for a $\mathrm{Rep}(H)$ SPT phase ${}_K \mathcal{M} \cong ({}_H \mathcal{M})_A$ are described by a right $K$-module $M^*$ and a left $K$-module $M$, where $K = A^{\mathrm{op}} \# H^{\mathrm{cop}}$.

It is instructive to consider the case of an ordinary finite group symmetry $G$.
A finite group symmetry $G$ is described by the category $\mathrm{Vec}_G$ of $G$-graded vector spaces, which is equivalent to the representation category of a dual group algebra $\mathbb{C}[G]^{*}$.
SPT phases with this symmetry are classified by the second group cohomology $H^2(G, \mathrm{U}(1))$ \cite{CGW2011a, CGW2011b, FK2011, SPGC2011, CGLW2013, KT2017, MS2006, Tur1999}. 
An SPT phase labeled by $\omega \in H^2(G, \mathrm{U}(1))$ is realized by the commuting projector Hamiltonian \eqref{eq: commuting projector Hamiltonian} when $A$ is a twisted group algebra $\mathbb{C}[G]^{\omega}$.
The edge modes $M^*$ of this model become a right $(\mathbb{C}[G]^{\omega})^{\mathrm{op}} \# (\mathbb{C}[G]^*)^{\mathrm{cop}}$-module, which is a left $\mathbb{C}[G]^{\omega}$-module in particular.
This implies that these edge modes have an anomaly $\omega$ of the finite group symmetry $G$.

\subsection{Generalization to anomalous fusion category symmetries}
\label{sec: A comment on a generalization to anomalous symmetries}
The most general unitary fusion category, which may or may not be anomalous, is equivalent to the representation category $\mathrm{Rep}(H)$ of a finite dimensional semisimple pseudo-unitary connected weak Hopf algebra $H$ \cite{Ostrik2003, NTV2003, Hayashi1999, Nikshych2004}.
As the case of Hopf algebras, any semisimple indecomposable module category over $\mathrm{Rep}(H)$ is given by the category ${}_K \mathcal{M}$ of left $K$-modules, where $K$ is an $H$-simple left $H$-comodule algebra \cite{Henker2011}.
We note that an $H$-simple left $H$-comodule algebra is semisimple \cite{Nikshych2004, Henker2011}.
Accordingly, we can construct all the TQFTs ${}_K \mathcal{M}$ with anomalous fusion category symmetry $\mathrm{Rep}(H)$ by pulling back the state sum TQFT with the input $K$ by a tensor functor $F_K: \mathrm{Rep}(H) \rightarrow {}_K \mathcal{M}_K$.
Moreover, the fact that $K$ is semisimple allows us to write down a commuting projector Hamiltonian in the same way as eq. \eqref{eq: commuting projector Hamiltonian}.
We can also define the action of $\mathrm{Rep}(H)$ on the lattice Hilbert space just by replacing a Hopf algebra with a weak Hopf algebra in \eqref{eq: Rep(H) action}.
One may expect that these Hamiltonians realize all the gapped phases with anomalous fusion category symmetries.
However, since our proof of the commutativity of the $\mathrm{Rep}(H)$ action \eqref{eq: Rep(H) action} and the commuting projector Hamiltonian \eqref{eq: commuting projector Hamiltonian} relies on the properties that are specific to a semisimple Hopf algebra, our proof does not work when $H$ is not a Hopf algebra, i.e. when the fusion category symmetry is anomalous.
Therefore, we need to come up with another proof that is applicable to anomalous fusion category symmetries.
We leave this problem to future work.

\section*{Acknowledgments}
I would like to thank Masaki Oshikawa and Ken Shiozaki for comments on the manuscript.
I also appreciate helpful discussions in the workshop ``Topological Phase and Quantum Anomaly 2021" (YITP-T-21-03) at Yukawa Institute for Theoretical Physics, Kyoto University.
I am supported by FoPM, WINGS Program, the University of Tokyo.

{\it Note added.} While this work was nearing completion, I became aware of a related paper by T.-C. Huang, Y.-H. Lin, and S. Seifnashri \cite{HLS2021}, in which correlation functions of 2d TQFTs with general fusion category symmetries are determined.
Our result provides lattice construction of these data.

\appendix

\section{State sum TQFTs on surfaces with interfaces}
\label{sec: State sum TQFTs on surfaces with interfaces}
In this appendix, we extend the state sum TQFTs to surfaces with interfaces following \cite{DKR2011}.
A TQFT on each region separated by interfaces is described by the state sum TQFT with non-anomalous fusion category symmetry, which we defined in section \ref{sec: Rep(H) symmetry of the state sum TQFT}.
We denote the state sum TQFT with the input $K$ as ${}_K \mathcal{M}$, whose ${}_K \mathcal{M}_K$ symmetry is pulled back to $\mathrm{Rep}(H)$ by a tensor functor $F_K: \mathrm{Rep}(H) \rightarrow {}_K \mathcal{M}_K$.
An interface between a $\mathrm{Rep}(H)$ symmetric TQFT ${}_K \mathcal{M}$ and a $\mathrm{Rep}(H^{\prime})$ symmetric TQFT ${}_{K^{\prime}} \mathcal{M}$ is labeled by a $K$-$K^{\prime}$ bimodule $M \in {}_K \mathcal{M}_{K^{\prime}}$.
The possible configurations of topological defects near an interface can be classified into the following cases:
\begin{equation}
\text{(I)} ~ \adjincludegraphics[valign = c, width = 2.0cm]{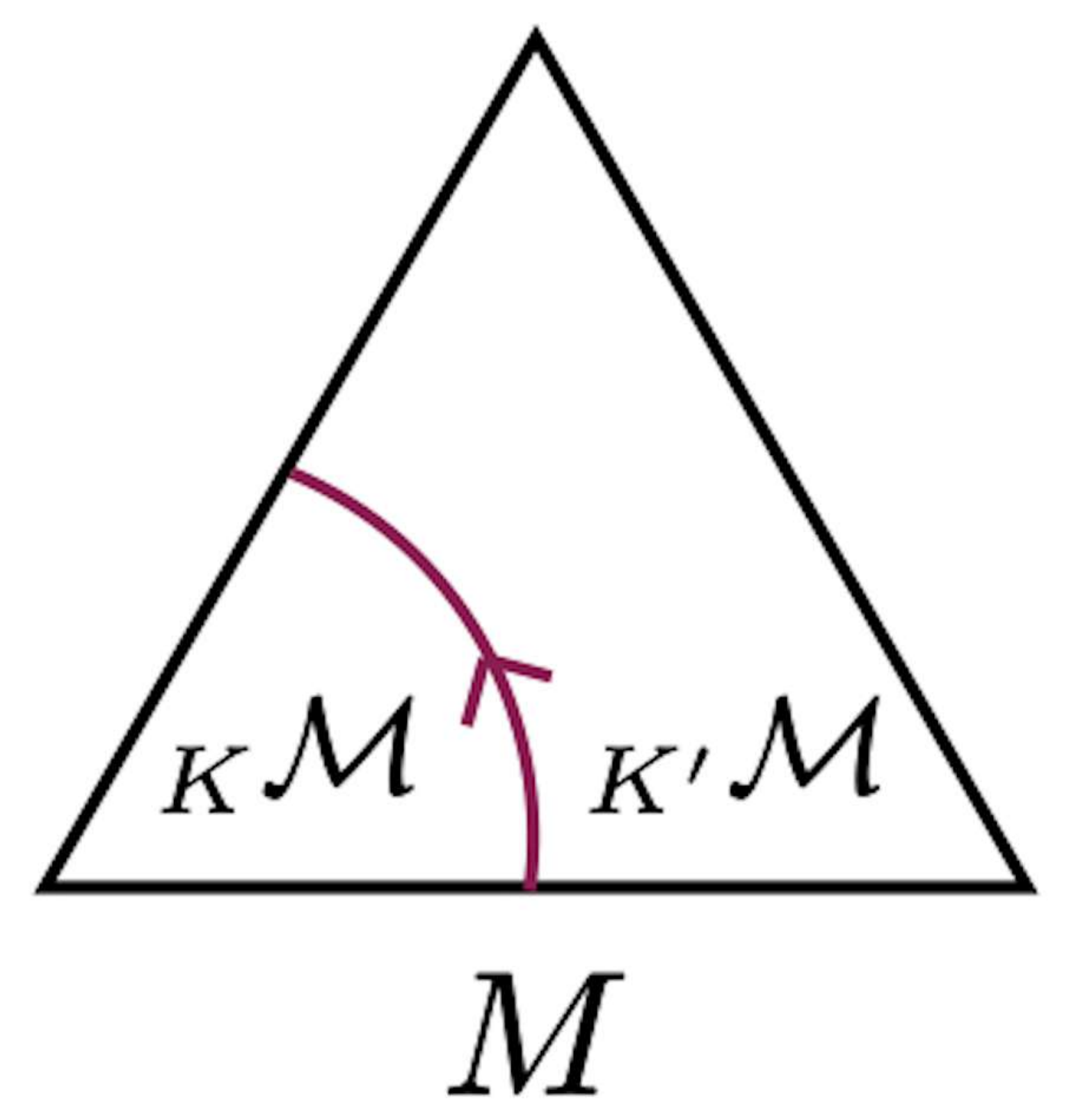}, \quad
\text{(II)} ~ \adjincludegraphics[valign = c, width = 2.1cm]{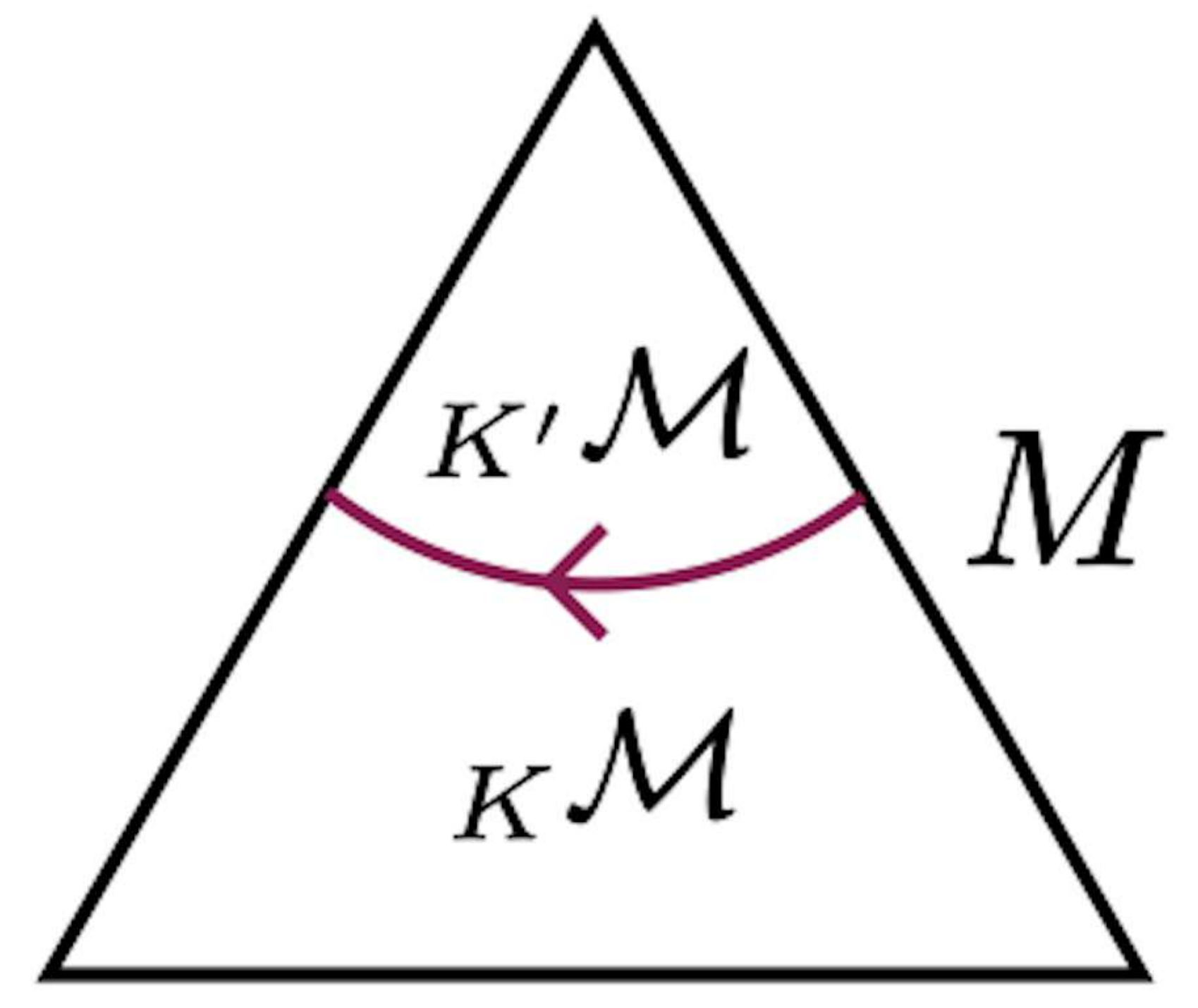}, \quad
\text{(III)} ~ \adjincludegraphics[valign = c, width = 2.1cm]{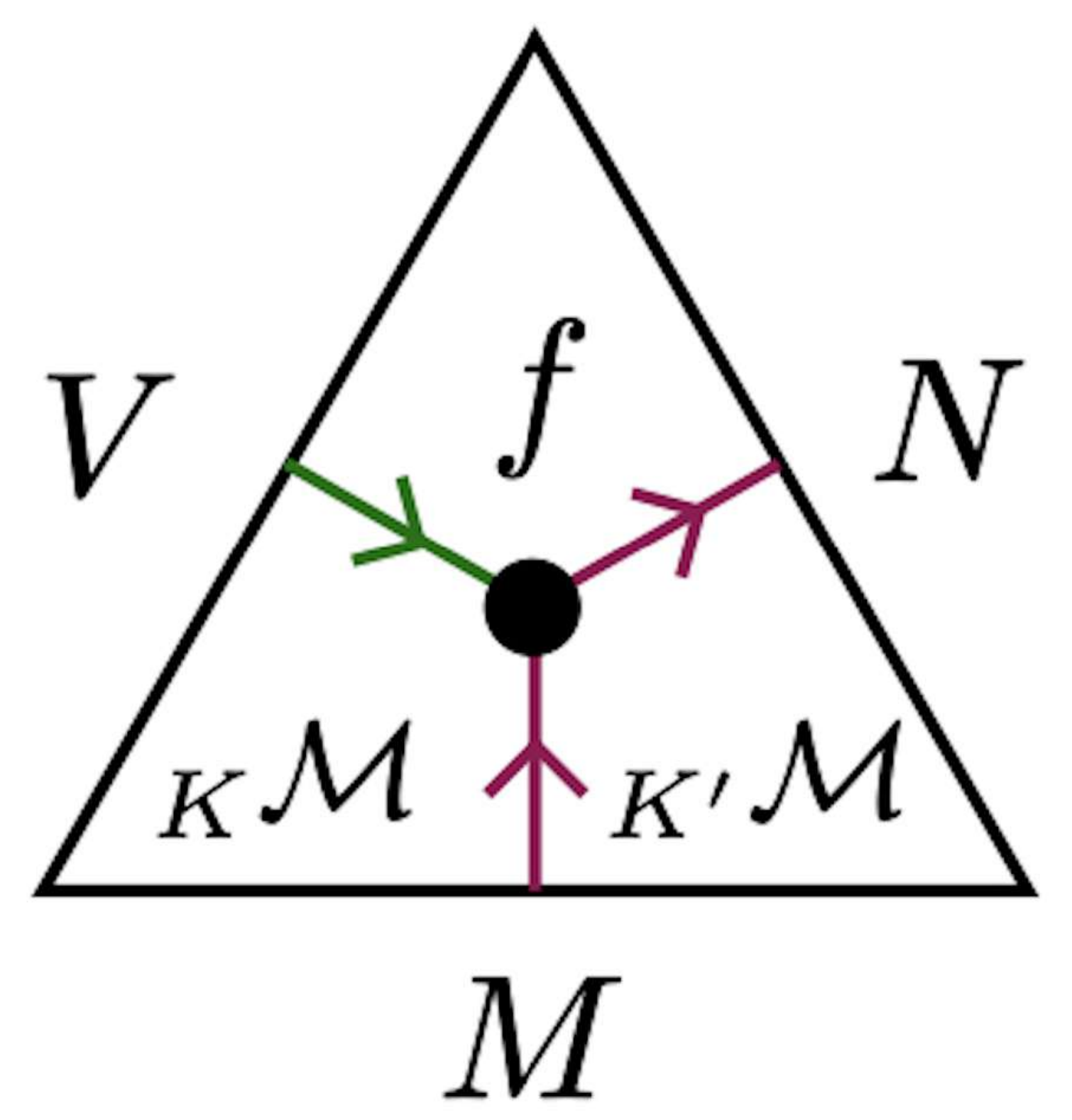}, \quad 
\text{(IV)} ~ \adjincludegraphics[valign = c, width = 2.1cm]{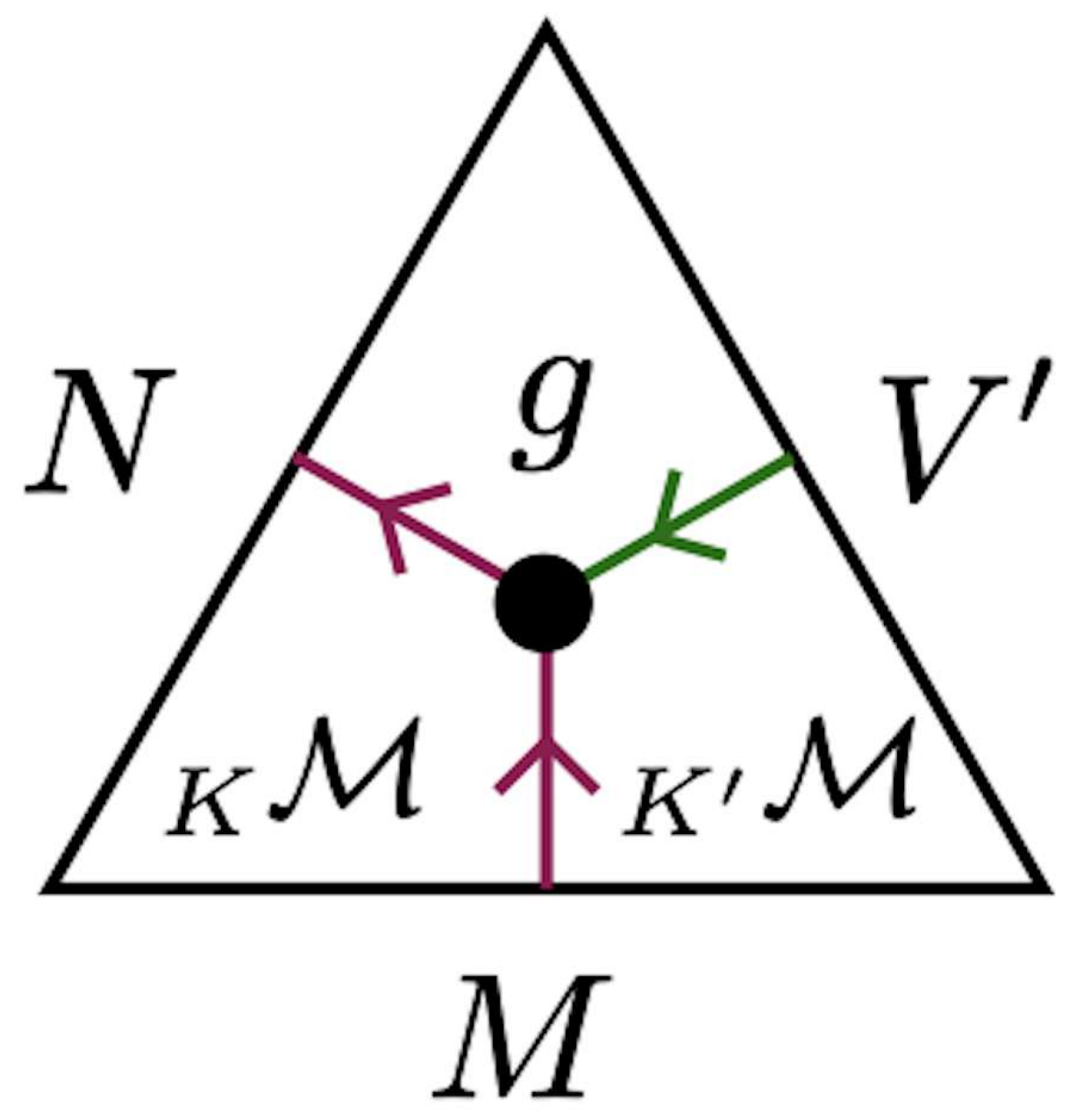}.
\label{eq: interface configurations}
\end{equation}
(I) and (II) represent an isolated interface $M$ between the state sum TQFTs ${}_K \mathcal{M}$ and ${}_{K^{\prime}} \mathcal{M}$.
(III) and (IV) represent topological defects $V \in \mathrm{Rep}(H)$ and $V^{\prime} \in \mathrm{Rep}(H^{\prime})$ that end on an interface.
The endpoints are labeled by $K$-$K^{\prime}$ bimodule maps $f \in \mathrm{Hom}_{K K^{\prime}}(V \overline{\otimes} M, N) = \mathrm{Hom}_{K K^{\prime}}(F_K(V) \otimes_K M, N)$ and $g \in \mathrm{Hom}_{K K^{\prime}}(M \overline{\otimes}^{\prime} V^{\prime}, N) = \mathrm{Hom}_{K K^{\prime}} (M \otimes_{K^{\prime}} F_{K^{\prime}}(V^{\prime}), N)$, where $\overline{\otimes}$ and $\overline{\otimes}^{\prime}$ are the left $\mathrm{Rep}(H)$ action and the right $\mathrm{Rep}(H^{\prime})$ action on ${}_K \mathcal{M}_{K^{\prime}}$.
We can also consider junctions of interfaces:
\begin{equation}
\text{(V)} ~~ \adjincludegraphics[valign = c, width = 2.4cm]{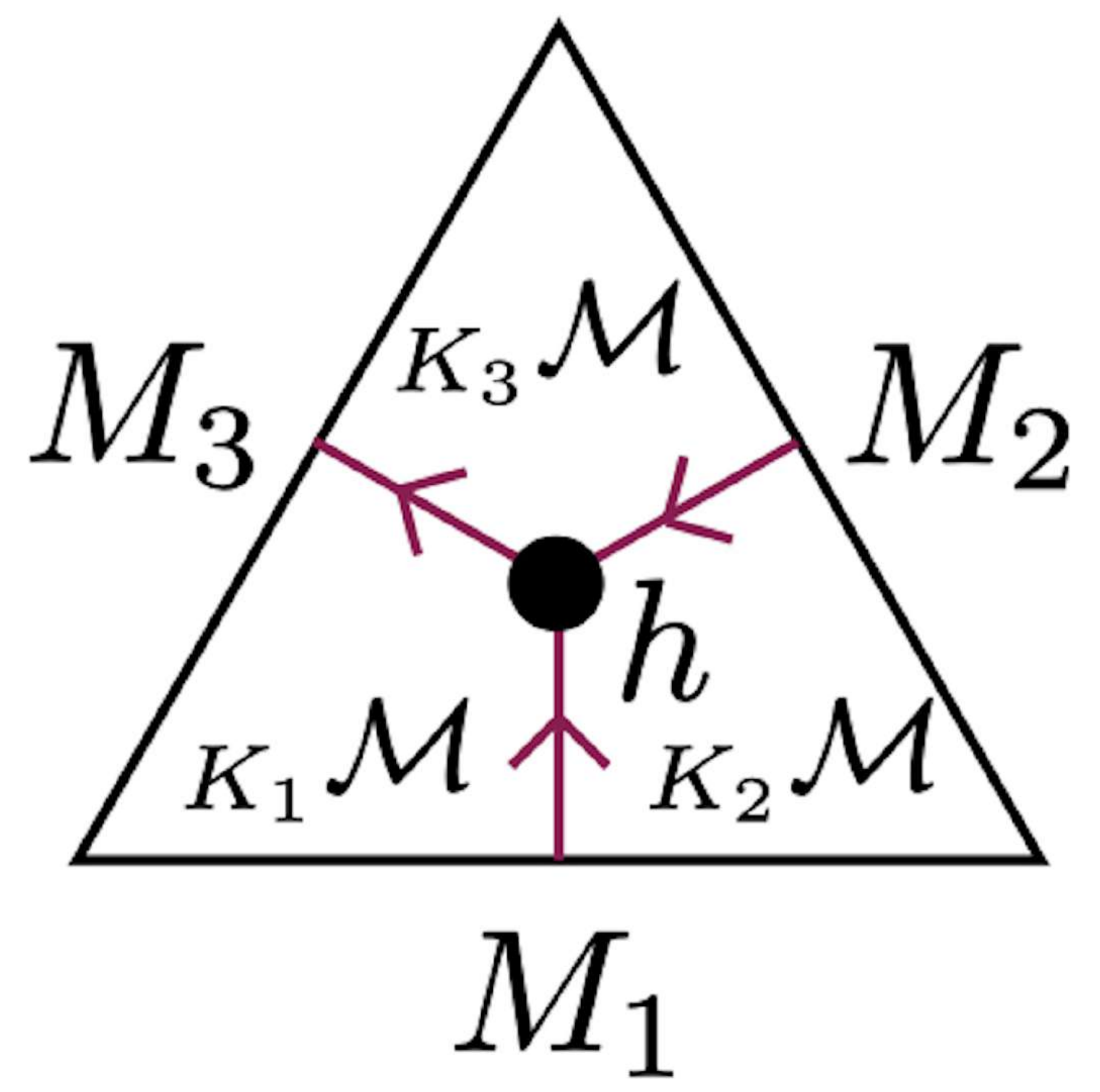}, \quad \quad
\text{(VI)} ~~ \adjincludegraphics[valign = c, width = 2.4cm]{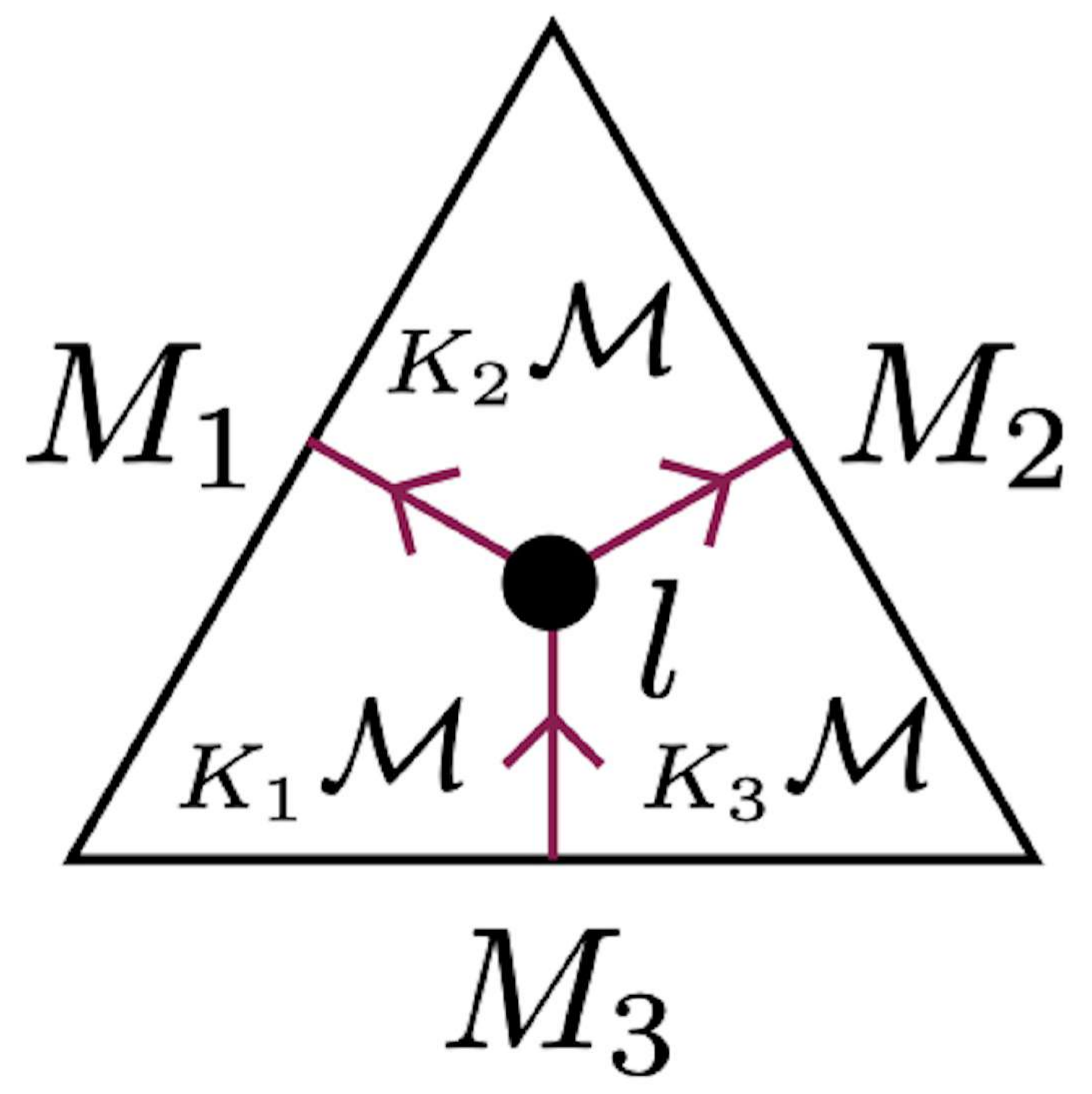},
\end{equation}
These junctions are labeled by $K_1$-$K_3$ bimodule maps $h \in \mathrm{Hom}_{K_1 K_3} (M_1 \otimes_{K_2} M_2, M_3)$ and $l \in \mathrm{Hom}_{K_1 K_3}(M_3, M_1 \otimes_{K_2} M_2)$, where $M_1 \in {}_{K_1} \mathcal{M}_{K_2}$, $M_2 \in {}_{K_2} \mathcal{M}_{K_3}$, and $M_3 \in {}_{K_1} \mathcal{M}_{K_3}$.

To incorporate these configurations, we need to extend the assignment of the vector spaces and the linear maps \eqref{eq: Re Rep(H)}--\eqref{eq: Ep Rep(H)}.
Specifically, we add the following vector spaces and linear maps:
\begin{align}
R_e & := M \quad \text{when $e$ intersects an interface $M$}. \label{eq: Re interface}\\
Q_{(p, e)} & := 
\begin{cases}
M \quad & \text{when $M$ goes into $p$ across $e$},\\
M^* \quad & \text{when $M$ goes out of $p$ across $e$}.
\end{cases}\\
P_e & := \mathrm{coev}_M: \mathbb{C} \rightarrow M \otimes M^* \quad \text{when $e$ intersects $M$}.\\
E_p & := 
\begin{cases}
\text{(I)} & \mathrm{ev}_{M} \circ (\mathrm{id}_{M^*} \otimes \rho_M^R): M^* \otimes M \otimes K^{\prime} \rightarrow \mathbb{C}, \\
\text{(II)} & \mathrm{ev}_M \circ (\mathrm{id}_{M^*} \otimes \rho_M^L): M^* \otimes K \otimes M \rightarrow \mathbb{C}, \\
\text{(III)} & \mathrm{ev}_N \circ (\mathrm{id}_{N^*} \otimes (f \circ \pi_{F_K(V), M})): N^* \otimes F_K(V) \otimes M \rightarrow \mathbb{C}, \\
\text{(IV)} & \mathrm{ev}_N \circ (\mathrm{id}_{N^*} \otimes (g \circ \pi_{M, F_{K^{\prime}}(V^{\prime})})): N^* \otimes M \otimes F_{K^{\prime}}(V^{\prime}) \rightarrow \mathbb{C}, \\
\text{(V)} & \mathrm{ev}_{M_3} \circ (\mathrm{id}_{M_3^*} \otimes (h \circ \pi_{M_1, M_2})): M_3^* \otimes M_1 \otimes M_2 \rightarrow \mathbb{C}, \\
\text{(VI)} & \mathrm{ev}_{M_1 \otimes M_2} \circ (\mathrm{id}_{(M_1 \otimes M_2)^*} \otimes (\iota_{M_1, M_2} \circ l)): (M_1 \otimes M_2)^* \otimes M_3 \rightarrow \mathbb{C}.
\end{cases}
\label{eq: Ep interface}
\end{align}
The above equations \eqref{eq: Re interface}--\eqref{eq: Ep interface} combined with eqs. \eqref{eq: Re Rep(H)}--\eqref{eq: Ep Rep(H)} give rise to a state sum TQFT on surfaces with interfaces.

\bibliographystyle{JHEP}
\bibliography{bibliography}

\end{document}